\documentclass[review]{elsarticle}

\pdfoutput=1
\usepackage{graphicx}
\usepackage{amsfonts, amsmath, amssymb}
\usepackage{booktabs,siunitx}
\usepackage[svgnames,table]{xcolor}
\usepackage[tableposition=above]{caption}
\usepackage{pifont}
\usepackage{bm}
\usepackage{cancel}
\usepackage{caption}
\usepackage{color}
\usepackage{enumerate}
\usepackage{float}
\usepackage{hyperref}
\usepackage[english]{babel}
\usepackage[margin=1in]{geometry}
\usepackage{mathtools}
\usepackage{setspace}
\usepackage{subfigure}
\usepackage{lineno}
\newcommand{\BigO}[1]{\ensuremath{\mathcal{O}\bigl(#1\bigr)}}

\newcommand \D [2]{\frac{\partial #1}{\partial #2}}

\newcommand \DDD [2]{\frac{{\rm D} #1}{{\rm D} #2}}
\renewcommand{\vec}[1]{\bm{\mathrm{#1}}}
\newcommand{\V}[1]{\bm{\mathrm{#1}}}
\newcommand{\norm}[1]{\left\lVert#1\right\rVert}

\def \div{\nabla \cdot \mbox{}}
\def \grad{\nabla}

\def \x{\vec{x}}
\def \y{\vec{y}}

\def \n{\vec{n}}

\def \u{\vec{u}}

\def \I{\vec{I}}

\def \N{\vec{N}}

\def \L{\vec{L}}

\def \A{\vec{A}}
\def \b{\vec{b}}
\def \bu{\vec{b_u}}
\def \bp{\vec{b_p}}
\def \B{\vec{B}}
\def \C{\vec{C}}
\def \vD{\vec{D}}

\def \epssub{\epsilon_\textrm{sub}}
\def \epsstokes{\epsilon_\textrm{Stokes}}
\def \G{\vec{G}}
\def \HG{H_\text{G}}
\def \I{\vec{I}}

\def \Lmu{\vec{L_{\mu}}}
\def \vrho{\vec{\rho}}
\def \vmu{\vec{\mu}}
\def \Lrho{\vec{L_{\wp}}}

\def \N{\vec{N}}
\def \Nx{N_x}
\def \Ny{N_y}

\def \Pproj{\vec{P}^{-1}_\textrm{proj}}

\def \R{\vec{R}}

\def \xb{\x_{\text{b}}}

\def \cC{\mathcal{C}}

\def \cC{{\mathcal{C}}}

\def \f{\vec{f}}
\def \fu{\vec{f_u}}

\def \fs{\f_{\text{s}}}

\def \half{\frac{1}{2}}
\def \3half{\frac{3}{2}}
\def \5half{\frac{5}{2}}

\def \n{\vec{n}}

\def \nref{n_{\text{ref}}}
\def \ncells{n_{\text{cells}}}
\def \ncycles{m}

\def \sgn{\textrm{sgn}}
\def \u{\vec{u}}

\def \up{\vec{u}_{\text{p}}}

\def \vtheta{\vec{\theta}}
\def \vsigma{\vec{\sigma}}

\def \x{\vec{x}}
\def \xu{\vec{x_u}}
\def \xp{\vec{x_p}}

\def \Re{\text{Re}}
\def \La{\text{La}}
\def \We{\text{We}}
\def \Fr{\text{Fr}}
\def \div{\nabla \cdot \mbox{}}
\def \grad{\nabla}

\def \dt{\Delta t}
\def \dx{\Delta x}
\def \dy{\Delta y}
\def \dz{\Delta z}

\def \dP{\Delta p}

\def \dt{\Delta t}
\def \dx{\Delta x}

\newcommand{\upperRomannumeral}[1]{\uppercase\expandafter{\romannumeral#1}}

\begin{document}

\begin{frontmatter}

%\title{A level set based, variable-coefficient and monolithic incompressible Navier-Stokes solver for high density ratio multiphase flows using a stable, consistent, and well-balanced discretization}
\title{A robust incompressible Navier-Stokes solver for high density ratio multiphase flows}
	
%\title{A combined level set immersed boundary method for high density ratio multiphase flows on staggered, Cartesian, and adaptively refined grids \upperRomannumeral{1} :  Monolithic variable-coefficient linear solvers and two-phase flows}

\author[Northwestern1]{Nishant Nangia}
\author[UNC]{Boyce E. Griffith}
\author[Northwestern1,Northwestern2]{Neelesh A. Patankar\corref{mycorrespondingauthor}}
\ead{n-patankar@northwestern.edu}
\author[SDSU]{Amneet Pal Singh Bhalla\corref{mycorrespondingauthor}}
\ead{asbhalla@sdsu.edu}

\address[Northwestern1]{Department of Engineering Sciences and Applied Mathematics, Northwestern University, Evanston, IL}
\address[UNC]{Departments of Mathematics, Applied Physical Sciences, and Biomedical Engineering, University of North Carolina, Chapel Hill, NC}
\address[Northwestern2]{Department of Mechanical Engineering, Northwestern University, Evanston, IL}
\address[SDSU]{Department of Mechanical Engineering, San Diego State University, San Diego, CA}
\cortext[mycorrespondingauthor]{Corresponding author}

\begin{abstract}
This paper presents a robust, adaptive numerical scheme for simulating high density ratio
and high shear multiphase flows on locally refined staggered Cartesian grids that adapt to the evolving interfaces and track regions of high vorticity. The algorithm combines the interface capturing level set method
with a variable-coefficient incompressible Navier-Stokes solver that is demonstrated to stably resolve material contrast ratios of up to six orders 
of magnitude. The discretization approach 
ensures second-order pointwise accuracy for both velocity and pressure with several physical boundary 
treatments, including velocity and traction boundary conditions. The paper includes several test cases that demonstrate the 
order of accuracy and algorithmic scalability of the flow solver.  To ensure the stability of the numerical scheme in the presence of high density and viscosity ratios, we employ a consistent treatment of mass and momentum transport in the conservative form of discrete equations. This consistency is achieved by solving an additional mass balance equation, which we approximate via a strong stability preserving Runga-Kutta time integrator 
and by employing the same mass flux (obtained from the mass equation) in the discrete momentum equation. 
The scheme uses higher-order \emph{total variation diminishing} (TVD) and \emph{convection-boundedness criterion} (CBC) satisfying limiter to avoid numerical fluctuations in the transported density field. 
The high-order bounded convective transport is done on a dimension-by-dimension basis, which makes the scheme simple to implement. We also demonstrate through several test cases that the lack of consistent mass 
and momentum transport in non-conservative formulations, which are commonly used in practice, or the use of non-CBC satisfying limiters can yield very large numerical error and very poor accuracy for convection-dominant high density ratio flows. Our numerical scheme also uses well-balanced surface tension and gravity force discretizations. In the hydrostatic limit, we show that the well-balanced formulation mitigates spurious flow currents and achieves discrete force-balance between the pressure gradient and surface tension or gravity.

\end{abstract}


\end{frontmatter}

%%%%%%%%%%%%%%%%%%%%%%%%%%%%%

\section{Introduction}
Incompressible two-phase flows involving liquid and gas at ambient conditions are prevalent 
in engineering and natural processes. The high density and high viscosity contrasts in multiphase flows 
often result in significant shear at the fluid-fluid interface. This poses a significant challenge in the 
numerical modeling of multicomponent flows because of the stiff system of equations arising from the discretization of the 
incompressible Navier-Stokes equations. In addition, numerical error accumulation near the highly 
deforming interface often leads to catastrophic failure of the scheme. 
Several multiphase flows are driven by interfacial forces like surface tension and volumetric forces like 
gravity; avoiding spurious currents near such interfaces requires a consistent and well-balanced discretization of the forces. Despite these difficulties, multiphase flow modeling 
has been the subject of extensive research for the past two decades owing to its importance in both industrial 
and natural processes. %The tremendous increase in the computational resources that allow 
%the detailed physical models to be executed at extreme scales.      

Several approaches to track the phases of a multiphase flow simulation exist in the literature, the most prominent being the volume-of-fluid (VOF) 
method of Hirt and Nichols~\cite{Hirt1981} and the level set method of Osher and Sethian~\cite{Osher1988}. % are the two main 
%categories. Each of these approaches have their own strengths and weaknesses. While it is not possible to list them 
%in detail, we briefly mention the salient features of the two approaches. 
The VOF method tracks the volume fraction 
of each phase in each computational cell and reconstructs the phase interface in a piecewise fashion from the 
volume fraction data. The advection of the VOF-scalar is done though geometric means to ensure strict mass conservation 
and boundedness of volume fraction. The disadvantages of VOF method include its non-smooth interface representation 
and tedious geometry implementation requirements. On the other hand, the level set method
captures the phase 
interface by computing a signed distance function. %on the Eulerian grid. 
The zero level set of the distance field 
implicitly defines the position of the interface. The level set method formulates the geometric problem of distance finding as 
a nonlinear hyperbolic partial differential equation (PDE) for which established numerical techniques can be applied. 
%``off-the-shelf".
Higher-order hyperbolic PDE discretization methods can enable smooth interface representation and accurate 
curvature calculation. The level set method can also lead to spurious mass changes resulting from the transport and reinitialization of the signed distance function~\cite{Osher2006}. However, this issue can be minimized by 
penalization techniques~\cite{Li2005} or by hybridizing the level set method with VOF-like methods, as done in the mass-conserving level set method (MCLS)~\cite{Van2005} and the coupled level set volume-of-fluid methods (CLSVOF~\cite{Sussman2000}, and VOSET~\cite{Sun2010}).
In this work, we use the level set method with sign~\cite{Son2005} and 
subcell~\cite{Min2010} fixes to mitigate spurious changes in the mass of each phase.

A challenging problem that has not received much attention in the multiphase flow literature is the treatment of large density ratios. The multiphase community has only recently begun to address the 
fundamental cause and resolution of numerical instabilities associated with flows with density ratios on the order of $100$--$1000$ 
and greater.
Over the past few years, many authors have proposed strategies for mitigating these instabilities
within a wide range of numerical contexts; for example,
Li et al.~\cite{Li2015} for a moment-of-fluid method,
Vaudor et al.~\cite{Vaudor2017} for a staggered grid CLSVOF method,
Le Chenadec and Pitsch~\cite{LeChenadec2013} and Owkes and Desjardins~\cite{Owkes2017}
for sharp interface VOF methods,
and Jemison et al.~\cite{Jemison2014} and Duret et al.~\cite{Duret2018} for compressible flow solvers, to name a few.
Interestingly, some earlier numerical schemes based on the VOF method, particularly that of 
Rudman~\cite{Rudman1998} and Bussman~\cite{Bussmann2002}, 
were not susceptible to flow instabilities with large density ratios because of their use of consistent transport schemes for the 
convective momentum and VOF-scalar in the discrete set of conservative equations.  In the context of the level set framework,
in which geometric information about interface transport is absent, it is difficult to maintain a discrete compatibility between mass and momentum transport. Raessi~\cite{Raessi2008} and Raessi and Pitsch~\cite{Raessi2012} 
introduced the first strategy for dealing with high density ratio flows within the 
level set framework by introducing geometric mass flux transport. This work has been limited to one- and two-dimensional problems because of the inherent difficulty 
of interface reconstruction and geometric transport with level sets. A more elegant solution to this problem that is easily extended to three spatial dimensions was proposed by Desjardins and Moureau~\cite{Desjardins2010} in the context of structured, staggered-grid Cartesian discretizations, and more recently by Ghods and Herrmann~\cite{Ghods2013} 
for collocated unstructured grids. Instead of constructing piecewise linear interfaces from the level set field, they suggest
solving an additional mass balance equation along with momentum and incompressibility constraint equations. The advantage 
of this approach is that the same (algebraic) mass flux can be used in both mass and momentum transport for a given time step,
yielding
a stable scheme for high density ratio flows. The level set is transported independently and is only used to synchronize the 
density field with level set field at the beginning of each time step. To prevent numerical oscillations in the density field near the interface, 
both Desjardins and Moureau~\cite{Desjardins2010} and Ghods 
and Herrmann~\cite{Ghods2013} used first-order upwinding for density and velocity transport. Doing so, however, produces diffusive 
flow features
and smeared vortices.  A key contribution of this work is to extend their approach to a second-order accurate, bounded scheme.
The resulting method is empirically demonstrated 
to remain stable even at very high density ratios of $10^6$. This is achieved in this work by employing a modified version of the third-order accurate 
Koren's limited \emph{Cubic Upwind Interpolation} (CUI) scheme described by Patel and Natarjan~\cite{Patel2015}. Koren's limited CUI 
is a total variation diminishing (TVD) and convection-boundedness criterion (CBC) satisfying convective limiter. %, generally expressed in normalized-variable form. 
The mass equation is integrated using a third-order strong-stability preserving Runge-Kutta time integrator (SSP-RK3)~\cite{Gottlieb2001}, and the mass flux from the penultimate (second) stage of the SSP-RK3 scheme is 
used in the 
convective operator to maintain discrete compatibility. We also compare the performance of the scheme using TVD but non-CBC
convective limiters, such as a version of the piecewise parabolic method (PPM), to CBC limiters like CUI, the Modified Gamma scheme (M-Gamma)~\cite{Patel2015}, and the Flux Blending Interface Capturing scheme (FCIBS)~\cite{Tsui2009} for convection-dominant high 
density ratio flows.

More recently Patel and Natarajan~\cite{Patel2017} used a consistent convective scheme for momentum and algebraic 
(i.e., geometric reconstruction-free) VOF transport. Their consistent transport scheme implicitly maintains the discrete compatibility of the density 
flux in the mass and momentum equations. Through several examples, they show that any mismatch between mass and 
momentum transport results in an inconsistent scheme and can result in extremely poor accuracy in certain flow scenarios. We remark that this is not the only way to achieve stabilization for high density ratio flows. For example, 
Desjardins and Moureau~\cite{Desjardins2010} and Sussman et al.~\cite{Sussman07} use the ghost fluid method to 
maintain sharp discontinuities across 
the interface within a hybridized level set framework. This approach also requires the use of some velocity extension algorithm 
across the phase interface to advect the level set field.  Specifically, Sussman et al.~\cite{Sussman07} use the higher density 
fluid's extrapolated velocity to advect the interface. 
%Maintaining sharp discontinuities across the interface stabilizes the multiphase 
%flow simulations even for the non-conservative form of momentum equation.  
We do not analyze the stability of sharp interface 
methods for high density ratio multiphase flows in this work.   

%As mentioned earlier, the discretization of multiphase flow equations with highly contrasting coefficients produces a stiff system 
%of equations.
A major challenge in incompressible flow simulation is the treatment of coupled velocity-pressure saddle-point system.
In this work, we solve this saddle-point system in a monolithic fashion (without any time-splitting approach) using a preconditioned flexible GMRES scheme.
We achieve high performance via a novel preconditioning strategy that combines a local-viscosity preconditioner~\cite{Furuichi2011} with a projection method-based  preconditioner introduced by Griffith in the context of constant coefficient problems \cite{Griffith2009} and extended by Cai et al.~\cite{Cai2014} to treat variable coefficient problems. This work extends the 
uniform grid variable-coefficient solvers of Cai et al.~to locally refined grids. % and also include non-trivial traction boundary 
%conditions (BCs) on the fluid domain. Cai et al. focussed mainly on the scalability of the preconditioner and lacked order of 
%accuracy results.
Another difference is that Cai et al.~\cite{Cai2014} considers the non-conservative form of equations, i.e, they compute 
velocity and pressure fields for a given density field. In this work, because we employ consistent mass and momentum transport 
with conservative form of equations, we solve an additional mass balance equation for density evolution. We empirically demonstrate that our 
spatio-temporal discretization achieves pointwise second-order accuracy for the velocity, pressure, and density fields for a variety 
of boundary conditions. %We remark that both conservative and non-conservative discretization yields the same linear operator, and so the 
%linear solver framework remains identical for them.
We note that most of the multiphase literature~\cite{Bussmann2002,Sussman1994,Ghasemi14,Pathak16,Bussmann99,Brackbill1992,Williams1998} treats the 
spatially varying viscous operator explicitly,
although an implicit treatment has been used by some authors as 
well~\cite{Sussman07,Lalanne2015,Mirzaii2012}. 
Because we use a solver strategy that allows us to treat the viscous operator implicitly, we are able to use a time step size based only on an advective CFL condition.
In the context of a level set framework, the implicit treatment of the viscous terms reduces the 
frequency of level set reinitialization without degrading accuracy, which reduces numerical perturbations in the interface 
configuration~\cite{Ervik2016}.

The remainder of the paper is organized as follows. We first introduce the continuous and discrete system of equations in 
Sections~\ref{sec_continuous} and~\ref{sec_discretized}, respectively. 
Next we discuss the projection preconditioner in 
Section~\ref{sec_solver}.
Software implementation is described in Section~\ref{sec_software}.
Accuracy and scalability results for the monolithic flow solver are presented thereafter for both conservative 
and non-conservative discretizations in Section~\ref{sec_convergence}. Finally, two-phase flow examples highlighting the 
importance of 
consistent and bounded mass and momentum transport are presented in Section~\ref{sec_examples}. We also contrast the 
consistent results against the 
results obtained from an inconsistent and non-conservative flow solver. Wherever possible, simulation results from 
locally refined grids are presented. Examples demonstrating the well-balanced discretization of the interfacial surface tension and 
volumetric gravitational force are also presented.  Well-balanced force formulations are particularly important in the hydrostatic limit,  
where it helps to mitigate spurious flow currents. This is achieved through a discrete force-balance between the pressure gradient 
and interfacial/external forces as shown by our test cases.

%%%%%%%%%%%%%%%%%%%%%%%%%%%%%%
\section{The continuous equations of motion}
\label{sec_continuous}

We follow the single fluid formulation~\cite{TryggvasonBook} for multiphase flows and 
consider a single viscous incompressible fluid with spatially and temporally varying density and viscosity occupying a fixed %rectangular 
region of space 
$\Omega \subset \mathbb{R}^d$. 
The equations of motion for the fluid are the incompressible Navier-Stokes equations, which in 
\emph{conservative} form are

\begin{align}
&\D{\rho \u}{t} + \div \rho\u\u = -\grad p + \div \left[\mu \left(\grad \u + \grad \u^T\right) \right]+ \f + \fs, \label{eqn_momentum}\\
&\div \u = 0, \label{eqn_continuity}
\end{align}
in which $\x = (x,y) \in \Omega$ are the fixed physical Eulerian coordinates, 
$\u(\x,t) = (u(\x,t),v(\x,t))$ is the fluid velocity, $p(\x,t)$ is the pressure, 
$\f(\x,t) = (f_1(\x,t), f_2(\x,t))$ is the momentum body force, $\fs(\x,t)$ is the continuum 
surface tension force, and $\rho(\x,t)$ and $\mu(\x,t)$ are the spatially and temporally varying fluid density 
and dynamic viscosity, respectively. Note that the continuity constraint 
Eq.~\eqref{eqn_continuity} follows directly from the conservation of 
mass equation over the entire domain,
\begin{align}
\D{\rho}{t} + \div \rho \u = \DDD{\rho}{t} + \rho \div \u = 0, \label{eqn_cons_of_mass}
\end{align}
 and the incompressible nature of the fluid, which in the Langrangian form can be expressed as  
$\DDD{\rho}{t} = 0$.  The equations of motion can also be cast to  \emph{non-conservative} form by combining 
Eqs.~\eqref{eqn_momentum},~\eqref{eqn_continuity} and~\eqref{eqn_cons_of_mass}, so that
\begin{align}
&\rho \left(\D{\u}{t} + \div \u\u \right) = -\grad p + \div \left[\mu \left(\grad \u + \grad \u^T\right) \right]+ \f + \fs, \label{eqn_nc_momentum}\\
&\div \u = 0. \label{eqn_nc_continuity}
\end{align}
Although these forms of the equations are equivalent, direct discretizations of these equations lead to numerical schemes with different properties, as discussed in later sections. Note that the
convective term $\div \u\u$ in Eq.~\eqref{eqn_nc_momentum} can also be expressed as 
$\u \cdot \grad \u$ using the continuity constraint.
 
To complete the description of the equations of motion, %one needs 
it is necessary to specify initial conditions for $\u$, $\rho$, and $\mu$ and 
boundary conditions on $\partial \Omega$. As described 
by Griffith~\cite{Griffith2009} for the constant-coefficient case, in this work we consider three 
types of boundary conditions: periodic, prescribed velocity, and prescribed traction. 
%A full description of these boundary conditions will be described
%in Section~\ref{sec_discretized}.

Finally, suppose that two fluids with differing densities and viscosities occupy the regions in the computational domain
$\Omega_0(t) \subset \Omega$ and $\Omega_1(t) \subset \Omega$, respectively. The codimension-$1$ interface
between the fluids $\Gamma(t) = \Omega_0 \cap \Omega_1$ can be tracked as the zero contour of a scalar
function $\phi(\x,t)$, which is the so-called level set function~\cite{Osher1988,Sethian2003,Sussman1994},

\begin{equation}
\Gamma(t) = \{\x \in \Omega \mid \phi(\x,t) = 0\}. \label{eq_interface}
\end{equation}
Level set methods are well-suited for tracking interfaces
undergoing complex topological changes and are relatively easy to implement in both two and three spatial dimensions. %Moreover, ``off-the-shelf" numerical advection schemes can be applied to them.
The density and viscosity in the two phases are determined as a function of this scalar field by 
\begin{align}
\rho (\x,t) &= \rho(\phi(\x,t)), \label{eq_rho_phi}\\
\mu (\x,t) &= \mu(\phi(\x,t)) \label{eq_mu_phi}.
\end{align}
The discretized form of Eqs.~\eqref{eq_rho_phi} and~\eqref{eq_mu_phi} are defined in Section~\ref{sec_level_set}
using a regularized Heaviside function.
The signed distance function is passively advected by the incompressible fluid velocity,
which in conservative form reads
\begin{equation}
\D{\phi}{t} + \div \phi \u = 0. \label{eq_ls_advection}
\end{equation}
One common choice of $\phi$ is the signed distance function; however, $\phi$ generally will not remain a signed
distance function under advection by Eq.~\eqref{eq_ls_advection}. A \emph{reinitialization}
or redistancing procedure is used to maintain the signed distance property of $\phi$ at every time step.
When the fluid properties are determined from $\phi$, we need initial conditions for $\phi$ but not for $\rho$ or $\mu$.

%%%%%%%%%%%%%%%%%%%%%%%%%%%%%%
\section{The discretized equations of motion}
\label{sec_discretized}
This section details the discretizations of the 
non-conservative and conservative forms of the governing 
equations. For notational simplicity, we present the discretized equations 
in two spatial dimensions. An extension to three spatial dimensions is 
straightforward with the exception of the discretization of the viscous term; the
treatment of this term in three dimensions is described in Appendix~\ref{app_3d_viscous}.
The treatment of physical boundary conditions in this work follows an approach similar to that of Griffith \cite{Griffith2009} and is detailed in Appendix~\ref{app_boundary_conditions}.

\subsection{Basic spatial discretization}
\label{sec_spatial_discretization}
\begin{figure}[]
  \centering
  \subfigure[Staggered grid]{
    \includegraphics[scale = 0.4]{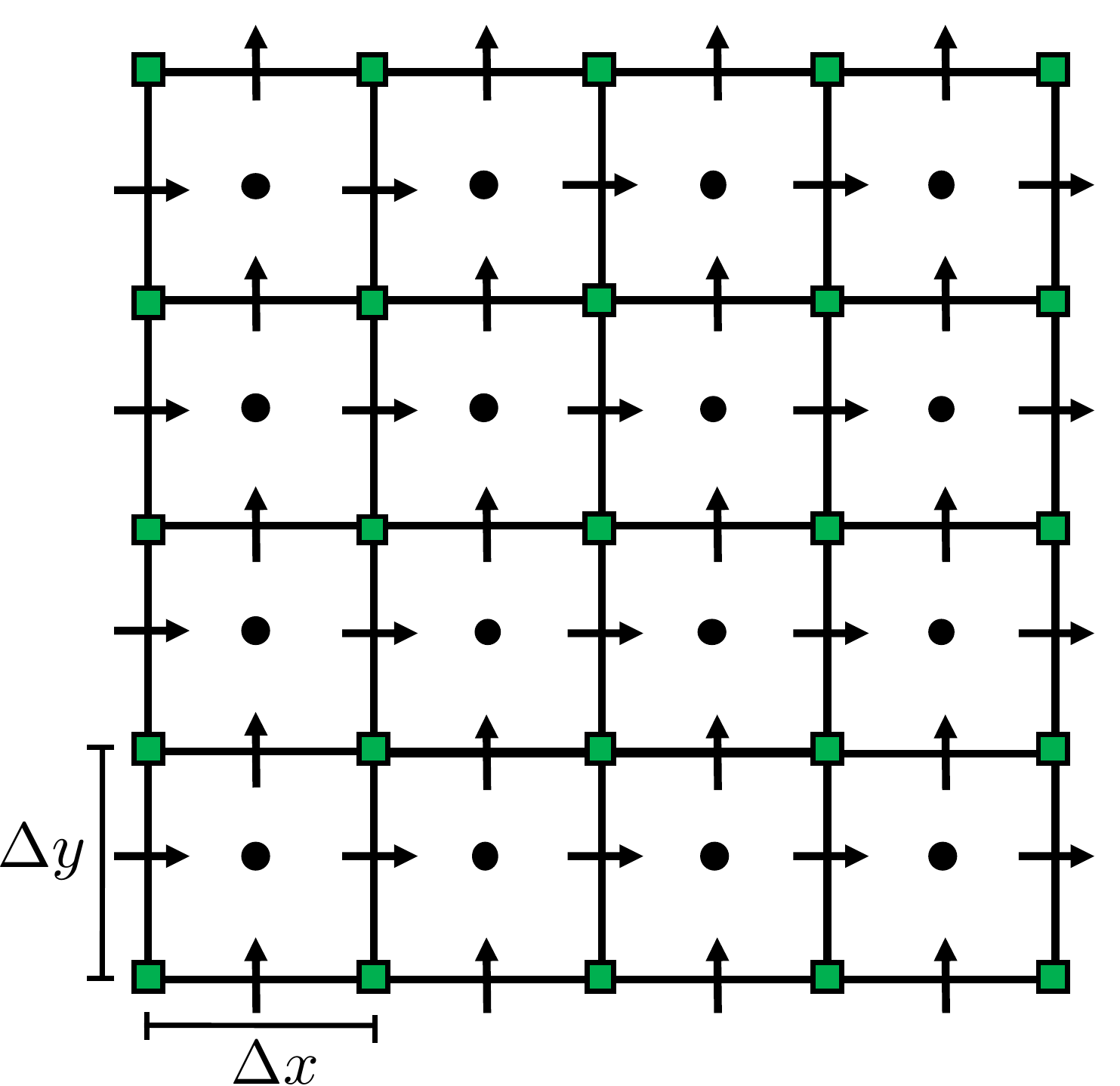}
    \label{staggered_grid}
  }
   \subfigure[Single grid cell]{
    \includegraphics[scale = 0.4]{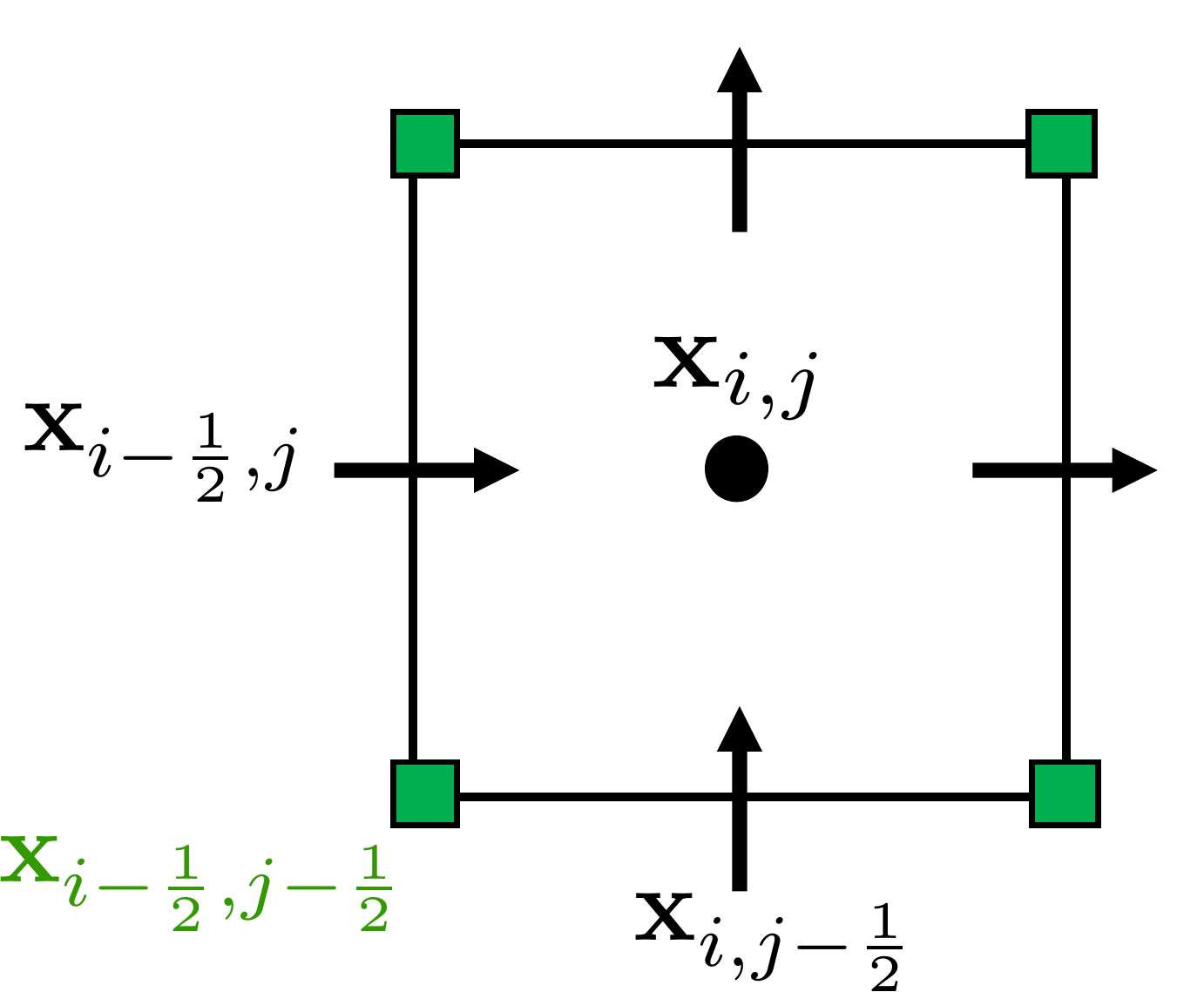}
    \label{grid_cell}
  }
  \caption{\subref{staggered_grid} A staggered-grid discretization of the computational domain $\Omega$.
  	       \subref{grid_cell} A single Cartesian grid cell on which the components of the velocity field $\u$ are approximated on the cell
	        faces ($\rightarrow$, black); the pressure $p$, and level set $\phi$ are approximated on the cell center
	        ($\bullet$, black); and the interpolated viscosity is approximated on cell nodes ($\blacksquare$, green).}
  \label{fig_staggered_grid}
\end{figure}

This work uses a staggered-grid discretization of the equations of incompressible
fluid flow on a rectangular domain $\Omega$. A $\Nx \times \Ny$ Cartesian grid covers the physical domain 
$\Omega$ with mesh spacing $\dx$ and $\dy$ in each direction.
Without loss of generality, the bottom left corner of the domain is assumed to be situated
at the origin $(0,0)$. The position of each grid cell center is then given by 
$\x_{i,j} = \left((i + \half)\dx,(j + \half)\dy\right)$ for $i = 0, \ldots, \Nx - 1$ and $j = 0, \ldots, \Ny - 1$. 
For a given cell $(i,j)$, $\x_{i-\half,j} = \left(i\dx,(j + \half)\dy\right)$ is the physical location of the cell face 
that is half a grid space away from $\x_{i,j}$ in the $x$-direction, and
$\x_{i,j-\half} =\left((i + \half)\dx,j\dy\right)$ is the physical location of the cell 
face that is half a grid cell away from  $\x_{i,j}$ in the $y$-direction.
The pressure is approximated at cell centers and is denoted by
$p_{i,j}^{n} \approx p\left(\x_{i,j},t^{n}\right)$, in which $t^n$ is the time at time step $n$.
Velocity components are defined at cell faces: $u_{i-\half,j}^{n} \approx u\left(\x_{i-\half,j}, t^{n}\right)$ and 
$v_{i,j-\half}^{n} \approx v\left(\x_{i,j-\half}, t^{n}\right)$. The components of the body force $\f = (f_1, f_2)$
are also approximated at $x$- and $y$-faces of the staggered grid cells, respectively. The density and viscosity
are approximated at cell centers of the staggered grid and are denoted by $\rho_{i,j}^{n} \approx \rho\left(\x_{i,j},t^{n}\right)$
and  $\mu_{i,j}^{n} \approx \mu\left(\x_{i,j},t^{n}\right)$.  In our numerical scheme, these values are interpolated onto the required degrees of freedom, as
needed. Similarly, the phase interface is tracked via the level set function, which is also defined
at cell centers and denoted by $\phi_{i,j}^{n} \approx \phi\left(\x_{ij}, t^n\right)$. See Fig.~\ref{fig_staggered_grid}.
Finally, we note that it is sometimes convenient to directly approximate the density field 
on faces of the staggered grid, i.e.  $\rho_{i-\half,j}^{n} \approx \rho\left(\x_{i-\half,j}, t^{n}\right)$ and 
$\rho_{i,j-\half}^{n} \approx \rho\left(\x_{i,j-\half}, t^{n}\right)$, despite it being a scalar quantity;
the reasoning behind this will be made apparent in Sec.~\ref{cons_discretization}.
%For these cases, the cell-centered approximation to density will never be formed.

The staggered-grid finite-difference approximations to the spatial differential operators
have been described in various prior studies~\cite{Griffith2009,Cai2014,Harlow1965,Guermond2006}.
We briefly summarize them here to complete the description of the spatial discretization. The divergence $\vD \cdot \u$ of the velocity field
$\u = (u,v)$ is approximated at cell centers by
\begin{align}
\label{eq_div_fd}
& (\vD \cdot \u)_{i,j} = (D^x u)_{i,j} + (D^y v)_{i,j}, \\
&(D^x u)_{i,j} = \frac{u_{i+\half, j} - u_{i-\half, j}}{\dx}, \\
&(D^y v)_{i,j} = \frac{v_{i, j+\half} - v_{i, j-\half}}{\dy}. 
\end{align}
The components of the gradient $\G p = (G^x p, G^y p)$ of cell-centered pressure $p$ are approximated at cell faces by
\begin{align}
\label{eq_grad_fd}
&(G^x p)_{i-\half,j} = \frac{p_{i,j} - p_{i-1,j}}{\dx}, \\
&(G^y p)_{i,j - \half} =\frac{p_{i,j} - p_{i,j-1}}{\dy}. 
\end{align}
The continuous strain rate tensor form of the viscous term is
\begin{equation}
\label{eq_visc_cont}
\div \left[\mu \left(\grad \u + \grad \u^T\right) \right] = 
\left[
\begin{array}{c}
 2 \D{}{x}\left(\mu \D{u}{x}\right) + \D{}{y}\left(\mu\D{u}{y}+\mu\D{v}{x}\right) \\
 2 \D{}{y}\left(\mu \D{v}{y}\right) + \D{}{x}\left(\mu\D{v}{x}+\mu\D{u}{y}\right) \\
\end{array}
\right],
\end{equation}
which leads to velocity coupling in the discretization for variable viscosity flows
\begin{equation}
\label{eq_visc_discrete}
\Lmu \u= 
\left[
\begin{array}{c}
 (\Lmu \u)^x_{i-\half,j} \\
 (\Lmu \u)^y_{i,j-\half}  \\
\end{array}
\right].
\end{equation}
The viscous operator is approximated using standard second-order, centered finite-differences
\begin{align}
 (\Lmu \u)^x_{i-\half,j} &= \frac{2}{\dx}\left[\mu_{i,j}\frac{u_{i+\half,j} - u_{i-\half,j}}{\dx} -
					        \mu_{i-1,j}\frac{u_{i-\half,j} - u_{i-\3half,j}}{\dx}\right] \nonumber \\ 
                    &+ \frac{1}{\dy}\left[\mu_{i-\half, j+\half}\frac{u_{i-\half,j+1} - u_{i-\half,j}}{\dy} - 
					         \mu_{i-\half, j-\half}\frac{u_{i-\half,j} - u_{i-\half,j-1}}{\dy}\right] \nonumber \\
	            &+ \frac{1}{\dy}\left[\mu_{i-\half, j+\half}\frac{v_{i,j+\half} - v_{i-1,j+\half}}{\dx} - 
					         \mu_{i-\half, j-\half}\frac{v_{i,j-\half} - v_{i-1,j-\half}}{\dx}\right] \label{eq_viscx_fd} \\				         
 (\Lmu \u)^y_{i,j-\half} &= \frac{2}{\dy}\left[\mu_{i,j}\frac{v_{i,j+\half} - v_{i,j-\half}}{\dy} -
					        \mu_{i,j-1}\frac{v_{i,j-\half} - v_{i,j-\3half}}{\dy}\right] \nonumber \\ 
                    &+ \frac{1}{\dx}\left[\mu_{i+\half, j-\half}\frac{v_{i+1,j-\half} - v_{i,j-\half}}{\dx} - 
					         \mu_{i-\half, j-\half}\frac{v_{i,j-\half} - v_{i-1,j-\half}}{\dx}\right] \nonumber \\
	            &+ \frac{1}{\dx}\left[\mu_{i+\half, j-\half}\frac{u_{i+\half,j} - u_{i+\half,j-1}}{\dy} - 
					         \mu_{i-\half, j-\half}\frac{u_{i-\half,j} - u_{i-\half,j-1}}{\dy}\right] \label{eq_viscy_fd},
\end{align}
in which we require approximations to the viscosity at both the \emph{centers} and \emph{nodes} of the Cartesian grid cells.
Node centered quantities are obtained via interpolation by either arithmetically averaging the neighboring cell centered quantities,
\begin{equation}
 \mu_{i-\half, j-\half}^{\text{A}} = \frac{\mu_{i, j} +  \mu_{i-1, j} +  \mu_{i, j-1} +  \mu_{i-1, j-1}}{4}, \label{eq_avg}
\end{equation}
or harmonically averaging those quantities,
\begin{equation}
\mu_{i-\half, j-\half}^{\text{H}} = \left(\frac{\frac{1}{\mu_{i, j}} +  \frac{1}{\mu_{i-1, j}} +  \frac{1}{\mu_{i, j-1}} +  \frac{1}{\mu_{i-1, j-1}}}{4}\right)^{-1}. \label{eq_har}
\end{equation}
We remark that in three spatial dimensions, the viscosity is required at the centers and \emph{edges} of the Cartesian grid cells.  The full three-dimensional discretization of the viscous term is detailed
in Appendix~\ref{app_3d_viscous}.
%The linear operators described above are needed to fully discretize the continuous equations of motion.
An additional approximation to the density-weighted, variable-coefficient Laplacian is required for the
projection preconditioner described in Sec.~\ref{sec_convergence},
\begin{align}
\label{eq_lap_fd}
(\Lrho p)_{i,j} =  \frac{1}{\dx}\left[\frac{1}{\rho_{i+\half,j}} \frac{p_{i+1,j} - p_{i,j}}{\dx} - \frac{1}{\rho_{i-\half,j}} \frac{p_{i,j} - p_{i-1,j}}{\dx}\right]
		     + \frac{1}{\dy}\left[\frac{1}{\rho_{i,j+\half}} \frac{p_{i,j+1} - p_{i,j}}{\dy} - \frac{1}{\rho_{i,j-\half}} \frac{p_{i,j} - p_{i,j-1}}{\dy}\right],
\end{align}
which requires density on faces of the Cartesian grid cells. These values also can be determined using either the arithmetic or harmonic averages of density from the two adjacent cell centers.
%In cases where we work directly with the face-centered density, these quantities are readily available.

Evaluating finite-difference operators near boundaries of the computational domain and locally refined
mesh boundaries requires specification of abutting ``ghost'' values, which will be described in Section~\ref{sec_amr} 
and Appendix~\ref{app_boundary_conditions}. 

\subsection{Discretization of the convective derivative}
\label{sec_convective}
In the present work, the nonlinear convective term, $\div \u \u$ for non-conservative form
and $\div \rho \u \u$ for conservative form, is computed using the third-order accurate
Koren's limited cubic upwind interpolation (CUI) scheme, first proposed by Roe and Baines~\cite{Roe1982}
and further investigated by Waterson  and Deconinck~\cite{Waterson2007} and by Patel and Natarajan~\cite{Patel2015}
for multiphase flows. CUI satisfies both a convection-boundedness criterion (CBC) as well as the total variation 
diminishing (TVD) property, both of which are essential to ensure a monotonic and bounded convective scheme.  
This specific method belongs to a class of \emph{nonlinear} upwind schemes (generally formulated in terms 
of flux limiters or normalized variables
\cite{Waterson2007}), which attempt to achieve higher than first-order accuracy while maintaining 
the monotonicity of the convected variable. The nonlinear monotonic schemes overcome the consequences of Godunov's order barrier theorem~\cite{Roe1986}, which states that no \emph{linear} convection scheme of second-order accuracy or higher
can be monotonic.
These schemes are generally described for cell-centered quantities, but can be formulated for face-centered
quantities after appropriate shifting and averaging operations \cite{Griffith2009}.  We will describe this next.

\begin{figure}[]
  \centering
  \subfigure[]{
    \includegraphics[scale = 0.4]{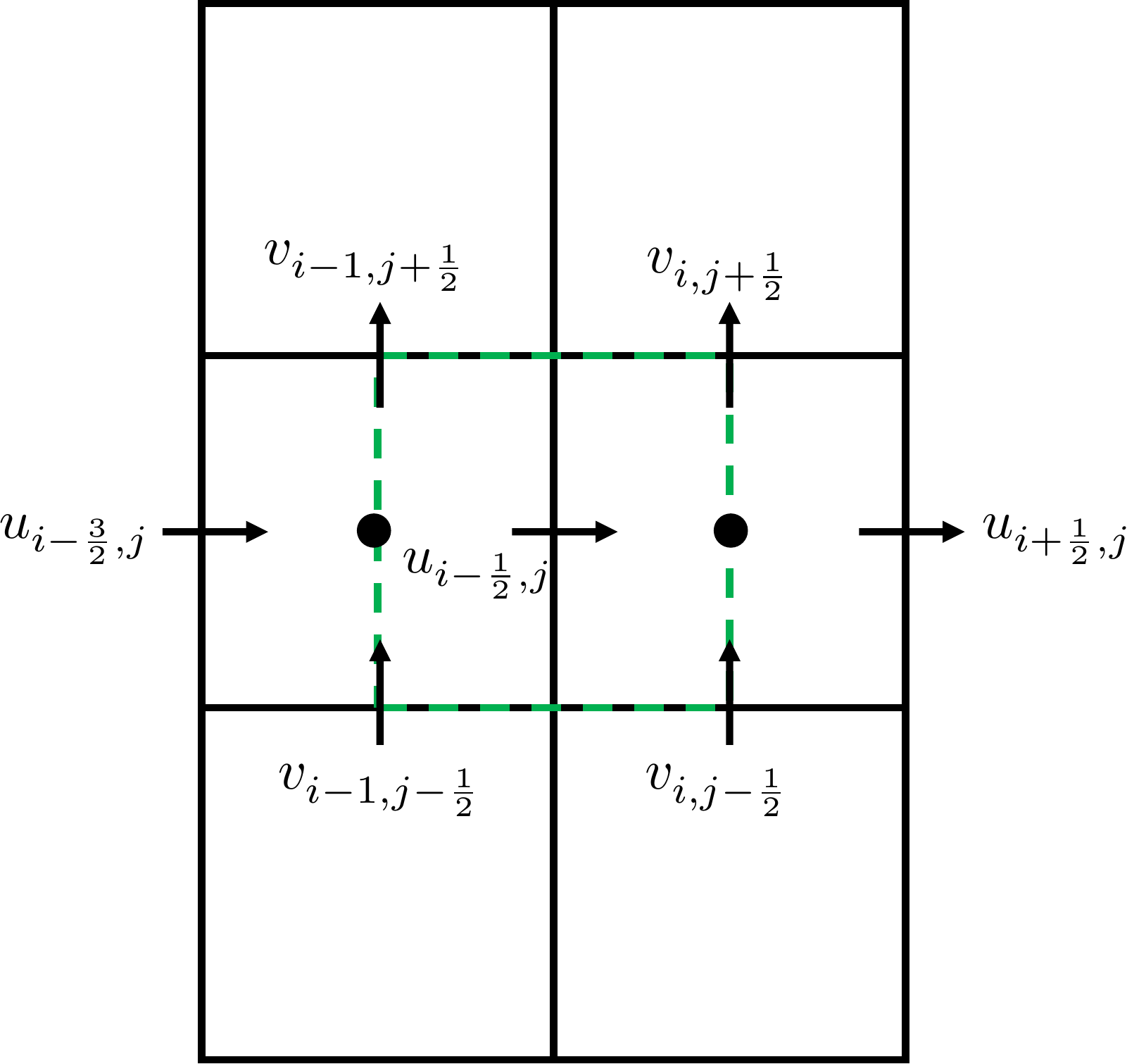}
    \label{convection_velocity}
  }
   \subfigure[]{
    \includegraphics[scale = 0.4]{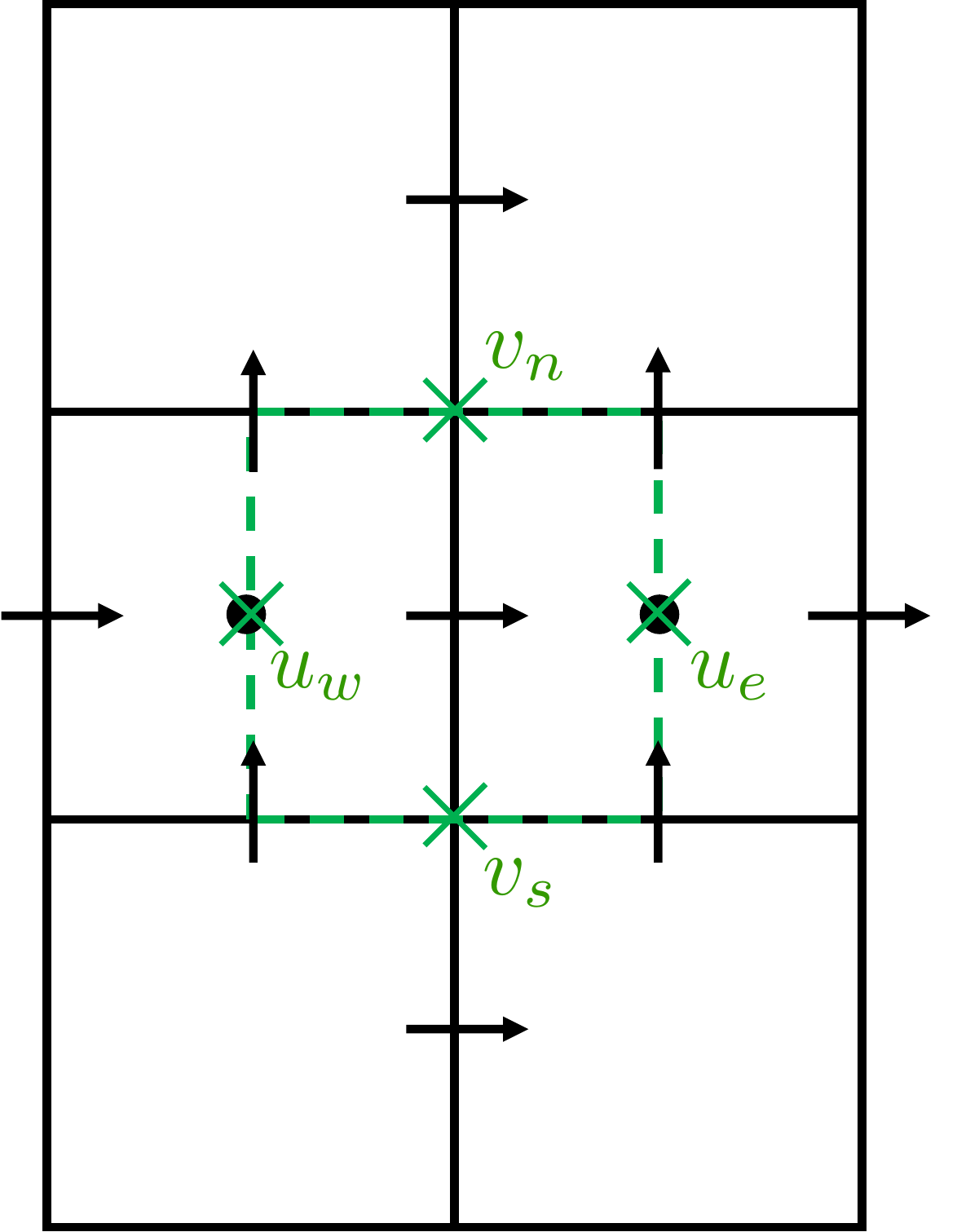}
    \label{convection_avg_velocity}
  }
  \caption{The shifted control volume (\texttt{---}, green) over which an approximation to $(\div \u \psi)_{i-\half,j}$ is computed.
  \subref{staggered_grid} The staggered grid velocities used to compute the advection velocity $\u_\text{adv}$.
  \subref{convection_avg_velocity} The location of the advection velocity on faces ($\times$, green) of the shifted
  control volume.}
  \label{fig_convection_velocity}
\end{figure}

For simplicity, we describe the CUI discretization of $(\div \u \psi)_{i-\half,j}$, in which $\psi$ is defined on $x$-faces
and is advected by a staggered grid velocity $\u$. Note that $\psi$ in this case could be the $x$-component of velocity
$u_{i-\half,j}$ or the density on $x$-faces $\rho_{i-\half,j}$. (The need for density advection will become apparent in Sec.~\ref{sec_temporal}.) As a first step, we construct control volumes centered about $\x_{i-\half,j}$
by shifting the computational grid by $\frac{\dx}{2}$ in the $x$-direction; see Fig.~\ref{fig_convection_velocity}. The
advection velocity $\u_\text{adv}$ on the faces of the shifted control volume is obtained by averaging the adjacent velocity components. 
As shown in Fig.~\ref{fig_convection_velocity}, these velocity components are
\begin{align}
u_w &= \frac{u_{i-\3half,j} + u_{i-\half,j}}{2}, \nonumber \\
u_e &= \frac{u_{i-\half,j} + u_{i+\half,j}}{2}, \nonumber \\
v_s &= \frac{v_{i-1,j-\half} + v_{i,j-\half}}{2}, \nonumber \\
v_n &= \frac{v_{i-1,j+\half} + v_{i,j+\half}}{2}. \label{eq_adv_vel}
\end{align}

\begin{figure}[]
  \centering
  \subfigure[CUI stencil for $\psi_e$ ($u_e \ge 0$)]{
    \includegraphics[scale = 0.4]{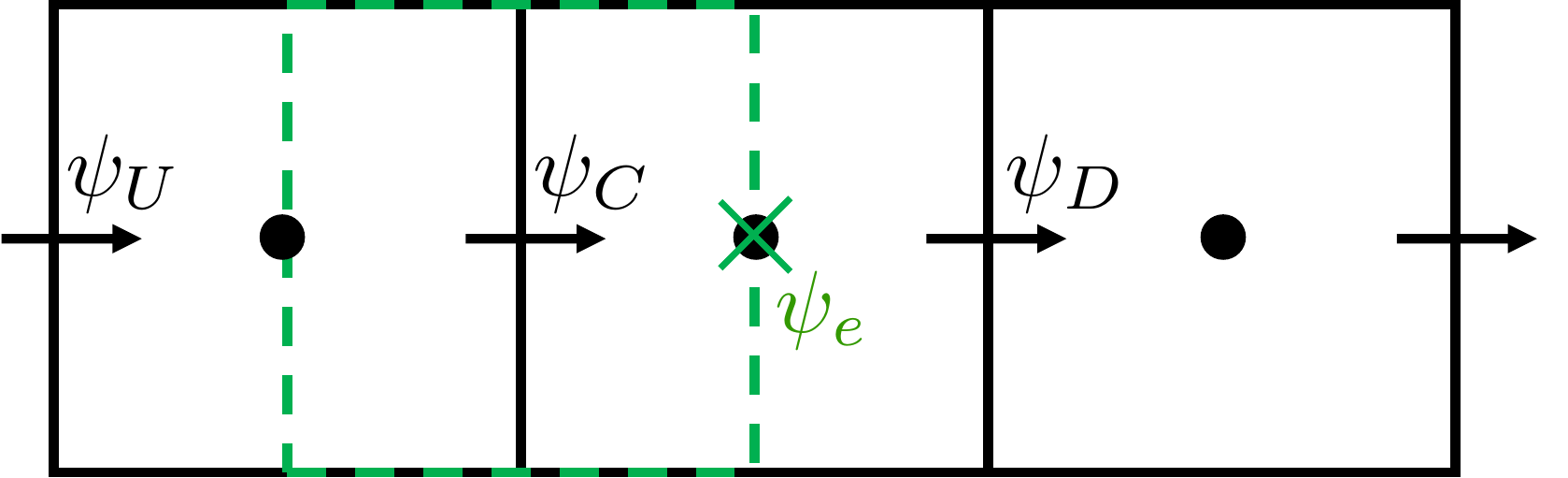}
    \label{convection_psi_e_right}
  }
   \subfigure[CUI stencil for $\psi_e$ ($u_e < 0$)]{
    \includegraphics[scale = 0.4]{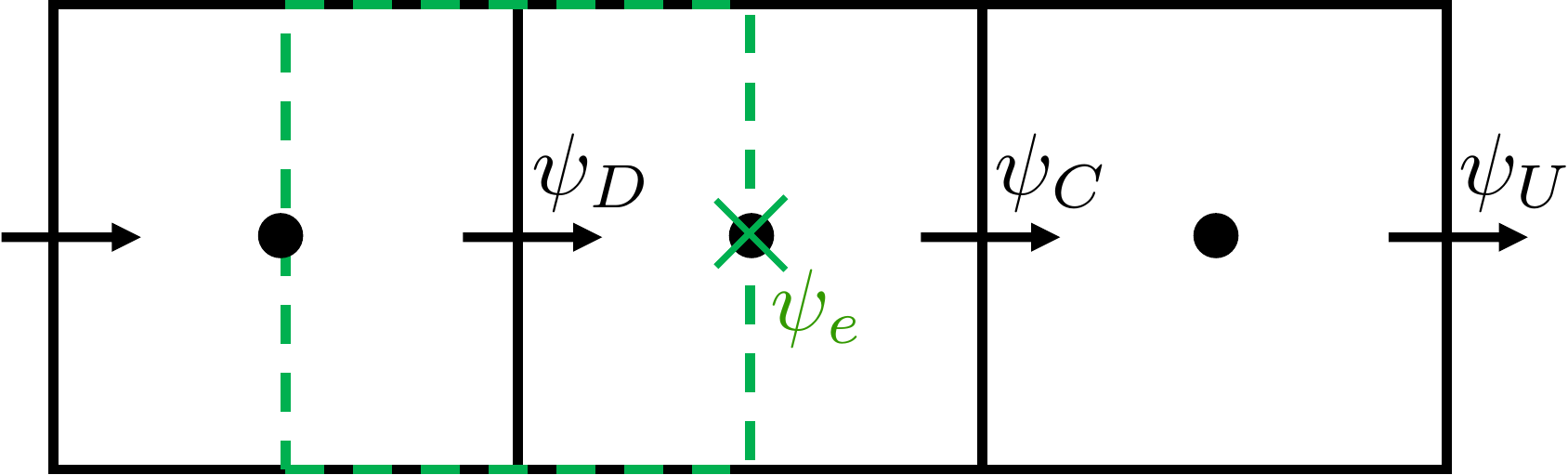}
    \label{convection_psi_e_left}
  }
   \subfigure[CUI stencil for $\psi_n$ ($v_n \ge 0$)]{
    \includegraphics[scale = 0.3]{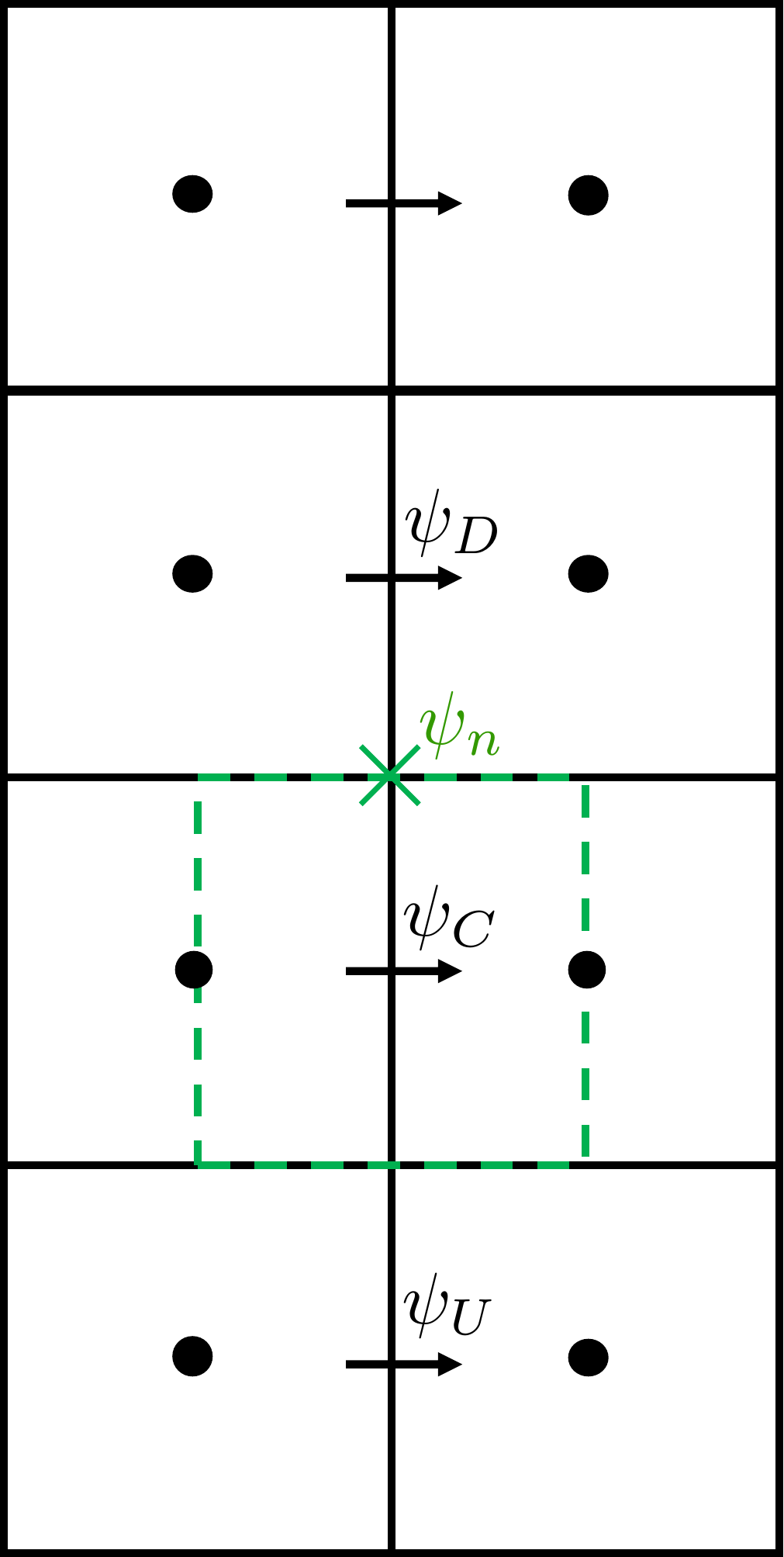}
    \label{convection_psi_n_top}
  }~~~~~~~~~~~~~~~~~~~~~~\subfigure[CUI stencil for $\psi_n $ ($v_n < 0$)]{
    \includegraphics[scale = 0.3]{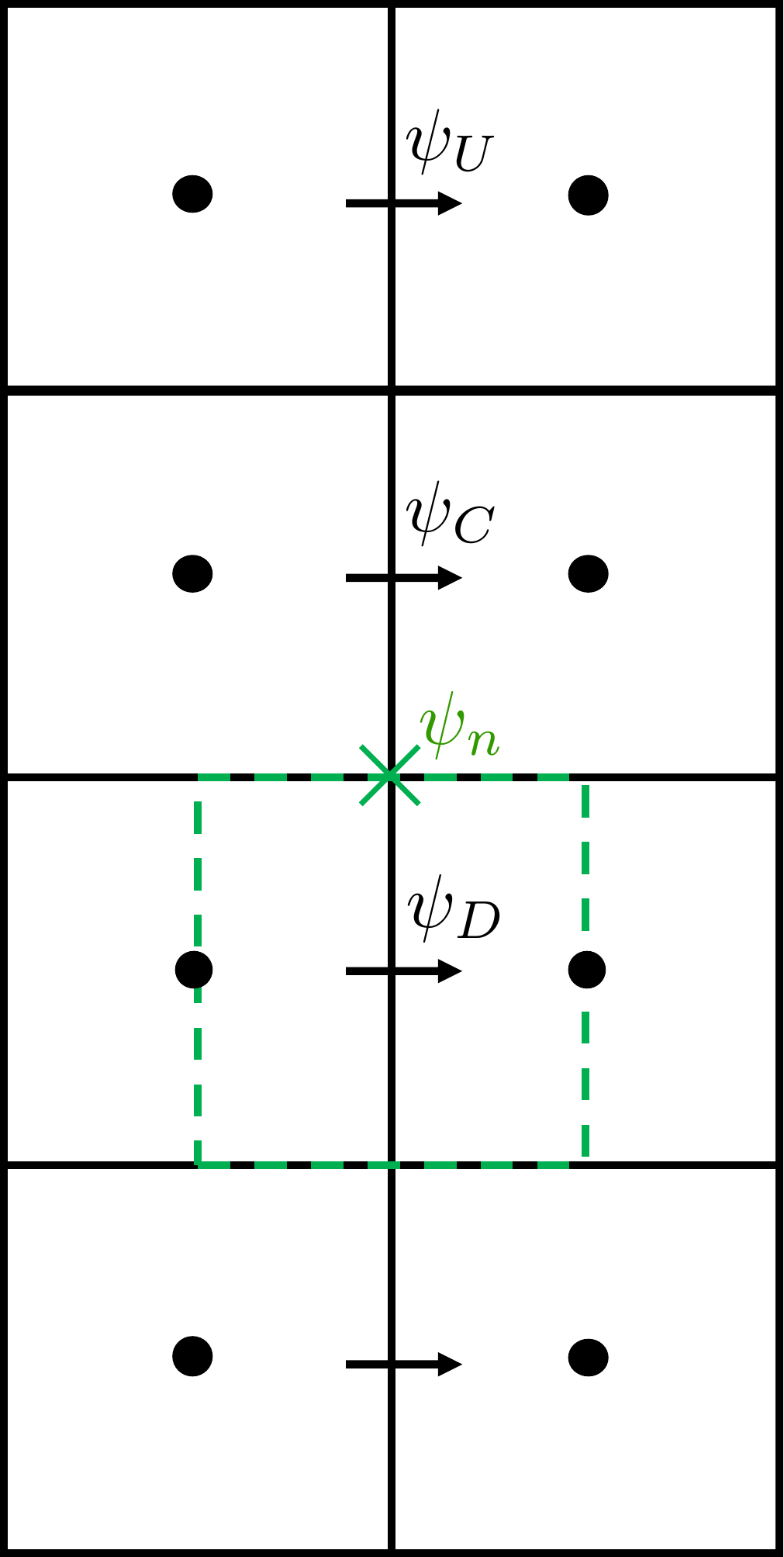}
    \label{convection_psi_n_bottom}
  }
  \caption{The shifted control volume (\texttt{---}, green) over which an approximation to $(\div \u \psi)_{i-\half,j}$ is computed.
  In each diagram, the control volume is surrounding the same face-centered degree of freedom located at $\x_{i-\half,j}$ and the upwind $\psi_C$, far upwind
  $\psi_U$, and downwind $\psi_D$ are labeled according to their usage in Eq.~\eqref{eq_cui_quantity}.
  \subref{convection_psi_e_right} The required degrees of freedom to compute the cubic upwind interpolation of $\psi_e$ when
  $u_e \ge 0$.
  \subref{convection_psi_e_left} The required degrees of freedom to compute the cubic upwind interpolation of $\psi_e$ when
  $u_e < 0$.
  \subref{convection_psi_n_top} The required degrees of freedom to compute the cubic upwind interpolation of $\psi_n$ when
  $v_n \ge 0$.
  \subref{convection_psi_n_bottom} The required degrees of freedom to compute the cubic upwind interpolation of $\psi_n$ when
  $v_n < 0$.}
  \label{fig_convection_psi}
\end{figure}

Next, we use the CUI scheme to obtain $\psi_w, \psi_e, \psi_s,$ and $\psi_n$ on the faces of the shifted control volume.
For a given shifted face $f \in \{e,w,n,s\}$, the upwind $\psi_C$, far upwind $\psi_U$ and downwind $\psi_D$ are
labeled depending on the direction of the advection velocity as shown in Fig.~\ref{fig_convection_psi}.
For instance, Fig.~\ref{convection_psi_e_right} depicts the case where $u_e \ge 0$ in which the face
right of $e$ is labeled as downstream ($\psi_D$) and the two faces counting leftward from $e$ are 
labeled upwind ($\phi_C$) and far upwind ($\psi_U$). Analogous three-point stencils are used for
the other shifted control volume faces.
The upwinded (limited) approximation of $\psi_{f,\text{lim}}$ on a shifted face can be  written in ``normalized variable" form as
\begin{equation}
\label{eq_cui_quantity}
\widetilde{\psi}_f  = 
\begin{cases} 
       3 \widetilde{\psi}_C,  & 0 < \widetilde{\psi}_C\ \le \frac{2}{13}\\
        \frac{5}{6}\widetilde{\psi}_C + \frac{1}{3} ,  & \frac{2}{13} < \widetilde{\psi}_C \le \frac{4}{5} \\
        1,  & \frac{4}{5} < \widetilde{\psi}_C \le 1 \\
        \widetilde{\psi}_C, & \textrm{otherwise},
\end{cases}
\end{equation}
in which the normalized value is defined by
\begin{equation}
\label{eq_normalized_value}
\widetilde{\psi} = \frac{\psi - \psi_U}{\psi_D - \psi_U}.
\end{equation}
Finally, we compute the approximation to $(\div \u \psi)_{i-\half,j}$ via
\begin{equation}
\label{eq_conv_fd}
(\div \u_\text{adv} \psi_\text{lim})_{i-\half,j} \approx \frac{u_e \psi_e - u_w \psi_w}{\dx} + \frac{v_n \psi_n - v_s\psi_s}{\dy}. 
\end{equation}

A similar procedure can be followed to compute a discretization for advected quantities on the $y$-faces.
It is also straightforward to extend this scheme to three spatial dimensions because all the computations
are performed on a dimension-by-dimension basis. We remark that there are other CBC and TVD satisfying limiters that can be used in place of CUI, including M-Gamma~\cite{Patel2015} and FBICS~\cite{Tsui2009}.
We refer readers to Waterson  and Deconinck~\cite{Waterson2007} and to Patel and Natarajan \cite{Patel2015} for further details. We also employ TVD but non-CBC satisfying high resolution limiters like xsPPM7~\cite{Griffith2009, Rider2007}, which is a version of the piecewise parabolic method (PPM), to advect scalars in this work. This method, however, can produce unphysical oscillations in the values of variables that have numerical bounds, as evidenced by numerical experiments reported in Sec.~\ref{sec_dense_droplet}.  Therefore, we use CUI (unless otherwise stated) for the convective discretization, and we use xsPPM7 to advect scalars such as the signed distance function only where noted. We remark that if density and viscosity are formulated in terms of a volume-fraction variable $\alpha$, which has bounds $0 \le \alpha \le 1$, a CBC satisfying limiter must be employed to advect and bound $\alpha$~\cite{Patel2015}.
We also remark that in the cases for which a cell-centered approximation to an advective derivative is needed (e.g., $(\div \u \phi)_{i,j}$ such as for evolving the level set function), the convective scheme remains the same (as described above), and the control volume is the \emph{unshifted} grid cell. As a result, the advective velocity for the cell-centered case would be the original staggered grid velocity $\u$ which is discretely divergence-free by construction. 
Finally, it is a simple exercise to show that if the staggered grid
velocity is discretely divergence-free, then its linear interpolation onto the staggered control volumes also satisfies the same discrete divergence-free property on those control volumes away from the boundaries of the computational domain~\cite{Griffith2012vol}.
Further, if a discrete divergence-free condition is also enforced in the ghost cells abutting the physical domain, as done in this work (see Appendix~\ref{app_boundary_conditions}), the interpolated advective velocity for the face-centered control volume is discretely divergence-free throughout the computational domain.     

\subsection{Interface tracking: the level set method}
\label{sec_level_set}
The interface between the two phases is represented by the zero level set of the scalar field $\phi(\x,t)$. It is convenient to initialize $\phi$ to be the signed distance from the interface $\Gamma^0 = \Gamma(0)$ (Eq.~\eqref{eq_interface}), i.e.
\begin{equation}
\label{eq_signed_distance}
%\phi\left(\x_{i,j}, 0\right) = 
\phi^{0}_{i,j} = 
\begin{cases} 
       \min\limits_{\y \in \Gamma^0} \|\x_{i,j}-\y\|,  & \x_{i,j} \in \Omega_0(0) \\
        -\min\limits_{\y \in \Gamma^0} \|\x_{i,j}-\y\|,  & \x_{i,j} \in \Omega_1(0),
\end{cases}
\end{equation}
which can be computed analytically for the simple initial interfaces considered in the present work.
Note that $\phi$ is not guaranteed to retain the signed distance property under linear advection, Eq.~\eqref{eq_ls_advection}.
Let $\widetilde{\phi}^{n+1}$ denote the level set function following an advection procedure after time stepping through the interval $\left[t^{n}, t^{n+1}\right]$. We aim to \emph{reinitialize} it to obtain a signed distance function ${\phi}^{n+1}$. This can be achieved by computing a steady-state
solution to the Hamilton-Jacobi equation
\begin{align}
&\D{\phi}{\tau} + \sgn\left(\widetilde{\phi}\right)\left(\|\grad \phi \| - 1\right) = 0, \label{eq_eikonal} \\
& \phi(\x, \tau = 0) = \widetilde{\phi}(\x), \label{eq_eikonal_init}
\end{align}
in which we have dropped the $n+1$ superscript because this process is agnostic to the particular time step under consideration.
At the end of a physical time step, Eq.~\eqref{eq_eikonal} is evolved in psuedo-time $\tau$, which, at steady state, produces
a signed distance function satisfying the Eikonal equation $\|\grad \phi \|  = 1$. Here, $\sgn$ denotes the sign of
$\widetilde{\phi}$, which is either $1$, $-1$, or $0$. The discretization of Eq.~\eqref{eq_eikonal} from the psuedo-time 
interval  $\left[\tau^{m}, \tau^{m+1} \right]$ yields

\begin{equation}
\label{eq_discretized_eikonal}
\frac{\phi^{m+1} - \phi^{m}}{\Delta \tau} +
\sgn\left(\widetilde{\phi}_{i,j}\right) \left[\HG\left(D^{+}_x \phi_{i,j}, D^{-}_x \phi_{i,j}, D^{+}_y \phi_{i,j}, D^{-}_y \phi_{i,j}\right) - 1 \right] = 0,
\end{equation}
in which $\HG$ denotes a discretization of $\|\grad \phi \|$ using the Godunov-Hamiltonian, and $D^{\pm}_x$ and $D^{\pm}_y$
denote one-sided discretizations of $\D{\phi}{x}$ and $\D{\phi}{y}$, respectively \footnote{In three spatial dimensions $\HG$ also includes $D^{\pm}_z$ terms.}. These are typically discretized using high-order
essentially non-oscillatory (ENO) or weighted ENO (WENO) schemes~\cite{Shu1998}.

It is well known that continually applying Eq.~\eqref{eq_discretized_eikonal} will cause the interface to shift as a function of
$\tau$~\cite{Russo2000}, which will eventually shrink closed interfaces and lead to substantial spurious changes in the volume of each phase. To mitigate
this numerical issue, we employ second-order ENO finite differences combined with a subcell-fix method described by Min~\cite{Min2010}.
Briefly, the subcell-fix method uses $\widetilde{\phi}$ to estimate the interface location (i.e., where $\widetilde{\phi} = 0$) by fitting 
a high-order polynomial and computing an improved estimate of the one-sided derivates $D^{\pm}_x$ and $D^{\pm}_y$ from the 
polynomial fit. A dimension-by-dimension approach is followed to fit the high-order polynomial. 
In addition, if $\widetilde{\phi}$ is already close to the desired distance function,
as is typically the case after advecting the level set field for only a single time step, further mitigation of spurious changes in mass is achieved by
enforcing an immobile boundary condition near the zero level set. This approach is described by Son~\cite{Son2005} and is easily implemented by fixing the nearest points to the interface, i.e.~Eq.~\ref{eq_discretized_eikonal} is not applied to $\phi_{i,j}$ satisfying $\phi_{i,j}\phi_{p,q} \le 0$ and $\left|\phi_{i,j}\right| \le \left|\phi_{p,q}\right|$ for $(p,q) = (i \pm 1,j)$ or $(p,q) = (i,j \pm 1)$. Notice that if Eq.~\ref{eq_discretized_eikonal} is not applied to $\phi_{i,j}$, the subcell-fix is also effectively omitted for $\phi_{i,j}$. After iterating Eq.~\ref{eq_discretized_eikonal} to some desired convergence criteria, the level set function $\phi^{n+1}$ is updated, and the next physical time step is carried out. In the present work, we always reinitialize the level set every time step and declare convergence when the $L^2$ norm between subsequent psuedo-time iterations is smaller than some tolerance (taken to be $10^{-6}$ in the present work) or when a maximum number of iterations have been carried out (taken to be the maximum grid size in one direction) --- whichever happens first.

We can use the signed distance property to define the material properties at cell-centers of the staggered grid.
For a mesh with uniform grid spacing $\dx = \dy$, we can define a smoothed
Heaviside function that has been regularized over $\ncells$ grid cells on either side of the interface,
\begin{equation}
\label{eq_heaviside}
\widetilde{H}_{i,j} = 
\begin{cases} 
       0,  & \phi_{i,j} < -\ncells \dx,\\
        \frac{1}{2}\left(1 + \frac{1}{\ncells \dx} \phi_{i,j} + \frac{1}{\pi} \sin\left(\frac{\pi}{ \ncells \dx} \phi_{i,j}\right)\right) ,  & |\phi_{i,j}| \le \ncells \dx,\\
        1,  & \textrm{otherwise}.
\end{cases}
\end{equation}
A given material property $\psi$ (such as $\rho$ or $\mu$) is then set in the whole domain via
\begin{equation}
\label{eq_material_set}
\psi_{i,j} = \psi_0 + (\psi_1 - \psi_0) \widetilde{H}_{i,j},
\end{equation}
in which $\psi_0$ and $\psi_1$ denote the material property value for phases occupying $\Omega_0$ and $\Omega_1$ and
we have assumed that $\Omega_0$ is represented by negative $\phi$ values (without loss of generality).

In all of the cases in the present work, we use either $\ncells = 1$ or $2$. In general we find that for
high inertia flows, $2$ grid cells of smearing leads to slightly more distorted interfaces than $\ncells = 1$. However, 
$n_\textrm{cells} = 2$ leads to slightly better convergence properties for the iterative solver described in
Sec.~\ref{sec_solver}. We would not recommend using $n_\textrm{cells} \ge 3$ because, in our experience,
larger smearing leads to diffuse interfaces and generates spurious vortex dynamics in the vicinity of those interfaces.

\subsection{Surface tension force}
We use the continuum surface tension model of  Brackbill et al.~\cite{Brackbill1992}  to define the volumetric
surface tension force in terms of the level set field

\begin{equation}
\label{eq_surface_tension}
\fs = \sigma \kappa \grad \widetilde{C} = -\sigma \div \left(\frac{\grad \phi}{\|\grad \phi\|}\right) \grad \widetilde{C},
\end{equation}
in which $\sigma$ is the uniform surface tension coefficient, $\kappa$ is the curvature of the interface computed directly from the signed distance function as \[\kappa = - \div \n = -\div \left(\frac{\grad \phi}{\|\grad \phi\|}\right),\] $\n$ is the unit normal to the surface, and 
$\widetilde{C}$ is a mollified version of the Heaviside function $\widetilde{H}$ that ensures the surface tension force is applied only near the zero level set.
In this work we use Peskin's four-point regularized delta function~\cite{Peskin02} to mollify the numerical Heaviside function,
although other functions may also be used to smooth the transition region as described by Williams et al.~\cite{Williams1998}.
Note that one could use a surface tension force function of the form $\fs = \sigma \kappa \n \widetilde{\delta}$, in which $\widetilde{\delta}$ is a regularized version of the Dirac delta function\footnote{This is because $ \grad \widetilde{C} \approx \n \widetilde{\delta}$.}. However, this would not yield a discrete balance between the surface tension force and the pressure gradient~\cite{Francois2006}, whereas Eq.~\eqref{eq_surface_tension} is discretely well-balanced with the pressure gradient because both $\grad \widetilde{C}$ and $\grad p$ are discretized in the same manner and at the same degrees of freedom.
We refer readers to the works of Brackbill et al.~\cite{Brackbill1992} and Williams et al.~\cite{Williams1998}
for more details on the continuum surface tension model and Francois et al.~\cite{Francois2006} for its well-balanced implementation.

%In the following section, we describe the overall temporal discretization of the equations of motion and the numerical schemes 
%to stabilize high density and high shear flows. We note that the time-stepping described next is agnostic to the particular
%level set reinitialization method used. Indeed, any equivalent (i.e. same order of accuracy) level set treatment could be 
%swapped out in place of the one described here, without affecting the consistency and accuracy of the multiphase flow calculation.

\subsection{Temporal discretization}
\label{sec_temporal}
Next, we describe the temporal discretization for both the non-conservative and conservative forms
of the equations of motion. Within one time step $\left[t^n, t^{n+1}\right]$, we employ $\ncycles$ cycles of fixed-point iteration to obtain
an approximate solution to the fully-coupled mass transport and fluid flow problem. In this approach, the
advective or convective terms and the body force are treated explicitly, and all other terms are treated
implicitly. To begin time stepping, we set $k = 0$ ($\u^{n+1,0} = \u^{n}$, $p^{n+\half,0} = p^{n-\half}$, and $\phi^{n+1,0} = \phi^{n}$)
and iterate until $k = \ncycles - 1$.  At the initial time step $n = 0$, these quantities are obtained using the prescribed initial conditions.
As described previously \cite{Griffith2009}, the initial value for pressure at the start of each time step $p^{n+\half,0}$ does not affect the flow dynamics nor the pressure solution at the end of the time step $p^{n+\half}$; rather it serves an initial guess to iterative solution of the linear system.

Although both solvers converge for a wide range of density and viscosity contrasts, %(to be shown in 
%Sec.~\ref{sec_convergence}, 
the non-conservative temporal discretization is only stable for
density ratios up to $\rho_1/\rho_0 \approx 100$. The fundamental cause for this is the 
discretely inconsistent transport of mass and momentum. As described by
Raessi~\cite{Raessi2008} and, more recently, by Ghods and Herrmann~\cite{Ghods2013}, inconsistencies in 
the the numerical mass and momentum fluxes used in the mass and momentum transport equations, respectively, can lead to
numerical instabilities at density ratios greater than $100$.
This problem is especially prevalent in the level set methods, in which the phase mass is transported via an auxiliary field
and no flux reconstruction is used. Numerical examples of instabilities for the non-conservative discretization
of two phase flows with high density ratios will be shown in Sec.~\ref{sec_examples}.

To prevent these instabilities for high density ratios, we extend an approach described by 
Desjardins and Moureau~\cite{Desjardins2010}, in which consistent transport was achieved
by solving an additional mass balance equation, Eq.~\eqref{eqn_cons_of_mass},
and by ensuring that the same numerical mass flux is used for both mass and momentum transport.
A discretization of the conservative form of Eqs.~\eqref{eqn_momentum} and~\eqref{eqn_continuity} 
is necessary to achieve discrete conservation and flux compatibility. The scheme described by Desjardins and Moureau
was first-order accurate and diffusive (see Appendix~\ref{app_first_order_density_update}).
We extend this scheme to achieve \emph{at least} second-order accuracy in velocity, pressure, and density.

\subsubsection{Non-conservative discretization}

During the time interval $\left[t^n, t^{n+1}\right]$, for the \emph{non-conservative} discretization the time stepping proceeds as follows:

\begin{enumerate}
	\item Advect the signed distance function $\phi$: \\
	The level set function is updated by discretizing Eq.~\eqref{eq_ls_advection} via
	\begin{equation}
	\frac{\phi^{n+1,k+1} - \phi^{n}}{\dt} + Q\left(\u^{n+\half,k}_\text{adv}, \phi^{n+\half,k}_\text{lim}\right) = 0,
	\end{equation}
	in which $Q\left(\u^{n+\half,k}_\text{adv}, \phi^{n+\half,k}_\text{lim}\right) \approx \left[\div \left(\u^{n+\half,k}_\text{adv} \phi^{n+\half,k}_\text{lim}\right)\right]_{i,j}$ is an explicit xsPPM7-limited approximation to the
	linear advection term on cell centers. The midpoint velocity and level set are given by $\u^{n+\half,k} = \half\left(\u^{n+1,k} + \u^{n}\right)$,
	and $\phi^{n+\half,k} = \half\left(\phi^{n+1,k} + \phi^{n}\right)$.
	 Here, the subscript ``adv'' indicates the staggered grid velocity on faces of cell-centered control volume, and the subscript ``lim'' indicates the limited value also defined on faces of the cell-centered control volume.
	\item Reset the material properties $\rho$ and $\mu$: \\
	The density and viscosity in the computational domain are determined from the signed distance function via
	\begin{align}
	&\rho^{n+1,k+1} = \rho_0 + (\rho_1 - \rho_0) \widetilde{H}\left(\phi^{n+1,k+1}\right), \label{eq_rho_nc_set}\\
	&\mu^{n+1,k+1} = \mu_0 + (\mu_1 - \mu_0) \widetilde{H}\left(\phi^{n+1,k+1}\right),
	\end{align}
	in which %$H\left(\phi^{n+1,k+1}\right)$ denotes a discrete approximation to the Heaviside function based on the level set function
	%and 
	$\rho_i$ and $\mu_i$ denote the density and viscosity for the two fluids, $i=0\text{ or }1$.
	The regularized Heaviside function
	is given by Eq.~\eqref{eq_heaviside}. Note that although the density and viscosity computed here are
	\emph{cell-centered}, the notations $\vrho^{n+1,k+1}$ and $\mu^{n+1,k+1}$ are also used to denote the interpolated
	material properties (to faces for density; to nodes in $2$D, or edges in $3$D for viscosity), as needed.
	% It should be understood by context
	%over which degrees of freedom these quantities are defined.
	
	\item Solve the incompressible Navier-Stokes equations for $\u$ and $p$: \\
	The velocity and pressure are computed from the discretization of the non-conservative fluid momentum and continuity equations
	\begin{align}
	&\vrho^{n+1,k+1} \left( \frac{\u^{n+1,k+1} - \u^n}{\dt} + \N\left(\u^{n+\half,k}_\text{adv},\u^{n+\half,k}_\text{lim}\right) \right) = -\G p^{n+\half, k+1} + \left(\L_{\mu} \u\right)^{n+\half, k+1} + \f^{n+\half}, \label{eq_nc_discrete_momentum}\\
	& \vD \cdot \u^{n+1,k+1} = 0 \label{eq_nc_discrete_continuity},
	\end{align}
	in which $\N\left(\u^{n+\half,k}_\text{adv},\u^{n+\half,k}_\text{lim}\right) \approx \left[\left(\div \left(\u^{n+\half,k}_\text{adv} u^{n+\half,k}_\text{lim}\right)\right)_{i-\half, j}, \left(\div \left(\u^{n+\half,k}_\text{adv} v^{n+\half,k}_\text{lim}\right)\right)_{i, j-\half}\right]$ is an explicit 
	CUI-limited approximation to the nonlinear
	convection term and 
	
	$\left(\L_{\mu} \u\right)^{n+\half, k+1} =  \half\left[\left(\L_{\mu} \u\right)^{n+1,k+1} + \left(\L_{\mu} \u\right)^n\right]$ is a semi-implicit approximation to the viscous
	strain rate. Here, the subscript ``adv'' indicates the interpolated advective velocity on the faces of face-centered control volume, and the subscript ``lim'' indicates the convective-limited value, as defined by Eq.~\eqref{eq_conv_fd}.	The above time-stepping scheme with $\ncycles = 2$ is similar to a combination of Crank-Nicolson for the viscous terms and explicit midpoint rule for the
	convective term, making it second-order accurate in time.
	Notice that this scheme is semi-implicit in that explicit approximations are used for $\rho$ and $\mu$ as well as the advective term.
\end{enumerate}

After $\ncycles$ cycles of fixed-point iteration, the final numerical solutions are given by $\u^{n+1} = \u^{n+1,\ncycles}$, 
$p^{n+\half} = p^{n+\half,\ncycles}$, and $\phi^{n+1} = \phi^{n+1,\ncycles}$. We employ $\ncycles = 2$ cycles of fixed-point 
iteration for all the numerical examples, which yields second-order spatio-temporal accuracy. %This is especially 
%important for the conservative form of equations where a coupled mass transport and incompressible momentum system are 
%solved, as described later in this section. Computationally, $\ncycles = 2$ means that the level set transport equation and the incompressible Navier-Stokes equations are numerically solved twice per time step. This is not a limitation of the numerical method nor the implementation; in practice one could use fewer or more cycles, or even a different number of cycles for the level set and fluid flow, and potentially obtain a different order of accuracy. 
Numerical experiments indicate that the largest stable advective CFL number for the above scheme with $\ncycles = 2$ is $0.5$.
% for $\ncycles = 2$; we expect lower (higher) stable CFL conditions when using fewer (more) cycles.

\subsubsection{Conservative discretization}
\label{cons_discretization}
During the time interval $[t^n, t^{n+1}]$, the time stepping proceeds as follows
for the \emph{discretely consistent, conservative} discretization:
\begin{enumerate}
	\item Advect the signed distance function $\phi$: \\
	The level set function is updated by discretizing Eq.~\eqref{eq_ls_advection} via
	\begin{equation}
	\frac{\phi^{n+1,k+1} - \phi^{n}}{\dt} + Q\left(\u^{n+\half,k}_\text{adv}, \phi^{n+\half,k}_\text{lim}\right) = 0,
	\end{equation}
	which is the same update used for the non-conservative discretization described earlier.
	
	\item Reset the material property $\mu$: \\
	The viscosity, but not the density, in the computational domain is determined from the signed distance function via
	\begin{align}
	&\mu^{n+1,k+1} = \mu_0 + (\mu_1 - \mu_0) \widetilde{H}\left(\phi^{n+1,k+1}\right).
	\end{align}
	
	\item Advect the \emph{face-centered} density and compute convective derivative $\C$: \\
	We next update the density and compute the nonlinear convective term in a way that ensures discrete consistency in the mass and momentum fluxes.
	First, we solve a discretized density update equation on \emph{faces} of the staggered grid using the third-order 
	accurate strong stability preserving Runge-Kutta (SSP-RK3) time integrator~\cite{Gottlieb2001}
	\begin{align}
	& \breve{\V \rho}^{(1)} = \breve{\V \rho}^{n} - \dt \R\left(\u^{n}_\text{adv}, \breve{\V \rho}^{n}_\text{lim}\right), \label{eq_rk1}\\
	&  \breve{\V \rho}^{(2)} = \frac{3}{4}\breve{\V \rho}^{n} + \frac{1}{4} \breve{\V \rho}^{(1)} - \frac{1}{4} \dt \R\left(\u^{(1)}_\text{adv}, \breve{\V \rho}^{(1)}_\text{lim}\right), \label{eq_rk2} \\
	& \breve{\V \rho}^{n+1, k+1} = \frac{1}{3} \breve{\V \rho}^n + \frac{2}{3} \breve{\V \rho}^{(2)} - \frac{2}{3} \dt \R\left(\u^{(2)}_\text{adv}, \breve{\V \rho}^{(2)}_\text{lim}\right) \label{eq_rk3},
	\end{align}
	in which  $\R\left(\u_{\text{adv}}, \breve{\V \rho}_{\text{lim}}\right) \approx \left[\left(\div \left(\u_\text{adv} \breve{\V \rho}_\text{lim}\right)\right)_{i-\half, j}, \left(\div \left(\u_\text{adv} \breve{\V \rho}_\text{lim}\right)\right)_{i, j-\half}\right]$ is an explicit CUI-limited approximation 
	to the linear density advection term. In contrast with the non-conservative form, the scalar density variable 
	is defined and directly evolved on faces of the staggered grid. Hence, we distinguish $\breve{\V \rho}$, 
	the face-centered density obtained via the SSP-RK3 integrator, from $\V \rho$, the face-centered density that is reset from the level set fields.
	Here, the subscript ``adv'' indicates the interpolated advective velocity on the faces of face-centered control volume, and the 
	subscript ``lim'' indicates the limited value, as defined by Eq.~\eqref{eq_conv_fd}.	
	Note that this time integration procedure is occurring \emph{within} the overall fixed-point iteration scheme.
	We have found it to be \emph{crucial} to use appropriately interpolated and extrapolated velocities to maintain the accuracy of the scheme.  To wit, for the first cycle ($k = 0$), the velocities are
	\begin{align}
		& \u^{(1)} = 2 \u^n - \u^{n-1}, \\
		& \u^{(2)} = \3half \u^n - \half \u^{n-1}.
	\end{align}
	For all remaining cycles ($k > 0$), the velocities are
	\begin{align}
		& \u^{(1)} = \u^{n+1,k}, \\
		& \u^{(2)} = \frac{3}{8} \u^{n+1,k} + \frac{3}{4} \u^{n} - \frac{1}{8} \u^{n-1}.
	\end{align}
	Notice that $\u^{(1)}$ is an approximation to $\u^{n+1}$, and $\u^{(2)}$ is an approximation to $\u^{n+\half}$. Similarly, $\breve{\V \rho}^{(1)}$ is an approximation to $\breve{\V \rho}^{n+1}$, and $\breve{\V \rho}^{(2)}$ is an approximation to $\breve{\V \rho}^{n+\half}$. 
	
	\item Solve the incompressible Navier-Stokes equations for $\u$ and $p$: \\
	Using the previously computed density and convective term, the velocity and pressure are computed
         from the discretization of the conservative fluid momentum and continuity equations
         \begin{align}
	&\frac{\breve{\V \rho}^{n+1,k+1} \u^{n+1,k+1} - \breve{\V \rho}^{n} \u^n}{\dt} + \C\left(\u^{(2)}_\text{adv}, \breve{\V \rho}^{(2)}_\text{lim} \u^{(2)}_\text{lim}\right) = -\G p^{n+\half, k+1} + \left(\L_{\mu} \u\right)^{n+\half, k+1} + \f^{n+\half}, \label{eq_c_discrete_momentum}\\
	& \vD \cdot \u^{n+1,k+1} = 0 \label{eq_c_discrete_continuity},
	\end{align}
	in which the approximation to the convective derivative is given by
	\begin{equation}
	\C\left(\u^{(2)}_\text{adv}, \breve{\V \rho}^{(2)}_\text{lim}\u^{(2)}_\text{lim}\right) \approx \left[\left(\div \left(\u^{(2)}_\text{adv} \breve{\V \rho}^{(2)}_\text{lim} u^{(2)}_\text{lim}\right)\right)_{i-\half,j}, \left(\div \left(\u^{(2)}_\text{adv} \breve{\V \rho}^{(2)}_\text{lim} v^{(2)}_\text{lim}\right)\right)_{i,j-\half}\right],
	\end{equation}
	and uses the same velocity $\u^{(2)}_{\text{adv}}$ and density $\breve{\V \rho}^{(2)}_{\text{lim}}$ used to update $\breve{\V \rho}^{n+1,k+1}$ in Eq~\eqref{eq_rk3}.
	This is the key requirement for consistent mass and momentum transport. 
\end{enumerate}
Results presented in Sec.~\ref{sec_examples} demonstrate that the consistent discretization is stable for density ratios of at 
least $10^6$ and produce significantly more accurate results than the inconsistent discretization for realistic 
two phase flow simulations.

For the conservative form as written, the density evolves along with the velocity and pressure at all times, with an initial 
value of $\breve{\V \rho}$ directly specified on the cell faces. 
We note the we can instead \emph{synchronize} the face-centered density via the signed distance function (averaged to faces) 
at time step $n$, i.e., 
$\V\rho^{n} = \rho_0 + (\rho_1 - \rho_0) \widetilde{\boldsymbol{H}}\left(\phi^{n}\right)$, while still maintaining numerical stability. When resetting the mass density in each time step,
the new density $\breve{\V \rho}^{n+1,k+1}$ (used in solving for $\u$ and $p$) is still obtained via the  SSP-RK3 update; however it is \emph{discarded} at the end of the time step.
\emph{When density synchronization is enabled, $\V \rho^{n}$ is used in place of $\breve{ \V \rho}^{n}$
wherever a density field is needed at time level $n$ in Eqs.~\eqref{eq_rk1} --~\eqref{eq_rk3} and Eq.~\eqref{eq_c_discrete_momentum}}.
Finally, we emphasize that using a level-set synchronized density at time level $n$ is recommended to
avoid significant distortions in the interface for high density ratio flows, which are generated by accumulation of errors in advecting $\breve{\V\rho}$ over the course of the simulation.\footnote{This justifies the additional computational cost incurred by level set and volume-of-fluid methods.}
Sec.~\ref{sec_cons_ms} investigates differences between evolving and resetting the density field, 
and all of the numerical examples in Sec.~\ref{sec_examples} use density 
synchronization.

Note that in both Eqs.~\eqref{eq_nc_discrete_momentum}  and~\eqref{eq_c_discrete_momentum} we only considered 
the volumetric body force term $\f^{n+\half}$. One can approximate the surface tension force $\fs^{n+\half}$ as a function 
of level set field by $\fs(\phi^{n+\half,k+1})$, in which $\phi^{n+\half,k+1} = \half \left( \phi^{n+1,k+1} + \phi^{n} \right)$. 
 
Finally, as discussed previously, there is no guarantee that the level set function will remain a signed distance function under advection
by an external velocity field. Thus, at the beginning of each time step, the reinitialization procedure described in 
Section~\ref{sec_level_set} is used.
Reinitializion is required to accurately evaluate regularized Heaviside functions
near the interface, as needed both to determine material properties and to evaluate interfacial forces related to surface tension.

\subsection{Adaptive mesh refinement}
\label{sec_amr}
Some cases presented in this work use a structured adaptive mesh refinement (SAMR) framework to discretize the equations of motion. 
These discretization approaches describe the computational domain as composed of
multiple grid levels, which together form a grid hierarchy.
Assuming uniform and isotropic mesh refinement, a grid hierarchy with $\ell$ levels and with a 
grid spacing $\dx_0$, $\dy_0$, and $\dz_0$ on the coarsest grid level
has grid spacings $\dx_\textrm{min} = \dx_0/\nref^{\ell-1}$, 
$\dy_\textrm{min} = \dy_0/\nref^{\ell-1}$, and $\dz_\textrm{min} = \dz_0/\nref^{\ell-1}$ on the finest grid level, in which
$\nref$ is the integer refinement ratio between levels.
% although the code can use different 
%values of $\nref$ between different levels. 
%In the present work, the refinement ratio is 
%taken to be the same in each direction (which is neither a limitation of the numerical method nor our implementation). 
(Although not considered here, both the numerical method and software implementation allow for general refinement ratios.)

The locally refined meshes can be static, in that they occupy a fixed region in the domain $\Omega$,
or adaptive, in that some criteria of interest is used to ``tag'' coarse cells for refinement. In our current implementation, cells are refined based on two criteria: 1)~if the local magnitude of vorticity 
$\|\omega\|_{i,j} = \|\grad \times \u\|_{i,j}$ exceeds a relative threshold and 2)~if the 
signed distance function $\phi_{i,j}$ is within some threshold of zero. This ensures that the important
dynamics (e.g., regions of high velocity gradients or the multiphase interfaces) are always approximated 
using appropriate mesh spacings.

Each time the grid hierarchy is regenerated, quantities must be transferred from the old grid hierarchy to the new grid hierarchy. In newly refined regions, the fluid velocity $\u$ is interpolated from the old coarse
grid using a conservative, discretely divergence- and curl-preserving interpolation scheme~\cite{Toth2002}.
Similarly, the level set $\phi$ and material properties $\rho$ and $\mu$ are interpolated from the old coarse
grid using a conservative linear interpolation scheme, which ensures that positivity is maintained for density and viscosity.
The pressure $p$ is interpolated using a simple non-conservative linear interpolation scheme as it is only used as 
an initial approximation to the updated pressure. In newly coarsened regions, all of the quantities are defined
as conservative averages of the old fine-grid data. These interpolations are used to define ghost cell
values at the coarse-fine interface, which enables composite-grid approximations to the linear operators
$\vD \cdot $, $\G $,  $\Lmu $, and $\Lrho$ described earlier in Sec.~\ref{sec_spatial_discretization}. We refer
readers to prior work by Griffith~\cite{Griffith2012} for additional details on the AMR discretization methods.

\section{Solution methodology}
\label{sec_solver}
This section describes the linear solvers required to compute
a solution to the fully-coupled, time-dependent incompressible 
Stokes system
\begin{equation}
\label{eq_full_stokes}
\left[
\begin{array}{cc}
 \frac{1}{\dt} \V \wp^{n+1,k+1} - \half \Lmu^{n+1,k+1} & \G\\
 -\vD\cdot & \mathbf{0} \\
\end{array}
\right]
\left[
\begin{array}{c}
 \u^{n+1,k+1}\\
 p^{n+\half,k+1} \\
\end{array}
\right] =
\left[
\begin{array}{c}
 \fu\\
\V0\\
\end{array}
\right],
\end{equation}
in which $\V \wp^{n+1,k+1}$ is a diagonal matrix of face-centered densities corresponding to each velocity
degree of freedom ($\V \wp \equiv \vrho$ as defined in Eq.~\eqref{eq_rho_nc_set} for the non-conservative discretization,
and $\V \wp \equiv \breve{\vrho}$ as defined in Eq.~\eqref{eq_rk3} for the conservative discretization),
and the right-hand side of the momentum equation is lumped into $\fu$, whose value
depends on the discretization type.
For non-conservative form, it is
\begin{equation}
\fu = \left(\frac{1}{\dt}  \vrho^{n+1,k+1} + \half \Lmu^{n}\right)\u^n - 
	\vrho^{n+1,k+1}\N\left(\u_\text{adv}^{n+\half,k}\u_\text{lim}^{n+\half,k}\right)+\f^{n+\half} + \fs^{n+\half},
\end{equation}
and for conservative form with density synchronization, it is
\begin{equation}
\fu = \left(\frac{1}{\dt}  \vrho^{n} + \half \Lmu^{n}\right)\u^n - 
	\C\left(\u_\text{adv}^{(2)},\breve{\V \rho}_\text{lim}^{(2)}\u_\text{lim}^{(2)}\right)+\f^{n+\half} +  \fs^{n+\half}.
\end{equation}
The operator on the left-hand side of Eq.~\eqref{eq_full_stokes} is the time-dependent
incompressible staggered Stokes operator. %, or \emph{Stokes operator} for short.
%In the next section, we will describe the solution of Eq.~\eqref{eq_full_stokes} via
%a preconditioned flexible GMRES (FGMRES) Krylov solver~\cite{Saad93}.
To solve this system of equations, we use a flexible GMRES (FGMRES) Krylov solver~\cite{Saad93} preconditioned 
by a variable-coefficient projection method solver that is hybridized with a local-viscosity solver. 
The efficient preconditioner enables rapid convergence of the variable-coefficient iterative Stokes solver; between $1$ and $20$ Krylov iterations are observed for all cases considered here. 
We briefly describe the hybrid preconditioner here and refer the readers to the work of Griffith~\cite{Griffith2009} and Cai et al.~\cite{Cai2014} for further details on the constant-coefficient and variable-coefficient preconditioners, respectively.

\subsection{Projection preconditioner}
\label{sec_proj_pc}
The Stokes system can be succinctly written as
\begin{equation}
\label{eq_stokes_system}
\B
\left[
\begin{array}{c}
 \xu\\
 \xp \\
\end{array}
\right] =
\left[
\begin{array}{cc}
 \A & \G\\
 -\vD\cdot\mbox{} & \mathbf{0} \\
\end{array}
\right]
\left[
\begin{array}{c}
  \xu\\
  \xp \\
\end{array}
\right] =
\left[
\begin{array}{c}
 \bu\\
 \bp \\
\end{array}
\right],
\end{equation}
in which $\B$ is the Stokes operator, $\A =  \frac{1}{\dt} \V \wp^{n+1,k+1} - \half \Lmu^{n+1,k+1}$ , %is
%the discretization of the temporal and viscous operator,
$\xu$ and $\xp$ denote the velocity and pressure
degrees of freedom, and $\bu$ and $\bp = \V 0$ denote the right-hand sides of the momentum and continuity equations, respectively.
%Krylov methods
%such as FGMRES require routines to compute the action of $\B$ on vectors, i.e. $\B \x$. Note that the matrix form
%of the operators $\A$, $\G$, and $\vD \cdot$ are merely a convenience. Rather these operators are ``applied" by
%direct evaluation of the finite difference operators defined on locally refined grids. 
%%described in Sec.~\ref{sec_spatial_discretization} at every grid face and cell.

Saddle-point problems such as Eq.~\ref{eq_stokes_system} are ill-conditioned, %which leads to slow iterative solver convergence rates in the absence of effective preconditioning strategies.
%of iterative solvers. 
and effective preconditioning strategies are needed to obtain scalable Krylov methods for such equations.
%for large, sparse systems. Preconditioned Krylov iterative methods require routines to compute the action of a
%preconditioner $\P^{-1}$ on vectors, i.e. $\P^{-1} \x$. 
The particular preconditioner used in the present work is based on the fractional-step \emph{projection method}, which is commonly used to  
solve the incompressible Navier-Stokes equations~\cite{Chorin1968, Chorin1969} in an operator-splitting manner. The conversion of the projection
method to a \emph{projection preconditioner} $\Pproj$ was described for constant material properties by Griffith \cite{Griffith2009},
and for variable material properties by Cai et al.~\cite{Cai2014}.

For theoretical completeness and to interpret solver scalability results of section~\ref{sec_cons_scale}, we briefly outline the derivation and form of 
$\Pproj$ and provide some practical considerations for the linear solvers towards the end of the section. 
To begin, we first note that for the Stokes preconditioner, $\xu$ and $\xp$ should be interpreted as error in velocity and pressure degrees of freedom, respectively, and $\bu$ and $\bp \ne \V0$ should be interpreted as residuals of the momentum and continuity constraint equations, respectively.
As is done in the conventional projection method, we first compute an intermediate approximation to
${\x}_{\u}$ by solving
\begin{equation}
\label{eq_frac_vel}
\A \widehat{\x}_{\u} = \bu.
\end{equation}
Note that this approximation does not in general satisfy the discrete continuity equation
i.e., $-\vD \cdot \widehat{\x}_{\u} \ne \bp$. This condition can be satisfied by introducing an auxiliary
scalar field $\vtheta$ and writing out a fractional timestep
\begin{align}
& \V \wp \frac{\x_{\u}-\widehat{\x}_{\u}}{\dt} = -\G \vtheta, \label{eq_frac_timestep} \\
& -\vD \cdot \xu = \bp \label{eq_frac_continuity}.
\end{align}
Multiplying Eq.~\eqref{eq_frac_timestep} by $\V \wp^{-1}$, taking the discrete divergence $\vD \cdot$, and substituting
in Eq.~\eqref{eq_frac_continuity} yields the density-weighted Poisson problem
\begin{equation}
\label{eq_frac_div}
-\vD \cdot \V \wp^{-1} \G \vtheta = -\Lrho \vtheta = -\frac{1}{\dt}\left(\bp + \vD \cdot \widehat{\x}_{\u}  \right).
\end{equation}
The updated velocity solution can be computed as
\begin{equation}
\label{eq_frac_up_vel}
\xu = \widehat{\x}_{\u} - \dt \V \wp^{-1} \G \vtheta.
\end{equation}
In the conventional projection method for constant viscosity, the pressure solution can
be computed as $\xp = \left(\I - \frac{\dt}{2} \mu \Lrho \right)\vtheta$. For preconditioning purposes,
it was shown in~\cite{Cai2014} that a reasonable approximation to the pressure solution in the presence of a spatially varying  
viscosity can be written as
\begin{equation}
\label{eq_frac_up_pressure}
\xp \approx \left(\I - \frac{\dt}{2} \Lrho 2 \vmu \right) \vtheta,
\end{equation}
in which $\vmu$ is a diagonal matrix of cell-centered viscosities corresponding to each pressure
degree of freedom. The above form of Eq.~\eqref{eq_frac_up_pressure} comes from an approximate 
Schur complement of the Stokes system Eq.~\eqref{eq_stokes_system} and we refer readers to~\cite{Cai2014}
for more details. Finally, the projection preconditioner can be written in matrix form as
\begin{equation}
\label{eq_proj_pc}
\Pproj =
\left[
\begin{array}{cc}
 \I & -\dt \V \wp^{-1} \G\\
 \mathbf{0} & \I - \frac{\dt}{2} \Lrho 2\vmu \\
\end{array}
\right]
\left[
\begin{array}{cc}
 \I & \mathbf{0}\\
 \mathbf{0} & (-\Lrho)^{-1} \\
\end{array}
\right]
\left[
\begin{array}{cc}
 \I & \mathbf{0}\\
-\frac{1}{\dt} \vD \cdot & -\frac{1}{\dt} \I \\
\end{array}
\right]
\left[
\begin{array}{cc}
 \A^{-1} & \mathbf{0}\\
 \mathbf{0} & \I \\
\end{array}
\right],
\end{equation}
in which $\vmu$ is a diagonal matrix of cell-centered viscosities corresponding to each pressure
degree of freedom.
This preconditioner can be obtained from an approximate 
block factorization of the Stokes system; see Cai et al.~\cite{Cai2014} for further details.

There are several advantages to using the projection method as a preconditioner rather than as a solver. We summarize them here:
\begin{itemize}
\item The projection method is an operator splitting approach that requires the specification of artificial boundary conditions for the velocity and pressure fields. This split affects the global  order of accuracy of the solution~\cite{Brown2001}. 
For example, consider the imposition of normal traction on domain boundary points $\xb \in \partial \Omega$, i.e, $\n \cdot \vsigma \cdot \n = -p+2 \mu \D{u_n}{n} = F(\xb,t)$ as detailed in Appendix~\ref{app_boundary_conditions}. Here, $u_n$ represents the normal velocity 
and $F(\xb,t)$ is the imposed function (possibly of time) defined on the boundary $\partial \Omega$. The normal traction boundary condition requires a linear combination of discretized 
pressure and velocity variables. This combination can be accounted for in the Stokes operator $\B$ directly. However, it is not possible to split the linear combination \emph{a priori} into the velocity operator $\A$ and pressure operator $\Lrho$ used by the projection solver. In contrast, this artificial split (or artificial boundary conditions) in the projection preconditioner~\footnote{Since preconditioners employ homogenous versions of the boundary conditions, one can use homogeneous Dirichlet conditions for pressure and homogeneous Neumann conditions for the normal velocity component. This treatment is followed in the current work for the projection preconditioner whenever normal traction boundary conditions are employed in the fluid solver.} does \emph{not} affect the final solution of velocity and pressure obtained from the outer Krylov solver. A preconditioner can only affect the convergence rate of the iterative solver, not the accuracy of the converged solution.     

\item Projection methods are derived by assuming that certain operators commute. Typically, this assumption is only satisfied in the case of constant-coefficient operators defined on periodic computational domains. For variable-coefficient operators, there is an unavoidable commutator error that is associated with using the projection method as a solver.

%\item The projection method as a solver introduces large errors for zero Reynolds number incompressible Stokes flow due to splitting of the velocity and pressure solutions; see the introduction section in~\cite{Cai2014} for a discussion. Stokes flow is a purely elliptic system that describes an instantaneous force equilibrium between pressure and viscous forces (and possibly other body forces), and therefore requires a simultaneous solution of velocity and pressure variables. As explained in Appendix~\ref{app_stokes_pc}, a modified projection preconditioner can be used to solve the incompressible Stokes equations simultaneously.

\item Finally, using the projection method as a preconditioner is \emph{no less efficient than using it as a solver}, as demonstrated by Griffith~\cite{Griffith2009} and Cai et al.~\cite{Cai2014}. Our tests in Sec.~\ref{sec_cons_scale} also support these previous findings.  
\end{itemize}

We remark that in contrast with the constant coefficient projection preconditioner~\cite{Griffith2009},
both the Stokes operator $\B$ and the preconditioner $\Pproj$ will change from time step to time step, and even cycle-to-cycle, because the density and viscosity generally will varying in both space and time. For the Krylov solver, it
is vital that $\B$ be updated every time the material properties are updated.  It is easy to achieve this in practice by using a matrix-free implementation of the present method.
In contrast, we find that there is no
need to update $\Pproj$ at the same frequency. Specifically, we find that in many cases, it is computationally efficient to update
$\Pproj$ only every $5$ to $50$ time steps to avoid impairing the convergence of the FGMRES iterations.
% (determined empirically), which does not lead to a significant increase in the number of FGMRES
%iterations required for convergence. This eliminates some of the computational cost required to reinitialize the linear
%solvers in the projection preconditioner.
% NN: don't think the below is true...
%We remark that we keep $\Pproj$ constant within a single time step, i.e, $\Pproj$ is not updated for $k > 0$ cycle number.
%In the context of the AMR framework, whenever the grid hierarchy is regenerated, both $\B$ and $\Pproj$ are updated to correspond to the new grid hierarchy. 

\subsection{Subdomain solvers}
\label{sec_subdomain}
Because Krylov methods require only the evaluation of the action of the preconditioning operator,  we do not actually form $\Pproj$, which would be a dense matrix.
Notice, however, that evaluating the action of the projection preconditioner $\Pproj$ requires the solution to momentum and pressure-Poisson equations.
In practice, these equations are only solved \emph{approximately} by so-called \emph{subdomain solvers}:
%The application of $\Pproj$ involves the solution of two distinct linear systems:
1) the velocity subdomain system, which requires the (approximate) application of $\A^{-1}$, and
2) the pressure subdomain system, which requires the (approximate) application of $\Lrho^{-1}$.
%$\left(-\Lrho\right)^{-1}\left(-\frac{1}{\dt}\bp + -\frac{1}{\dt}\vD \cdot \widehat{\x}_{\u}  \right)$.
We again employ iterative methods for these inner subdomain solvers to produce approximate solutions up
to a specified relative residual tolerance $\epssub$. It is neither recommended nor required that tight
tolerances be used. We find that for the cases considered in this work, $\epssub = 10^{-2}$ is sufficient
to produce a convergent outer solver that achieves a relative residual tolerance of $\epsstokes = 10^{-12}$
for Eq.~\eqref{eq_stokes_system} within $1$ to $20$ FGMRES iterations, even for highly contrasting material properties, or on highly refined grids.

For both the velocity and pressure subdomain problems, we employ a Richardson solver with a single
multigrid V-cycle of an FAC (fast adaptive composite) 
preconditioner~\cite{Mccormick1986}, respectively. For each multigrid level,
for both the velocity and 
pressure problems, $3$ iterations of Gauss-Seidel smoothing are used. The solvers chosen here are shown to 
converge with density and viscosity contrasts up to $10^6$ in Sec.~\ref{sec_convergence}.
%We note that other iterative methods could be used to approximately solve the subdomain problems
%and could be more efficient that those chosen for this work.
Homogenous Dirichlet or Neumann boundary conditions, similar to those used in a projection method-based solver, are prescribed for the subdomain problems.
We remark that choosing boundary conditions for the inner solvers that are incompatible with the physical boundary conditions imposed in the outer solver can cause the outer FGMRES iterations to fail to converge. More details on the choice of compatible subdomain solver boundary conditions are provided by Griffith \cite{Griffith2009}.
Unlike the case of a projection method-based solver, however, the choice of these boundary conditions do not affect the final solution or the accuracy of the outer solver so long as that outer solver converges.

%%%%%%%%%%%%%%%%%%%%%%%%%%%%%%
\section{Software implementation}
\label{sec_software}

The algorithms and fluid solver described here are implemented
in the IBAMR library~\cite{IBAMR-web-page}, which is open-source C++
simulation software focused on immersed boundary methods with
adaptive mesh refinement. All of the numerical examples presented here
are publicly available via \url{https://github.com/IBAMR/IBAMR}.
IBAMR relies on SAMRAI \cite{HornungKohn02, samrai-web-page} for Cartesian grid 
management and the AMR framework. Linear and nonlinear solver support in IBAMR is provided by 
the PETSc library~\cite{petsc-efficient, petsc-user-ref, petsc-web-page}.

%All of the example cases in the present work made use of distributed-memory
%parallelism using the Message Passing Interface (MPI) library. Various computing environments were used to
%execute the simulations, including:
%\begin{itemize}
%\item a personal desktop computer comprised of 16 cores (processors) and 24 GB total memory; 
%\item the Quest cluster at Northwestern University, which is comprised of 679 compute nodes (16,028 cores) interconnected by an InfiniBand network and 128 GB memory per node;
%\item the XSEDE Bridges cluster at the Pittsburgh Supercomputing Center, which contains 752 Regular Shared Memory nodes (21,056 cores) and 128 GB memory per node;
%\end{itemize}
%Between $1$ and $128$ processors were used in all the cases described here.

%%%%%%%%%%%%%%%%%%%%%%%%%%%%%%%%%%%%%%
\section{Solver accuracy and performance}

\label{sec_convergence}
This section investigates the accuracy and convergence rates for the multiphase flow solver, in both non-conservative and conservative form, with various choices of boundary 
conditions. In all cases, the computed solution from the solver is compared against manufactured solutions on uniform 
and locally refined grids. 
We also demonstrate the robustness and scalability of the flow solver for highly contrasting densities and viscosities 
and also under grid refinement in this section.
We employ $\ncycles = 2$ cycles of fixed-point iteration for all cases considered in this section.
To assess the order of accuracy, errors in the cell centered pressure $p$ are computed using standard formulae for the $L^1$ and $L^{\infty}$ norms
\begin{align}
&\|p\|_1 = \sum_{i = 0}^{\Nx-1} \sum_{j = 0}^{\Ny - 1} \left| p_{i,j}\right| \dx \dy, \\
&\|p\|_{\infty} = \max_{\x_{i,j} \in \Omega} \left|p_{i,j}\right|.
\end{align}
The $L^1$ error in the face centered velocity $\u$ is computed using a simple modification for the $L^1$ norms
described in~\cite{Griffith2009}, which ensures that $\|(1,0)\|_1 = \|(0,1)\|_1 = 1$; the $L^1$ norm is computed as
$\|\u\|_1 = \|(u,v)\|_1 = \|u\|_1 + \|v\|_1$, with
\begin{align}
&\|u\|_1 = \frac{1}{2} \sum_{j = 0}^{\Ny-1} \left|u_{-\half,j}\right| \dx \dy + \sum_{i = 1}^{\Nx - 1} \sum_{j = 0}^{\Ny - 1} \left|u_{i-\half,j}\right| \dx\dy + \frac{1}{2} \sum_{j = 0}^{\Ny-1} \left|u_{\Nx-\half,j}\right| \dx\dy, \\
&\|v\|_1 = \frac{1}{2} \sum_{i = 0}^{\Nx-1} \left|v_{i,-\half}\right| \dx \dy + \sum_{i = 0}^{\Nx - 1} \sum_{j = 1}^{\Ny - 1} \left|v_{i,j-\half}\right| \dx\dy + \frac{1}{2} \sum_{i = 0}^{\Nx-1} \left|v_{i,\Ny-\half}\right| \dx\dy.
\end{align}
Finally, the $L^{\infty}$ norm of the face centered velocity is computed as $\|\u\|_{\infty} = \|(u,v)\|_{\infty} = \max\left(\|u\|_{\infty}, \|v\|_{\infty}\right)$, with
\begin{align}
&\|u\|_{\infty} = \max_{\x_{i-\half,j} \in \Omega} \left|u_{i-\half,j}\right|, \\
&\|v\|_{\infty} = \max_{\x_{i,j-\half} \in \Omega} \left|v_{i,j-\half}\right|.
\end{align}
The presented convergence rates are essentially the same independent of the time at which
these errors are evaluated.
Unless otherwise stated, a grid size of $N \times N$ is used to discretize the computational domain and a relative convergence tolerance of $\epsstokes = 10^{-12}$ is specified for the FGMRES solver in all of these cases.

For many of the manufactured solutions considered in this section, a
smoothing parameter $\delta$ is used to ensure that the density and viscosity fields are transitioned
from low values to high values over a spatial region of constant width. The specific values of $\delta$
are chosen to ensure that the coarsest grid in each convergence study is able to resolve this region over at 
least $1.5$ grid cells. Moreover, a constant spatial transition width ensures that the same exact analytical solution 
is considered on all grids, 
which makes the order-of-accuracy results consistent.

\subsection{Non-conservative form: Effect of boundary conditions}
\label{sec_non_cons_ms}

\begin{figure}[]
  \centering
  \subfigure[$\rho(\x)$]{
    \includegraphics[scale = 0.255]{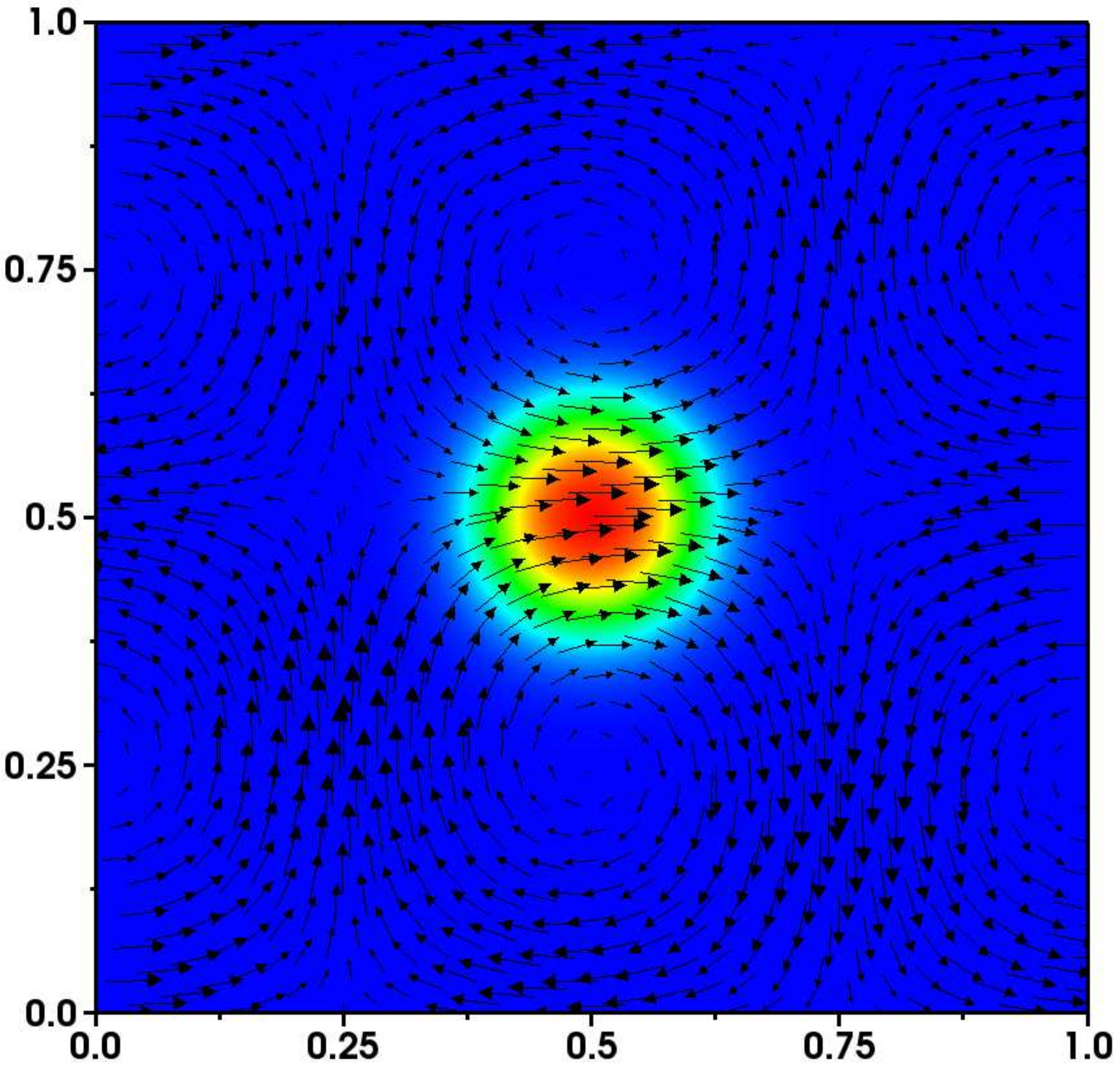}
    \label{fig_NC_Density_Init_Schematic}
  }
   \subfigure[$\mu(\x)$]{
    \includegraphics[scale = 0.255]{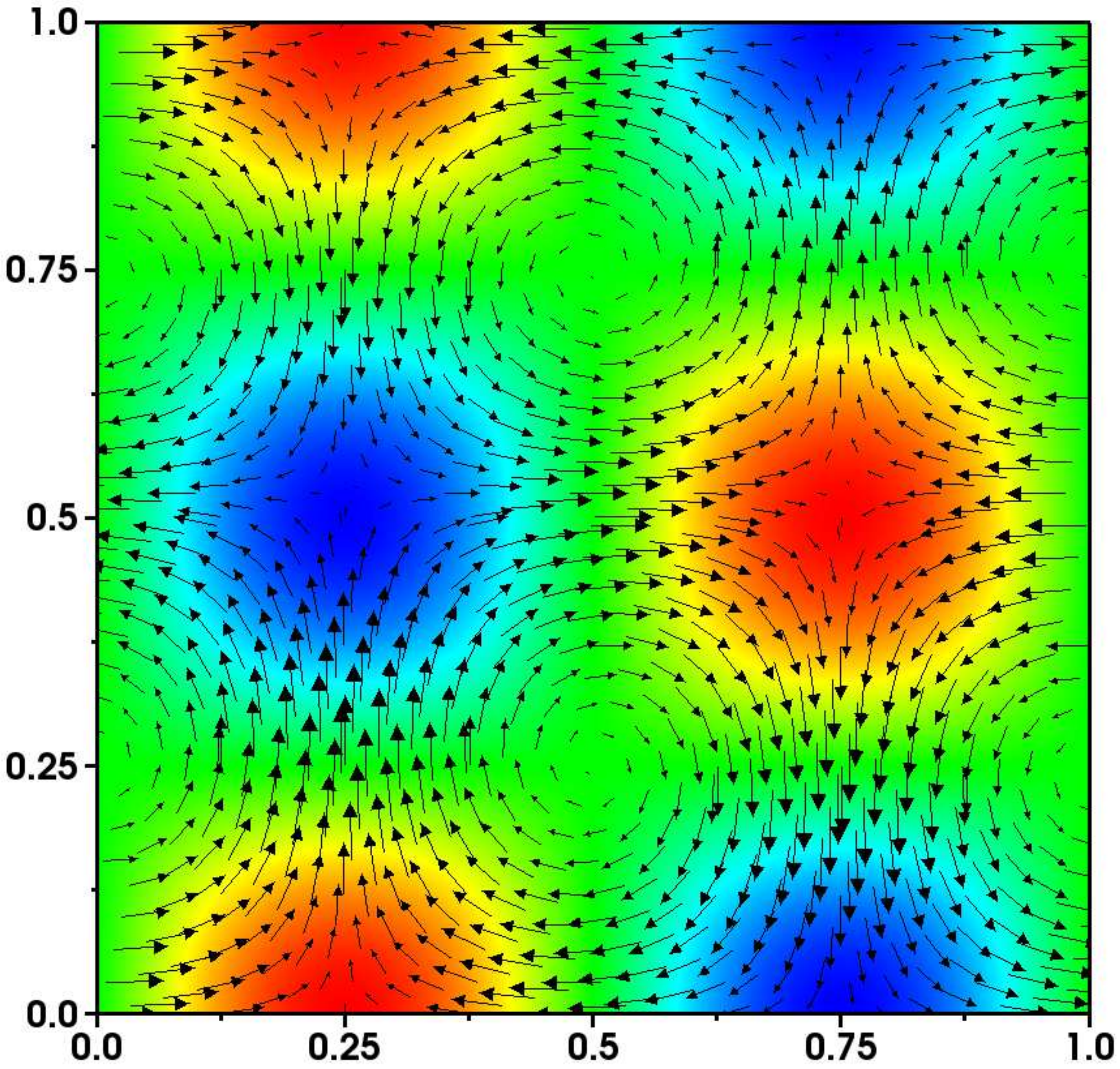}
    \label{fig_NC_Viscosity_Init_Schematic}
  }
   \caption{The \subref{fig_NC_Density_Init_Schematic} density and
    \subref{fig_NC_Viscosity_Init_Schematic} viscosity fields, along with the initial velocity vectors
    for the manufactured solution described
   in Sec.~\ref{sec_non_cons_ms}}.
  \label{fig_non_cons_ms_init}
\end{figure}

We begin by considering the non-conservative set of equations, \eqref{eqn_nc_momentum}--\eqref{eqn_nc_continuity}.
A manufactured solution for velocity and pressure is given by
\begin{align}
u(\x,t) &= 2 \pi  \cos (2 \pi  x) \cos (2 \pi  t-2 \pi  y), \label{eq_non_cons_ms_u}\\
v(\x,t) &= -2 \pi  \sin (2 \pi  x) \sin (2 \pi  t-2 \pi  y)-\sin (2 \pi  t-2 \pi  x), \label{eq_non_cons_ms_v}\\
p(\x,t) &= -2 \pi  \sin (2 \pi  t-2 \pi  x) \cos (2 \pi  t-2 \pi  y), \label{eq_non_cons_ms_p}
\end{align}
together with time-independent density and viscosity fields of the form   
\begin{align}
\rho(\x) &= \rho_0 + \frac{\rho_1}{2}  \left(\tanh \left(\frac{0.1\,
   -\sqrt{(x-0.5)^2+(y-0.5)^2}}{\delta }\right)+1\right), \label{eq_non_cons_ms_rho}\\
\mu(\x) &= \mu_0+\mu_1+\mu_1 \sin (2 \pi  x) \cos (2 \pi  y), \label{eq_non_cons_ms_mu}
\end{align}
The density, viscosity, and initial velocity fields are shown in Fig.~\ref{fig_non_cons_ms_init}.
%In the above, $\rho_0, \rho_1, \mu_0, \mu_1 \in \mathbb{R}$ 
%and $\delta \in \mathbb{R}$ is a smoothing parameter. 
Plugging Eqs.~\eqref{eq_non_cons_ms_u}-\eqref{eq_non_cons_ms_mu} 
into the non-conservative momentum equation~\eqref{eqn_nc_momentum} yields a forcing term $\f(\x,t)$ that produces the specified 
solution. % given by Eqs.~\eqref{eq_non_cons_ms_u}-\eqref{eq_non_cons_ms_p}. 
We set $\rho_0 = 1$, $\rho_1 = -\rho_0 + 10^3$, $\mu_0 = 10^{-4}$, and $\mu_1 = -\mu_0 + 10^{-2}$,
which yields variations in density and viscosity are set to be similar to that of air and water.
The computational domain is the unit square, $\Omega = [0,L]^2 = [0,1]^2$.
The smoothing parameter is set to $\delta = 0.05L$. %Note that for the non-conservative equations, the density and viscosity fields are given, and therefore we only need to test the order of accuracy of the computed velocity and pressure fields. 

\begin{figure}[]
  \centering
  \subfigure[Velocity]{
    \includegraphics[scale = 0.255]{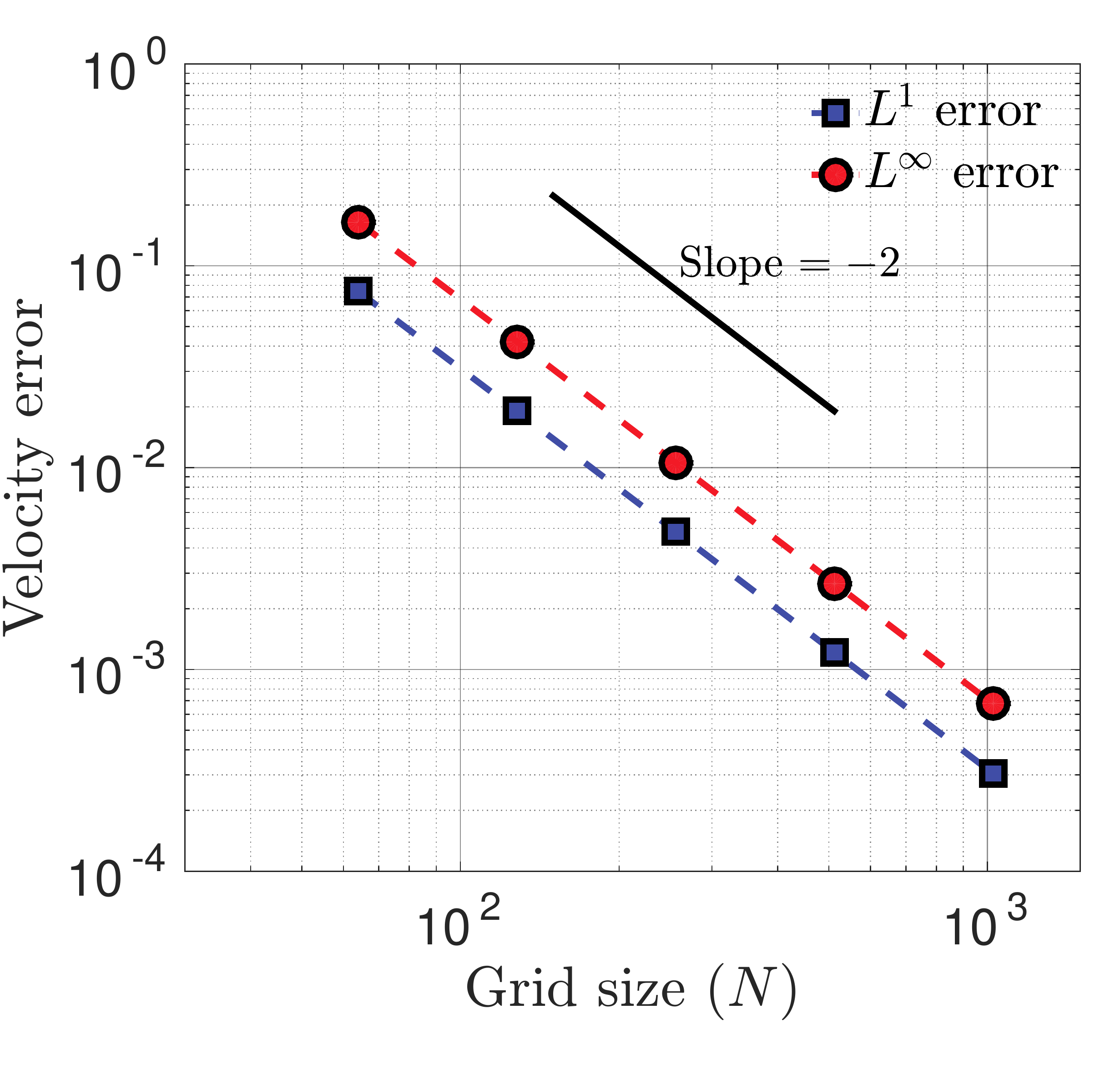}
    \label{fig_nc_U_err_periodic}
  }
   \subfigure[Pressure]{
    \includegraphics[scale = 0.255]{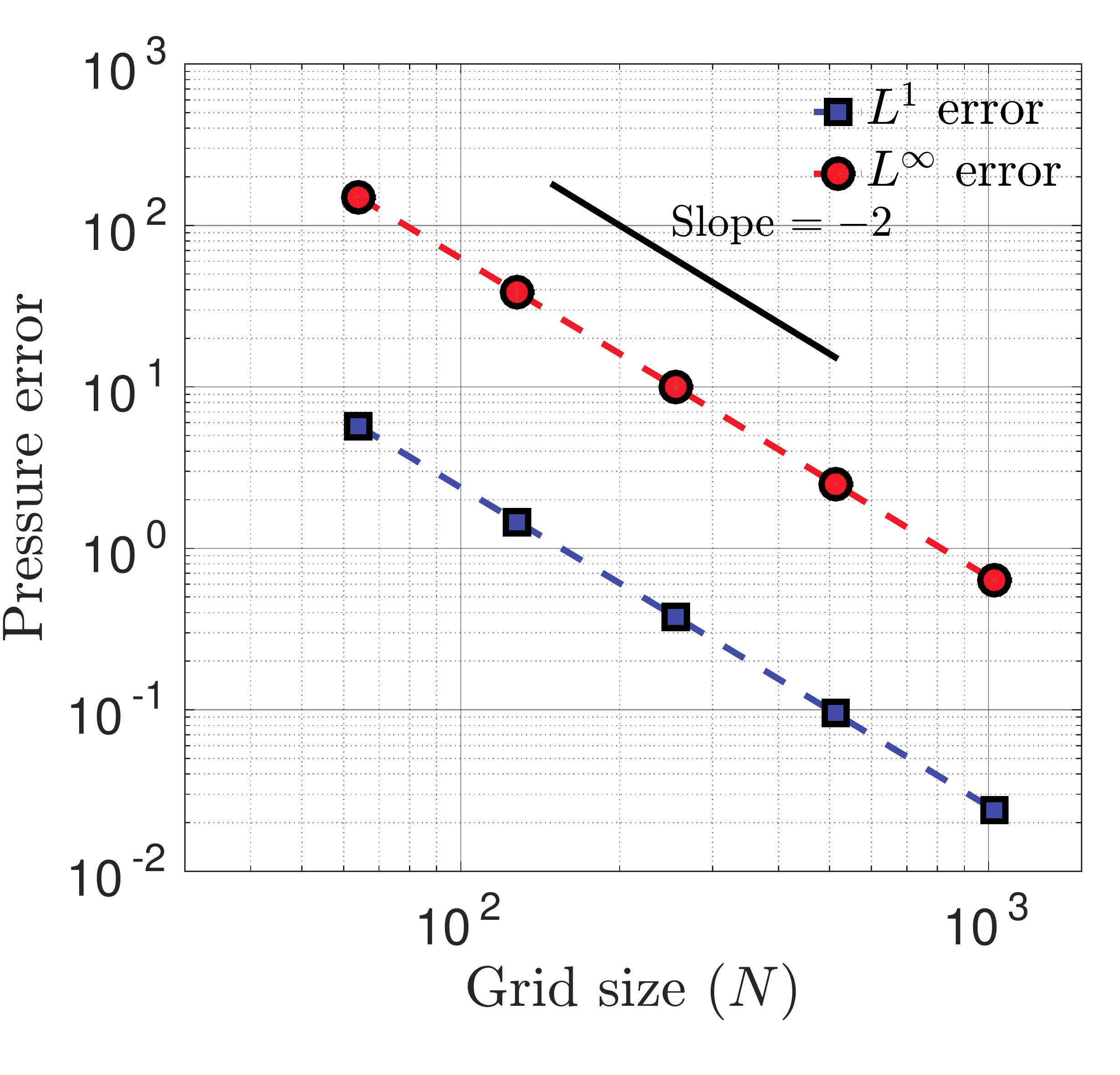}
    \label{fig_nc_P_err_periodic}
  }
  \caption{ 
  $L^1$ ($\blacksquare$, blue) and $L^\infty$ ($\bullet$, red) errors as a function of grid size $N$ for the non-conservative
  manufactured solution with periodic boundary conditions:
  \subref{fig_nc_U_err_periodic}
  convergence rate for $\u$;
  \subref{fig_nc_P_err_periodic}
   convergence rate for $p$.
   }
  \label{fig_nc_err_periodic}
\end{figure}

\begin{figure}[]
  \centering
  \subfigure[Velocity]{
    \includegraphics[scale = 0.255]{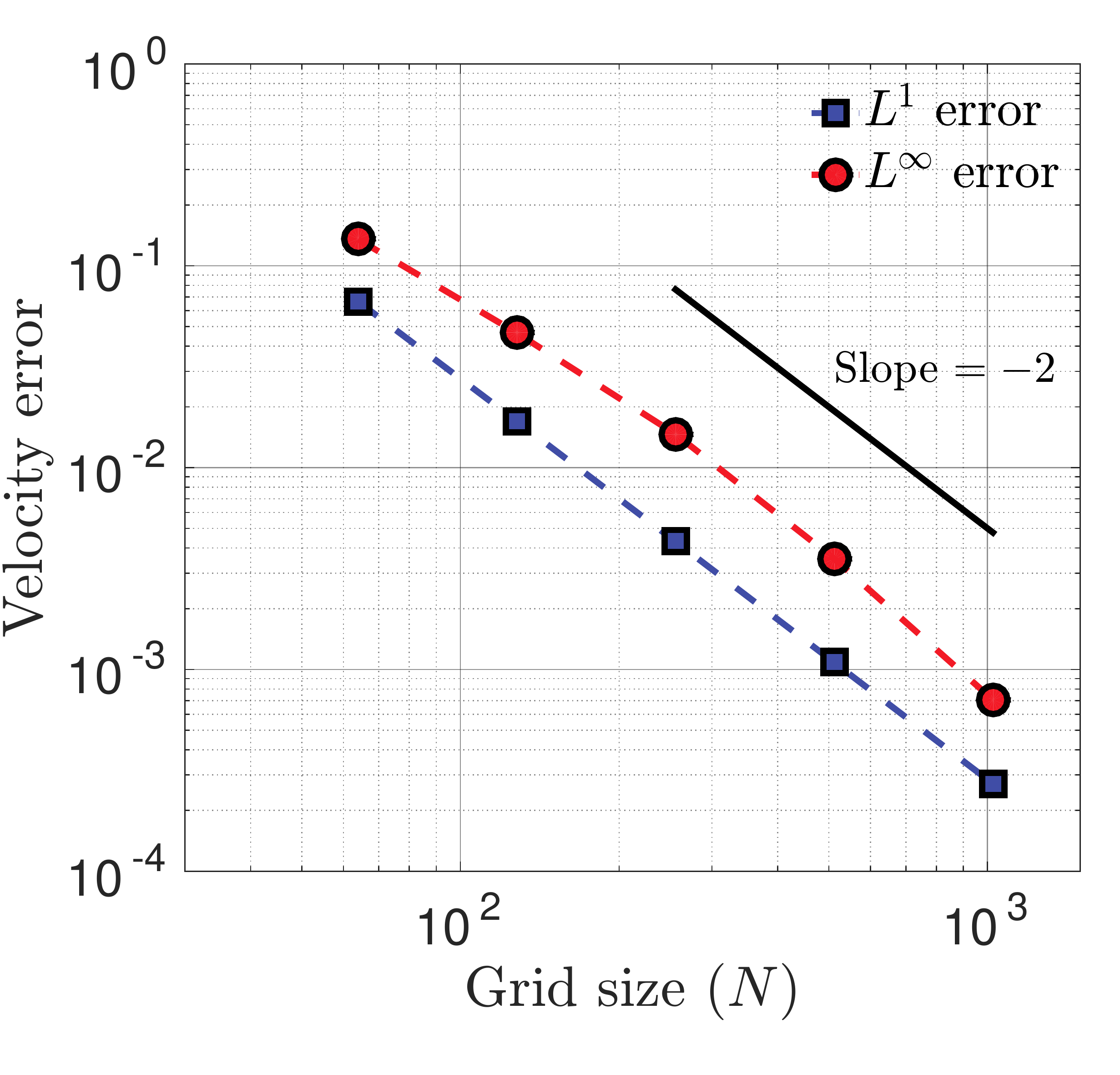}
    \label{fig_nc_U_err_vel_vel}
  }
   \subfigure[Pressure]{
    \includegraphics[scale = 0.255]{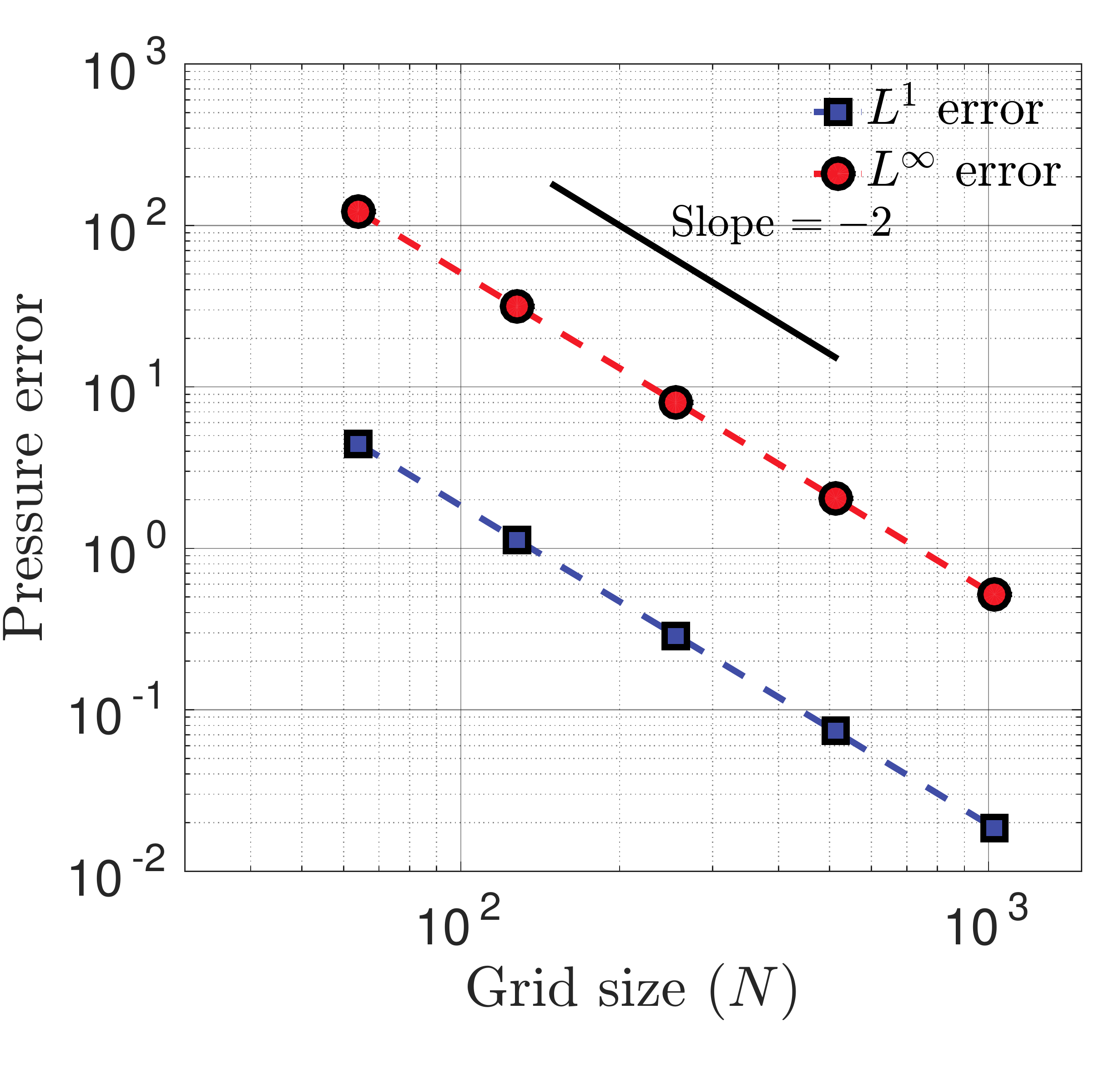}
    \label{fig_nc_P_err_vel_vel}
  }
  \caption{ 
  $L^1$ ($\blacksquare$, blue) and $L^\infty$ ($\bullet$, red) errors as a function of grid size $N$ for the non-conservative
  manufactured solution with specified normal and tangential velocity (vel-vel) boundary conditions:
  \subref{fig_nc_U_err_vel_vel}
  convergence rate for $\u$;
  \subref{fig_nc_P_err_vel_vel}
   convergence rate for $p$.
   }
  \label{fig_nc_err_vel_vel}
\end{figure}

\begin{figure}[]
  \centering
  \subfigure[Velocity]{
    \includegraphics[scale = 0.255]{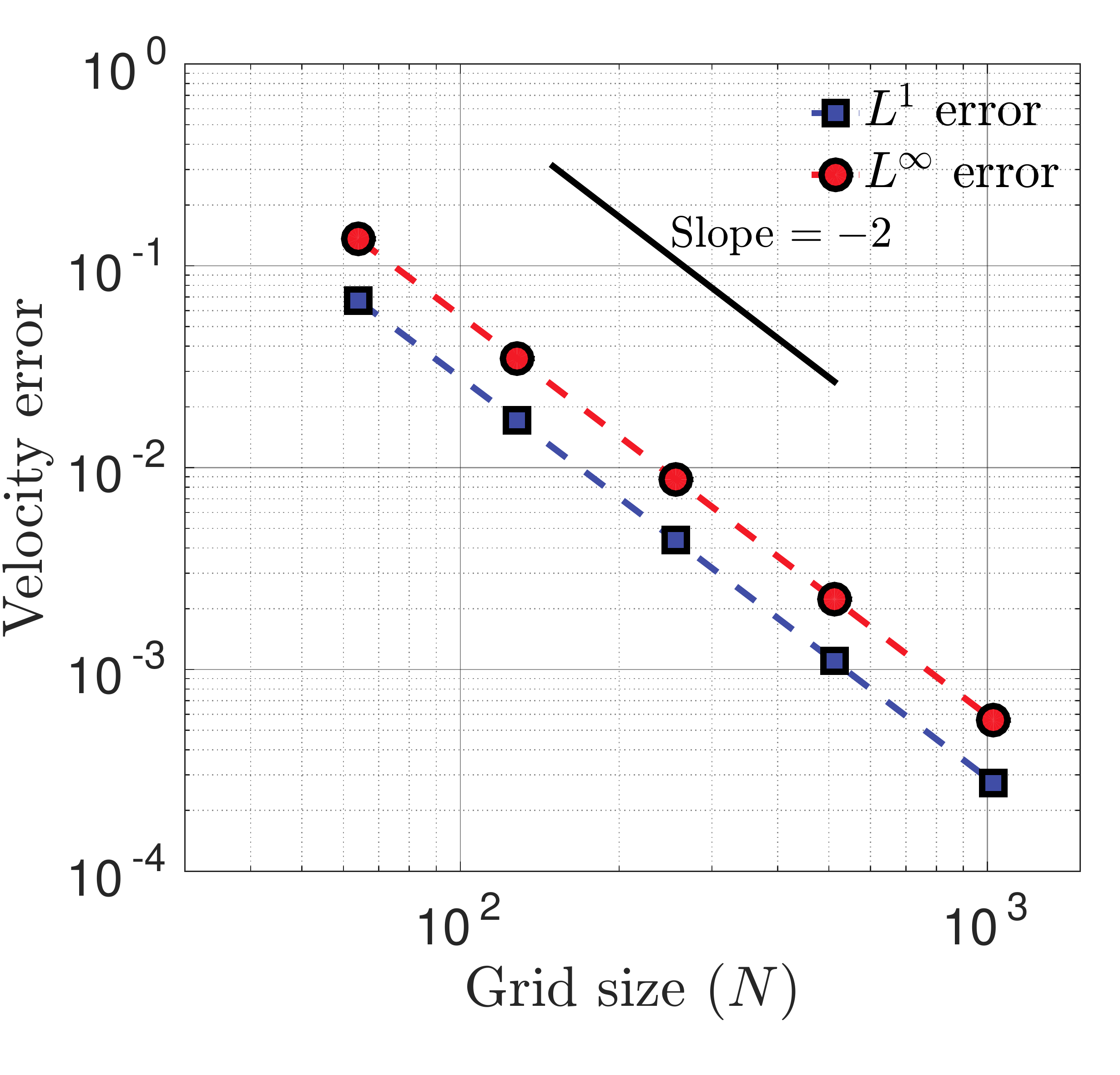}
    \label{fig_nc_U_err_vel_tra}
  }
   \subfigure[Pressure]{
    \includegraphics[scale = 0.255]{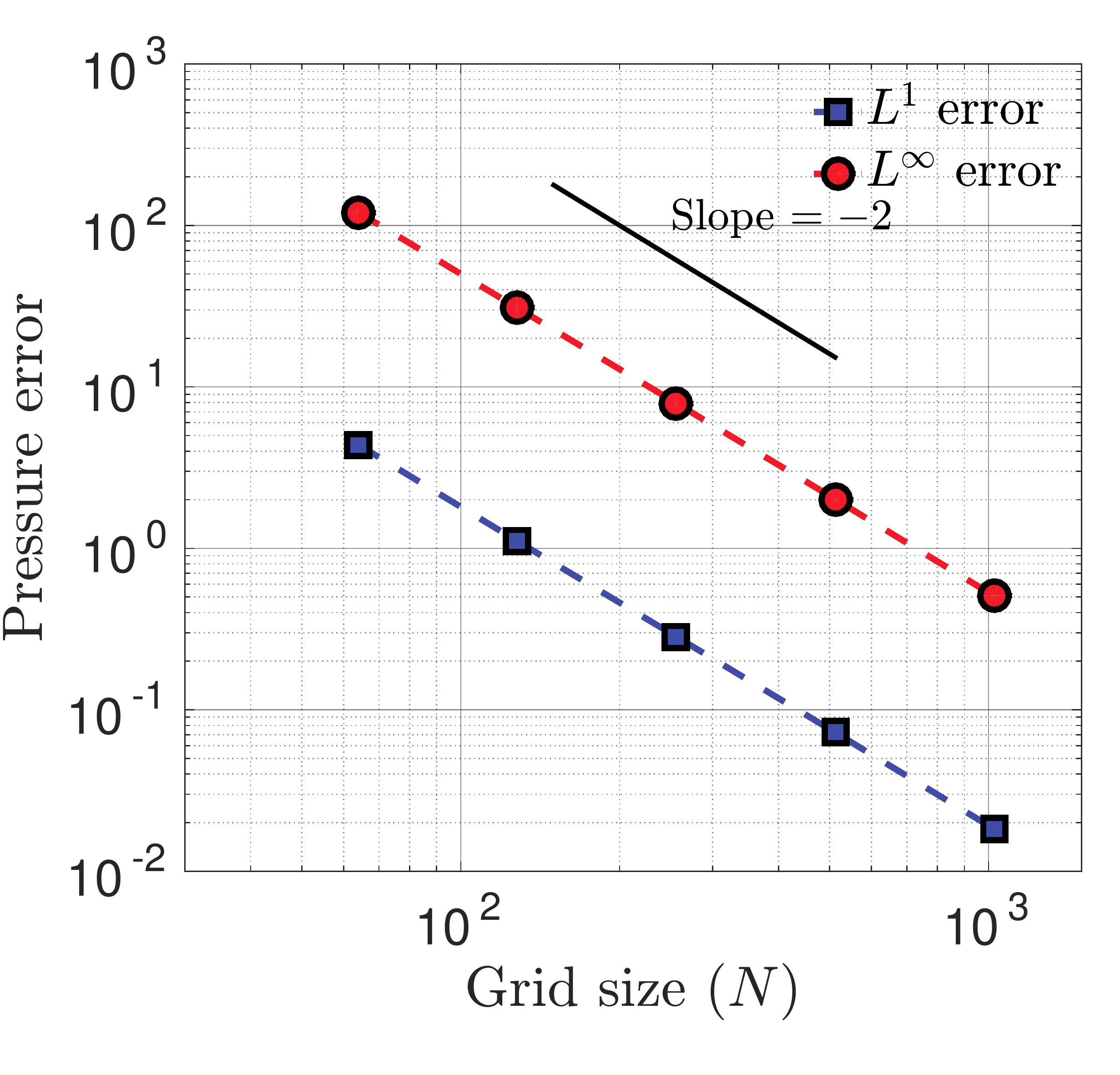}
    \label{fig_nc_P_err_vel_tra}
  }
  \caption{ 
  $L^1$ ($\blacksquare$, blue) and $L^\infty$ ($\bullet$, red) errors as a function of grid size $N$ for the non-conservative
  manufactured solution with specified normal velocity and tangential traction (vel-tra) boundary conditions:
  \subref{fig_nc_U_err_vel_tra}
  convergence rate for $\u$;
  \subref{fig_nc_P_err_vel_tra}
   convergence rate for $p$.
   }
  \label{fig_nc_err_vel_tra}
\end{figure}

\begin{figure}[]
  \centering
  \subfigure[Velocity]{
    \includegraphics[scale = 0.255]{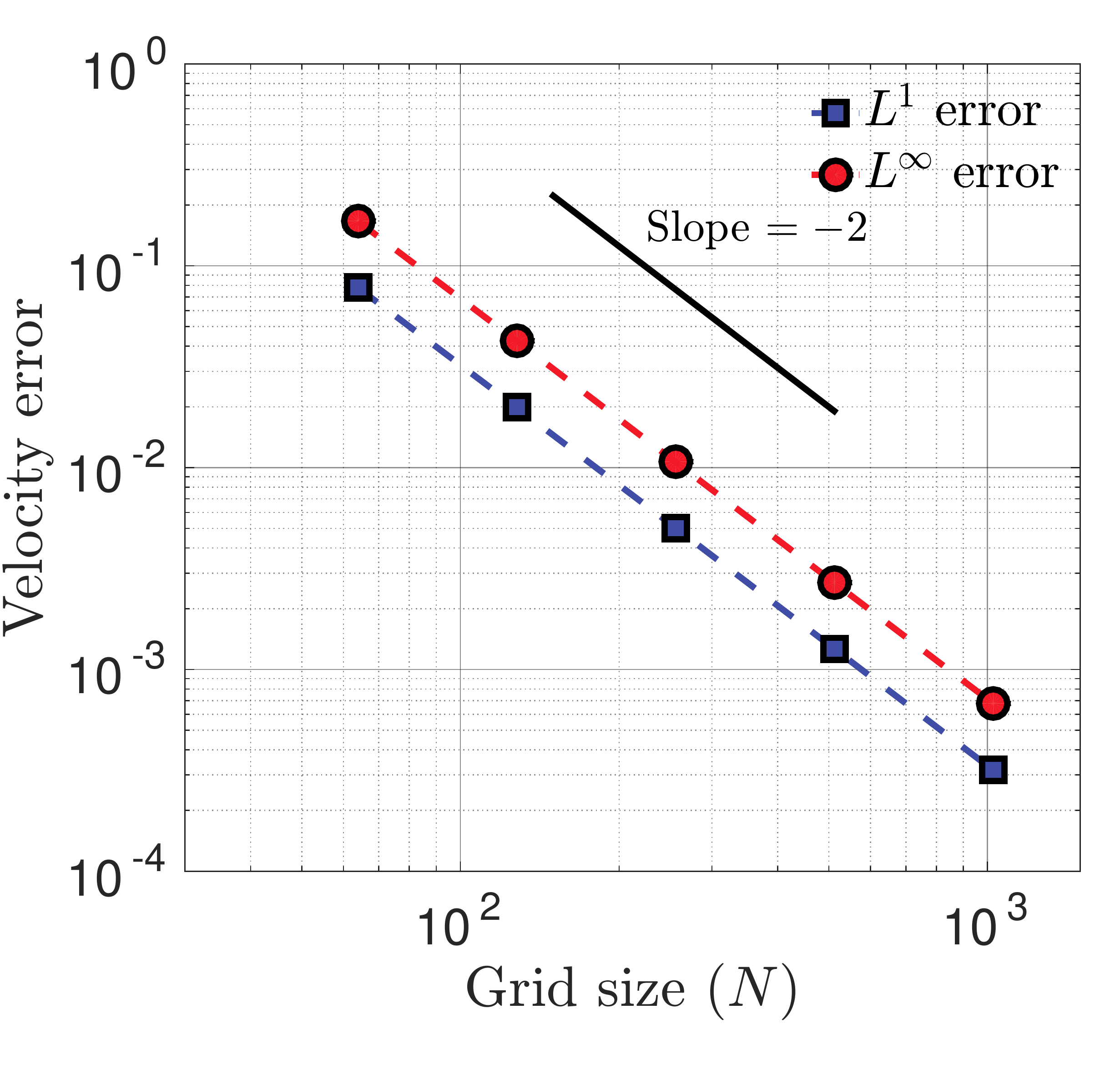}
    \label{fig_nc_U_err_tra_vel}
  }
   \subfigure[Pressure]{
    \includegraphics[scale = 0.255]{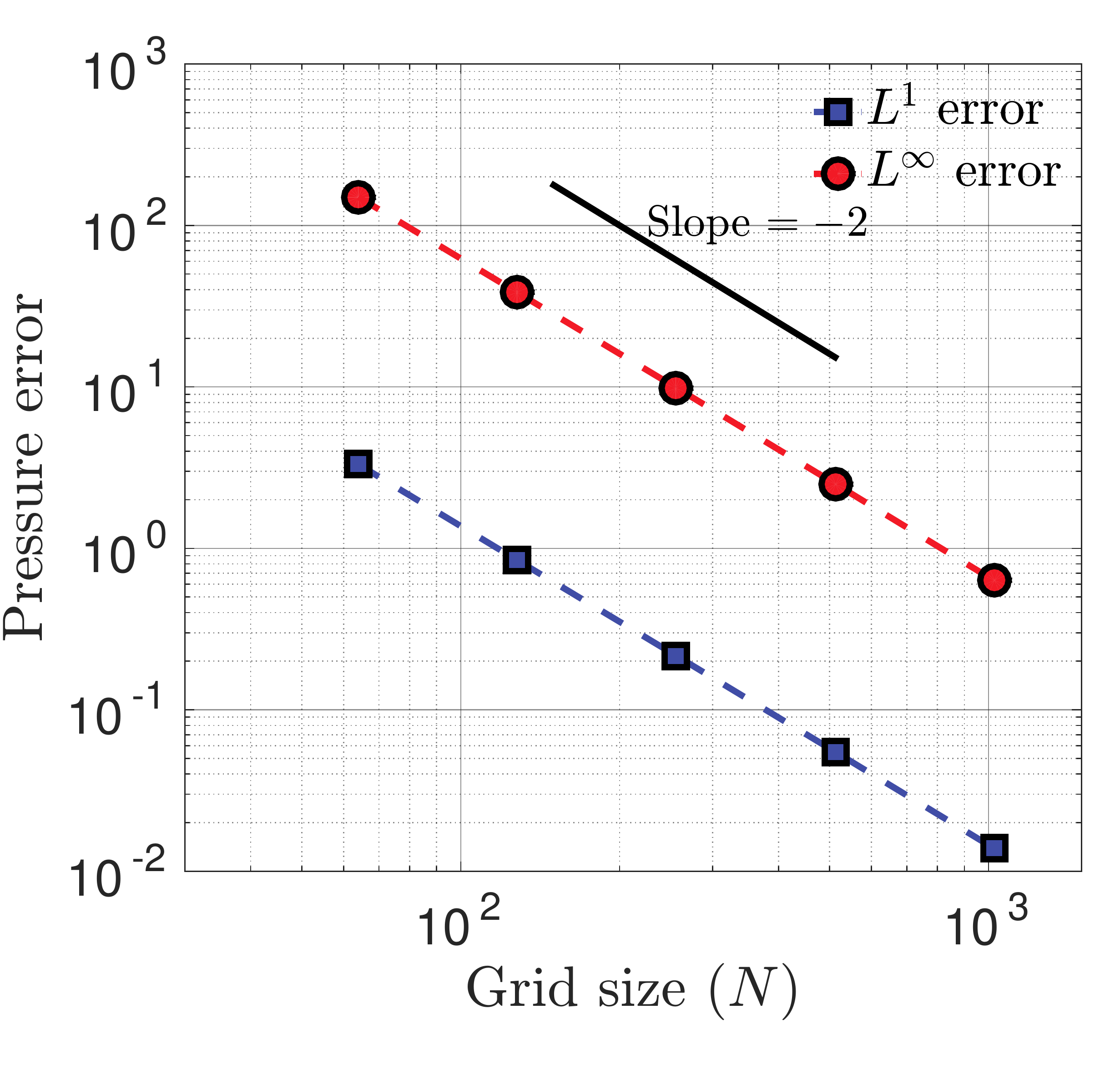}
    \label{fig_nc_P_err_tra_vel}
  }
  \caption{ 
  $L^1$ ($\blacksquare$, blue) and $L^\infty$ ($\bullet$, red) errors as a function of grid size $N$ for the non-conservative
  manufactured solution with specified normal traction and tangential velocity (tra-vel) boundary conditions:
  \subref{fig_nc_U_err_tra_vel}
  convergence rate for $\u$;
  \subref{fig_nc_P_err_tra_vel}
   convergence rate for $p$.
   }
  \label{fig_nc_err_tra_vel}
\end{figure}

\begin{figure}[]
  \centering
  \subfigure[Velocity]{
    \includegraphics[scale = 0.255]{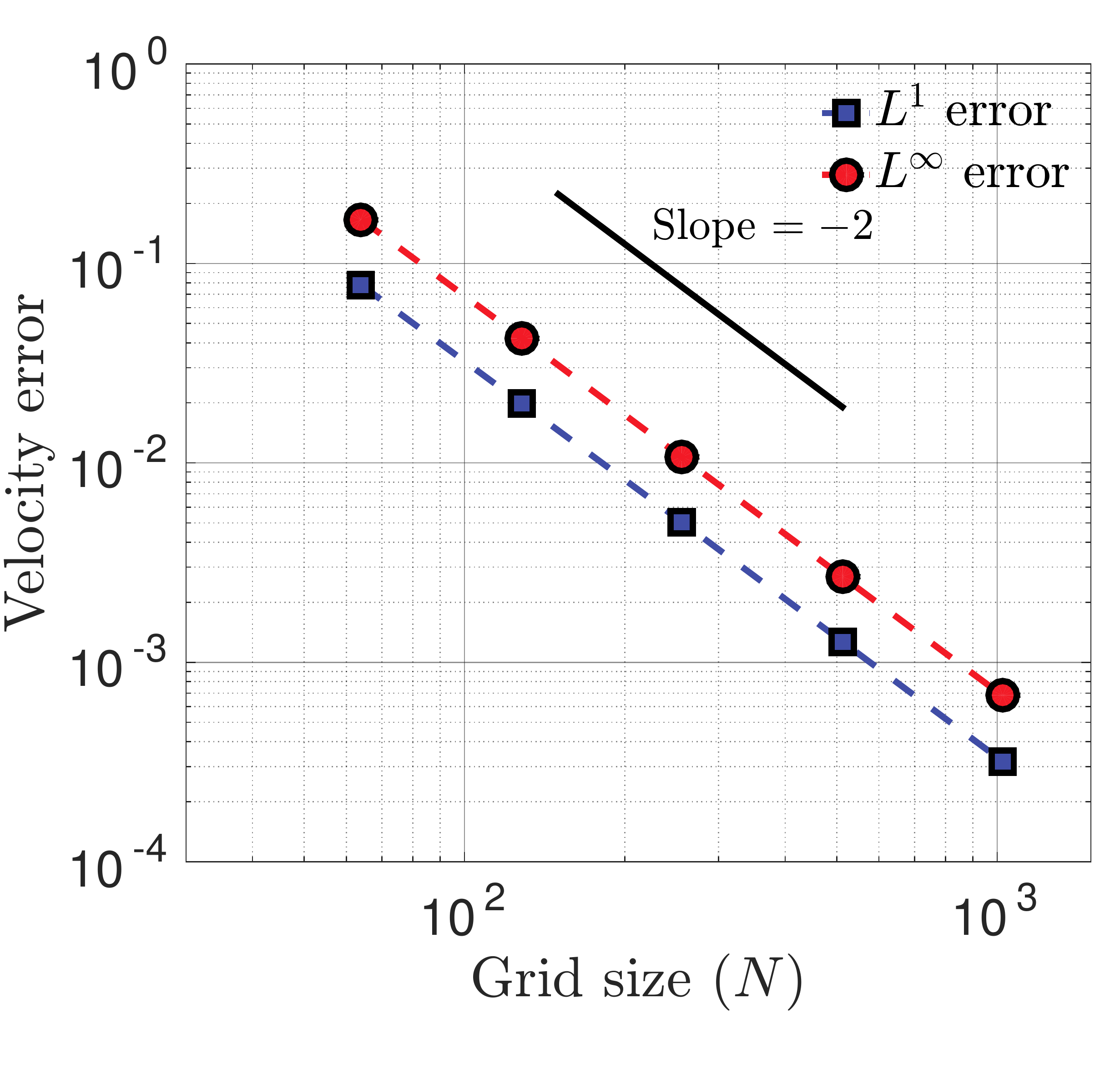}
    \label{fig_nc_U_err_tra_tra}
  }
   \subfigure[Pressure]{
    \includegraphics[scale = 0.255]{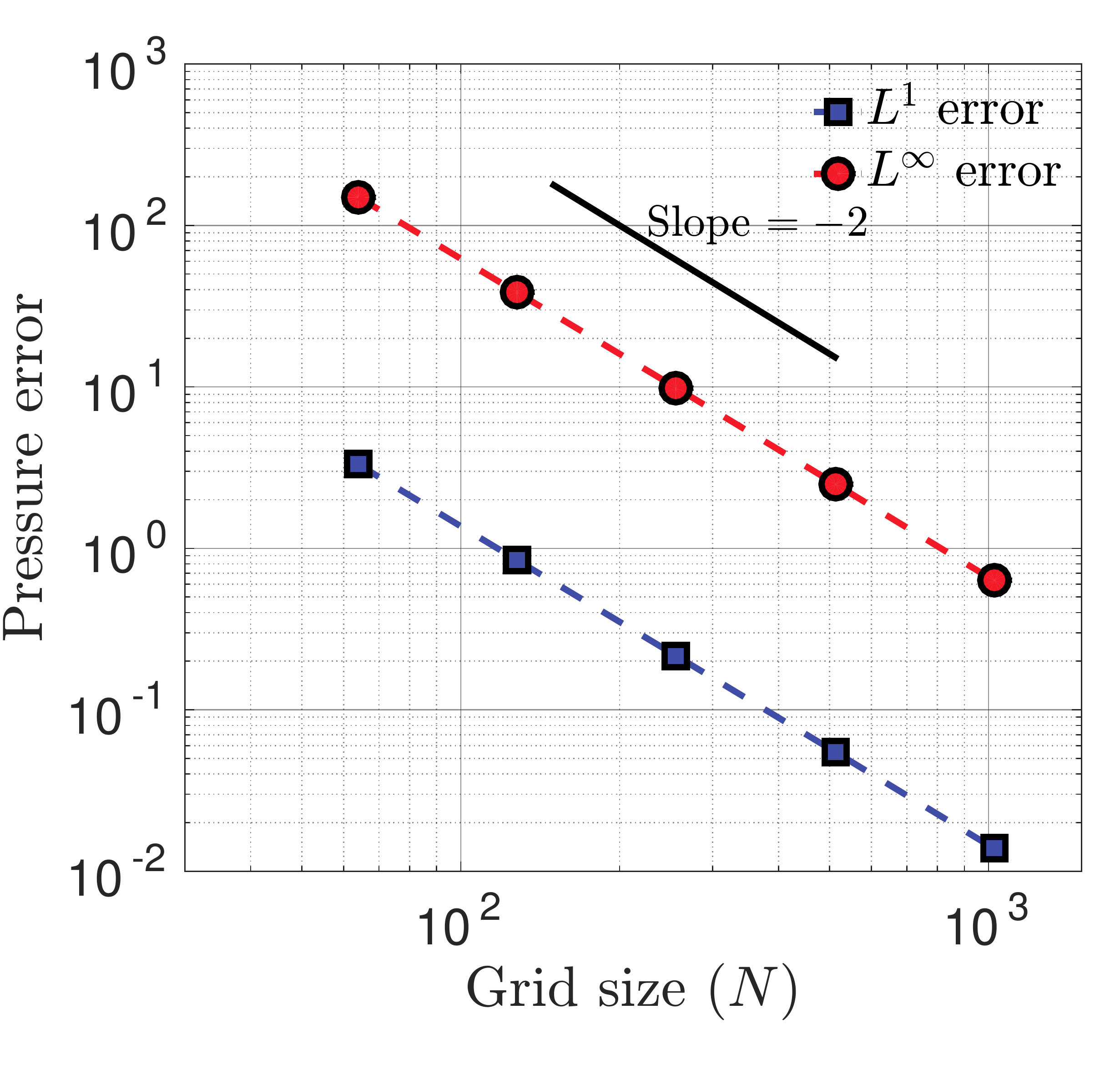}
    \label{fig_nc_P_err_tra_tra}
  }
  \caption{ 
  $L^1$ ($\blacksquare$, blue) and $L^\infty$ ($\bullet$, red) errors as a function of grid size $N$ for the non-conservative
  manufactured solution with specified normal and tangential traction (tra-tra) boundary conditions:
  \subref{fig_nc_U_err_tra_tra}
  convergence rate for $\u$;
  \subref{fig_nc_P_err_tra_tra}
   convergence rate for $p$.
   }
  \label{fig_nc_err_tra_tra}
\end{figure}

%Following Griffith~\cite{Griffith2009} for a constant coefficient monolithic flow solver, 
We impose periodic boundary conditions 
in the $x$-direction and various boundary conditions in the $y$-direction. The boundary conditions in the $y$-direction are: 
\begin{itemize}
\item specified normal and tangential velocities, denoted as ``vel-vel"; 
\item specified normal velocity with specified tangential traction, denoted as ``vel-tra"; 
\item specified normal traction with specified tangential velocity, denoted as ``tra-vel"; 
\item specified normal and tangential tractions, denoted as ``tra-tra";
\item periodic boundary conditions denotes as ``periodic".
\end{itemize}
The maximum velocities in the domain for this manufactured solution are $\BigO{1}$, hence a relevant  time scale is $L/U$ with $U = 1$. Errors in the velocity and pressure are computed at time $T = tU/L = 0.1$ with a uniform time step $\dt = 1/(15.625 N)$, which
yields an approximate CFL number of $0.5$. Figs.~\ref{fig_nc_err_periodic}--\ref{fig_nc_err_tra_tra} show the discrete $L^1$ and $L^{\infty}$ errors for velocity and pressure
as a function  of grid size. Second-order convergence rates are observed for both velocity and pressure in both
norms. We note that at lower resolutions, the pointwise convergence rates at coarser resolutions are less than two for
vel-vel boundary conditions (see Fig.~\ref{fig_nc_U_err_vel_vel}), although second-order convergence rates are ultimately 
obtained on finer grids. We remark that reduction in accuracy at lower resolution is observed for small base viscosity; at higher 
(base) viscosities, we see second-order convergence rates at all the resolutions considered for the vel-vel boundary conditions (data not shown). This is also consistent with what was observed for the unsplit discretization of the constant coefficient incompressible Navier-Stokes equations in Griffith~\cite{Griffith2009}. These tests show that the present numerical discretization and boundary treatment maintains the pointwise second-order accuracy for a variety of physical boundary conditions.

\subsection{Non-conservative form: Effect of density and viscosity ratios}
\label{sec_non_cons_bubble}
\begin{figure}[]
  \centering
  \subfigure[$\rho(\x)$]{
    \includegraphics[scale = 0.255]{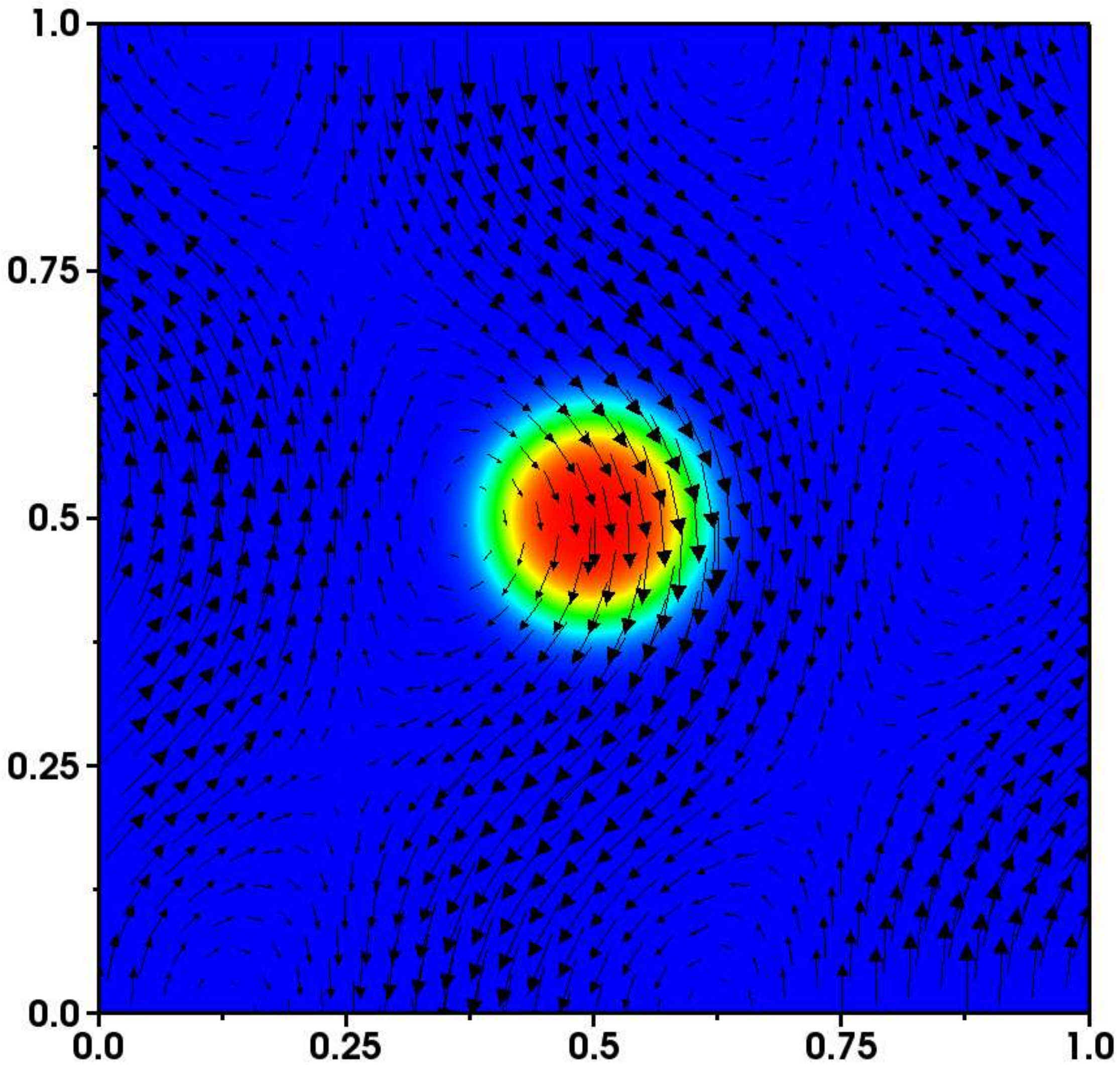}
    \label{fig_NC_Bubble_Density_Init_Schematic}
  }
   \subfigure[$\mu(\x)$]{
    \includegraphics[scale = 0.255]{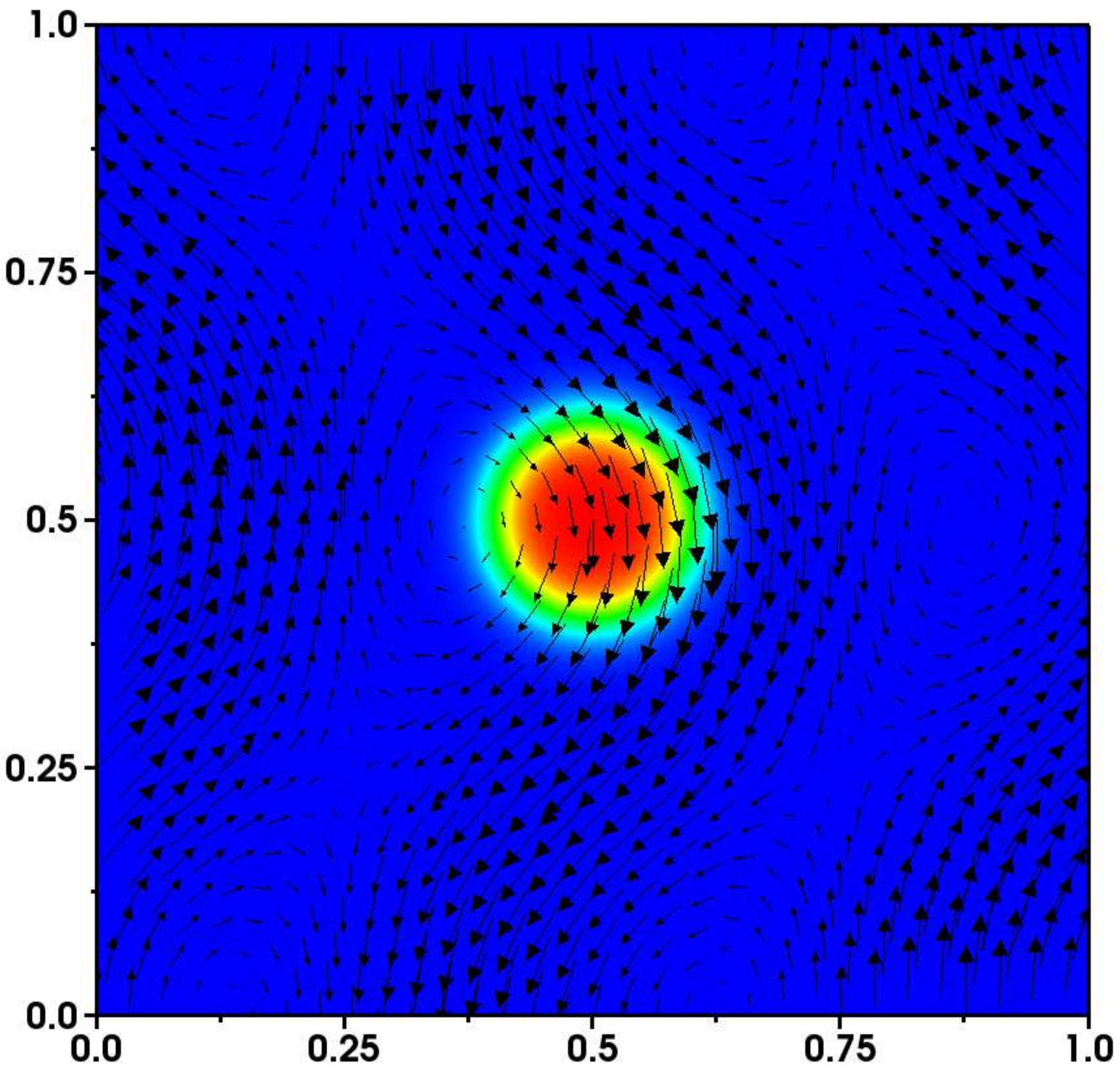}
    \label{fig_NC_Bubble_Viscosity_Init_Schematic}
  }
   \caption{The \subref{fig_NC_Bubble_Density_Init_Schematic} density and
    \subref{fig_NC_Bubble_Viscosity_Init_Schematic} viscosity fields, along with the initial velocity vectors
    for the manufactured solution described
   in Sec.~\ref{sec_non_cons_bubble}.}
  \label{fig_non_cons_bubble_init}
\end{figure}

Next, we verify the accuracy of the solver/discretization for a wide range of density and viscosity ratios.
A manufactured solution for the non-conservative form is taken to be
\begin{align}
u(\x,t) &= -\cos (2 \pi  t-2 \pi  x) \sin (2 \pi  t-2 \pi  y), \label{eq_non_cons_bubble_u}\\
v(\x,t) &= \sin (2 \pi  t-2 \pi  x) \cos (2 \pi  t-2 \pi  y)+\cos (t-2 \pi  x), \label{eq_non_cons_bubble_v}\\
p(\x,t) &= \sin (2 \pi  t-2 \pi  x) \sin (2 \pi  t-2 \pi  y), \label{eq_non_cons_bubble_p}
\end{align}
with time-independent density and viscosity fields,
\begin{align}
\rho(\x) &= \rho_0 \left(\frac{R_\rho-1}{2}  \tanh \left(\frac{0.1\,
   -\sqrt{(x-0.5)^2+(y-0.5)^2}}{\delta }\right)+\frac{R_\rho+1}{2}\right), \label{eq_non_cons_bubble_rho}\\
\mu(\x) &=\mu_0 \left(\frac{R_\mu-1}{2}  \tanh \left(\frac{0.1\,
   -\sqrt{(x-0.5)^2+(y-0.5)^2}}{\delta }\right)+\frac{R_\mu+1}{2}\right). \label{eq_non_cons_bubble_mu}
\end{align}
Physically, this describes a ``bubble'' centered in the computational domain $\Omega = [0,L]^2 = [0,1]^2$.
The density (viscosity) in the domain varies smoothly between $\rho_0$ ($\mu_0$) outside of the domain
to $R_{\rho} \cdot \rho_0$ ($R_{\mu} \cdot \mu_0$), indicating a density (viscosity) ratio of $R_{\rho}$ ($R_{\mu}$).
The smoothing parameter is set to $\delta = 0.03L$.
The density, viscosity, and initial velocity fields are shown in Fig.~\ref{fig_non_cons_bubble_init}.

%Plugging Eqs.~\eqref{eq_non_cons_bubble_u}--\eqref{eq_non_cons_bubble_mu} into the 
%non-conservative momentum equation~\eqref{eqn_momentum} yields a forcing term 
%$\f(\x,t)$ that will produce the desired manufactured solution. % given by Eqs.~\eqref{eq_non_cons_bubble_u}-\eqref{eq_non_cons_bubble_p}.

The effect of varying density ratio $R_\rho$ is considered first. The outer viscosity is set to $\mu_0 = 10^{-4}$
with viscosity ratio $R_\mu = 10^1$, while the outer density is set to $\rho_0 = 10^0$.
Density ratios of $R_\rho = 10^1, 10^2, 10^3, 10^4, 10^5, 10^6$ are considered.
For \emph{all} physical boundaries, specified normal velocity 
and tangential traction boundary conditions are used.
The maximum velocities in the domain for this manufactured solution are $\BigO{1}$, hence a relevant  time scale is $L/U$ with $U = 1$.
Errors in the velocity and pressure are computed at
time $T = tU/L =  0.1$ using a uniform time step $\dt = 1/(6.25 N)$, which yields an approximate CFL
number of $0.3$.

\begin{figure}[]
  \centering
  \subfigure[Velocity $L^1$ error]{
    \includegraphics[scale = 0.255]{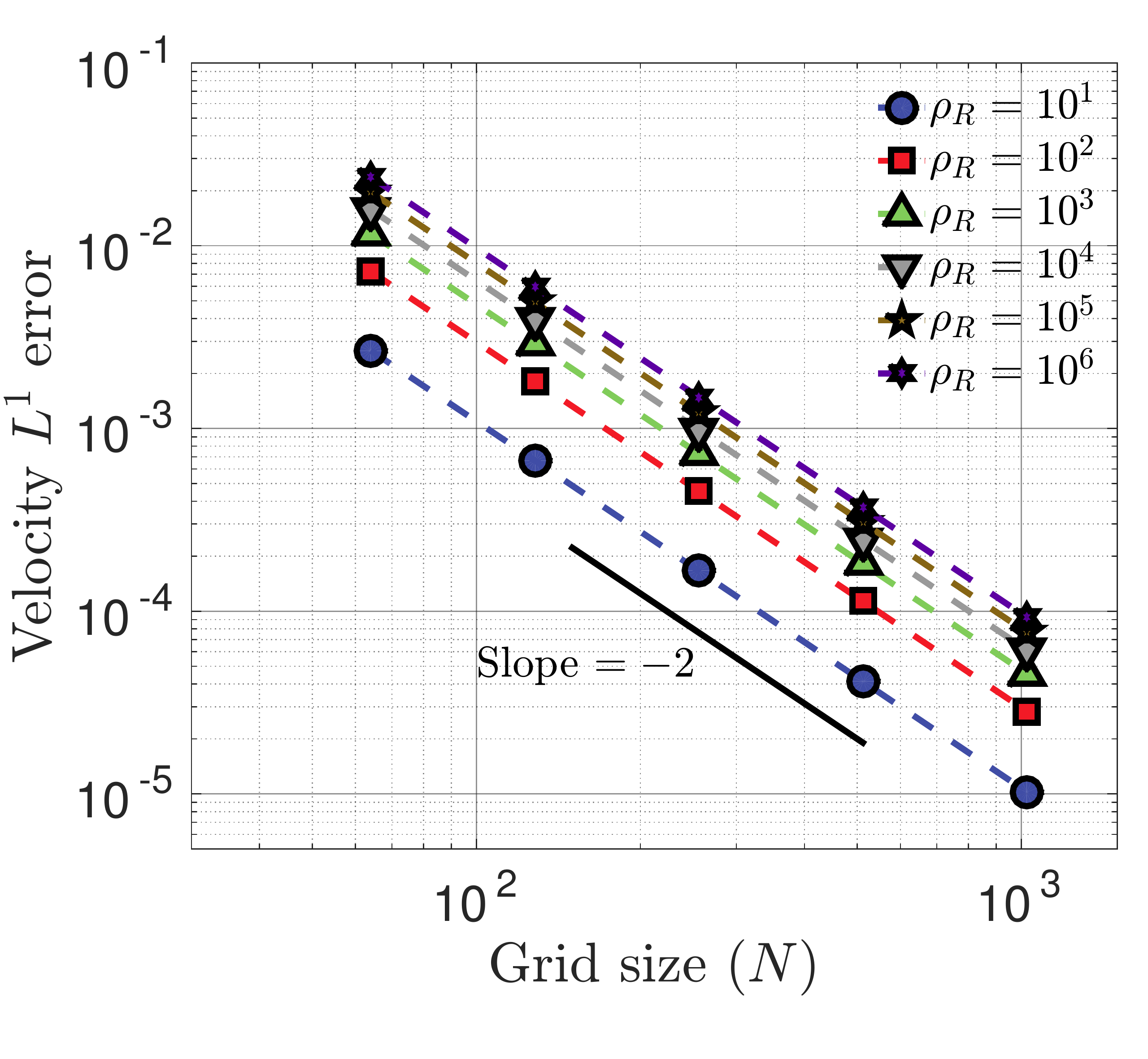}
    \label{fig_rho_bubble_UL1}
  }
  \subfigure[Velocity $L^\infty$ error]{
    \includegraphics[scale = 0.255]{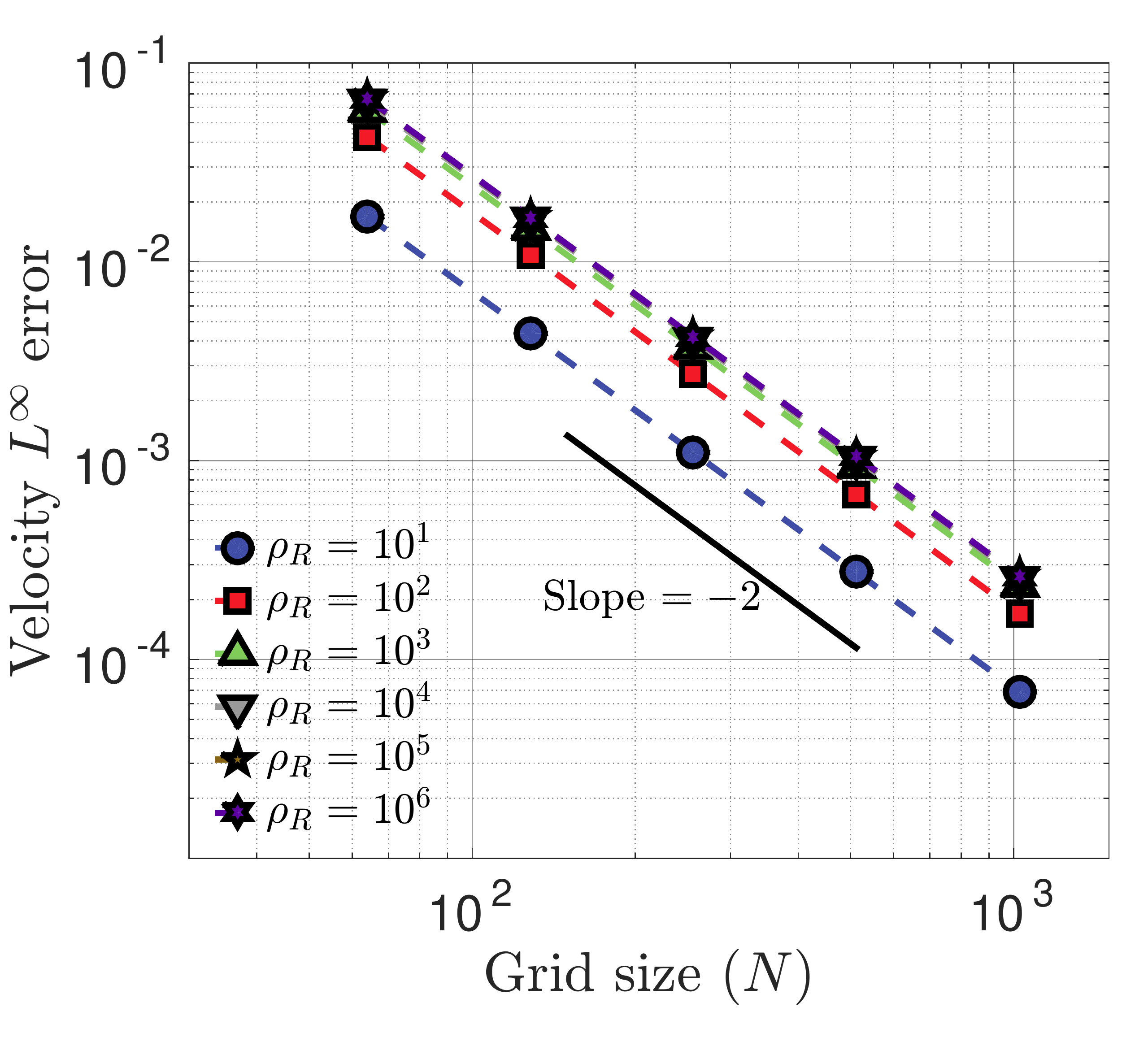}
    \label{fig_rho_bubble_ULInf}
  }
   \subfigure[Pressure $L^1$ error]{
    \includegraphics[scale = 0.255]{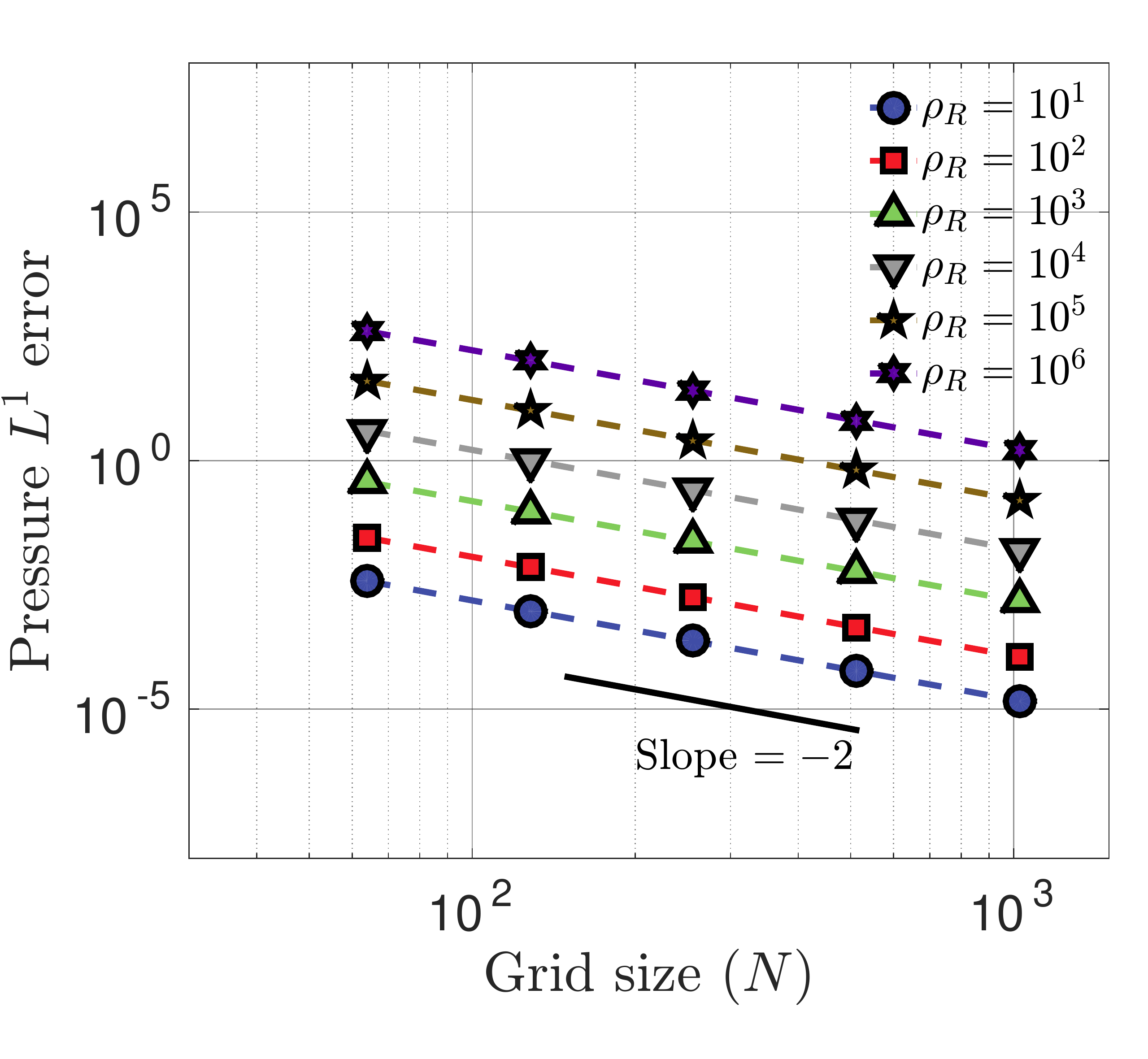}
    \label{fig_rho_bubble_PL1}
  }
  \subfigure[Pressure $L^\infty$ error]{
    \includegraphics[scale = 0.255]{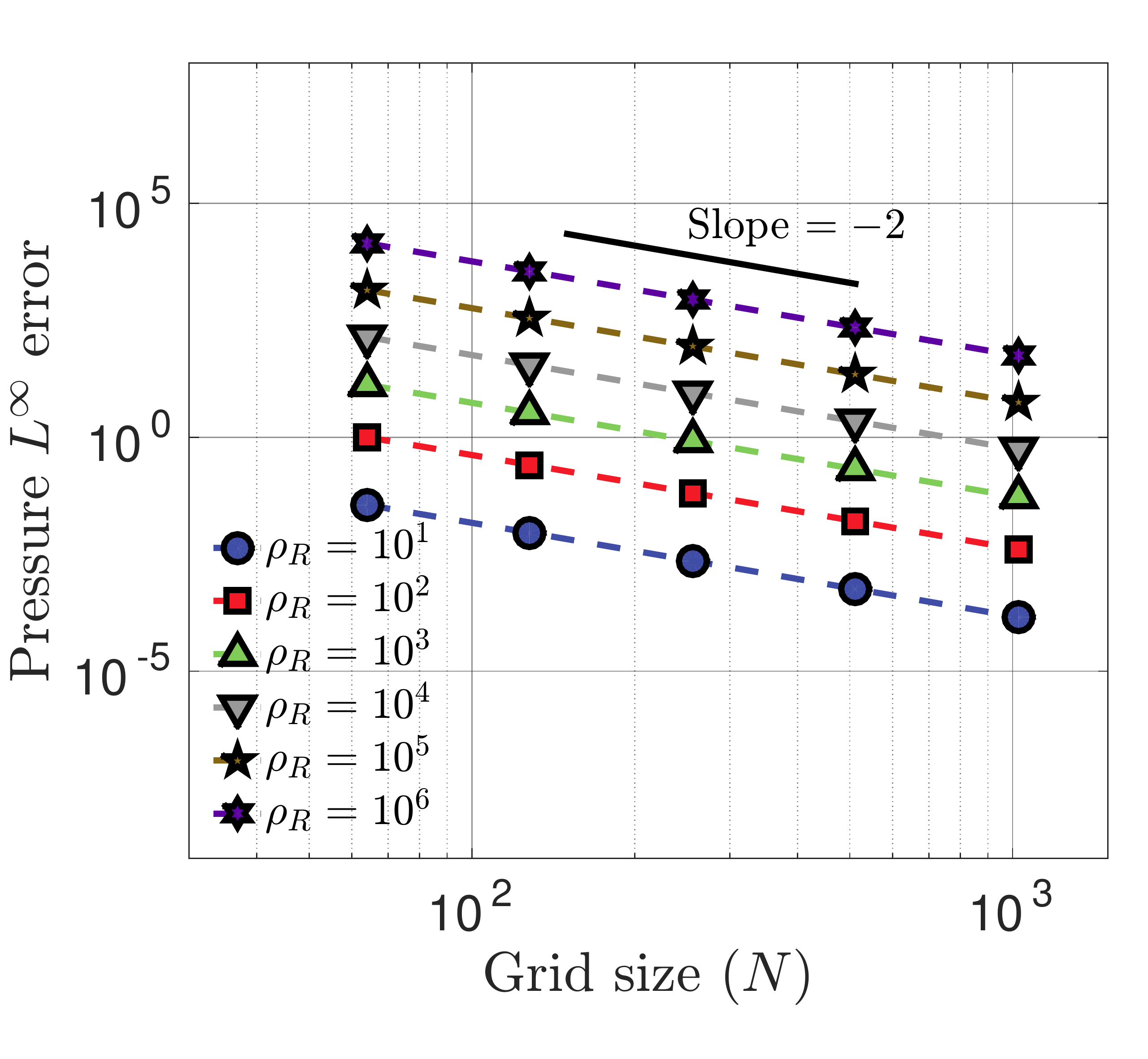}
    \label{fig_rho_bubble_PLInf}
  }
  \caption{ 
  Errors as a function of grid size $N$ for varying density ratios $\rho_R = 10^1 - 10^6$ with $\rho_0 = 10^0$
  and constant viscosity ratio $\mu_R = 10^1$ with $\mu_0 = 10^{-4}$ for a non-conservative manufactured solution.
  Normal velocity and tangential traction boundary conditions are specified at all physical boundaries in every case.
  \subref{fig_rho_bubble_UL1}
  $L^1$ convergence rate for $\u$;
  \subref{fig_rho_bubble_ULInf}
  $L^\infty$ convergence rate for $\u$;
  \subref{fig_rho_bubble_PL1}
   $L^1$ convergence rate for $p$;
   \subref{fig_rho_bubble_PLInf}
   $L^\infty$ convergence rate for $p$.
   }
  \label{fig_rho_bubble}
\end{figure}

Fig.~\ref{fig_rho_bubble} shows the $L^1$ and $L^\infty$ errors for velocity and pressure as a function
of grid size. Second-order convergence rates are achieved for velocity and pressure in both norms.
It is clear that for a given grid size, the error increases as a function of density ratio. Additionally, this increase
in error is more pronounced in the pressure than in the velocity. This is unsurprising: high density ratio simulations
need to be adequately refined in order to accurately resolve the pressure jump across a dense region. We remark 
that for many air-water interface impact problems that are prevalent in ocean and marine engineering, 
pressure forces dominate viscous traction. Therefore, high accuracy in the pressure is desirable
for this class of applications. %fluid-structure interaction (FSI) problems. We will consider such FSI applications in future studies. 

Next, we consider the effect of varying viscosity ratio. The outer density is set to $\rho_0 = 10^{0}$
with density ratio $R_\rho = 10^1$, while the outer viscosity is set to $\mu_0 = 10^{-4}$.
Viscosity ratios of $R_\mu = 10^1, 10^2, 10^3, 10^4, 10^5, 10^6$ are considered. The same grid sizes and
time step from the varying density case are used.

\begin{figure}[]
  \centering
  \subfigure[Velocity $L^1$ error]{
    \includegraphics[scale = 0.255]{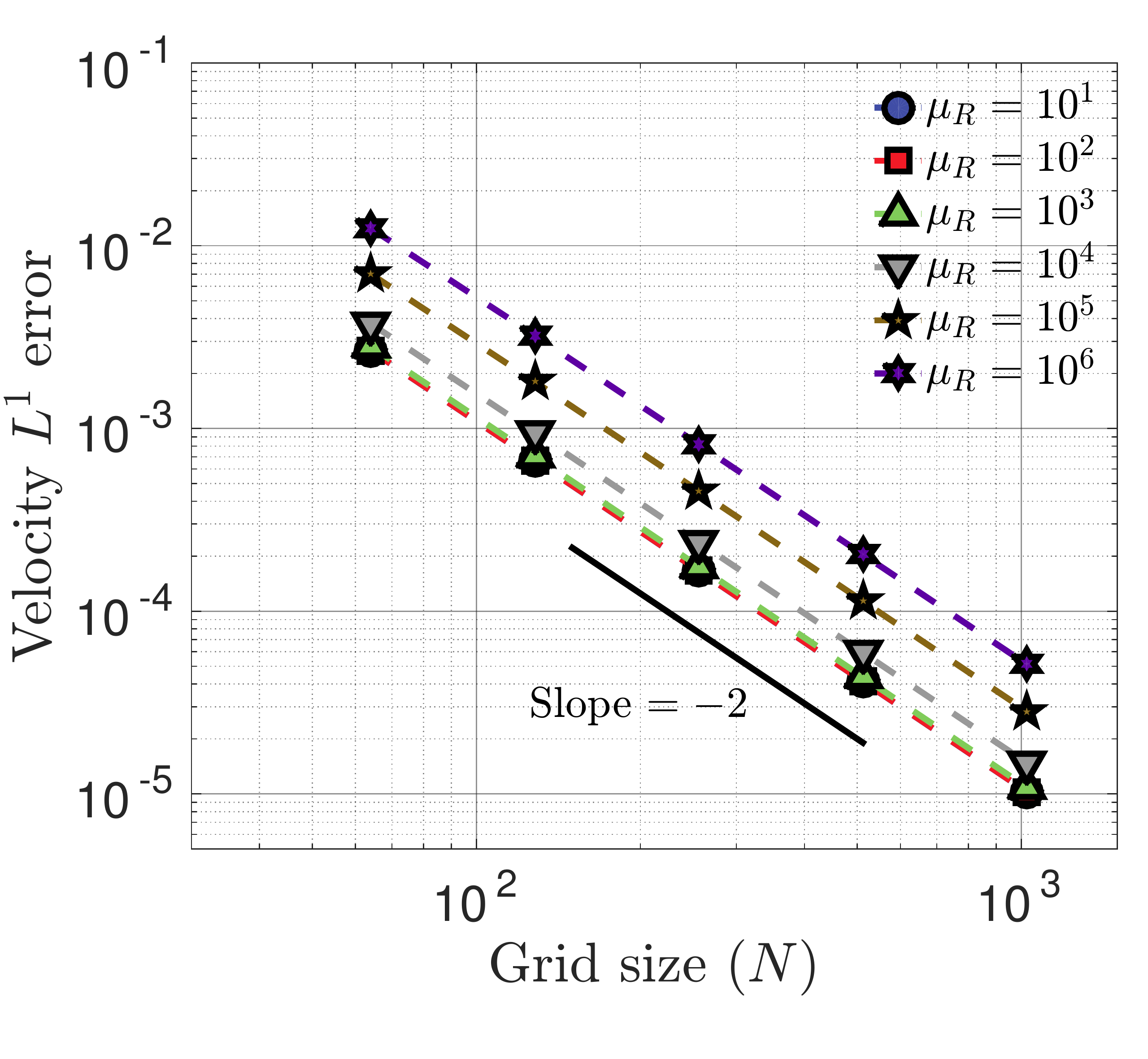}
    \label{fig_mu_bubble_UL1}
  }
  \subfigure[Velocity $L^\infty$ error]{
    \includegraphics[scale = 0.255]{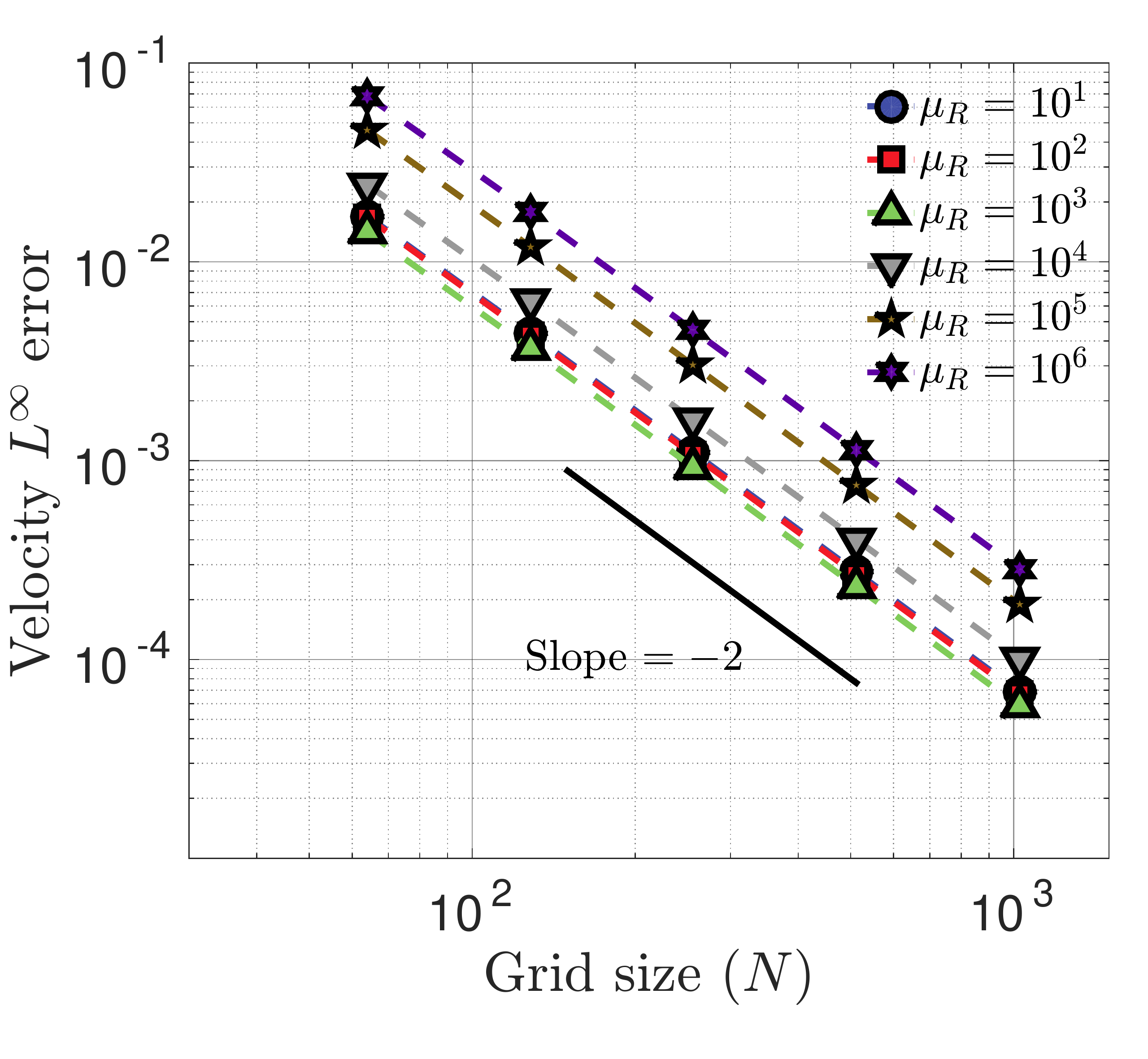}
    \label{fig_mu_bubble_ULInf}
  }
   \subfigure[Pressure $L^1$ error]{
    \includegraphics[scale = 0.255]{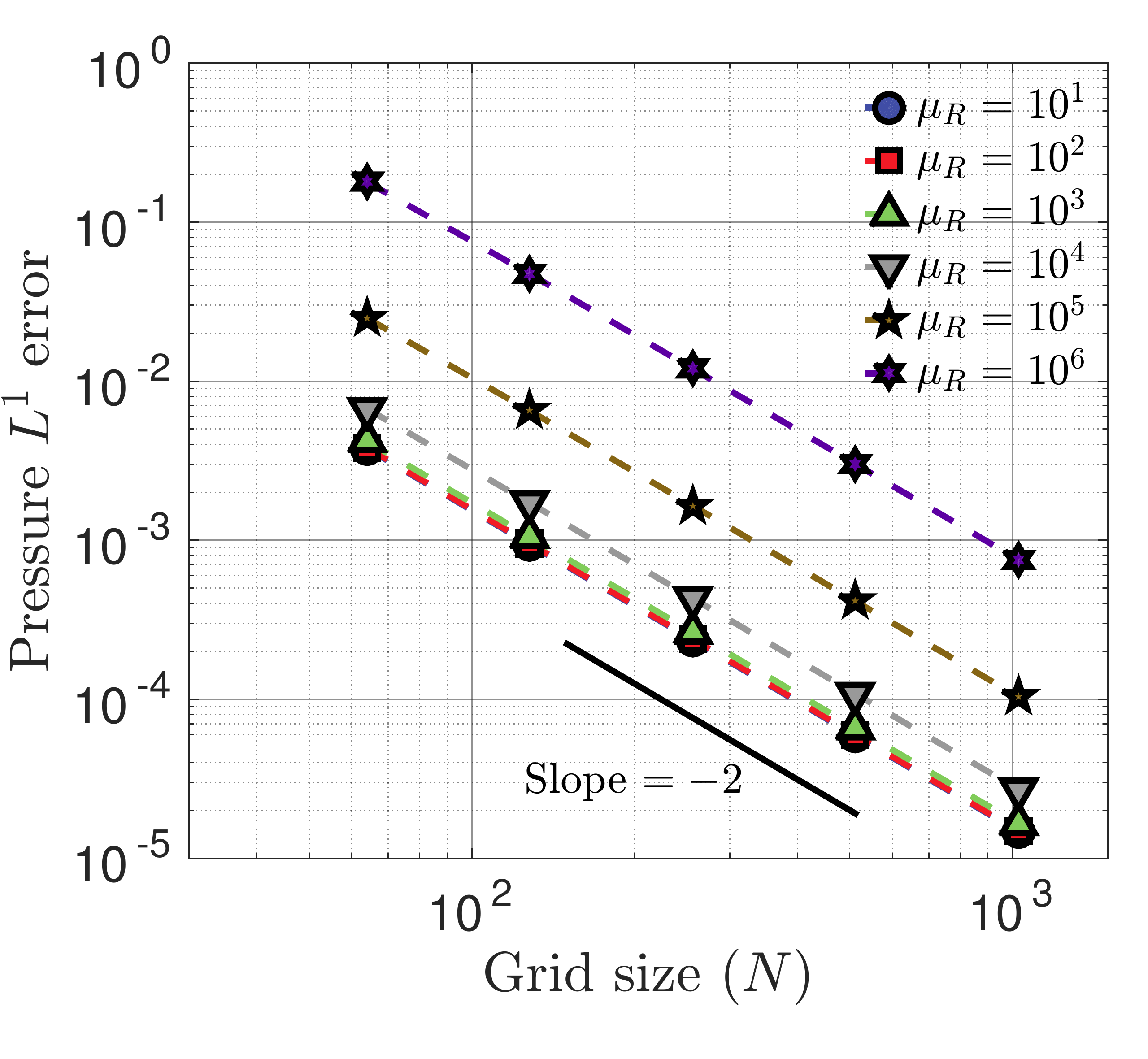}
    \label{fig_mu_bubble_PL1}
  }
  \subfigure[Pressure $L^\infty$ error]{
    \includegraphics[scale = 0.255]{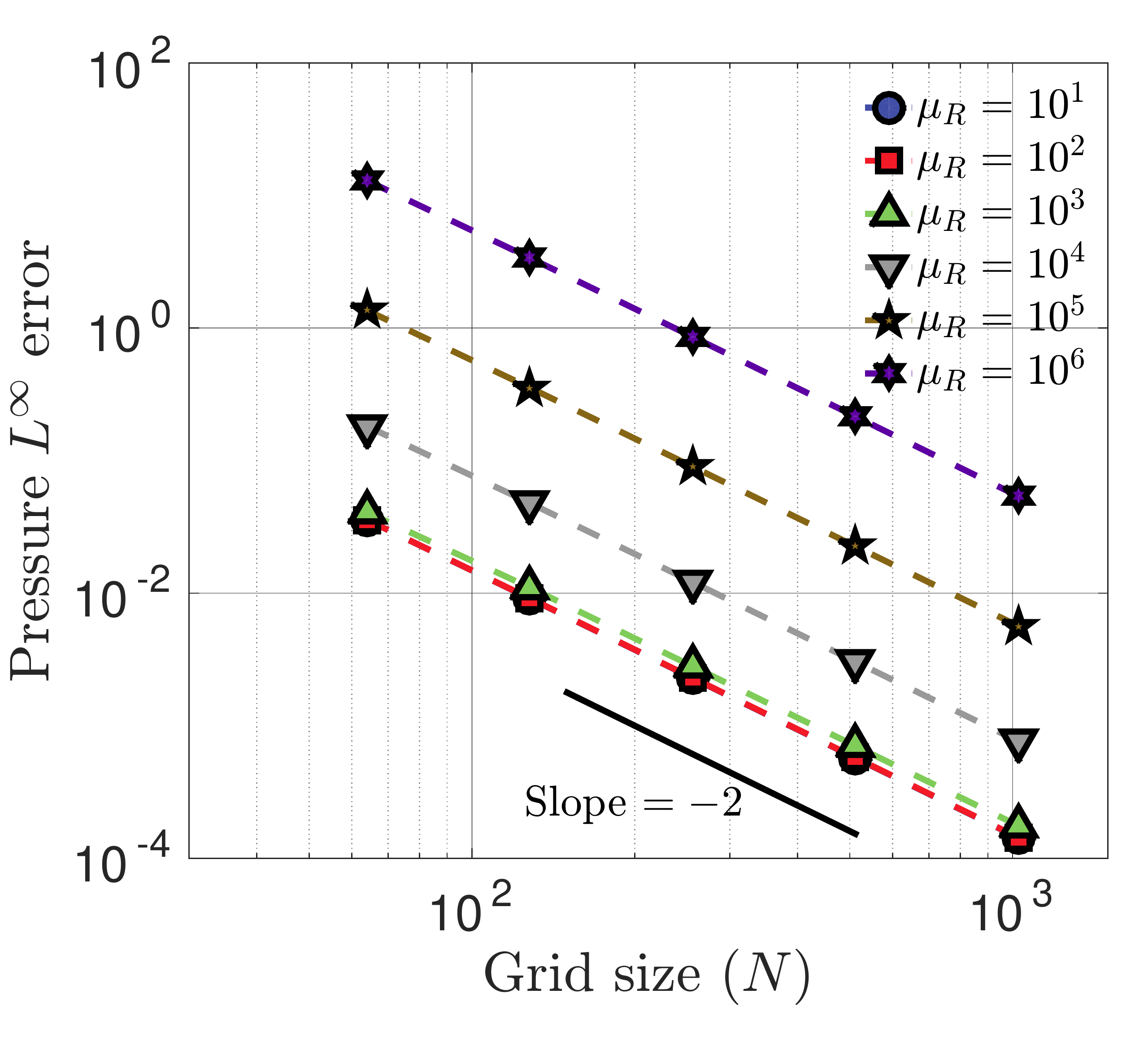}
    \label{fig_mu_bubble_PLInf}
  }
  \caption{ 
 Errors as a function of grid size $N$ for varying viscosity ratios $\mu_R = 10^1 - 10^6$ with $\mu_0 = 10^{-4}$
  and constant density ratio $\rho_R = 10^1$ with $\rho_0 = 10^0$ for a non-conservative manufactured solution.
  Normal velocity and tangential traction boundary conditions are specified at all physical boundaries in every case.
  \subref{fig_mu_bubble_UL1}
  $L^1$ convergence rate for $\u$;
  \subref{fig_mu_bubble_ULInf}
  $L^\infty$ convergence rate for $\u$;
  \subref{fig_mu_bubble_PL1}
   $L^1$ convergence rate for $p$;
   \subref{fig_mu_bubble_PLInf}
   $L^\infty$ convergence rate for $p$.
   }
  \label{fig_mu_bubble}
\end{figure}

Fig.~\ref{fig_mu_bubble} shows the $L^1$ and $L^\infty$ errors for velocity and pressure as a function
of grid size. We again observe second-order convergence rates for both velocity and pressure.
At a given grid size, the errors increase as a function of viscosity ratio, although the growth
is not as significant as in the varying density case. These tests show that the present numerical discretization
maintains the desired order of accuracy for a wide range of fluid properties.

\subsection{Conservative form: Effect of density evolution and synchronization}
\label{sec_cons_ms}
\begin{figure}[]
  \centering
  \subfigure[$\rho(\x,0)$]{
    \includegraphics[scale = 0.255]{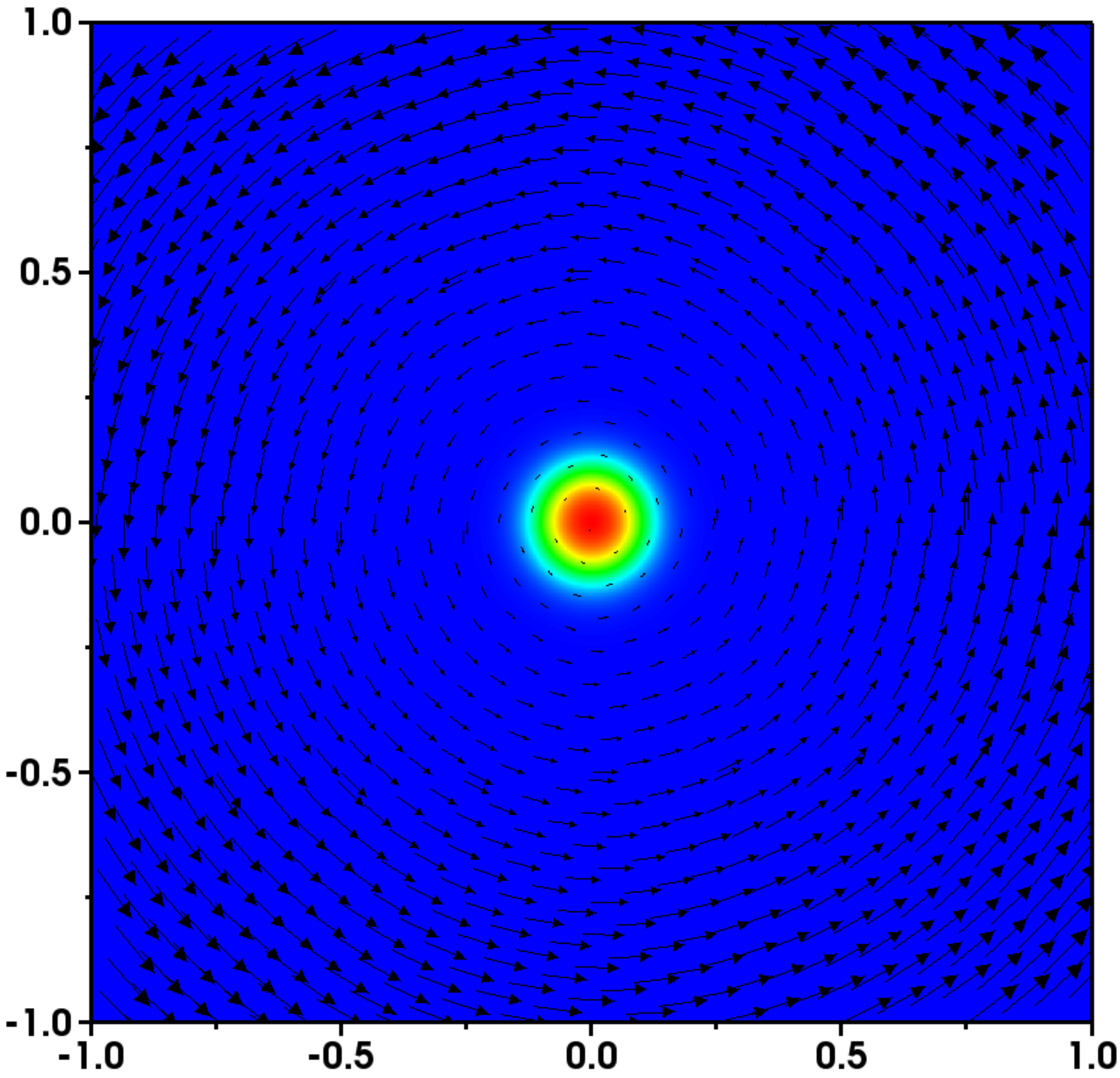}
    \label{fig_C_Density_Init_Schematic}
  }
   \subfigure[$\mu(\x)$]{
    \includegraphics[scale = 0.255]{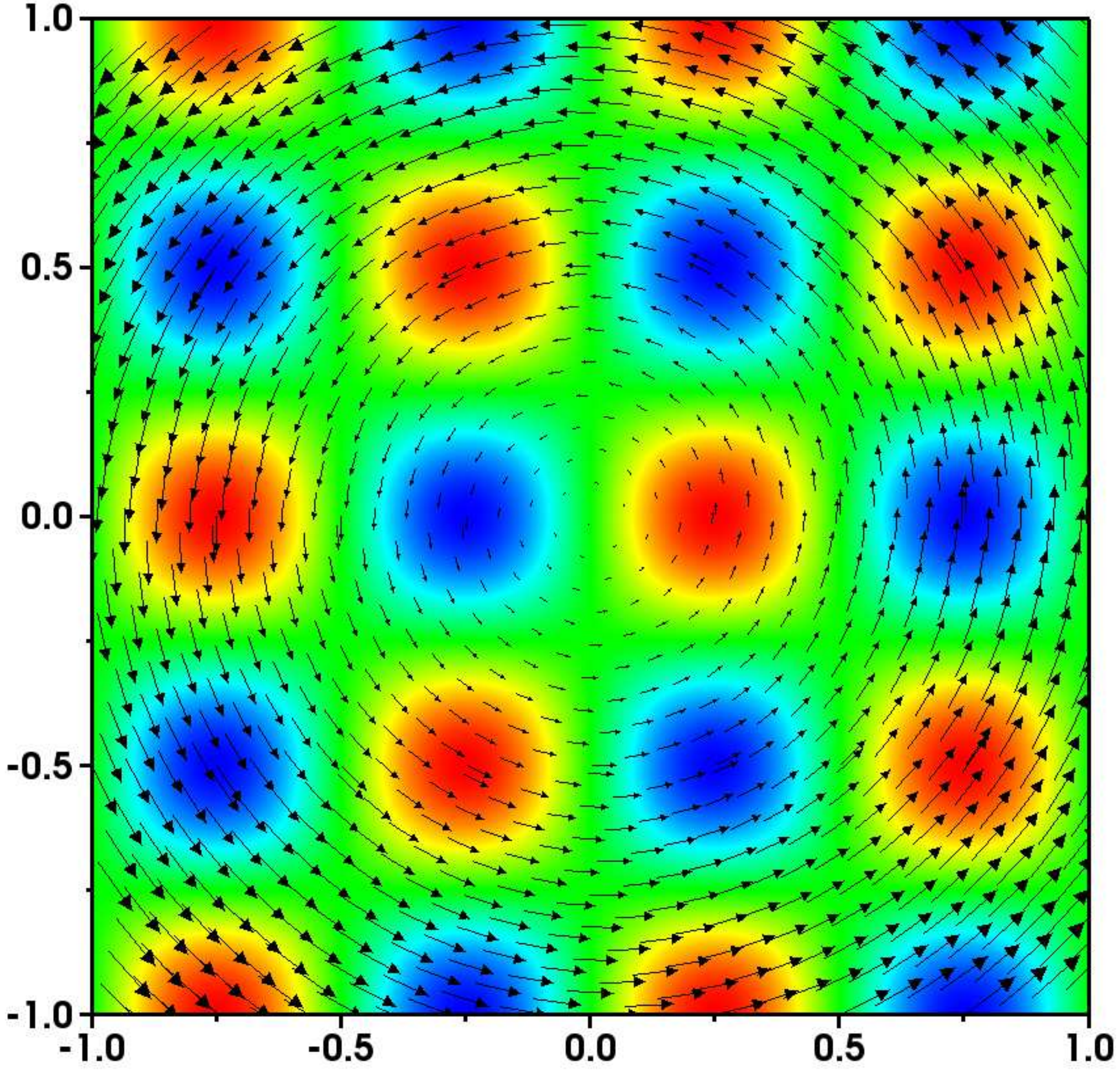}
    \label{fig_C_Viscosity_Init_Schematic}
  }
   \caption{The \subref{fig_C_Density_Init_Schematic} initial density and
    \subref{fig_C_Viscosity_Init_Schematic} viscosity fields, along with the initial velocity vectors
    for the manufactured solution described
   in Sec.~\ref{sec_cons_ms}.}
  \label{fig_cons_ms_init}
\end{figure}

In this next case, we verify the effect of consistent mass density update and conservative discretization 
on the order of accuracy of the computed solution. For this test case we take the velocity and pressure 
solutions to be
\begin{align}
u(\x,t) &= -y \cos (t), \label{eq_cons_ms_u}\\
v(\x,t) &= x \cos (t), \label{eq_cons_ms_v}\\
p(\x,t) &= \sin (t) \sin (x) \sin (y), \label{eq_cons_ms_p}
\end{align}
and we prescribe
\begin{align}
\rho(\x,t) &= \rho_0 + \frac{\rho_1}{2}  \left(\tanh \left(\frac{0.1\,
   -\sqrt{x^2+y^2}}{\delta }\right)+1\right)+x \cos (\sin
   (t))+y \sin (\sin (t)) + 2, \label{eq_cons_ms_rho}\\
\mu(\x) &= \mu_0+\mu_1+\mu_1 \sin (2 \pi  x) \cos (2 \pi  y). \label{eq_cons_ms_mu}
\end{align}
Plugging Eqs.~\eqref{eq_non_cons_ms_u}--\eqref{eq_cons_ms_mu} into the conservative momentum 
equation~\eqref{eqn_momentum} yields a forcing term $\f(\x,t)$ that produces the desired conservative manufactured 
solution. Moreover, it can be verified that the time-dependent density function in Eq.~\eqref{eq_cons_ms_rho} satisfies
the conservative mass balance equation~\eqref{eqn_cons_of_mass}. Also 
notice that density function in Eq.~\eqref{eq_cons_ms_rho}
is only $C^0$ continuous, and, in particular, that it has singular spatial derivatives at the origin. We remark that unlike the non-conservative 
form of equations, in which an arbitrary density function can be selected for the manufactured solution, constructing manufactured solutions for the conservative form 
is non-trivial because the velocity and density fields must satisfy both the mass balance and divergence-free condition.
The viscosity and initial velocity and density fields are shown in Fig.~\ref{fig_cons_ms_init}.

For this case, the variations in density are set to $\rho_0 = 1$ and $\rho_1 = 10^3 - \rho_0$,
and the variations in viscosity are set to $\mu_0 = 10^{-2}$ and $\mu_1 = 1 - \mu_0$.
The computational domain is
$\Omega = [-L,L]^2 =[-1,1]^2$.
The smoothing parameter is set to $\delta = 0.05L$.  %For \emph{all physical boundaries, 
Normal velocity and tangential traction boundary conditions are used at all physical boundaries.
The maximum velocities in the domain for this manufactured solution are $\BigO{1}$, hence a relevant  time scale is $L/U$ with $U = 1$.
Errors in the velocity and pressure
are computed at time $T = tU/L = 0.6$ with a uniform time step size $\dt = 1/(1.042N)$, which yields an approximate CFL
number of $0.5$.

\begin{figure}[]
  \centering
  \subfigure[Velocity]{
    \includegraphics[scale = 0.255]{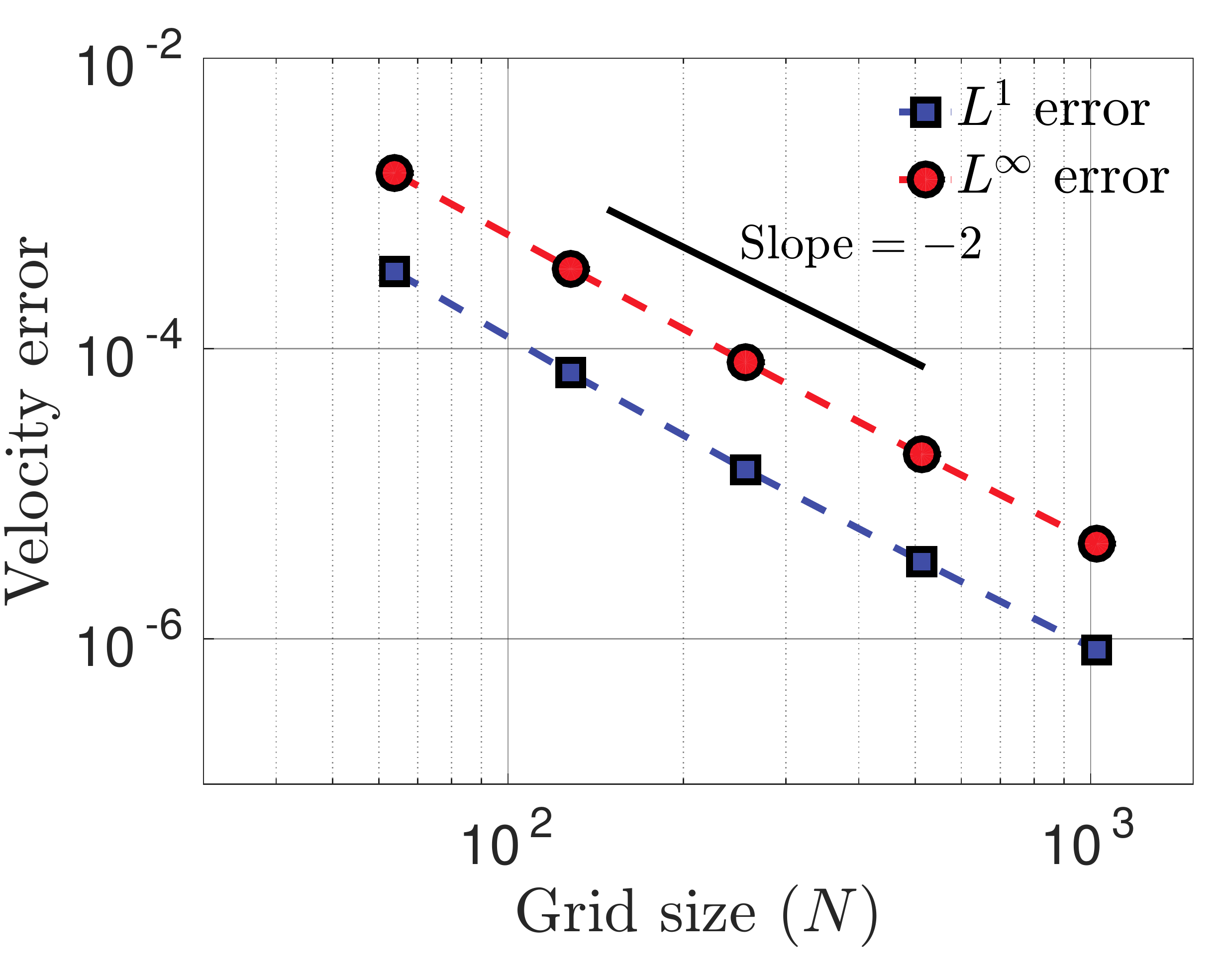}
    \label{fig_c_U_err_vel_tra}
  }
   \subfigure[Pressure]{
    \includegraphics[scale = 0.255]{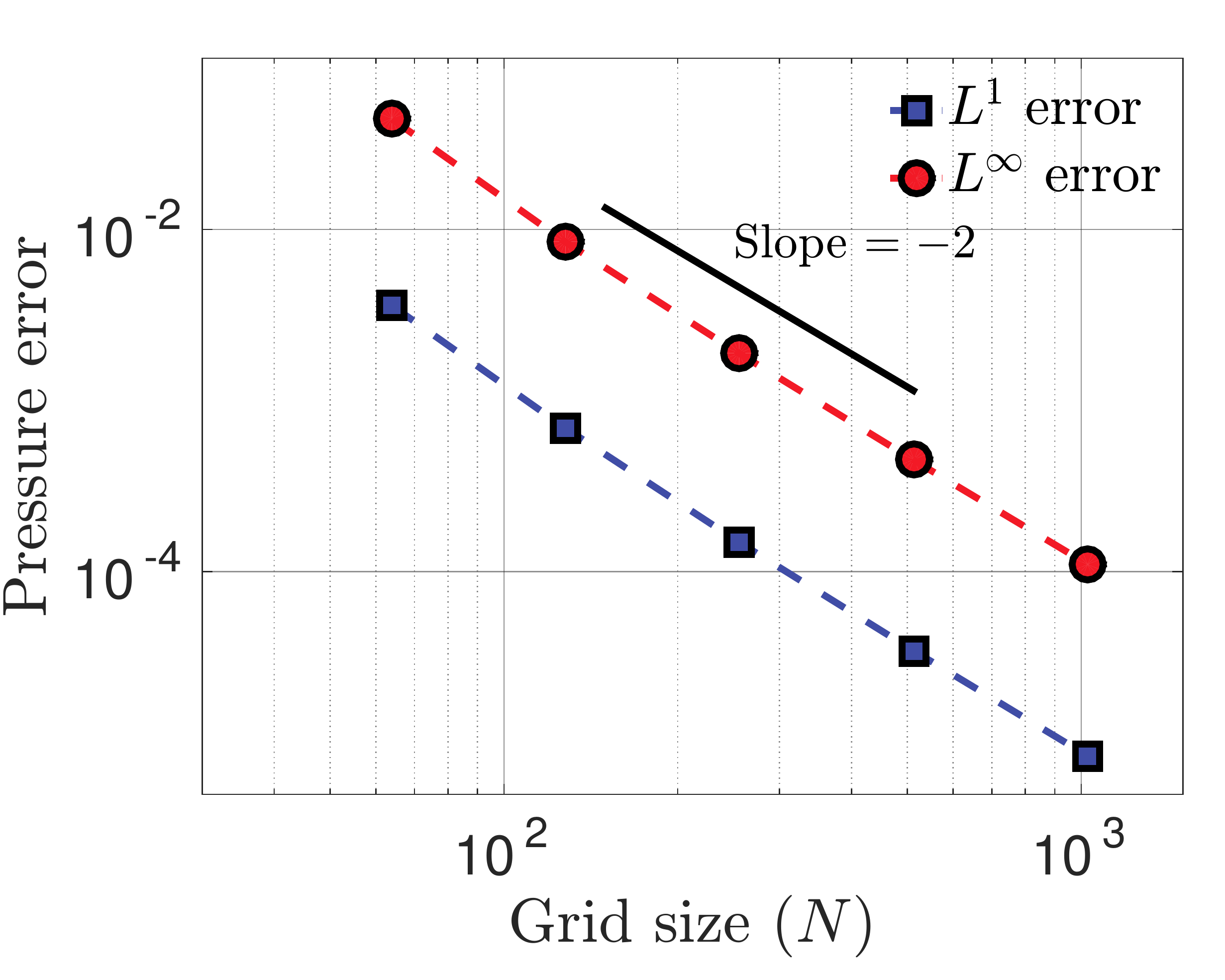}
    \label{fig_c_P_err_vel_tra}
  }
   \subfigure[Density]{
    \includegraphics[scale = 0.255]{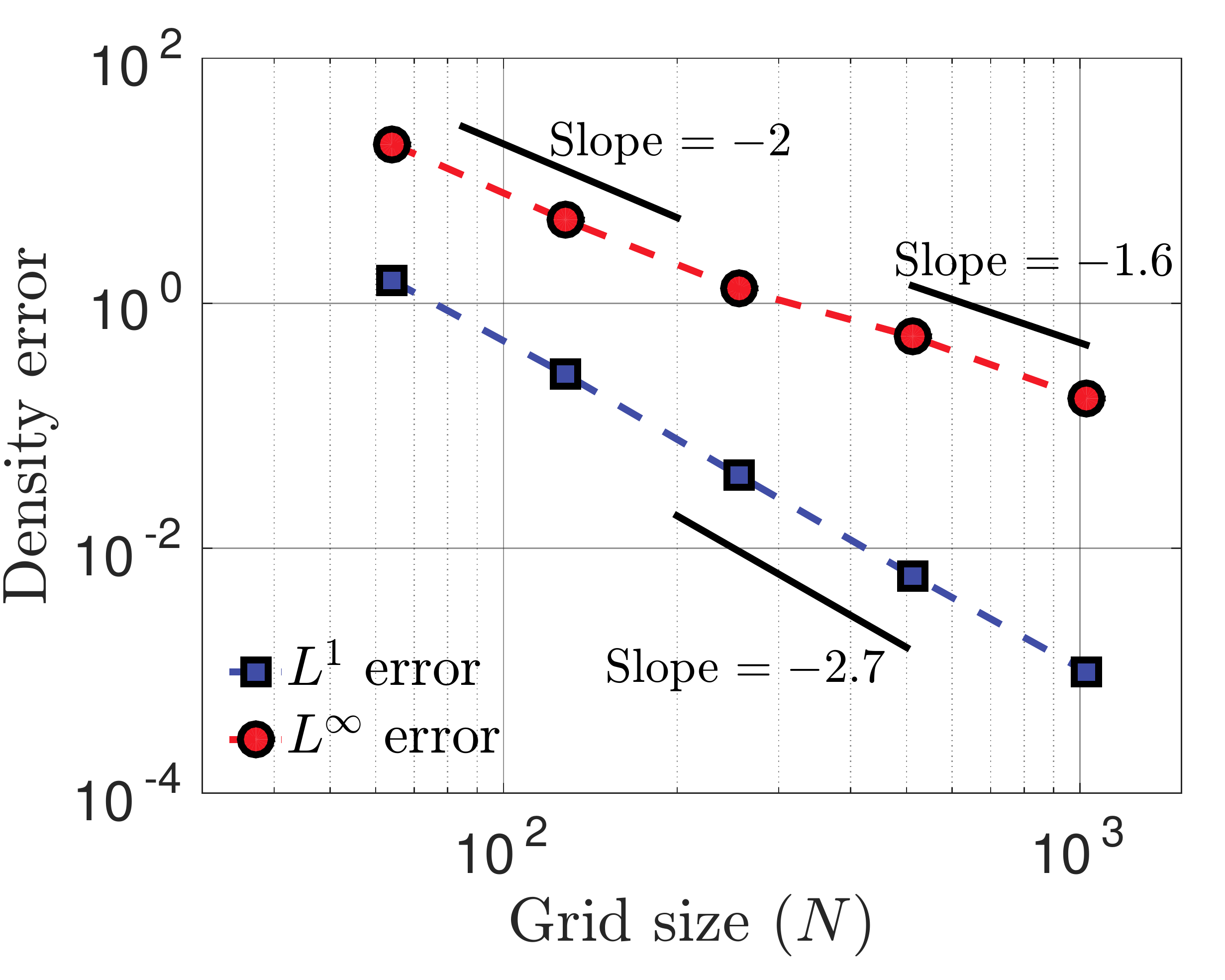}
    \label{fig_c_RHO_err_vel_tra}
  }
  \caption{ 
  $L^1$ ($\blacksquare$, blue) and $L^\infty$ ($\bullet$, red) errors as a function of grid size $N$ for the conservative
  manufactured solution with specified normal velocity and tangential traction boundary conditions on all boundaries. In these cases,
  the density $\rho$ is \emph{not} reset between time steps.
  \subref{fig_c_U_err_vel_tra}
  convergence rate for $\u$;
  \subref{fig_c_P_err_vel_tra}
   convergence rate for $p$;
   \subref{fig_c_RHO_err_vel_tra}
   convergence rate for $\rho$.
   }
  \label{fig_c_err_vel_tra}
\end{figure}

As a first test, we consider the evolution of density along with velocity and pressure. The time-independent viscosity is set
at the initial time and is not evolved during the simulation. Fig.~\ref{fig_c_err_vel_tra} shows the $L^1$
and $L^{\infty}$ errors for velocity, pressure, and density as a function of grid size. Second-order convergence rates
are achieved for velocity and pressure in both norms. The convergence rate for the $L^1$ density error
is at least second order. Local reductions in pointwise convergence rates (less than 
second-order) are seen for density. This can be attributed to the $\cC^0$ spatial continuity of $\rho(\x,t)$. 
We do indeed obtain full second-order convergence rates for all norms when considering a smooth density 
manufactured solution (see Appendix \ref{app_smooth_density}).

In our next test, we consider the same conservative manufactured solution as above, but we
instead \emph{reset} the density field at the beginning of each time step by computing the face-centered
$\vrho^n$ directly from Eq.~\eqref{eq_cons_ms_rho} at time $t^n$.
Hence, within each time step, $\vrho^n$ is numerically evolved to
$\breve{\V \rho}^{n+1, k+1}$ using the SSP-RK3 update described in Sec.~\ref{cons_discretization}. To wit, the density evolution reads as

\begin{align}
	& \breve{\V \rho}^{(1)} = \vrho^n - \dt \R\left(\u^{n}_\text{adv}, \vrho^n_\text{lim}\right),\\
	&  \breve{\V \rho}^{(2)} = \frac{3}{4}\vrho^{n} + \frac{1}{4} \breve{\V \rho}^{(1)} - \frac{1}{4} \dt \R\left(\u^{(1)}_\text{adv}, \breve{\V \rho}^{(1)}_\text{lim}\right), \\
	& \breve{\V \rho}^{n+1, k+1} = \frac{1}{3} \vrho^n + \frac{2}{3} \breve{\V \rho}^{(2)} - \frac{2}{3} \dt \R\left(\u^{(2)}_\text{adv}, \breve{\V \rho}^{(2)}_\text{lim}\right).
	\end{align}
This evolved quantity is only used in the conservative discretization of the momentum
equation~\eqref{eq_c_discrete_momentum}. Upon numerically solving for the velocity and pressure,
this density approximation is discarded and we begin the next time step with
$\vrho^{n+1}$ computed from Eq.~\eqref{eq_cons_ms_rho} at time $t^{n+1}$. On the final time step, the density
error norms are computed with respect to the evolved $\breve{\V \rho}$ in order to determine the order of
accuracy of a single time step of SSP-RK3 integration. This density resetting procedure emulates the level
set synchronization approach described in Sec.~\ref{cons_discretization}, which is used in all 
of the numerical examples presented in Sec.~\ref{sec_examples}. This methodology is commonly used in
other interface capturing models as well, such as
volume of fluid~\cite{Hirt1981} and phase field~\cite{Boettinger2002} approaches,
in which the material properties are set via an auxiliary indicator function. 

%This resetting procedure for 
%the density is not arbitrary.  Instead, it is a commonly used approach in interface capturing approaches such as level set~\cite{Sethian2003}, 
%volume of fluid~\cite{Hirt1981}, and phase field~\cite{Boettinger2002} methods, in which the material properties are set 
%via an auxiliary indicator function. 

\begin{figure}[]
  \centering
  \subfigure[Velocity]{
    \includegraphics[scale = 0.255]{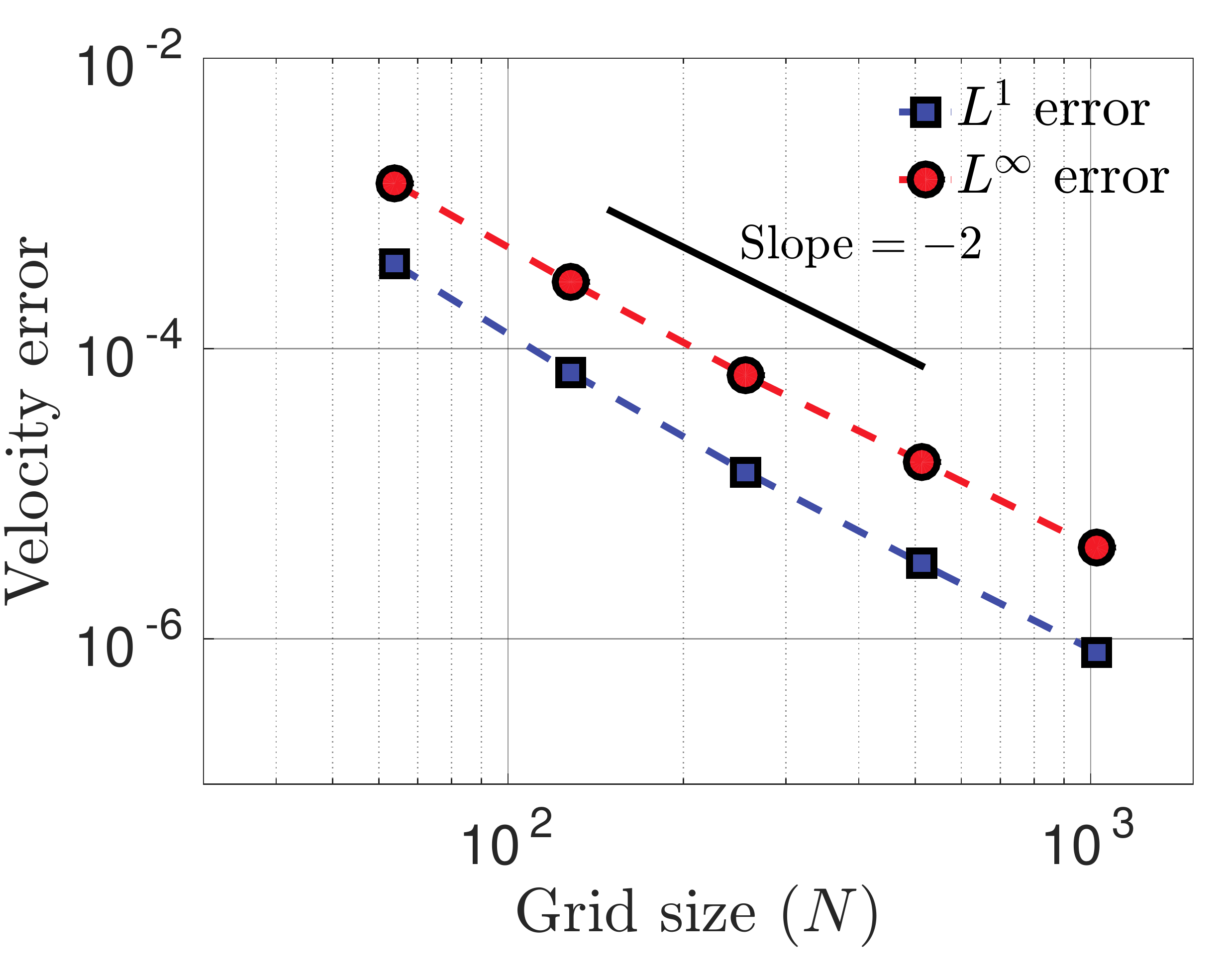}
    \label{fig_c_U_err_vel_tra_reset_rho}
  }
   \subfigure[Pressure]{
    \includegraphics[scale = 0.255]{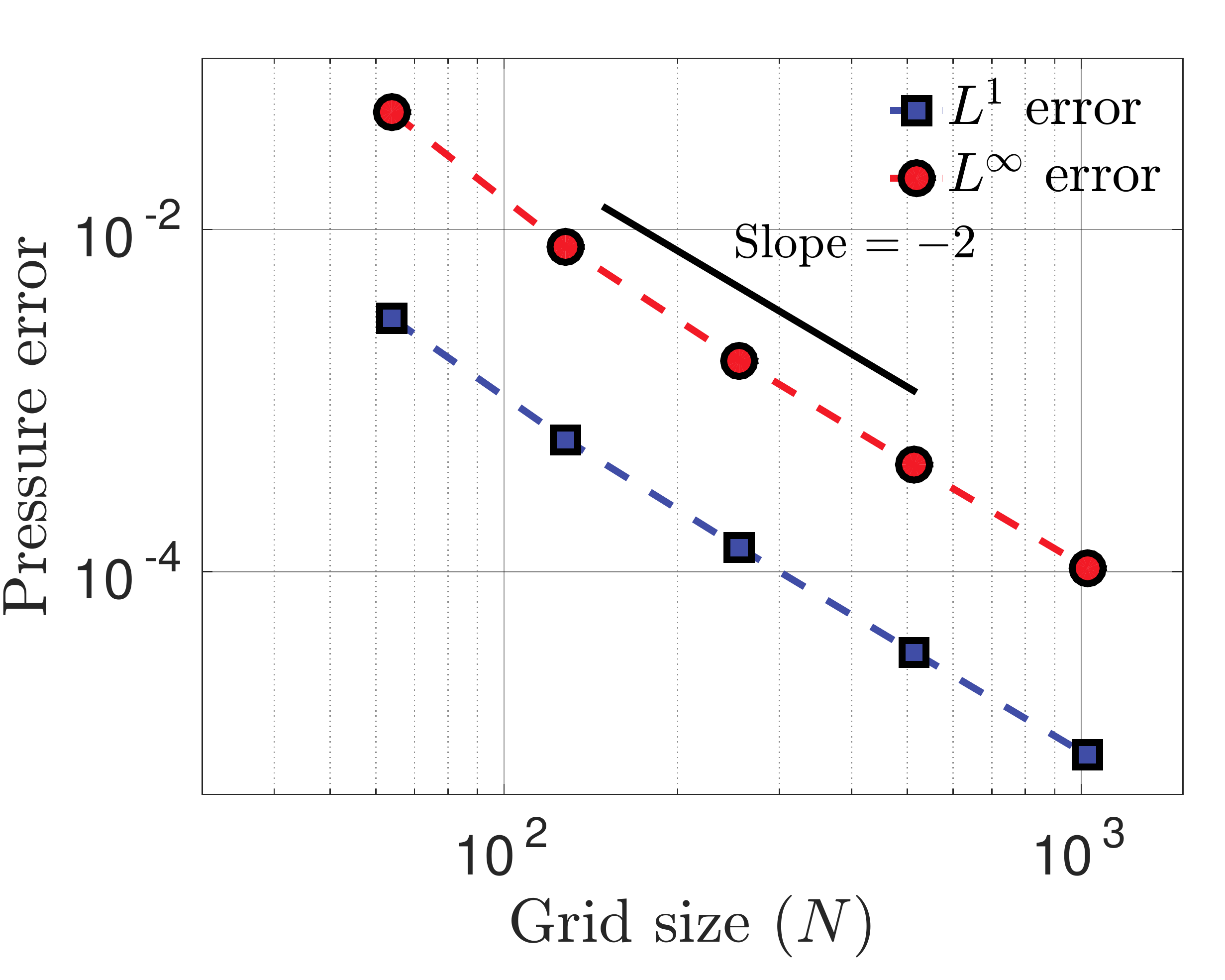}
    \label{fig_c_P_err_vel_tra_reset_rho}
  }
   \subfigure[Density]{
    \includegraphics[scale = 0.255]{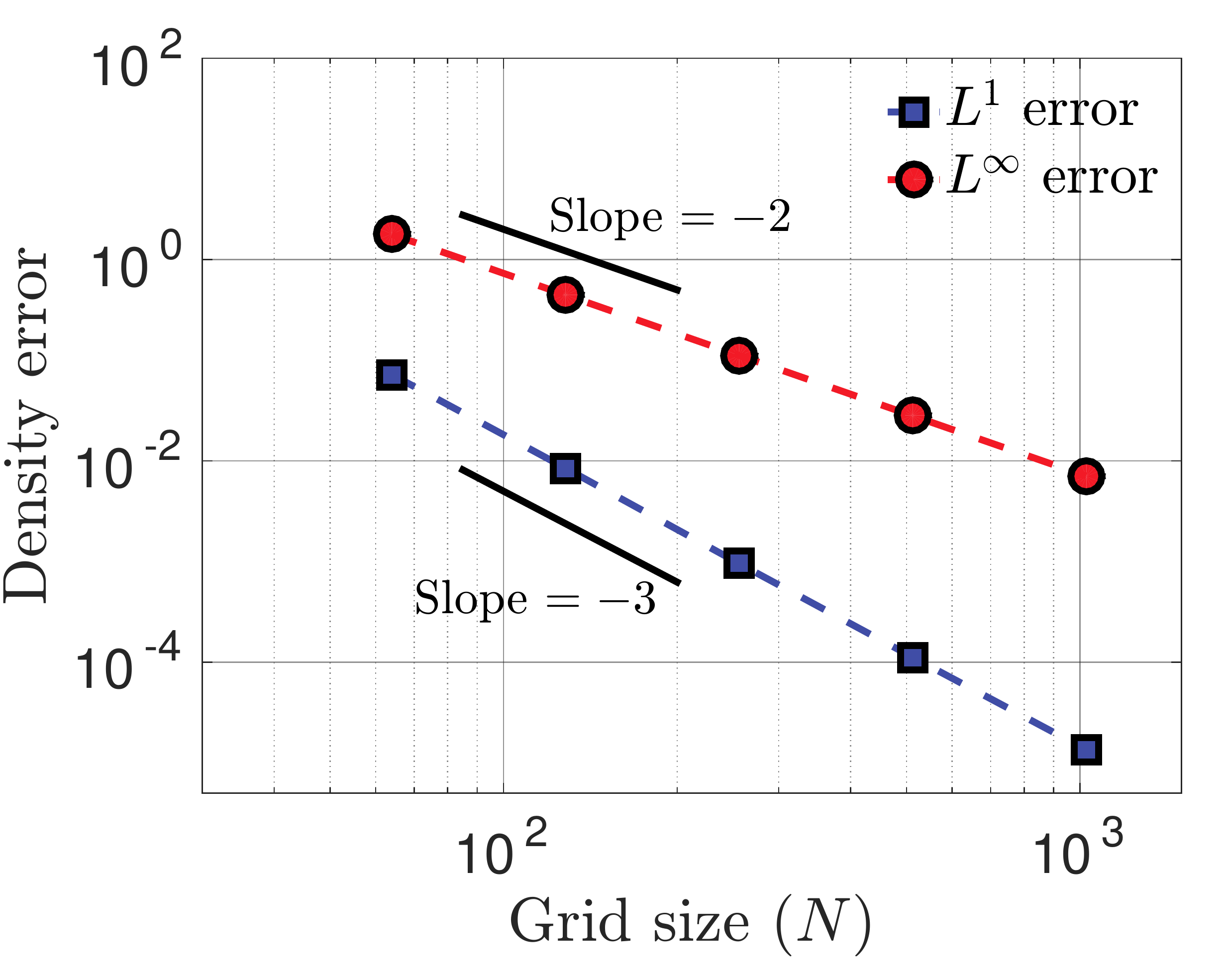}
    \label{fig_c_RHO_err_vel_tra_reset_rho}
  }
  \caption{ 
  $L^1$ ($\blacksquare$, blue) and $L^\infty$ ($\bullet$, red) errors as a function of grid size $N$ for the conservative
  manufactured solution with specified normal velocity and tangential traction boundary conditions on all boundaries. In these cases,
  the density $\rho$ is reset between time steps.
  \subref{fig_c_U_err_vel_tra_reset_rho}
  convergence rate for $\u$;
  \subref{fig_c_P_err_vel_tra_reset_rho}
   convergence rate for $p$;
   \subref{fig_c_RHO_err_vel_tra_reset_rho}
   convergence rate for $\rho$.
   }
  \label{fig_c_err_vel_tra_reset_rho}
\end{figure}

Fig.~\ref{fig_c_err_vel_tra_reset_rho} shows the $L^1$ and $L^{\infty}$ errors for 
velocity, pressure, and density as a function of grid size. We again observe full second-order convergence rates for 
velocity and pressure. We also no longer see a reduction in the order of accuracy for density, even for the $\cC^0$ density field; this is because the error is essentially computed after a single time step of density evolution.
%Therefore, the computed error in 
%density is essentially computed after a single time step of density evolution. %Nevertheless, 
These tests therefore show that the present numerical discretization maintains the desired order of accuracy for a conservative formulation with consistent mass update.

\subsection{Non-conservative/Conservative form: Effect of local mesh refinement}
\begin{figure}[]
  \centering
  \subfigure[Non-conservative]{
    \includegraphics[scale = 0.255]{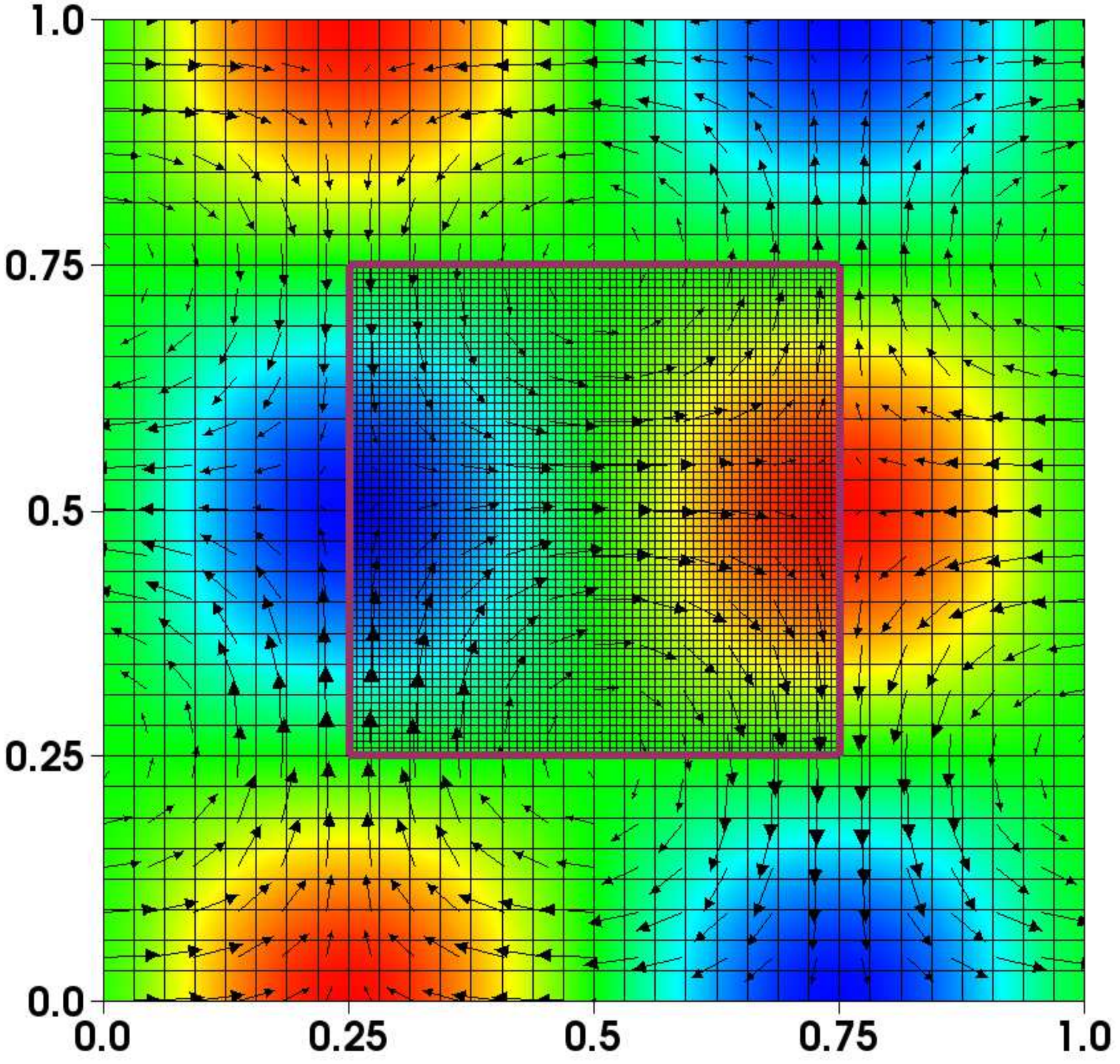}
    \label{fig_nc_amr_grid}
  }
   \subfigure[Conservative]{
    \includegraphics[scale = 0.255]{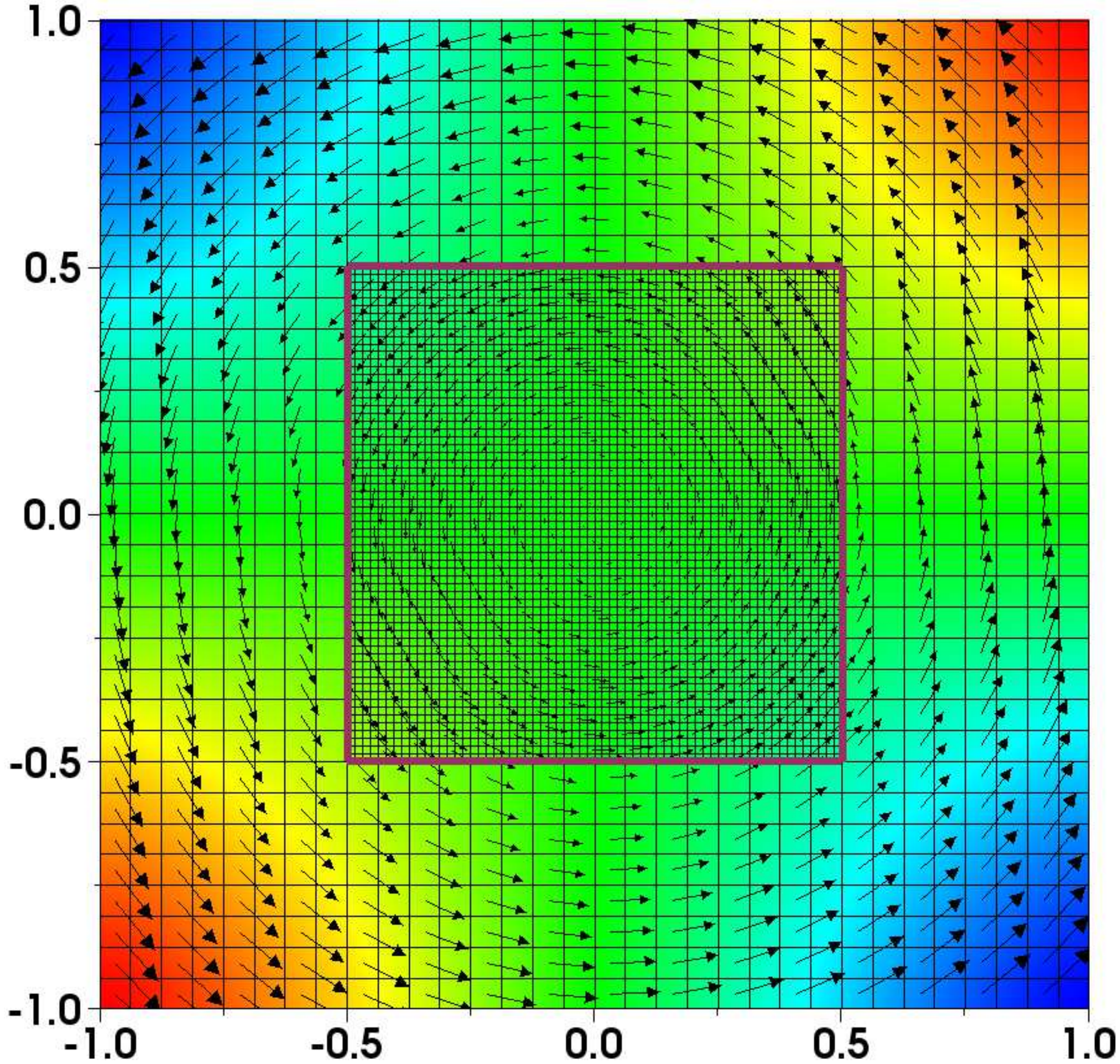}
    \label{fig_c_amr_grid}
  }
   \caption{Velocity vectors and pressure field plotted on the grid hierarchy used for the locally refined mesh
   convergence study. In both figures, the purple box indicates a locally refined grid with grid spacing
   $\dx_\textrm{min} = \dx_0/4$.}
  \label{fig_amr_grid}
\end{figure}

We now consider the non-conservative and conservative manufactured solutions of the previous sections,~\ref{sec_non_cons_ms} and~\ref{sec_cons_ms}, on locally refined grids and analyze the error convergence rate. In all of the cases considered here, we use $\ell = 2$ mesh levels with a refinement ratio of $\nref = 4$. The refined region is always assigned to the center of the domain, which contains the dense region, as shown in
Fig.~\ref{fig_amr_grid}. It should be noted that for practical multiphase simulations, we always restrict the fluid-gas interface
to the finest mesh level in the grid hierarchy.
%as currently done in this example.

\begin{figure}[]
  \centering
  \subfigure[Velocity]{
    \includegraphics[scale = 0.255]{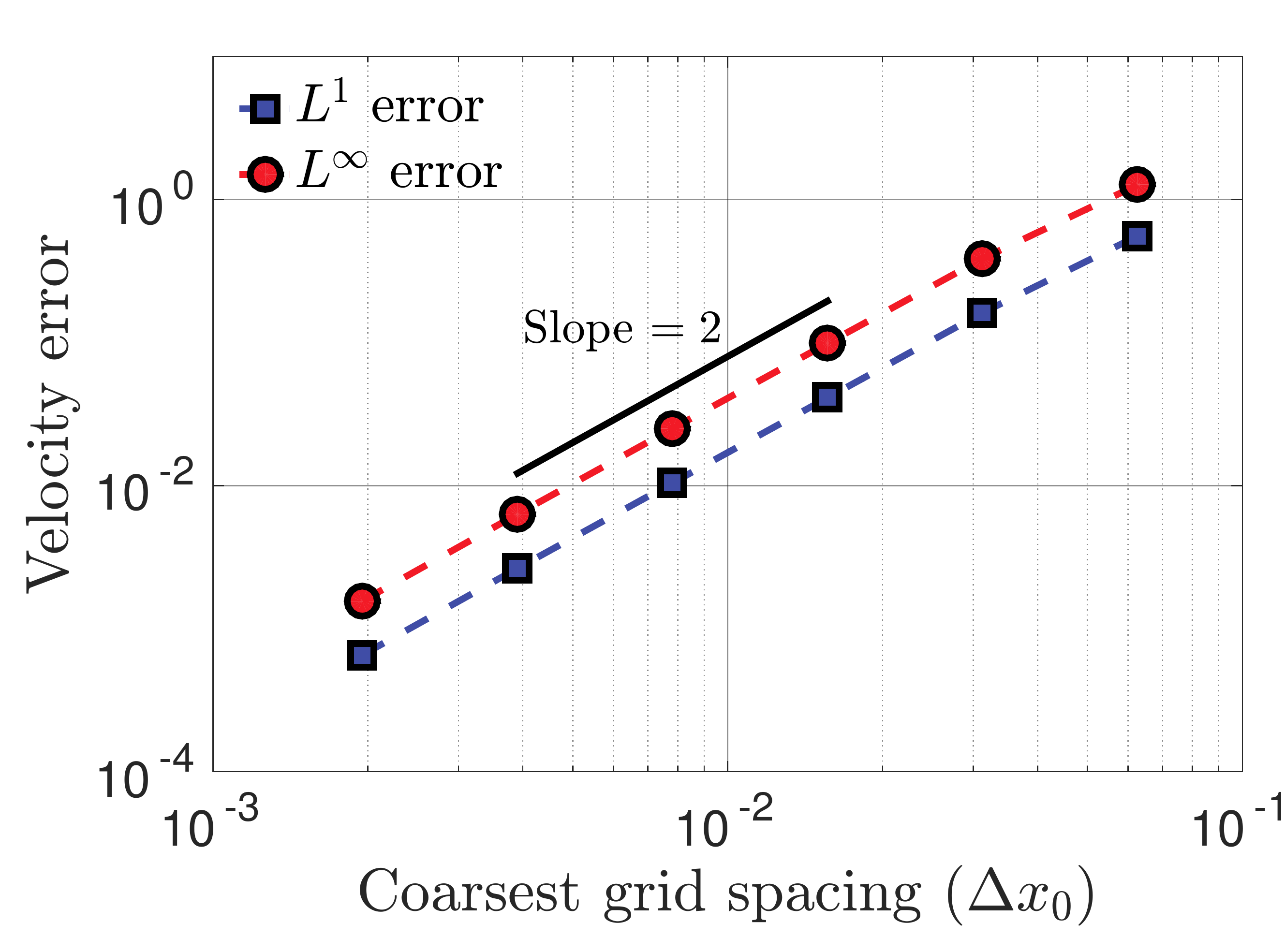}
    \label{fig_nc_U_err_amr}
  }
   \subfigure[Pressure]{
    \includegraphics[scale = 0.255]{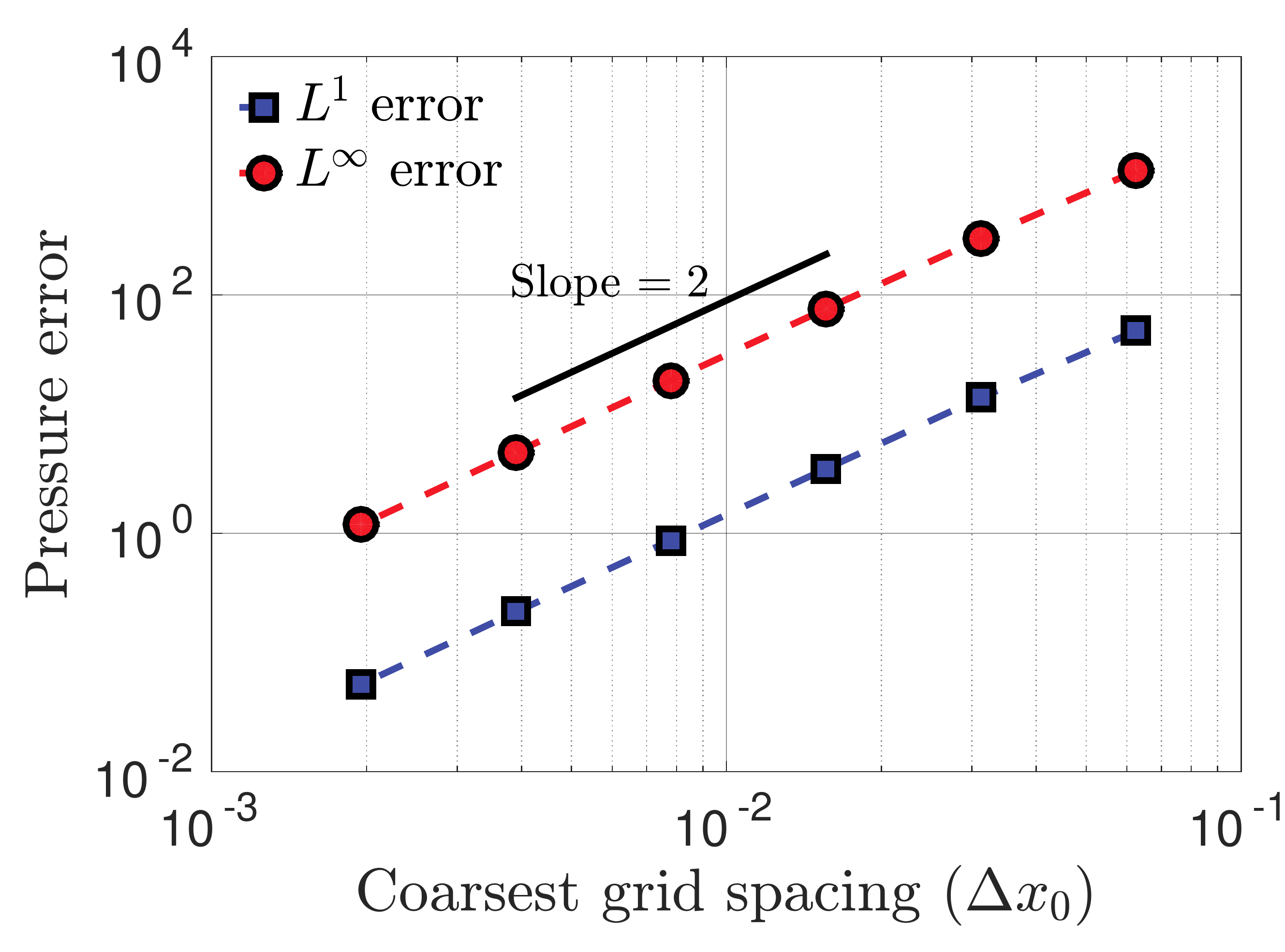}
    \label{fig_nc_P_err_amr}
  }
  \caption{ 
  $L^1$ ($\blacksquare$, blue) and $L^\infty$ ($\bullet$, red) errors as a function of coarsest grid spacing $\dx_0$
  for the non-conservative manufactured solution described in Sec.~\ref{sec_non_cons_ms} on a locally refined mesh.
  The specified boundary conditions are normal velocity and tangential traction (vel-tra).
  \subref{fig_nc_U_err_amr}
  convergence rate for $\u$;
  \subref{fig_nc_P_err_amr}
   convergence rate for $p$.
   }
  \label{fig_nc_err_amr}
\end{figure}

In our first test, we consider the non-conservative manufactured solution detailed in Sec.~\ref{sec_non_cons_ms}.
We consider vel-tra boundary conditions, i.e., periodic boundary conditions in the $x$-direction and specified
normal velocity and tangential traction in the $y$-direction. Fig.~\ref{fig_nc_err_amr} shows the $L^1$ and
$L^{\infty}$ errors for velocity and pressure as a function of coarsest grid spacing. We again obtain second-order 
convergence rates for velocity and pressure on the locally refined grid.

\begin{figure}[]
  \centering
  \subfigure[Velocity]{
    \includegraphics[scale = 0.255]{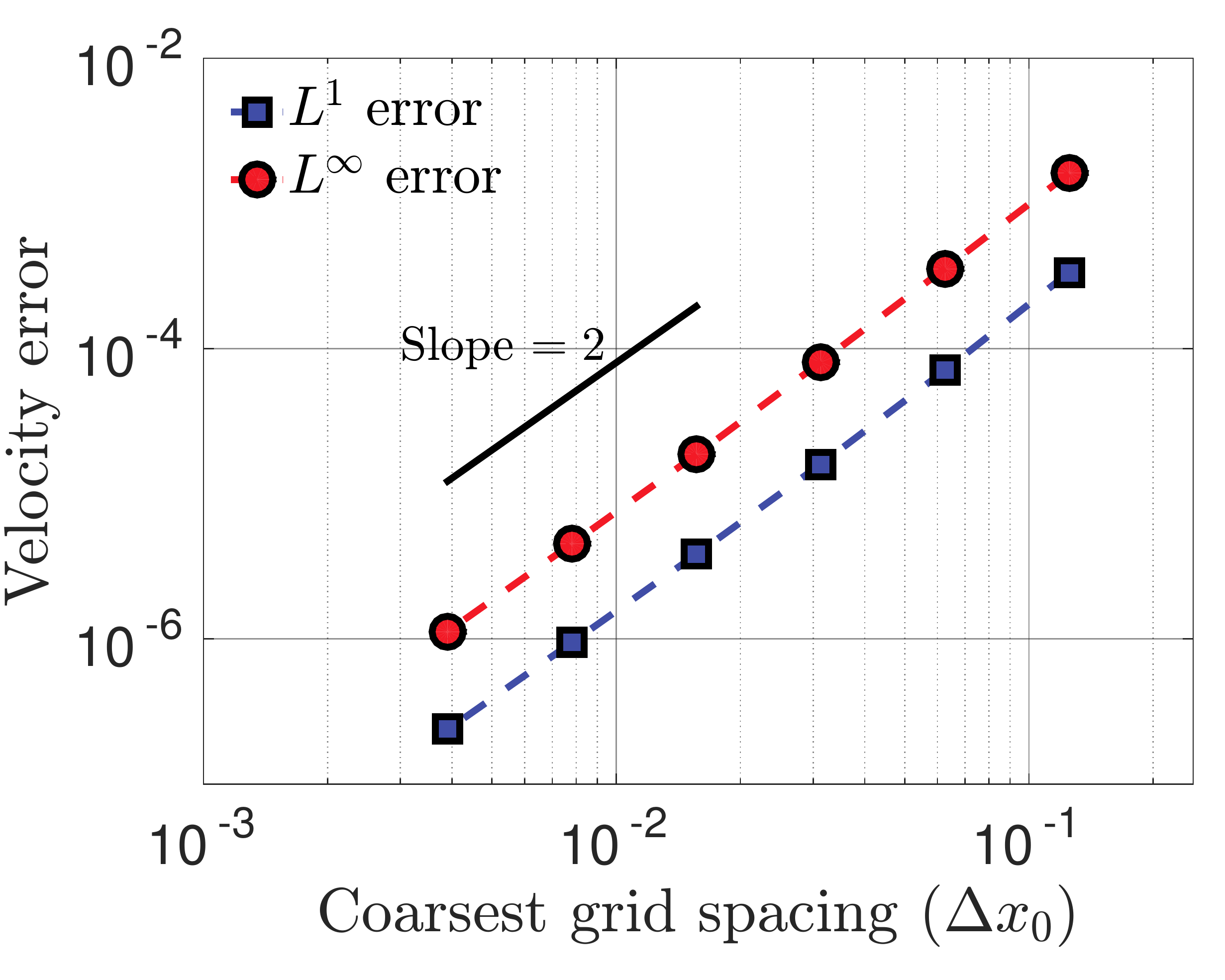}
    \label{fig_c_U_err_amr}
  }
   \subfigure[Pressure]{
    \includegraphics[scale = 0.255]{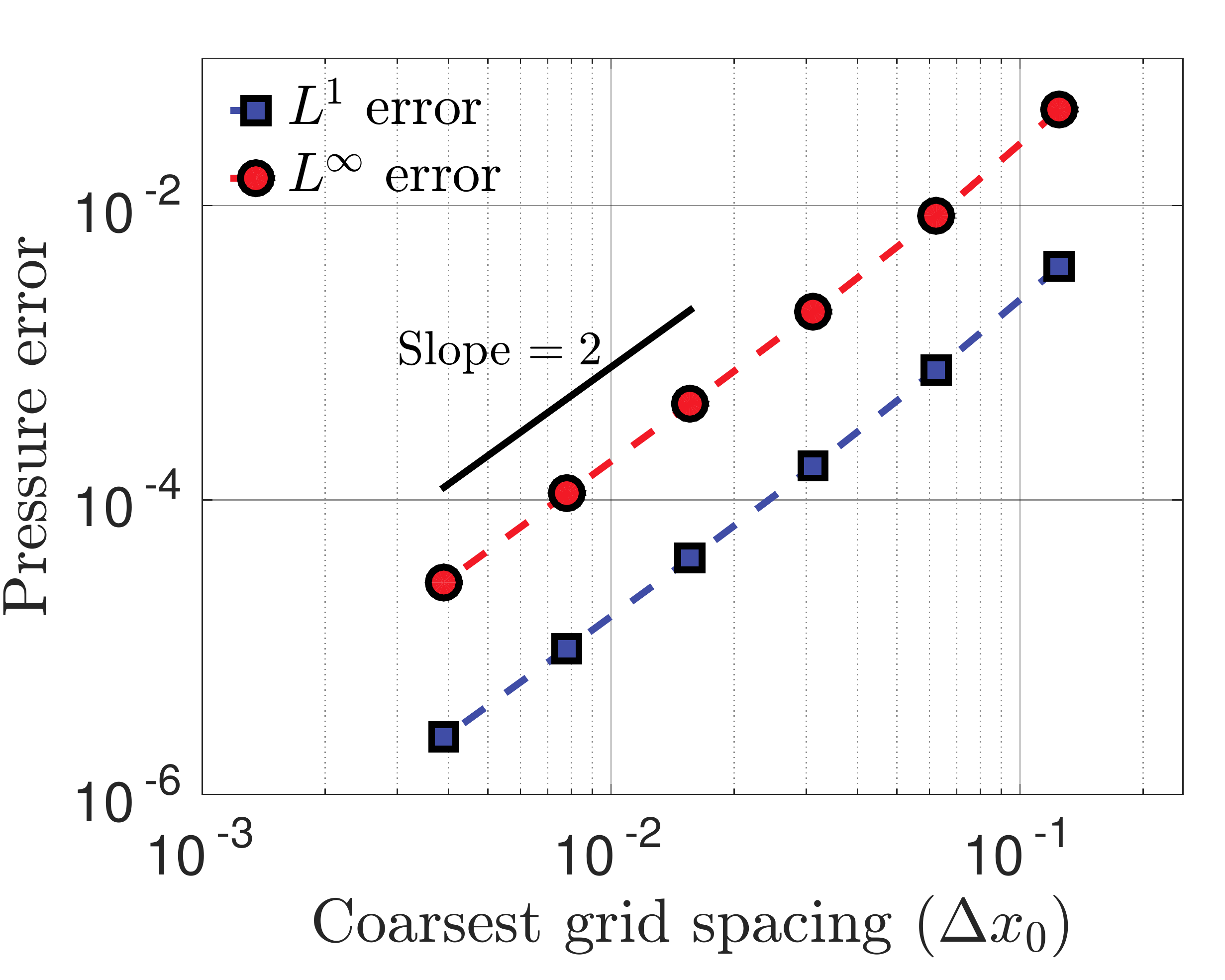}
    \label{fig_c_P_err_amr}
  }
   \subfigure[Density]{
    \includegraphics[scale = 0.255]{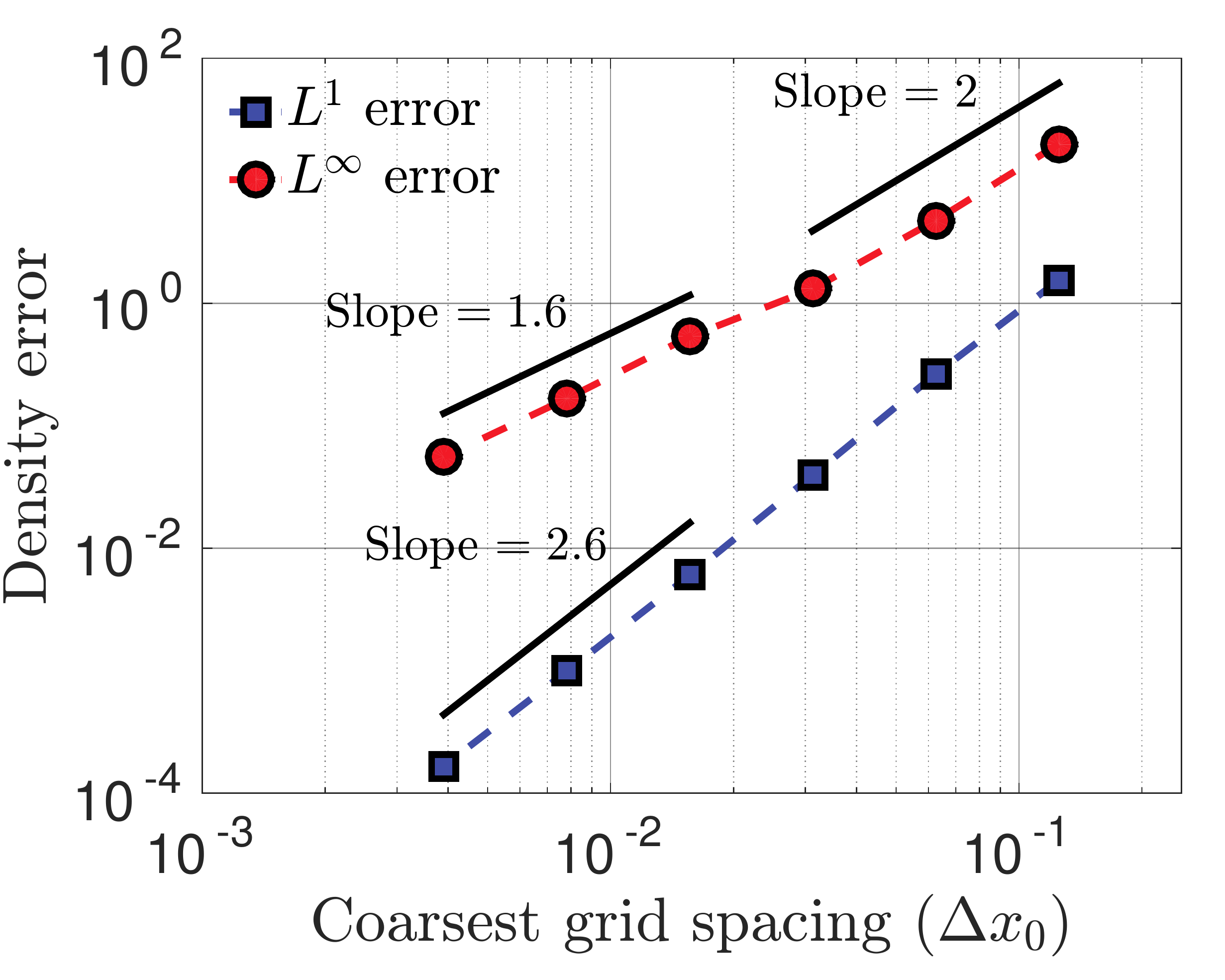}
    \label{fig_c_RHO_err_amr}
  }
  \caption{ 
  $L^1$ ($\blacksquare$, blue) and $L^\infty$ ($\bullet$, red) errors as a function of coarsest grid spacing $\dx_0$
  for the conservative manufactured solution with specified normal velocity and tangential traction boundary conditions on all boundaries. In these cases,
  the density $\rho$ is \emph{not} reset between time steps.
  \subref{fig_c_U_err_amr}
  convergence rate for $\u$;
  \subref{fig_c_P_err_amr}
   convergence rate for $p$;
   \subref{fig_c_RHO_err_amr}
   convergence rate for $\rho$.
   }
  \label{fig_c_err_amr}
\end{figure}

Next, we consider the conservative manufactured solution described in Sec.~\ref{sec_cons_ms}.
In this case, we do not reset the density at the beginning of time steps.
Fig.~\ref{fig_c_err_amr} shows the $L^1$ and $L^{\infty}$ errors for velocity, pressure,
and density as a function of coarsest grid spacing. The observed convergence rates observed
on this locally refined mesh are nearly identical to those shown for the uniform mesh case.

%Somewhat surprisingly, the pointwise convergence rates for these locally refined meshes
%are in excellent agreement with the uniform mesh cases presented in earlier sections.
%Presumably, the $L^{\infty}$ convergence rates would be reduced under sufficient spatial resolution.

\subsection{Conservative form: Solver scalability}
\label{sec_cons_scale}
Finally, we investigate the scalability of the preconditioned FGMRES solver for the staggered 
Stokes system given by Eq.~\eqref{eq_stokes_system} and also show the linear solver iteration results for the
velocity %(Eq.~\eqref{eq_frac_vel}) 
and pressure %(Eq.~\eqref{eq_frac_div}) 
subdomain problems.
A bubble is placed in a computational domain $\Omega = [0, L]^2$ with $L = 1$,
which is discretized by an $N \times N$ grid. The radius of the bubble is $R = 0.2L$ 
with initial center position $(X_0, Y_0) = (L/2, L/2)$. The bubble has density $\rho_\text{i}$ and
viscosity $\mu_\text{i}$ and is placed within ambient fluid of density $\rho_\text{o}$ and $\mu_\text{o}$.
The density and viscosity jump is smeared over $\ncells = 2.5$ grid cells on each side of the interface and harmonic averaging
is used to average viscosity from cell-centers to nodes of the staggered grid. The initial velocity
and pressure are set to
\begin{align}
u(\x,0) &= \cos (2 \pi  x) \sin (2 \pi  y), \\
v(\x,0) &= -\sin (2 \pi  x) \cos (2 \pi  y),\\
p(\x,0) &= 0.
\end{align}
Homogenous normal velocity and homogeneous tangential traction boundary conditions are used at all sides of the domain, and homogenous Neumann 
boundary conditions for $\phi$, $\rho$, and $\mu$ are specified on $\partial \Omega$. No additional
body forces are applied to the momentum equations (i.e., $\f(\x,t) = 0$).
The maximum velocities in the domain for this manufactured solution are $\BigO{1}$, hence a relevant  time scale is $L/U$ with $U = 1$.
Each case is run until $T = t U/L = 0.1$ with
a uniform time step $\dt = 1/(3.125 N)$, which yields an approximate CFL number of $0.3$.
Residual values are presented at the final time from the second fixed-point cycle.

\begin{figure}[]
  \centering
    \includegraphics[scale = 0.255]{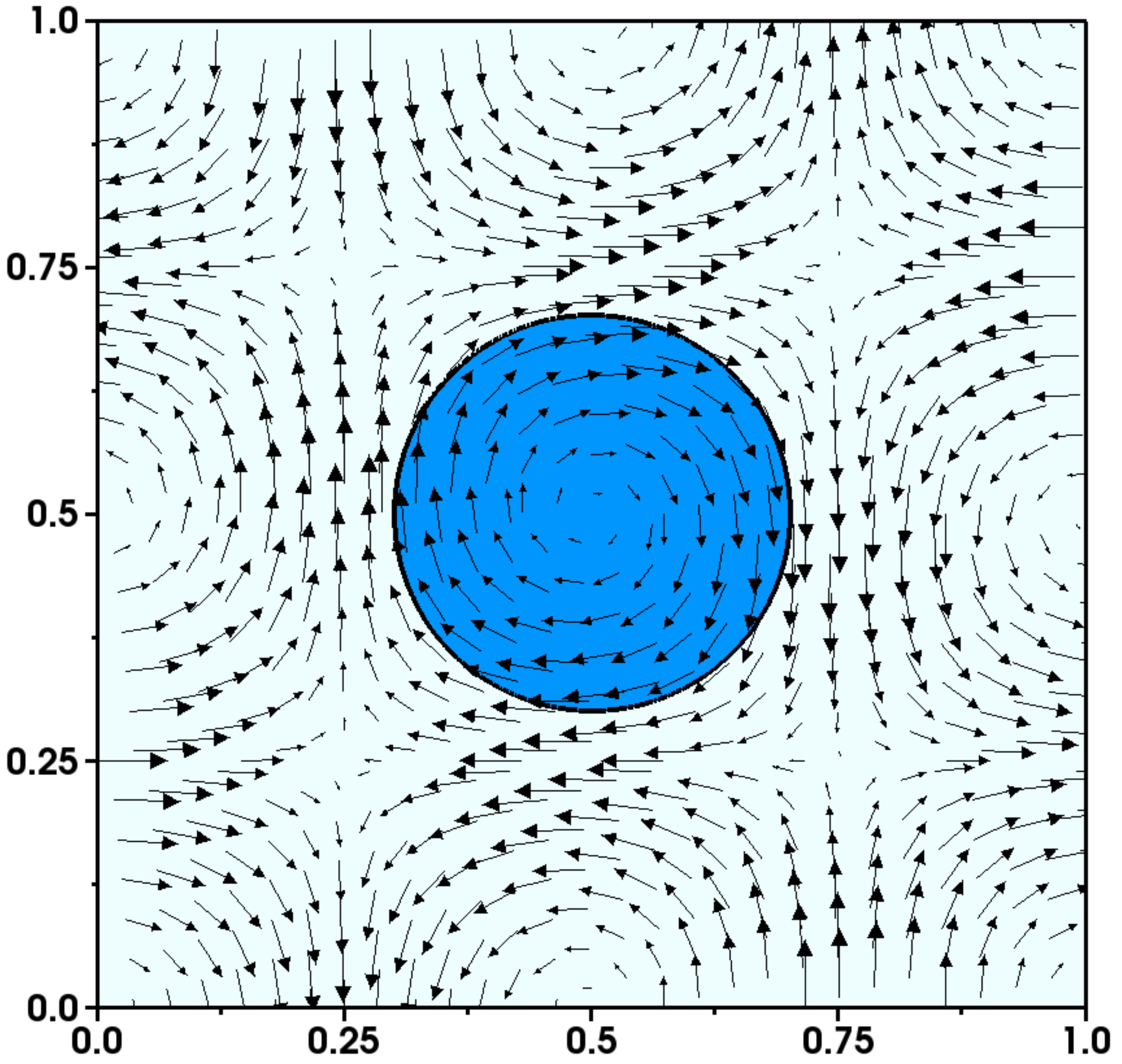}
   \caption{The initial location of the air-suspended water droplet, and the initial velocity
   vectors for the problem set up described in Sec.~\ref{sec_cons_scale}.}
  \label{Bubble_Residual_Init_Schematic}
\end{figure}

\begin{figure}[]
  \centering
  \subfigure[Velocity subdomain residuals]{
    \includegraphics[scale = 0.27]{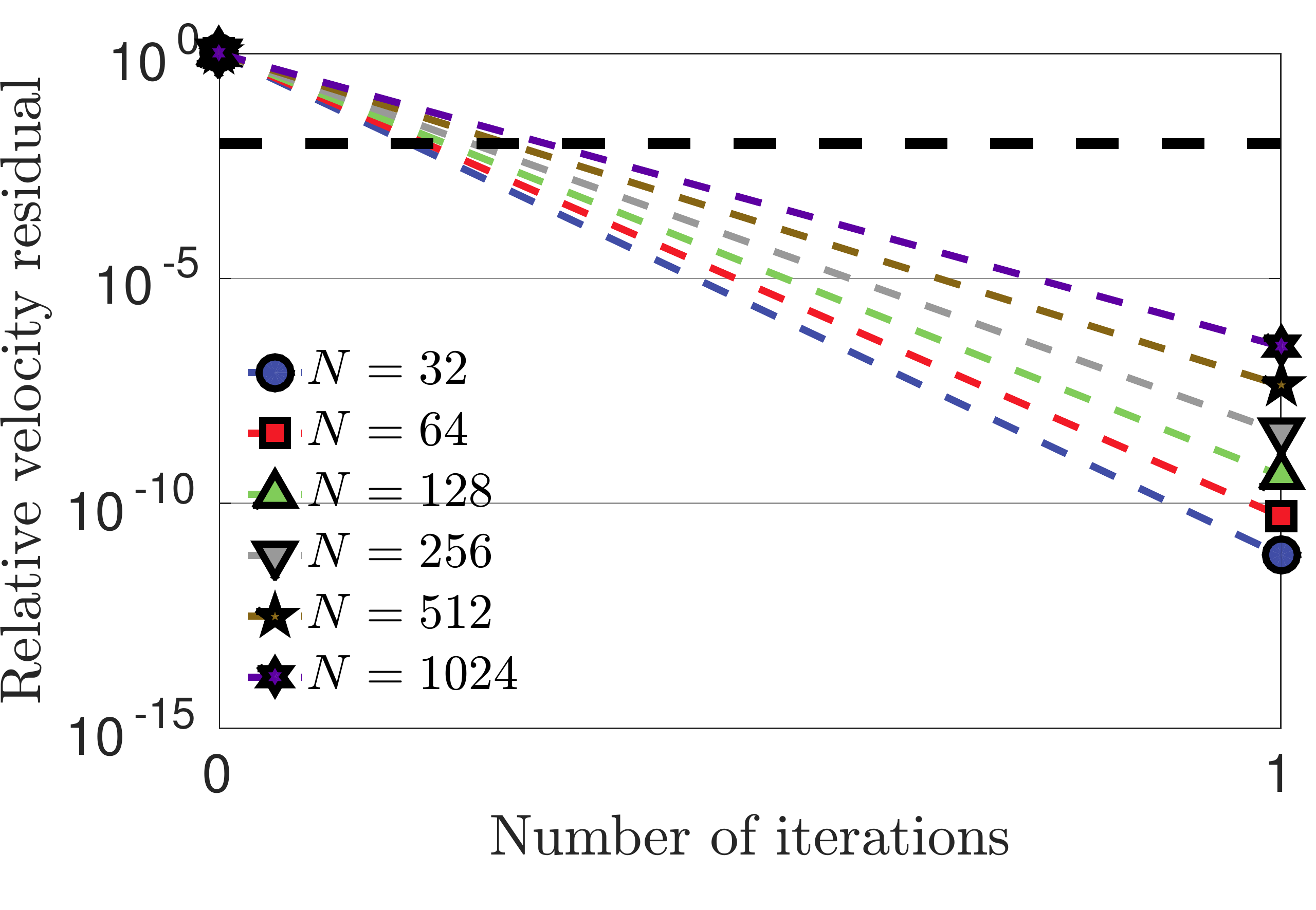}
    \label{MKS_Velocity_Residuals}
  }
   \subfigure[Pressure subdomain residuals]{
    \includegraphics[scale = 0.27]{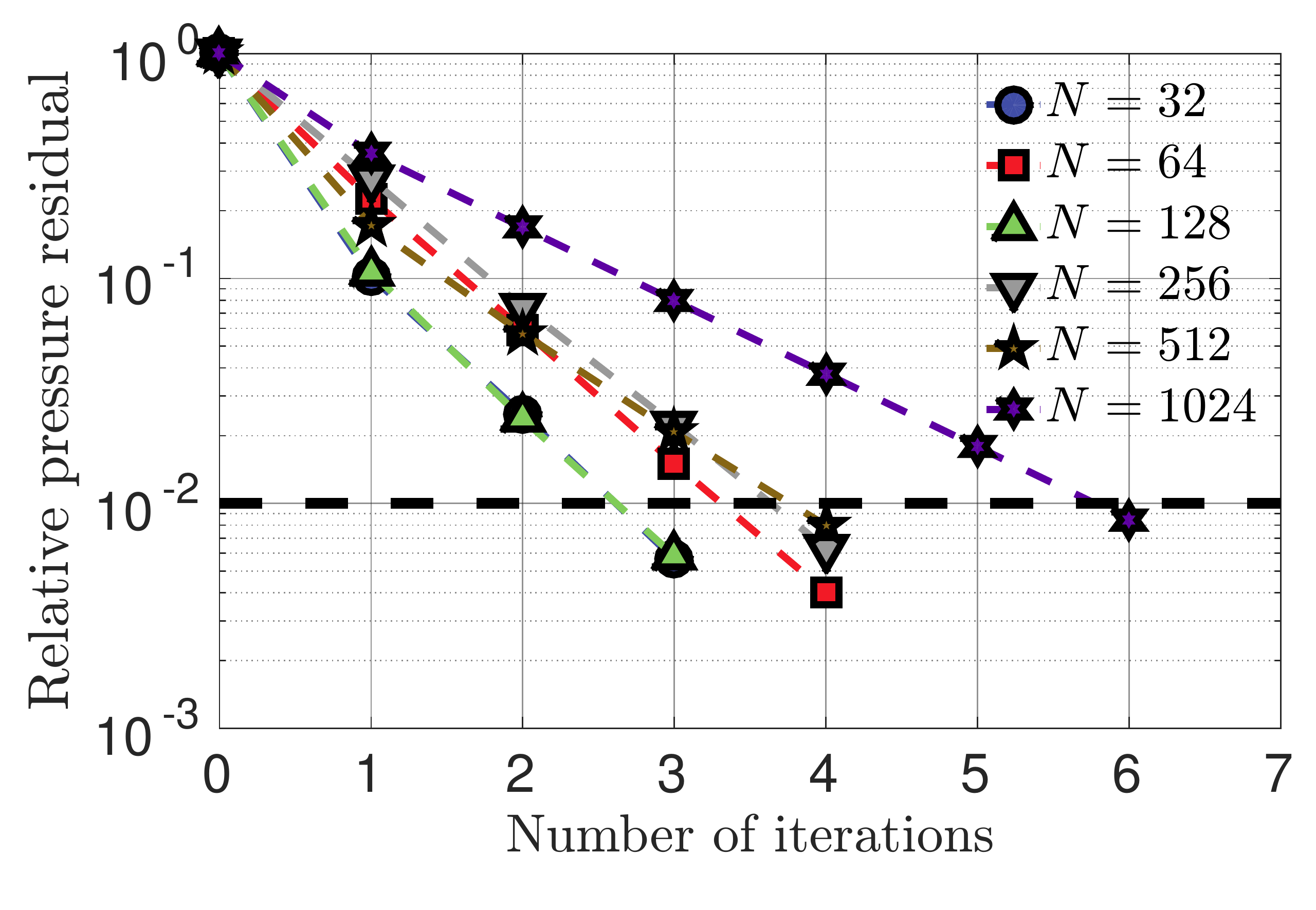}
    \label{MKS_Pressure_Residuals}
  }
   \subfigure[Stokes solver residuals]{
    \includegraphics[scale = 0.27]{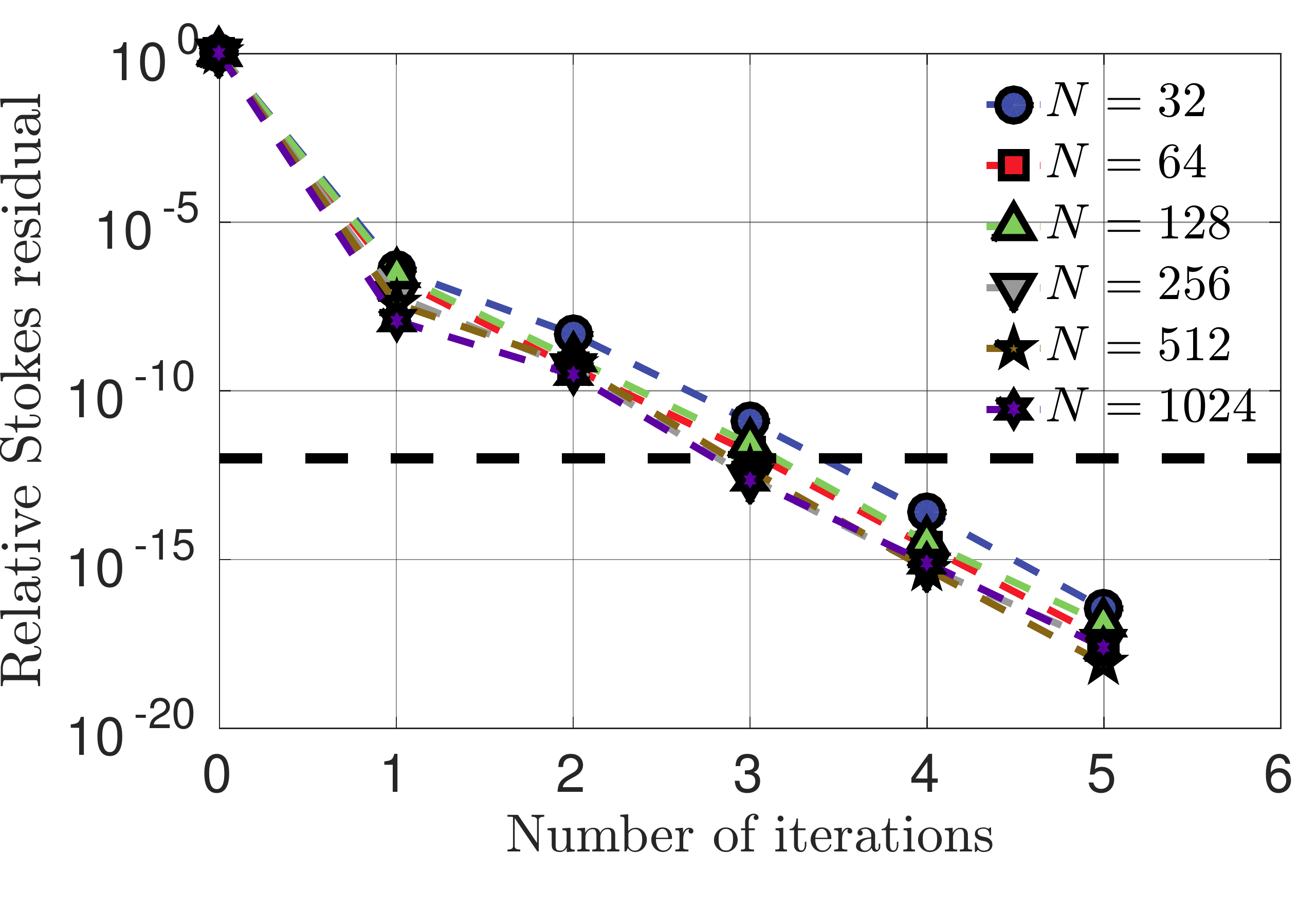}
    \label{MKS_Stokes_Residuals}
  }
  \caption{Relative residual as a function of iteration count for the final solves of the water bubble problem described in
  Sec.~\ref{sec_cons_scale}, for varying grid sizes $N = 32 - 1024$. MKS units are used to specify the
  material properties of both water and air, yielding a Reynolds number of $\Re = 2.71\times 10^4$.
  \subref{MKS_Velocity_Residuals}
  Velocity subdomain solver relative residual $\norm{\A \widehat{\x}_{\u} - \bu}_2/\norm{\bu}_2$ vs. number of iterations.
  \subref{MKS_Pressure_Residuals}
   Pressure subdomain solver relative residual $\norm{-\Lrho \vtheta + \frac{1}{\dt}\left(\bp + \vD \cdot \widehat{\x}_{\u}  \right)}_2/\norm{-\frac{1}{\dt}\left(\bp + \vD \cdot \widehat{\x}_{\u}  \right)}_2$ vs. number of iterations.
   \subref{MKS_Stokes_Residuals}
   Staggered Stokes solver relative residual $\norm{\B \x - \b}_2/\norm{\b}_2$ vs. number of iterations. 
   }
  \label{fig_MKS_residuals}
\end{figure}

In the first case, the bubble is set to have $\rho_\text{i} = 10^3$ and $\mu_\text{i} = 8.9\times10^{-4}$, and the exterior
fluid is set to have $\rho_\text{o} = 1.225$ and $\mu_\text{o} = 1.81 \times 10^{-5}$. This corresponds to the material properties
of water and air, respectively, in meter-kilogram-second (MKS) units.
The initial problem set up is shown in Fig.~\ref{Bubble_Residual_Init_Schematic}.
The Reynolds number based on the properties of the exterior fluid, the diameter of the bubble, and the maximum initial
velocity in the domain is $\Re = 2.71\times 10^4$.
The subdomain solvers with loose relative convergence tolerance of $\epssub = 10^{-2}$ are used for this problem. %; see Sec.~\ref{sec_subdomain} for details. 
For the outer FGMRES solver,  a tight relative convergence tolerance of $\epsstokes = 10^{-16}$ is employed. 
Fig.~\ref{fig_MKS_residuals} summarizes iteration results for the three
linear solvers considered. We only show residual data for the velocity and pressure solves
from the \emph{final} FGMRES iteration of each simulation, although we note that each outer iteration
requires essentially the same number of inner solver iterations. Hence, the number of inner solver V-cycles
per Stokes cycle can be approximated as the number of subdomain solver iterations multiplied by the number of FGMRES 
iterations shown in Fig.~\ref{MKS_Stokes_Residuals}. For the velocity subdomain solver at all grid sizes,
we see that it only takes a single iteration to reduce the residual by several orders of magnitude, 
well below the desired tolerance. This is expected for this particular case because the system is 
strongly diagonally dominant. For the pressure subdomain solver, we see that it takes between three
and six iterations to reduce the relative residual below $\epssub$. Although the required number of iterations 
increases as a function of grid size, it does so at only a modest rate. For the full staggered Stokes system, the 
FGMRES solver converges to machine precision within five iterations for all the grid sizes $N$.
To converge the Stokes system for the wide range of grid sizes shown here to a relative residual tolerance of
$\epsstokes = 10^{-12}$, we would expect to carry out around $3$--$4$ velocity V-cycles
and $12$--$18$ pressure V-cycles. To obtain accurate results for practical multiphase fluid flow problems, 
it is reasonable to converge the Stokes system to a relative residual of around $10^{-6}$, requiring only a single FGMRES
iteration, a single V-cycle for the velocity and around $3$--$6$ V-cycles for the pressure.

\begin{figure}[]
  \centering
  \subfigure[Velocity subdomain residuals]{
    \includegraphics[scale = 0.27]{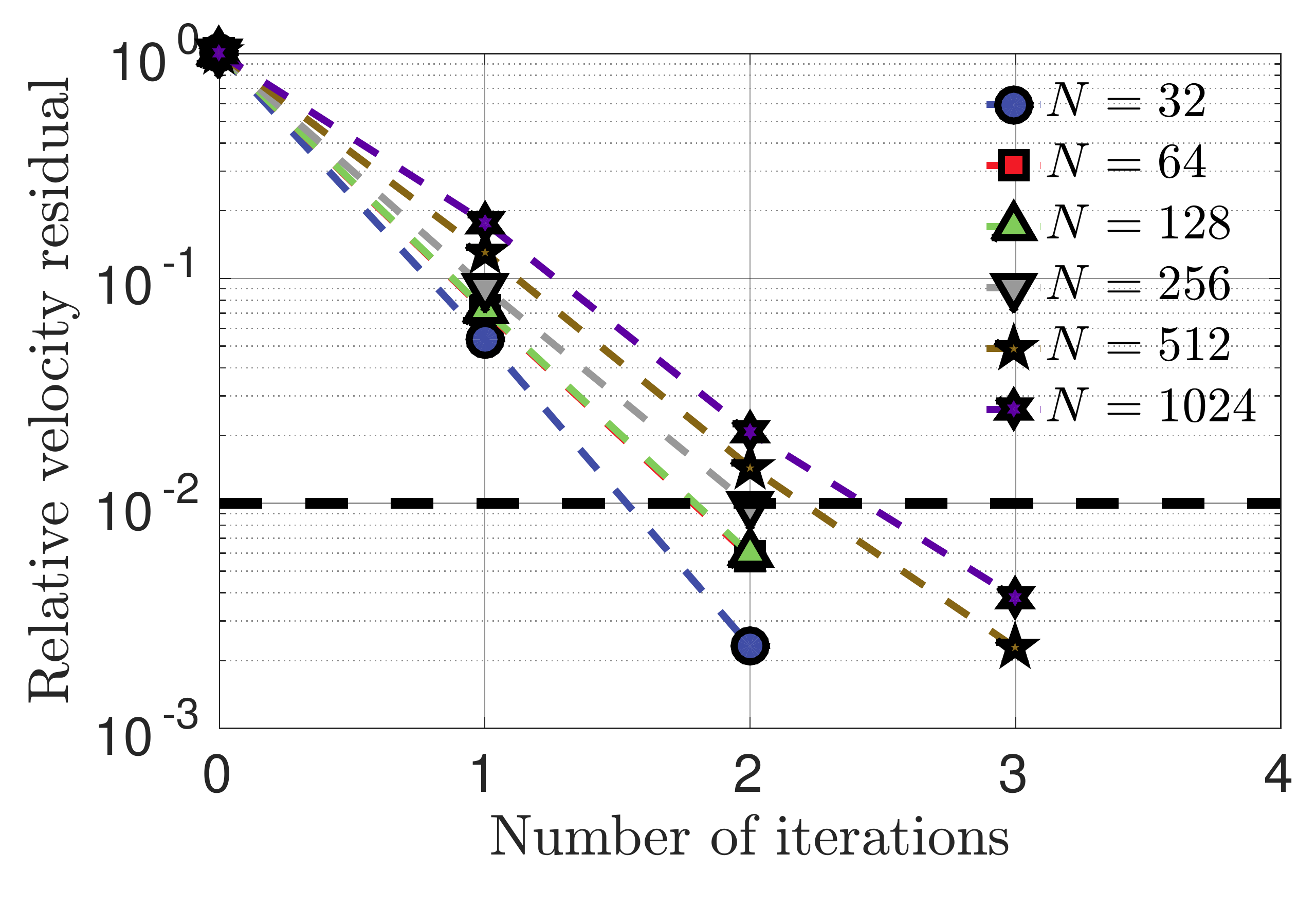}
    \label{CGS_Velocity_Residuals}
  }
   \subfigure[Pressure subdomain residuals]{
    \includegraphics[scale = 0.27]{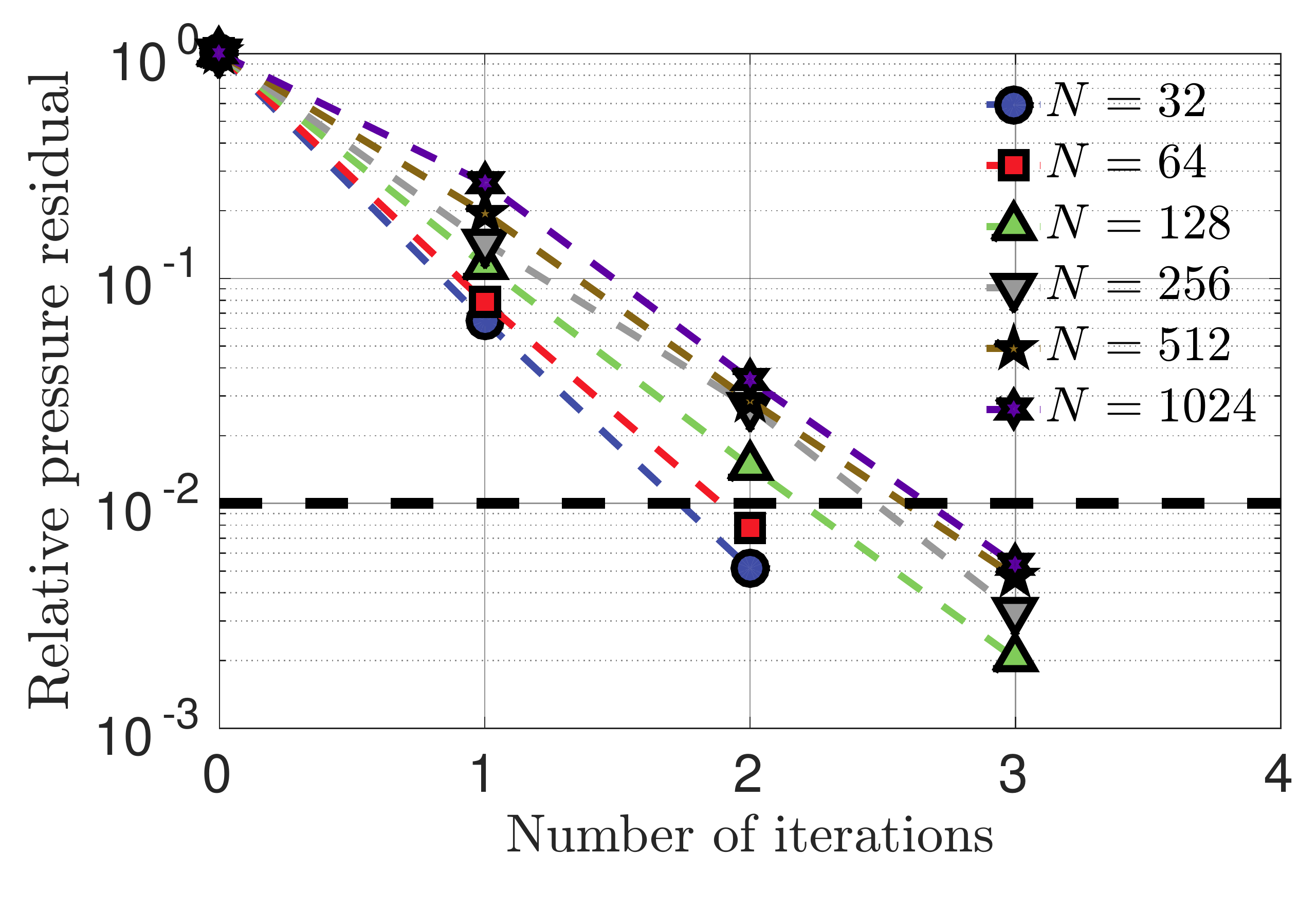}
    \label{CGS_Pressure_Residuals}
  }
   \subfigure[Stokes solver residuals]{
    \includegraphics[scale = 0.27]{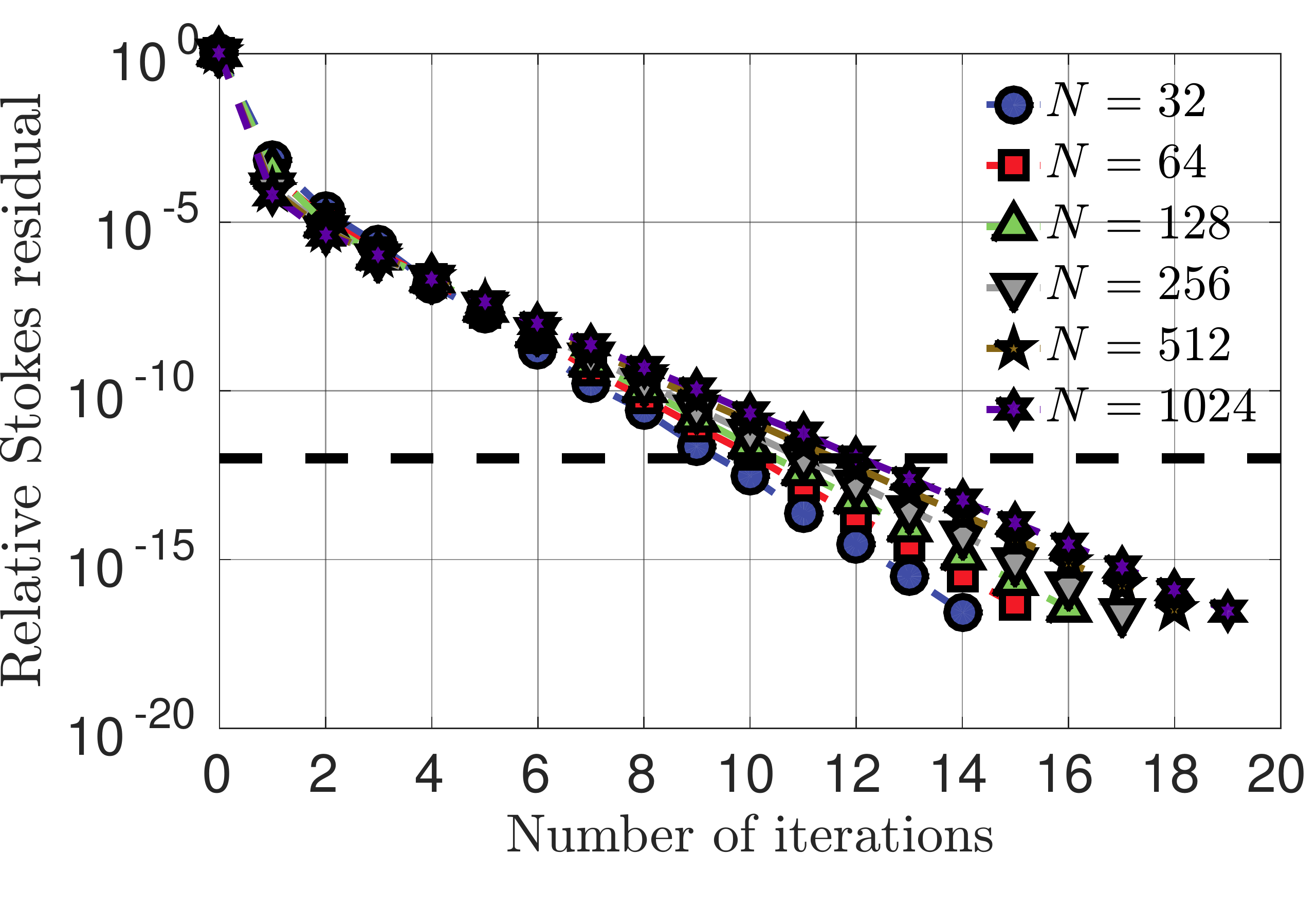}
    \label{CGS_Stokes_Residuals}
  }
  \caption{Relative residual as a function of iteration count for the final solves of the water bubble problem described in
  Sec.~\ref{sec_cons_scale}, for varying grid sizes $N = 32 - 1024$. CGS units are used to specify the
  material properties of both water and air, yielding a Reynolds number of  $\Re = 2.71\times 10^0$.
  \subref{CGS_Velocity_Residuals}
  Velocity subdomain solver relative residual $\norm{\A \widehat{\x}_{\u} - \bu}_2/\norm{\bu}_2$ vs. number of iterations.
  \subref{CGS_Pressure_Residuals}
   Pressure subdomain solver relative residual $\norm{-\Lrho \vtheta + \frac{1}{\dt}\left(\bp + \vD \cdot \widehat{\x}_{\u}  \right)}_2/\norm{-\frac{1}{\dt}\left(\bp + \vD \cdot \widehat{\x}_{\u}  \right)}_2$ vs. number of iterations.
   \subref{CGS_Stokes_Residuals}
   Staggered Stokes solver relative residual $\norm{\B \x - \b}_2/\norm{\b}_2$ vs. number of iterations. 
   }
  \label{fig_CGS_residuals}
\end{figure}

In the second case, the bubble is set to have $\rho_\text{i} = 1$ and $\mu_\text{i} = 8.9\times10^{-3}$, whereas the outside
fluid is set to have $\rho_\text{o} = 1.225 \times 10^{-3}$ and $\mu_\text{o} = 1.81 \times 10^{-4}$. This corresponds to the
material properties of water and air, respectively, in centimeter-gram-second (CGS) units. The Reynolds number based on the properties of the exterior fluid, the diameter of the bubble, and the maximum initial
velocity in the domain is $\Re = 2.71$.
Fig.~\ref{fig_CGS_residuals} 
summarizes iteration results for the three linear solvers considered. In contrast with the higher Reynolds number case, it generally takes
between two and three iterations for the velocity subdomain solver to converge below the desired tolerance. Again,
this is expected, because at this Reynolds number, $\A$ is not as diagonally dominant. Similarly, the pressure
subdomain solver requires between two and three iterations for all grid sizes. The full staggered Stokes system requires
between $14$ and $19$ iterations to converge to machine precision. Hence, to converge the Stokes system to a relative
residual tolerance of $\epsstokes = 10^{-12}$, we would expect to carry out around $20$--$39$ velocity V-cycles and a comparable number of pressure V-cycles.
To achieve a relative residual tolerance of $10^{-6}$, we would expect to carry out around $6$--$9$ V-cycles for both the velocity and pressure.

The results here suggest that the preconditioned iterative solver is scalable for high density and high viscosity ratio flows, including practical air-water interfacial flows at both low and high Reynolds numbers. The number of iterations required to converge the outer and inner solvers are relatively insensitive to grid size. Similar convergence
behaviors are observed for other combinations of physical boundary conditions (results not shown). In the following sections, we 
mainly focus on multiphase flows involving air-water interaction, for which the solver parameters discussed here are
reasonably efficient.

\section{Numerical examples}
\label{sec_examples}
This section investigates several multiphase flow problems to verify the
accuracy, consistency, and stability of the present numerical method. Both conservative and 
non-conservative formulations are used, and differences between the formulations are examined.
We also compare 
our results to benchmark problems drawn from the multiphase flow literature.
We employ $\ncycles = 2$ cycles of fixed-point 
iteration for all the examples in this section.
We note that errors lower than machine precision may be truncated during computation,
and we only report errors lower than $10^{-16}$ as $\le 10^{-16}$.

\subsection{Static bubble with surface tension}
We first demonstrate that our treatment of surface tension is \emph{well-balanced}
with respect to the treatment of the pressure gradient in the hydrostatic limit. A bubble is placed in a computational domain
$\Omega = [0,L]^2$ with $L = 1$, which is discretized by a uniform $N \times N$ grid. The radius of the bubble
is $R = 0.25L$ with initial center position $(X_0, Y_0) = (L/2, L/2)$. Both the fluids inside and outside the
bubble are initially at rest. Zero normal and tangential velocity boundary conditions are imposed on
the boundaries of the computational domain. The only momentum body force  considered for this case is the surface
tension force; other body forces such as gravity are neglected. The conservative discretization is used for all cases considered here,
but the non-conservative discretization yields similar results because these flows are not
dominated by convection (data not shown). A relative convergence tolerance of $\epsstokes = 10^{-15}$ is
specified for the FGMRES solver, which can corresponds to absolute tolerances in the range of $10^{-10}-10^{-15}$.
Two grid cells ($\ncells = 2$) of smearing are used on either side of the interface.

As a first test case, we consider an inviscid flow with inner and outer viscosities $\mu_{\text{i}} = \mu_{\text{o}} = 0$.
The grid size is taken to be $N = 80$, and a constant time step size $\dt = 1/(12.5N)$ is used.
Density ratios of $\rho_{\text{i}}/\rho_{\text{o}} = 1, 10^3, $ and $10^6$ are considered, and each case is run for
$50$ time steps. For this particular case, the surface tension should exactly balance the pressure
difference across the interface, resulting in a stationary bubble with zero velocity everywhere in the domain.
The exact pressure difference is analytically given by the Young-Laplace equation
\begin{equation}
\dP_{\textrm{exact}} = \sigma \kappa = \frac{\sigma}{R},
\end{equation}
in which the surface tension coefficient is set to $\sigma = 73$, and the exact curvature of the
bubble is $\kappa = 1/R$. This case has been extensively investigated by a number of previous studies,
including Williams et al.~\cite{Williams1998}, Francois et al.~\cite{Francois2006}, and Patel and Natarajan~\cite{Patel2017}.
If an unbalanced treatment of surface tension and pressure gradient is used, significant parasitic currents
will be generated during the first  time step, and the exact pressure jump will not be captured.
As in previous studies, we use the exact curvature for the surface tension force calculation.
Because the fluid should remain at rest, any nonzero velocities are the result of solver error.
We assess the accuracy of the solver in terms of the $L^1$ and $L^\infty$ norms of the velocity along with the relative error in the pressure jump
\begin{equation}
E(\dP) = \frac{\left| \dP - \dP_{\textrm{exact}}\right|}{\left|\dP_{\textrm{exact}}\right|},
\end{equation}
in which $\dP = p_{\text{i}} - p_{\text{o}}$ is the numerically computed pressure difference across the interface
and $\dP_{\textrm{exact}} = 73/0.25 = 292$. Tables~\ref{tab_static_droplet_dt} and~\ref{tab_static_droplet_50dt}
show the errors in velocity and pressure at $t = \dt$ and $t = 50 \dt$, respectively. For all density ratios, we see that
the errors are close to machine precision, indicating negligible spurious currents and a numerical balance between
pressure and surface tension forces.

\begin{table}
    \centering
    \caption{Errors in velocity and pressure after a single time step for the two-dimensional static and inviscid bubble.
    The exact curvature $\kappa = 1/R$ is used for the surface tension force calculation.}
    \rowcolors{3}{}{gray!10}
    \begin{tabular}{*4c}
        \toprule
        & \multicolumn{3}{c}{Errors at $t = \dt$} \\
        \cmidrule(lr){2-4}
        $\rho_{\text{i}}/\rho_{\text{o}}$  & $\|\u\|_1$ & $\|\u\|_{\infty}$ & $E(\dP)$ \\
        \midrule
        $1$ & $3.76 \times 10^{-14}$ & $1.11 \times 10^{-13}$ & $9.95 \times 10^{-14}$\\
        $10^3$ & $2.03 \times 10^{-15}$ & $1.20 \times 10^{-13}$ & $\le 10^{-16}$\\
        $10^6$ & $8.44 \times 10^{-13}$ & $4.73 \times 10^{-11}$ & $5.84 \times 10^{-16}$\\
        \bottomrule
    \end{tabular}
    \label{tab_static_droplet_dt}
\end{table}

\begin{table}
    \centering
    \caption{Errors in velocity and pressure after a $50$ time steps for the two-dimensional static and inviscid bubble.
    Exact curvature $\kappa = 1/R$ is used for the surface tension force calculation.}
    \rowcolors{3}{}{gray!10}
    \begin{tabular}{*4c}
        \toprule
        & \multicolumn{3}{c}{Errors at $t = 50 \dt$} \\
        \cmidrule(lr){2-4}
        $\rho_{\text{i}}/\rho_{\text{o}}$  & $\|\u\|_1$ & $\|\u\|_{\infty}$ & $E(\dP)$ \\
        \midrule
        $1$ & $4.17 \times 10^{-16}$ & $9.36 \times 10^{-15}$ & $5.84 \times 10^{-16}$\\
        $10^3$ & $1.95 \times 10^{-14}$ & $5.16 \times 10^{-13}$ & $\le 10^{-16}$\\
        $10^6$ & $3.61 \times 10^{-12}$ & $1.81 \times 10^{-10}$ & $\le 10^{-16}$\\
        \bottomrule
    \end{tabular}
    \label{tab_static_droplet_50dt}
\end{table}

\begin{figure}[]
  \centering
  \subfigure[Pressure difference relative error]{
    \includegraphics[scale = 0.25]{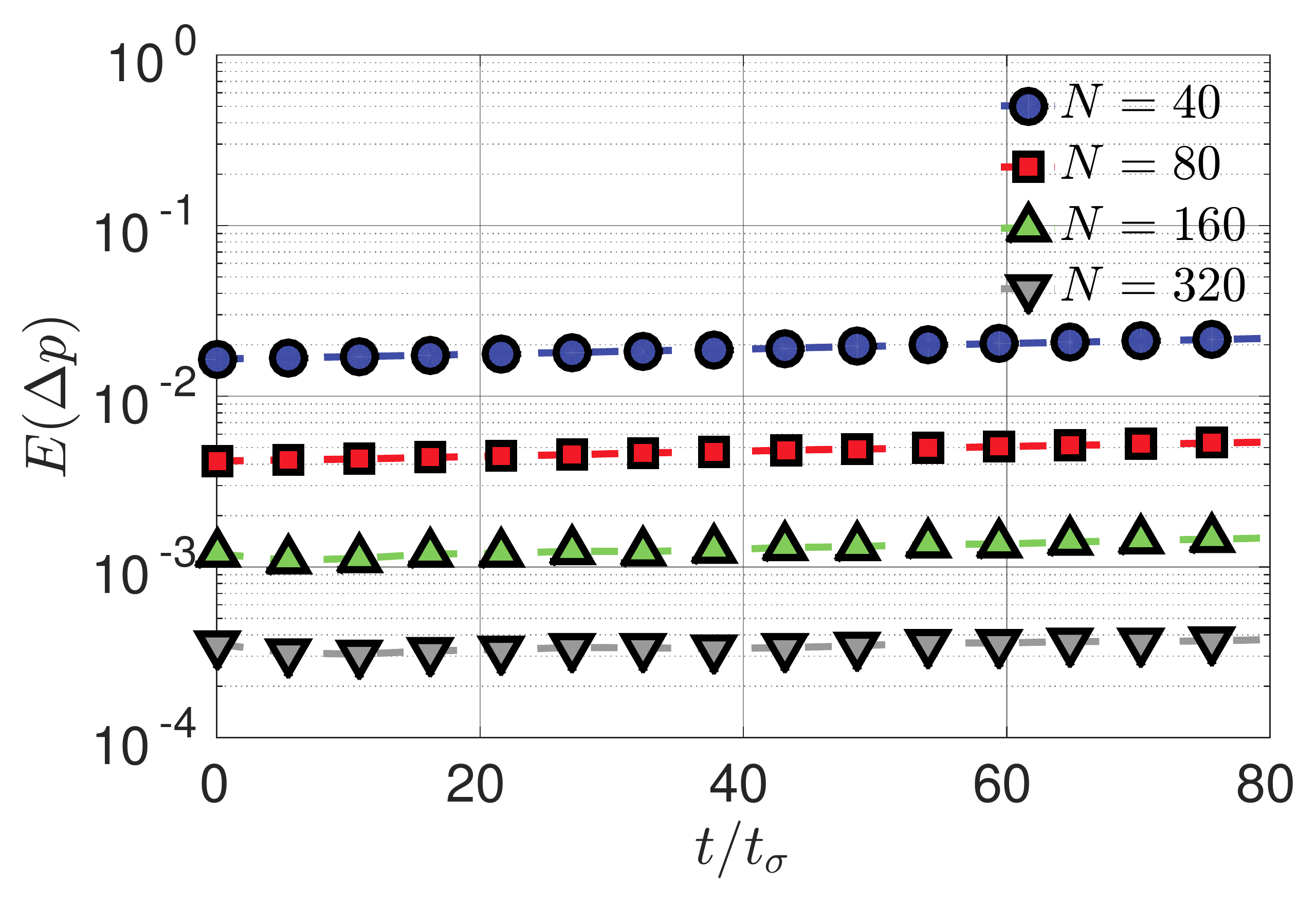}
    \label{Static_Bubble_Pressure_Error}
  }
   \subfigure[Velocity $L^1$ norm]{
    \includegraphics[scale = 0.25]{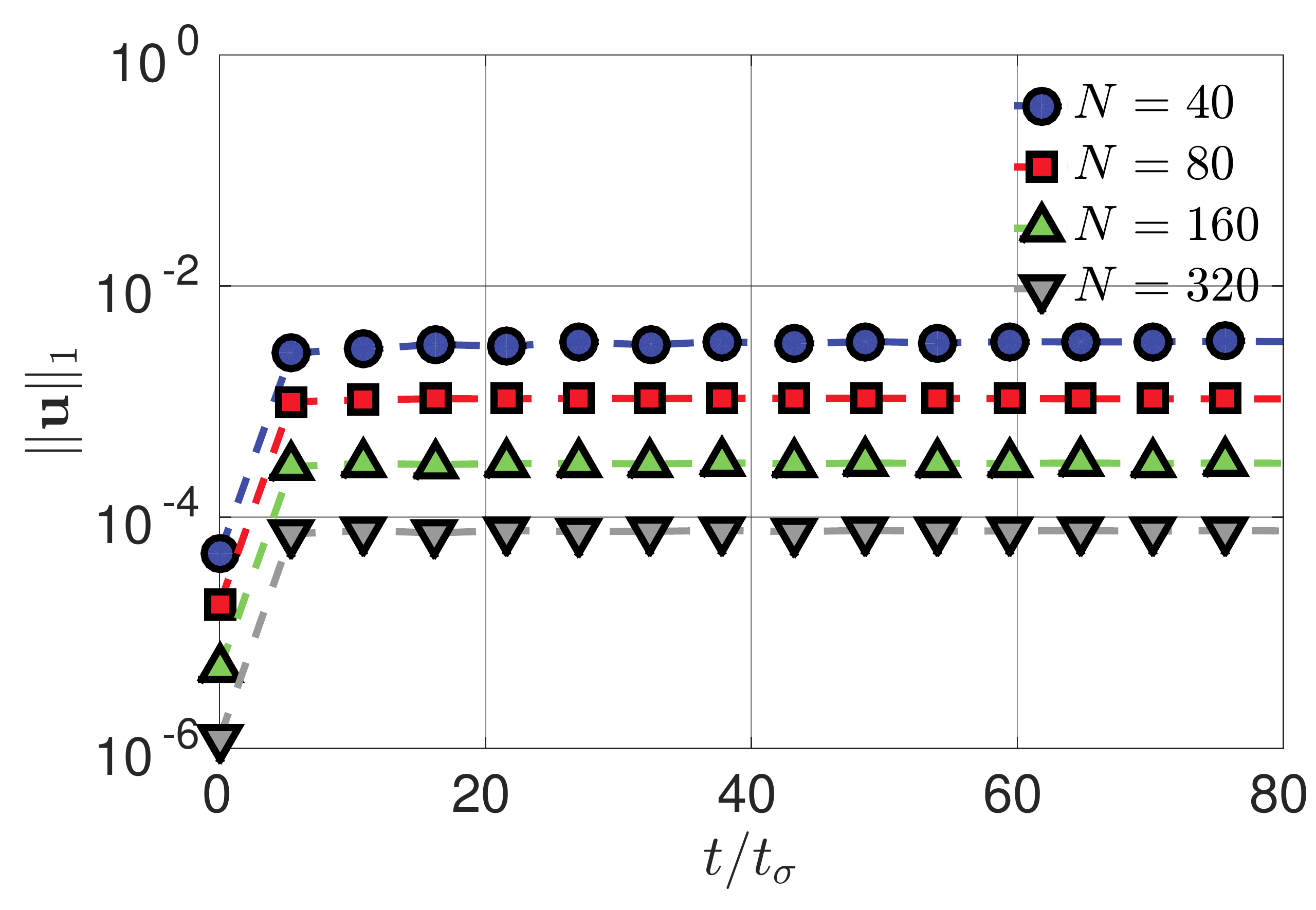}
    \label{Static_Bubble_L1Velocity_Error}
  }
   \subfigure[Velocity $L^{\infty}$ norm]{
    \includegraphics[scale = 0.25]{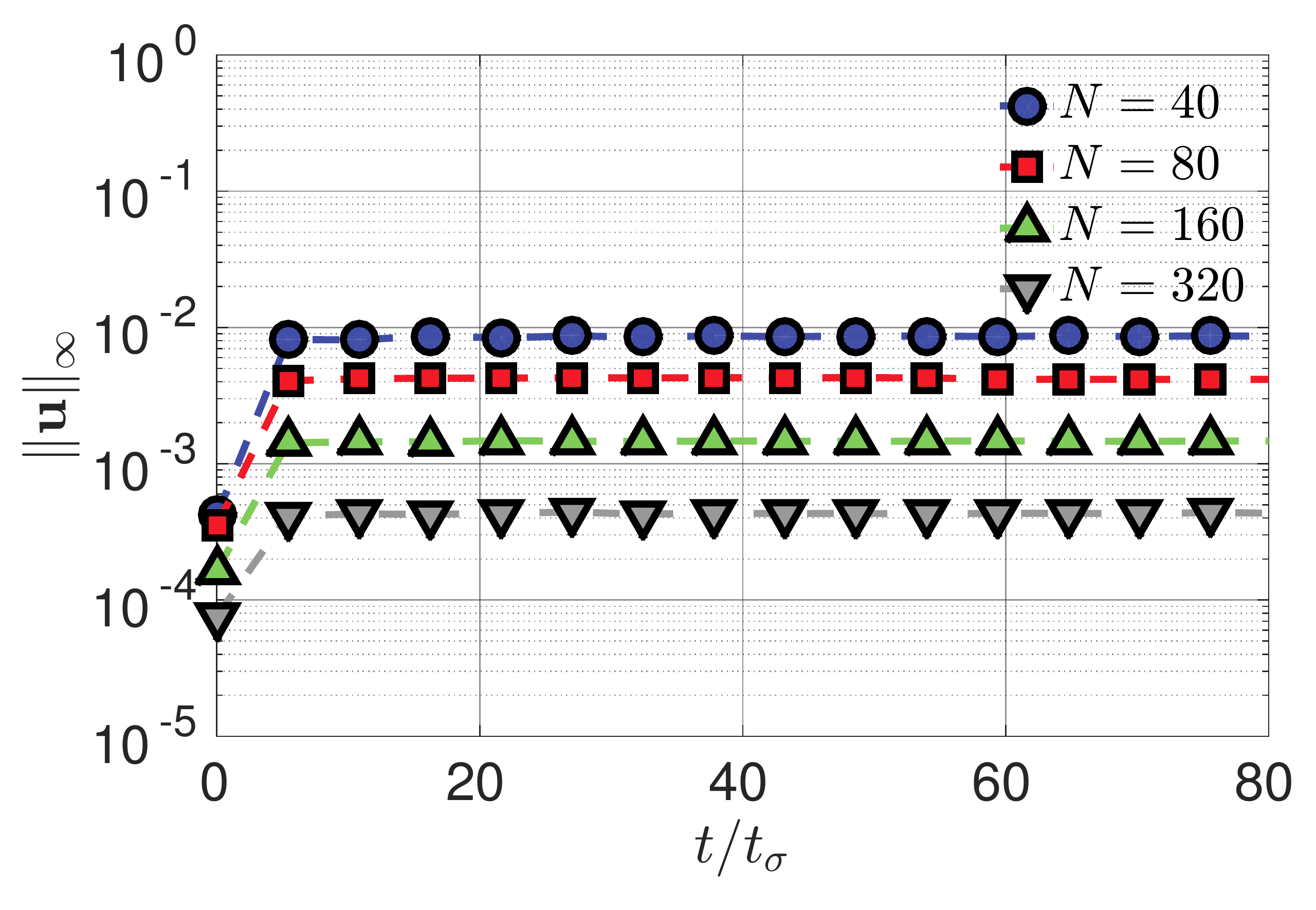}
    \label{Static_Bubble_LinfVelocity_Error}
  }
  \caption{Temporal evolution of errors for a viscous bubble with $\rho_{\text{i}}/\rho_{\text{o}} = 10^3$ and $\La = 12000$,
  for grid sizes $N = 40$ -- $320$. Here, curvature is computed numerically from the level set function.
  \subref{Static_Bubble_Pressure_Error}
  Pressure difference relative error $E(\dP)$ vs. nondimensional time.
  \subref{Static_Bubble_L1Velocity_Error}
   $L^1$ norm of velocity vs. nondimensional time.
   \subref{Static_Bubble_LinfVelocity_Error}
    $L^{\infty}$ norm of velocity vs. nondimensional time.
   }
  \label{fig_Static_Bubble_Error}
\end{figure}

Next, we consider the temporal evolution of velocity and pressure errors for a viscous droplet
suspended in fluid. We consider the same geometric parameters as the previous example
for the bubble and the domain size. The density ratio is held fixed at $\rho_{\text{i}}/\rho_{\text{o}} = 10^3$,
and we consider grids with $N = 40, 80, 160,$ and $320$. A constant time step size of $\dt = 1/(50 N)$
is used, and the surface tension coefficient is again set to $\sigma = 73$. The viscosity in each fluid
is determined using the dimensionless Laplace number $\La = \rho D \sigma/\mu^2 = 12000$, in which $D = 2R$ is the diameter of the bubble.
Using the parameters described here yields a viscosity ratio of $\mu_{\text{i}}/\mu_{\text{o}} = 31.62$, and both fluids are initially
at rest. In contrast with the previous example, here the curvature is computed directly from the level set
function. Time is nondimensionalized using the capillary time scale
$t_{\sigma} = \sqrt{\rho_{\text{o}} D^3/\sigma}$. Similar cases have been numerically investigated by Popinet~\cite{Popinet2009}
and by Abadie et al.~\cite{Abadie2015}.

\begin{figure}[]
  \centering
    \includegraphics[scale = 0.3]{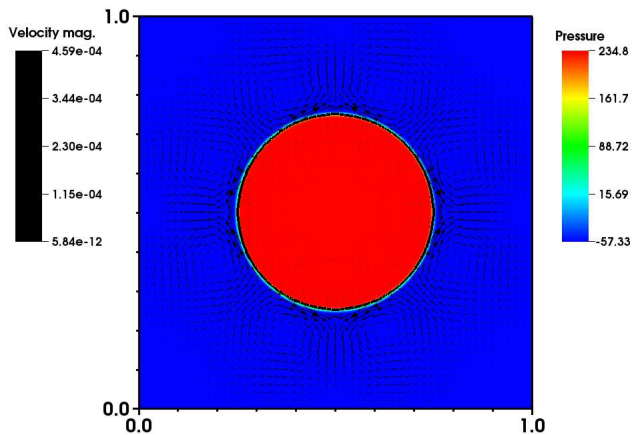}
  \caption{Velocity vectors and pressure of the static bubble with $\rho_{\text{i}}/\rho_{\text{o}} = 10^3$ and $\La = 12000$,
  for grid size $N = 320$ at time $t/t_\sigma = 60.41$.
  Curvature is computed numerically from the level set function.}
  \label{fig_La12000_bubble_t2.5}
\end{figure}

Fig.~\ref{fig_Static_Bubble_Error} shows the time evolution of $E(\dP)$, $\|\u\|_1$, and $\|\u\|_\infty$.  
We observe that the errors do not grow substantially as a function of time, and that each computation maintains
stability.
These results are consistent with the results of other level set based multiphase flow solvers.
In particular, it is known that the reinitialization procedure will slightly shift the interface every time step,
which can prevent the spurious velocities from decaying to machine precision~\cite{Abadie2015}. In contrast, certain 
geometric VOF methods exhibit exponential decay of the spurious velocities because of the absence of any redistancing process~\cite{Popinet2009}.
Nonetheless, the errors shown in shown in Fig.~\ref{fig_Static_Bubble_Error} decrease as the resolution increases, yielding convergence rates of $1.98$ for $E(\dP)$,
$1.94$ for $\|\u\|_1$, and $1.75$ for $\|\u\|_\infty$ between the two finest cases $N = 160$ and
$N = 320$. Fig.~\ref{fig_La12000_bubble_t2.5} shows the velocity vectors and pressure for the finest
case $N = 320$ at $t/t_\sigma = 60.41$.
Although we do see some spurious velocities across
the computational domain, the large internal pressure required to maintain the bubble's shape is
accurately resolved.
These results indicate that the surface tension treatment described here is well-balanced and does not significantly reduce
the expected order of accuracy of the fluid solver.

\subsection{Gas/liquid tank}
This section demonstrates that our treatment of the gravitational body force is
well-balanced with respect to the treatment of the pressure gradient.
The computational domain $\Omega = [0,L]^2$ with $L = 1$
is filled halfway from $y = 0$ to $y = L/2$ with a heavy fluid of density $\rho_{\text{h}}$.
The remainder of the tank, from $y = L/2$ to $y = L$, is occupied by a lighter
fluid with density $\rho_{\text{l}} = 1$, and the density ratio $\rho_{\text{h}}/\rho_{\text{l}}$ is varied.
The viscosity is set to zero for both fluids, $\mu_{\text{l}} = \mu_h = 0$.
The domain is discretized by a uniform $N \times N$ grid, and both the fluids
are initially at rest. A gravitational acceleration $\mathbf{g} = (0, -g_y) = (0, -9.81)$ is 
specified, and surface tension is neglected.
The conservative discretization is used for all cases considered here, but we
found that the non-conservative discretization yields similar results because these
flows are not dominated by convection (data not shown).
A relative convergence
tolerance of $\epsstokes = 10^{-15}$ is specified for the FGMRES solver, and
two grid cells of smearing ($\ncells = 2$) are used on either side of the interface.

As a first test case, we prescribe homogenous normal velocity conditions
on the left, bottom, and right boundaries and homogenous normal traction
conditions on the top boundary along with homogenous tangential velocity conditions on all boundaries of the computational domain.
The grid is held fixed with size $N = 100$ and a constant
time step size $\dt = 1/(10 N)$ is used. Density ratios of $\rho_{\text{h}}/\rho_{\text{l}} = 10^1, 10^3, $
and $10^6$ are considered and each case is run for $50$ time steps. For this particular
case, the gravitational force should exactly balance the pressure difference across
the interface, resulting in a stationary tank of fluid with zero velocity everywhere in the
domain. Similar cases has been computationally investigated by Montazeri et al.~\cite{Montazeri2012}
and by Patel and Natarajan~\cite{Patel2017}. %We note that in contrast with these previous studies, we
%employ a purely staggered-grid discretization of the pressure and velocity degrees of freedom.
Because the pressure is defined on cell-centers of the staggered grid, the analytical pressure jump
for this problem can be computed by integrating the momentum equation (ignoring velocity terms):
\begin{align}
&\int_{\dy/2}^{L-\dy/2} \grad p(y) \,dy = \int_{\dy/2}^{L-\dy/2} \rho(y) \mathbf{g}  \,dy
= -g_y \left(\int_{L/2}^{L-\dy/2} \rho_{\text{l}}  \,dy + \int_{\dy/2}^{L/2} \rho_{\text{h}}  \,dy\right) \nonumber \\
\Rightarrow &\dP_{\textrm{exact}} = p\bigg\rvert^{L-\dy/2}_{\dy/2} =  - \frac{1}{2} g_y \left(\rho_{\text{h}} + \rho_{\text{l}}\right)\left(L-\dy\right) \label{eq_column_pressure_jump},
\end{align}
in which the pressure is evaluated at the cell centers adjacent to the top and bottom computational
boundaries and $\dy = L/N$ is the vertical grid spacing for this particular problem.
If an unbalanced treatment of gravity and pressure gradient is used, significant parasitic currents
will be generated, and the exact pressure jump will not be captured.
Similar to the static bubble case in the previous section, two quantitative measurements are used to 
assess the accuracy of the simulation: the $L^{\infty}$ norm in velocity, and the relative error in pressure 
jump $E(\dP)$ using the numerically computed pressure difference between $y = \dy/2$ and $y = L-\dy/2$,
and the exact pressure difference $\dP_{\textrm{exact}} = -9.81/2 \times (\rho_{\text{h}} + \rho_{\text{l}}) \times (1-0.01)$.

Tables~\ref{tab_static_column_dt} and~\ref{tab_static_column_50dt}
show the errors in velocity and pressure at $t = \dt$ and $t = 50 \dt$, respectively. For all density ratios considered, we see that
the errors are close to machine precision, indicating negligible spurious currents and a numerical balance between
pressure and gravitational body force.

\begin{table}
    \centering
    \caption{Errors in velocity and pressure after a single time step for the two-dimensional static and inviscid fluid column.}
    \rowcolors{3}{}{gray!10}
    \begin{tabular}{*4c}
        \toprule
        & \multicolumn{2}{c}{Errors at $t = \dt$} \\
        \cmidrule(lr){2-3}
        $\rho_{\text{h}}/\rho_{\text{l}}$ & $\|\u\|_\infty$ & $E(\dP)$ \\
        \midrule
        $10^1$ & $2.53 \times 10^{-16}$ & $5.45 \times 10^{-15}$\\
        $10^3$ & $7.21 \times 10^{-15}$ & $9.32 \times 10^{-14}$\\
        $10^6$ & $4.40 \times 10^{-12}$ & $5.42 \times 10^{-12}$\\
        \bottomrule
    \end{tabular}
    \label{tab_static_column_dt}
\end{table}

\begin{table}
    \centering
    \caption{Errors in velocity and pressure after a $50$ time steps for the two-dimensional static and inviscid fluid column.}
    \rowcolors{3}{}{gray!10}
    \begin{tabular}{*4c}
        \toprule
        & \multicolumn{2}{c}{Errors at $t = 50 \dt$} \\
        \cmidrule(lr){2-3}
        $\rho_{\text{h}}/\rho_{\text{l}}$  & $\|\u\|_\infty$ & $E(\dP)$ \\
        \midrule
        $10^1$ & $\le 10^{-16}$ & $1.33 \times 10^{-16}$\\
        $10^3$ & $\le 10^{-16}$ & $\le 10^{-16}$\\
        $10^6$ & $1.62 \times 10^{-16}$ & $1.92 \times 10^{-16}$\\
        \bottomrule
    \end{tabular}
    \label{tab_static_column_50dt}
\end{table}

Next, we consider the dynamic case of a tank being filled with fluid. The same numerical 
parameters from the previous case are used, except that the normal velocity at the bottom
boundary is set to be unity. Over time, the fluid will displace the gas out of the domain
until the tank consists only of the fluid phase. Only the case for density ratio of $\rho_{\text{h}}/\rho_{\text{l}} = 10^3$
is presented here, although the results are similar for a wide range of density ratios.
%(data from non-conservative formulation not shown).
Montazeri et al.~demonstrated that the use of an unbalanced formulation will produce
significant interface deformation resulting from the spurious currents, which eventually destabilizes 
the computation~\cite{Montazeri2012}.

\begin{figure}[]
  \centering
  \subfigure[t = 0.0]{
    \includegraphics[scale = 0.18]{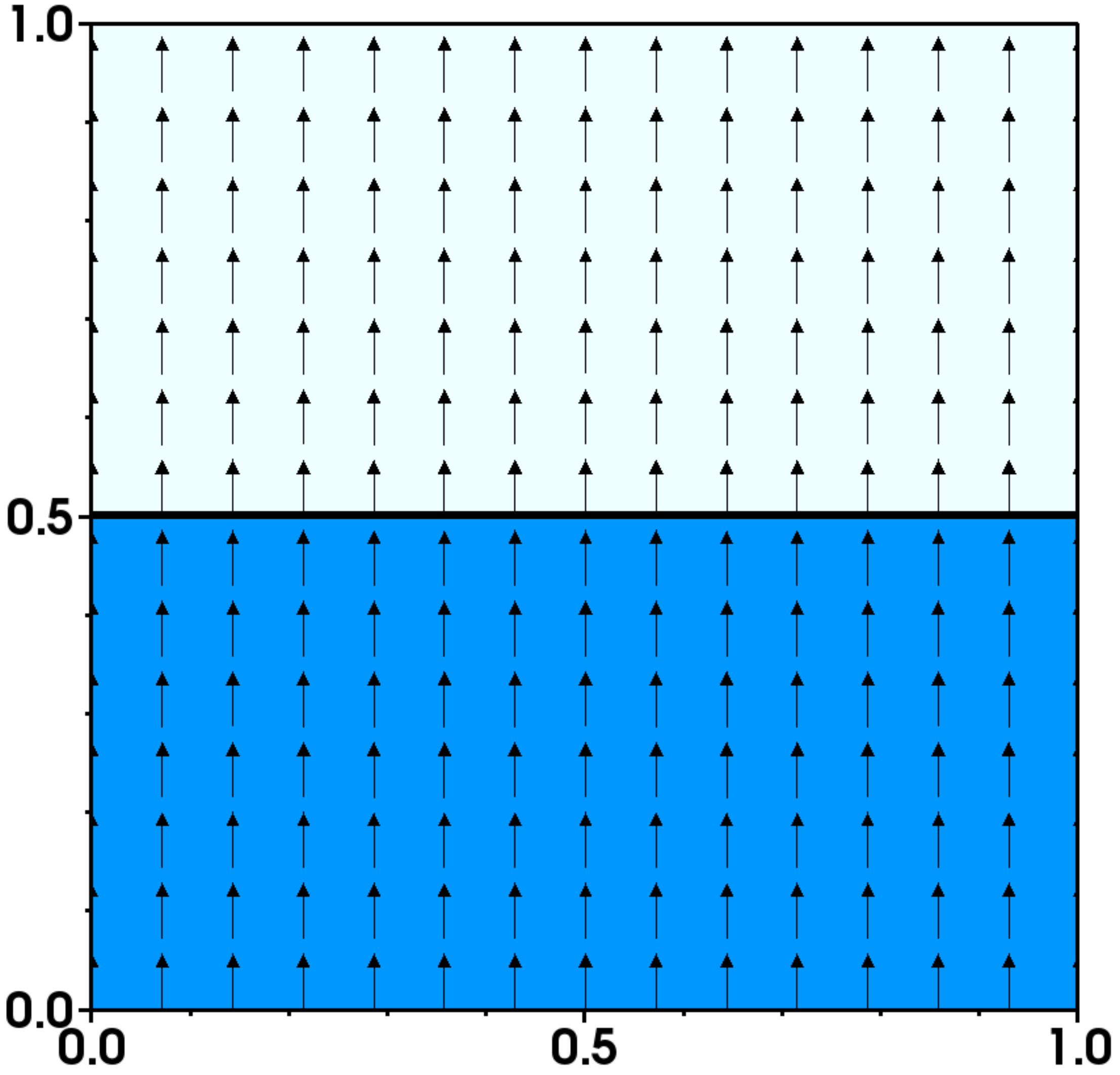}
    \label{liquid_filling_t0}
  }
   \subfigure[t = 0.2]{
    \includegraphics[scale = 0.18]{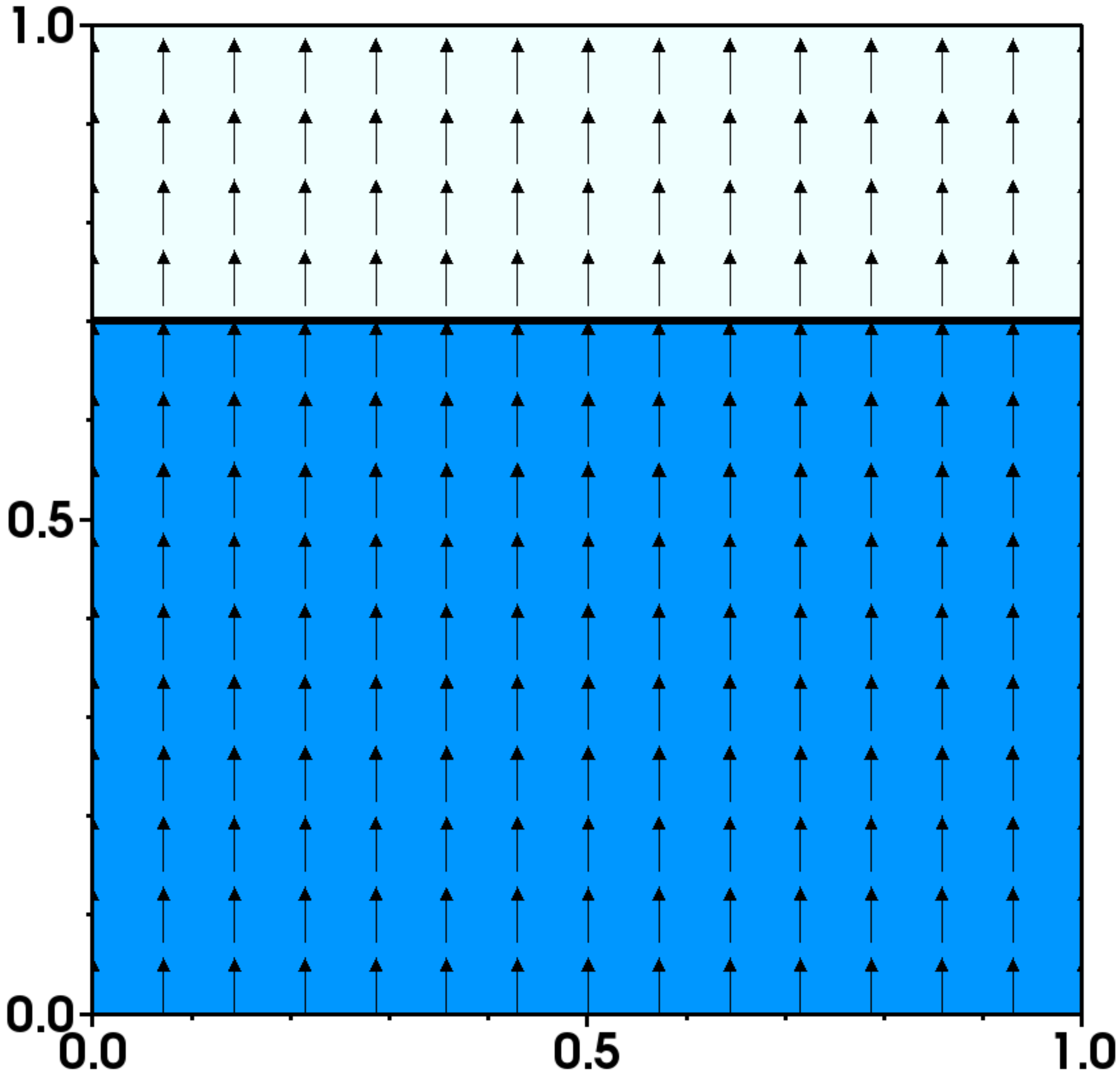}
    \label{liquid_filling_t02}
  }
   \subfigure[t = 0.4]{
    \includegraphics[scale = 0.18]{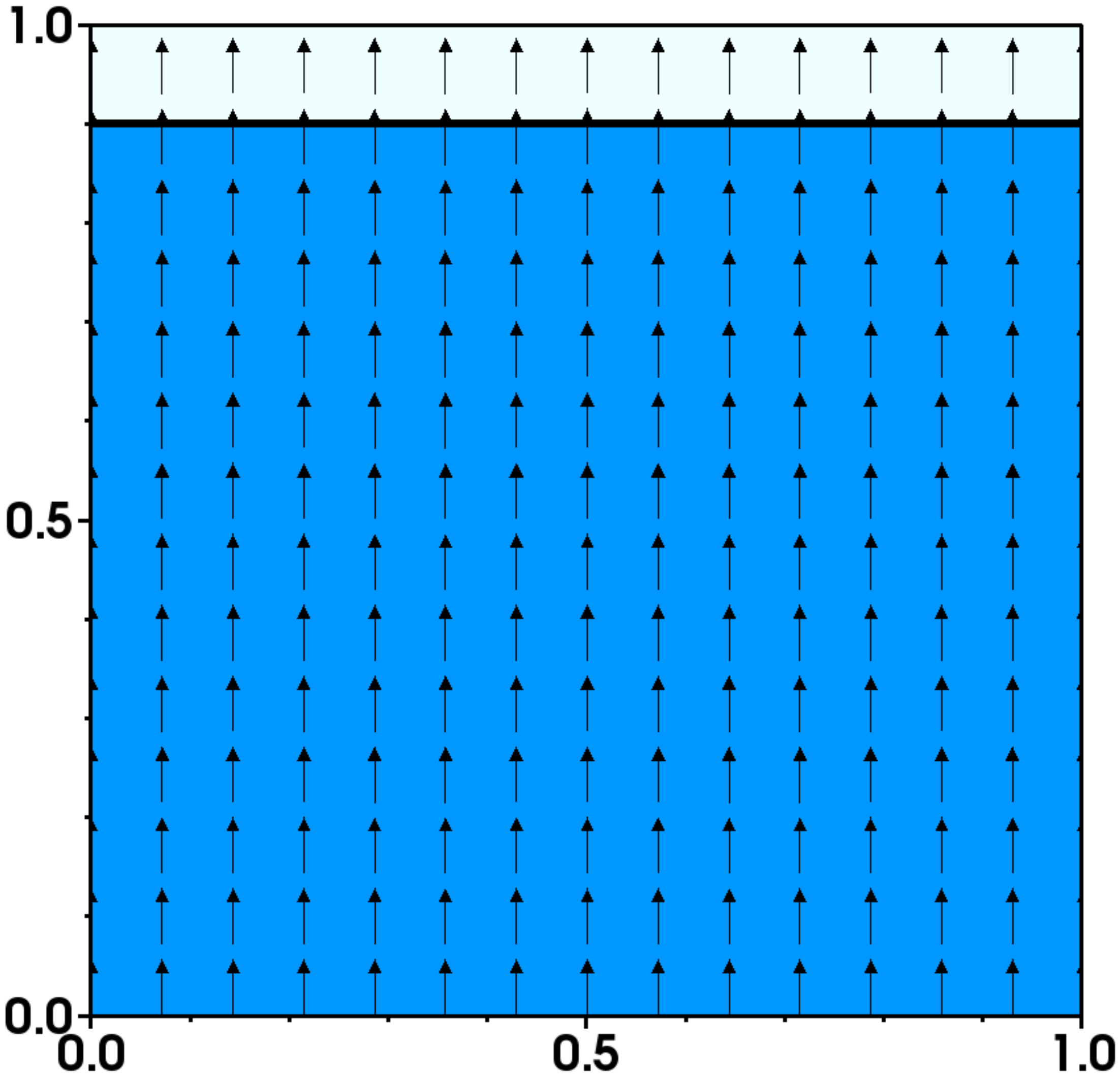}
    \label{liquid_filling_t04}
  }
  \caption{Temporal evolution of a tank being filled with heavy liquid for $\rho_{\text{h}}/\rho_{\text{l}} = 10^3$ and $N = 100$. Velocity vectors are denoted by arrows ($\rightarrow$, black).
   }
  \label{fig_Filling_Column}
\end{figure}

Fig.~\ref{fig_Filling_Column} shows three snapshots in time for the evolution of the filling
water tank. We see the fluid-gas interface evolve upward with no deformation because
negligible spurious horizontal velocities generated.
These cases demonstrate that even though we are treating gravity as a momentum
body force acting throughout the entire domain (and \emph{not} as an interfacial force as done in
by Montazeri et al.~\cite{Montazeri2012}), the staggered-grid discretization provides a well-balanced
pressure gradient and gravity forcing. The reason is attributed to the fact that the pressure gradient and the gravity
forces are defined at the same spatial location, namely, the Cartesian grid faces.
%for the staggered grid configuration at which gravity force and pressure gradient are defined.
It was also recently recognized by Patel and Natarajan~\cite{Patel2017} that it is \emph{not} necessary to express gravity force in terms of a gradient operator  to obtain a well-balanced formulation.

\subsection{Convection of high density droplet}
\label{sec_dense_droplet}
In this section, we demonstrate the importance of consistent mass and momentum transport to achieve
stability in high density ratio flows. A dense bubble is placed in a fully periodic computational domain
$\Omega = [0, L]^2$ with $L = 1$, which is discretized by an $N \times N$ grid.
The radius of the bubble is $R = 0.2L$ with initial center position $(X_0, Y_0) = (L/4, L/2)$.
The density ratio between the bubble and the outer fluid is $\rho_{\text{i}}/\rho_{\text{o}} = 10^6$, and the viscosity
is set to $\mu = 0$ in the entire domain. The initial velocity is set via a smoothed Heaviside function defined on faces
%defined on the faces of the grid cells
\begin{equation}
\label{eq_bubble_init_velocity}
\widetilde{H}^f_{i-\half,j} = 
\begin{cases} 
       0,  & \phi_{i-\half,j} < -\dx\\
        \frac{1}{2}\left(1 + \frac{1}{\dx} \phi_{i-\half,j} + \frac{1}{\pi} \sin\left(\frac{\pi}{\dx} \phi_{i-\half,j}\right)\right) ,  & |\phi_{i-\half,j}| \le \dx\\
        1,  & \textrm{otherwise}
\end{cases}
\end{equation}
in which $\phi(\x, 0) = \sqrt{\left(x-X_0 \right)^2 + \left(y-Y_0 \right)^2} - R$ is the initial signed distance
function away from the circular interface and the face centered level set is obtained by averaging $\phi$
in the two adjacent cell centers $\phi_{i-\half,j} = \frac{1}{2} \left(\phi_{i-1,j} +\phi_{i,j}\right)$. The initial horizontal 
velocity $u(\x,0)$ is set to be unity inside the bubble and zero outside the bubble, i.e.
$u_{i-\half,j} = 1 - \widetilde{H}^f_{i-\half,j}$. The vertical velocity $v$ is initialized to be identically zero.

Notice that this initial velocity profile is not discretely divergence-free. Because this is an inviscid case,
numerical instabilities and errors that are generated during the first time step can persist for the duration of the 
simulation. Hence, we apply a density-weighted projection to construct a discretely mass-conserving initial
velocity $\up(\x,0)$. This can be achieved by numerically solving the Poisson problem
\begin{equation}
\label{eq_bubble_proj_poisson}
-\vD \cdot \left(\frac{1}{\rho(\x,0)} \G{\psi(\x)} \right) = - \vD \cdot \u(\x,0),
\end{equation}
for $\psi(\x)$ and then adding its gradient to the initial velocity field,
\begin{equation}
\label{eq_bubble_proj_velocity}
\up(\x,0) = \u(\x,0) - \frac{1}{\rho(\x,0)} \G \psi(\x).
\end{equation}
This problem has been studied numerically by Bussman et al.~\cite{Bussmann2002},
Desjardins and Moureau~\cite{Desjardins2010}, Ghods and Herrmann~\cite{Ghods2013}, 
and Patel and Natarajan~\cite{Patel2017}.
For all of the following cases, the density is set via the level set function at the beginning
of each time step with one grid cell of smearing ($\ncells = 1$) on either side of the interface. 
We find that smearing over two or more cells leads to a more distorted interface for this particular 
problem. This is expected because analytically the dense bubble mimics
a thick solid blob, but numerically, as the interface of the blob thickens, it acts more like a fluidic blob whose interface tends 
to ``flow" under the pressure drag. Each case is run until
$t = 1$ with a constant time step size of $\dt = 1/(31.25 N)$. 
Under the prescribed conditions, the dense rigid blob should maintain its circular shape and return exactly to its initial position, analytically.

\begin{figure}[]
  \centering
  \subfigure[$N = 32, t = 1$]{
    \includegraphics[scale = 0.2]{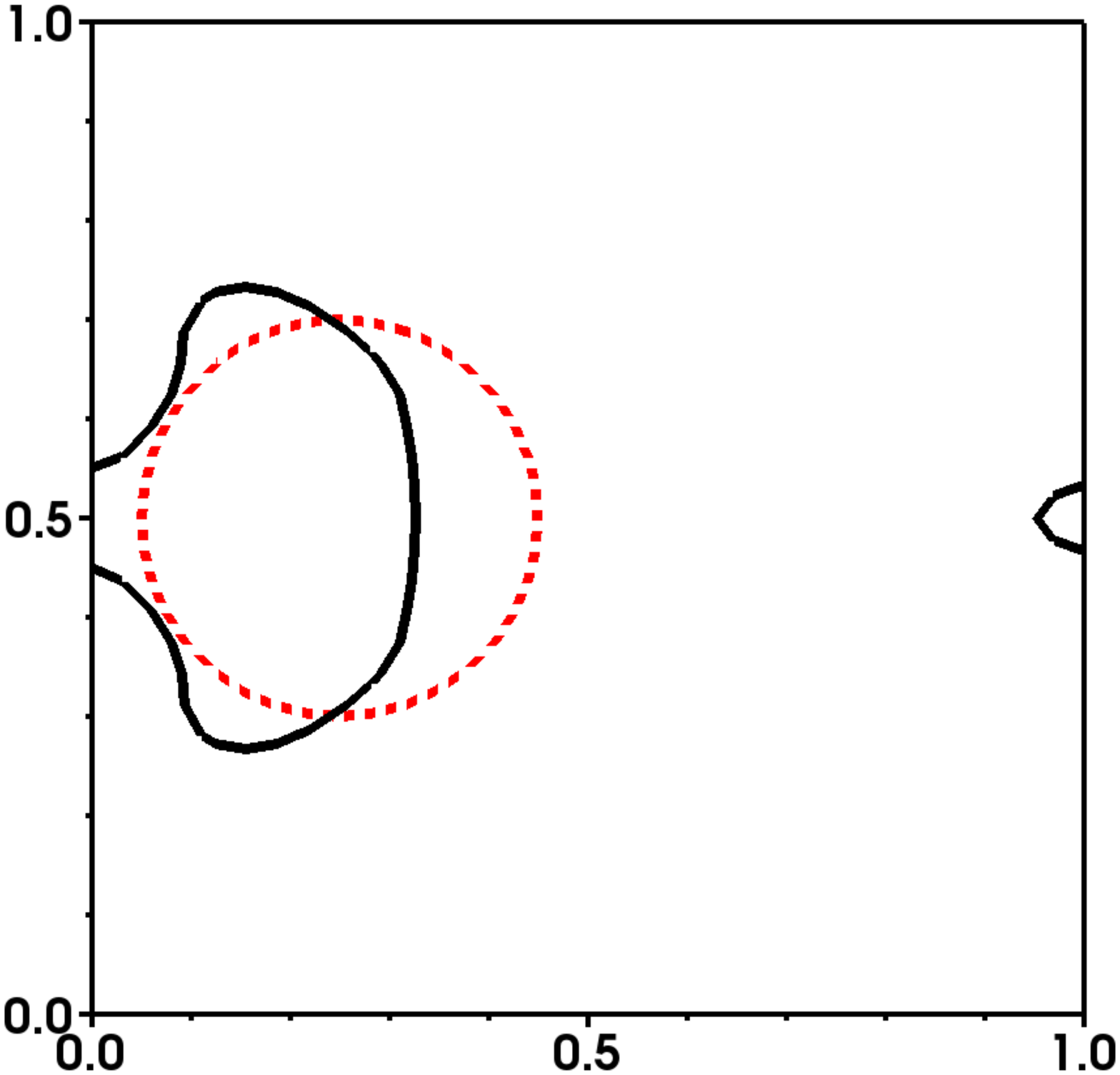}
    \label{nc_PPM_bubble_N32}
  }
   \subfigure[$N = 64, t = 1$]{
    \includegraphics[scale = 0.2]{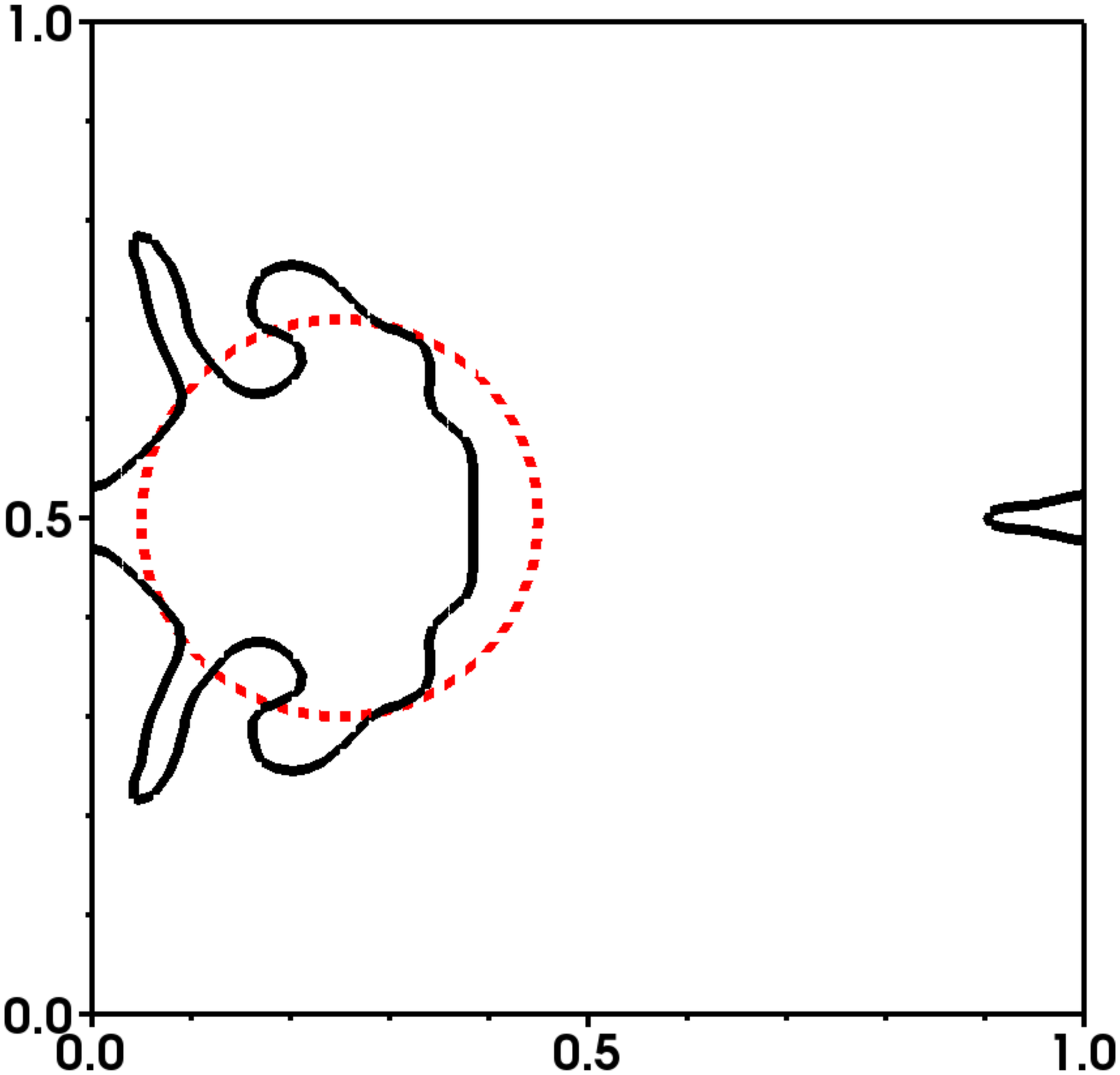}
    \label{nc_PPM_bubble_N64}
  }
   \subfigure[$N = 128, t = 0.5$]{
    \includegraphics[scale = 0.2]{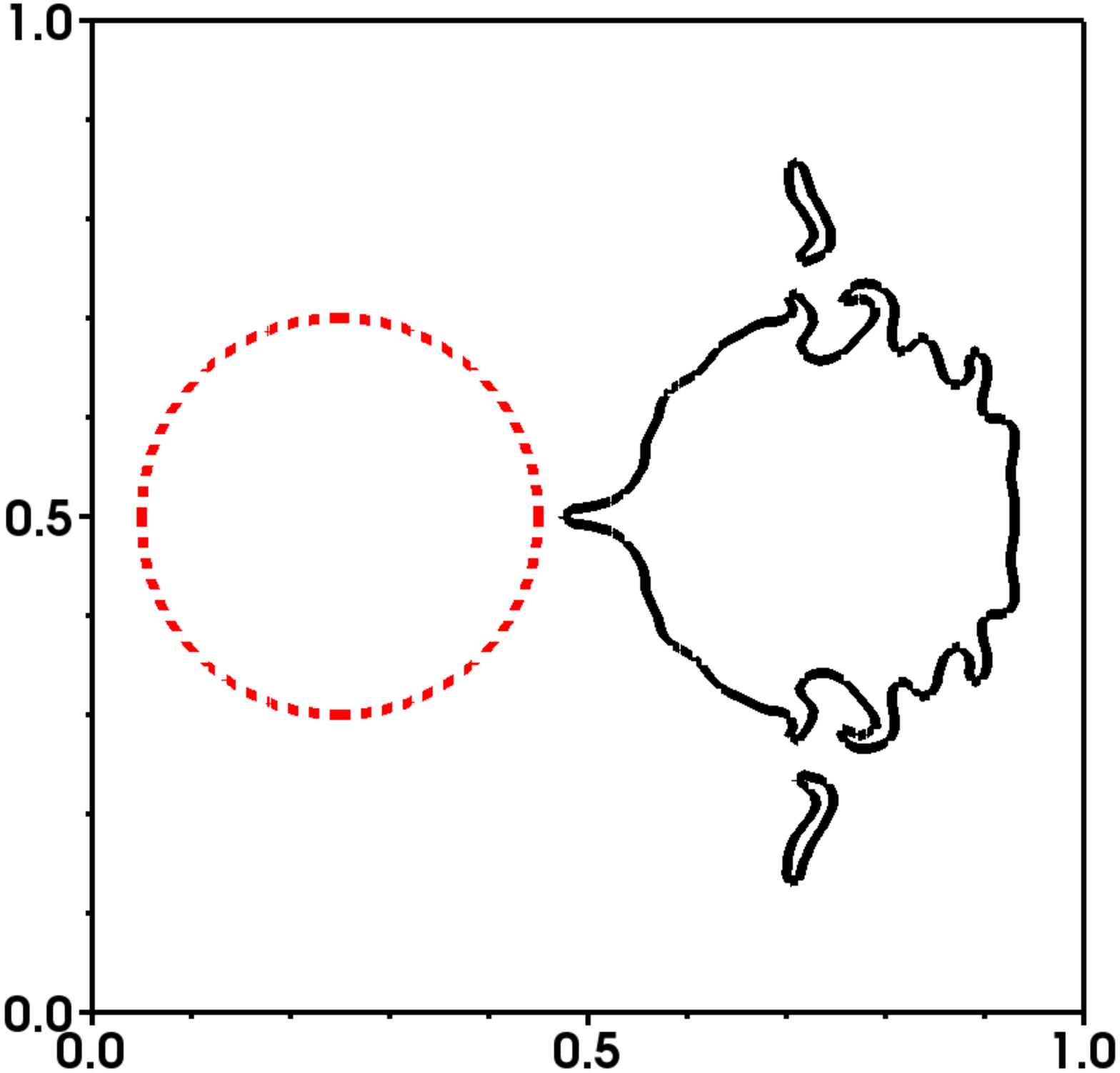}
    \label{nc_PPM_bubble_N128}
  }
  \subfigure[$N = 256, t = 0.5$]{
    \includegraphics[scale = 0.2]{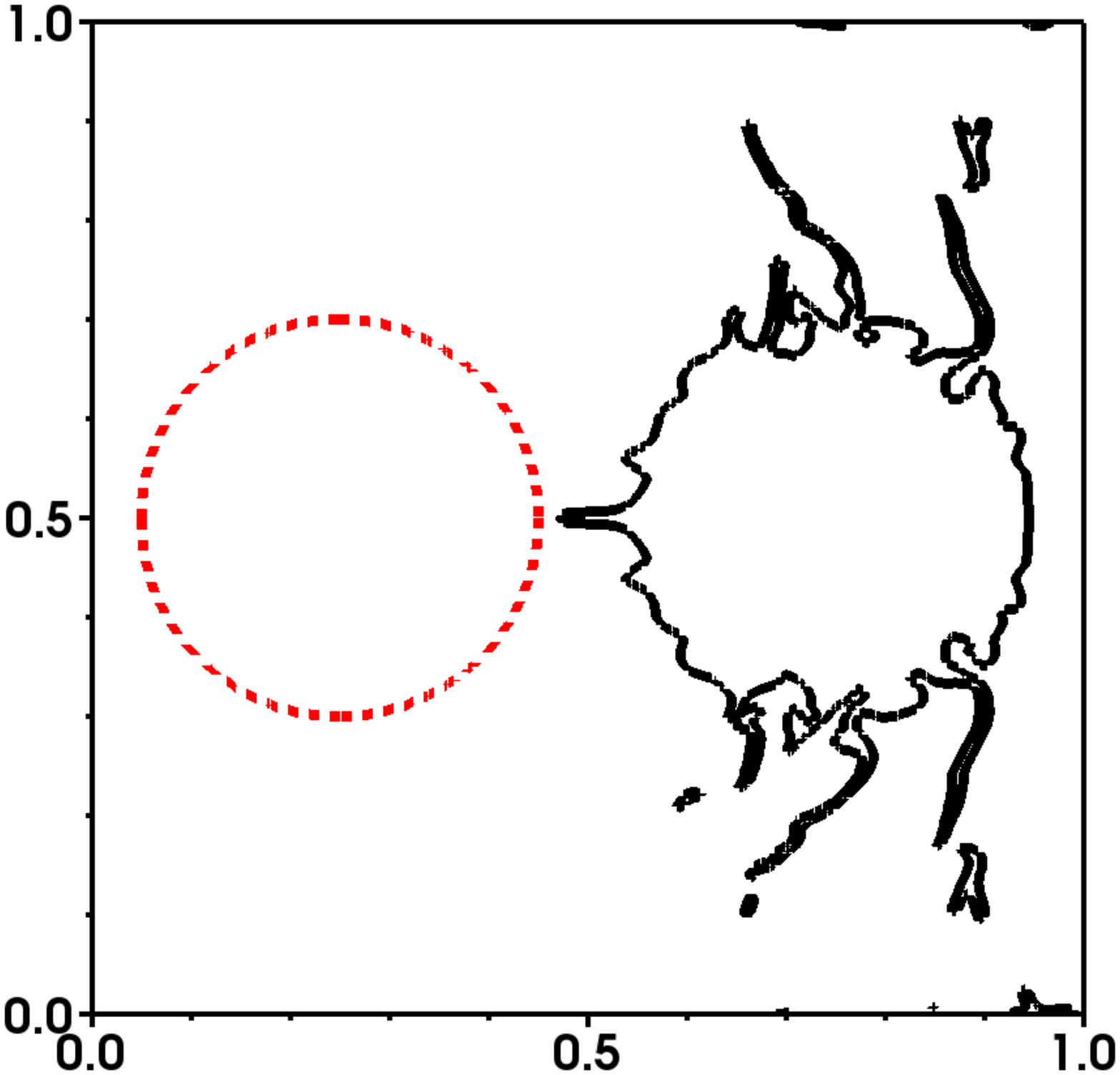}
    \label{nc_PPM_bubble_N256}
  }
  \caption{Initial condition/exact solution (\texttt{---}, red) and numerical solution (---, black) 
  for convection of a high density droplet with density ratio $\rho_{\text{i}}/\rho_{\text{o}} = 10^6$ and initial
  velocity $\up(\x,0)$. Inconsistent transport of mass and momentum occurs while employing 
  a non-conservative momentum integrator.
  PPM is used for advection of the level set and CUI limiter is used for the convective operator.
  \subref{nc_PPM_bubble_N32}
  Grid size $N = 32$ at time $t = 1$
  \subref{nc_PPM_bubble_N64}
  Grid size $N = 64$ at time $t = 1$
  \subref{nc_PPM_bubble_N128}
  Grid size $N = 128$ at time $t = 0.5$; the simulation becomes unstable shortly after the shown time
  \subref{nc_PPM_bubble_N256}
  Grid size $N = 256$ at time $t = 0.5$; the simulation becomes unstable shortly after the shown time}
  \label{fig_nc_PPM_bubble}
\end{figure}

\begin{figure}[]
  \centering
  \subfigure[$N = 32, t = 1$]{
    \includegraphics[scale = 0.22]{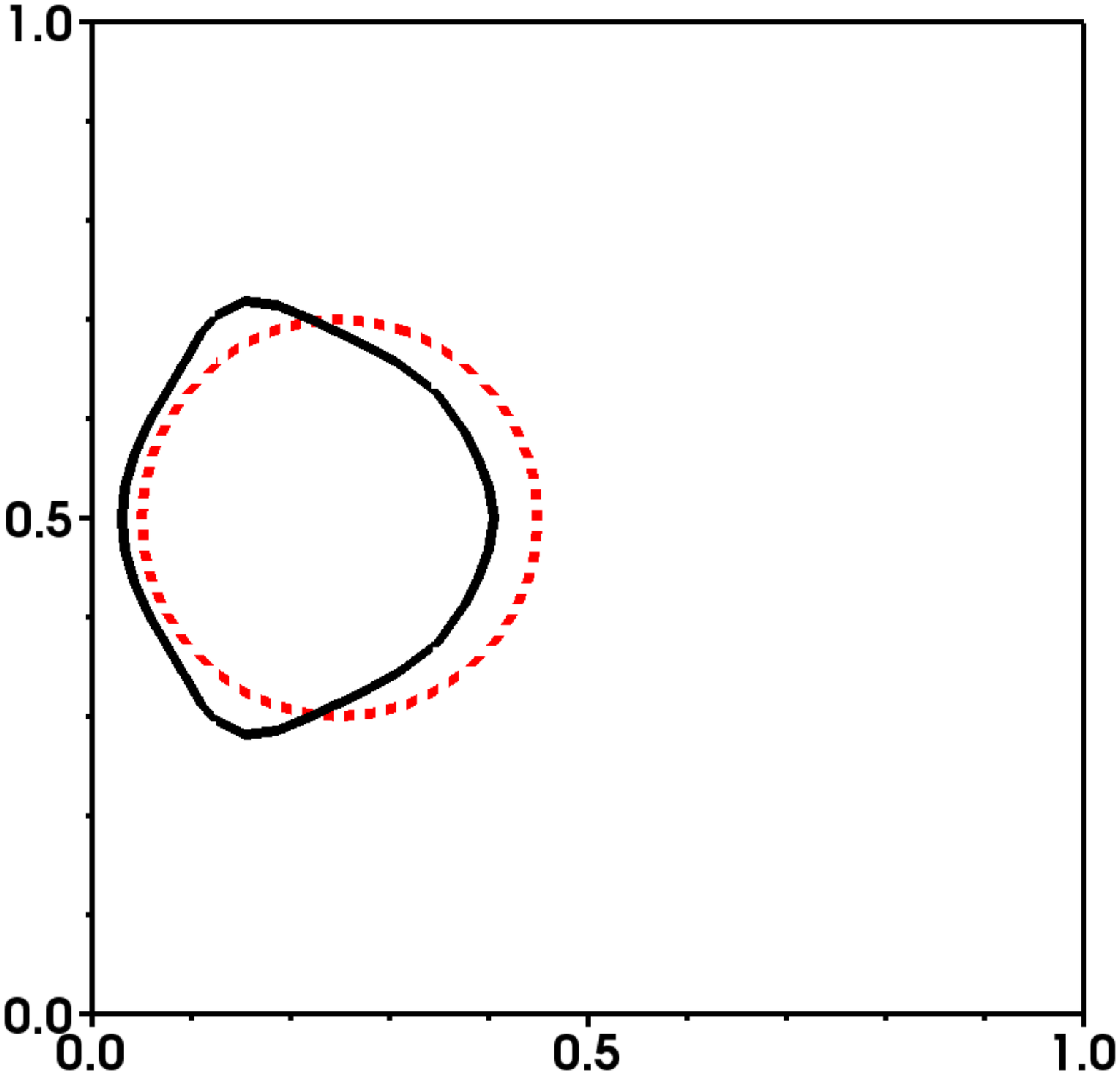}
    \label{c_PPM_bubble_N32}
  }
   \subfigure[$N = 64, t = 1$]{
    \includegraphics[scale = 0.22]{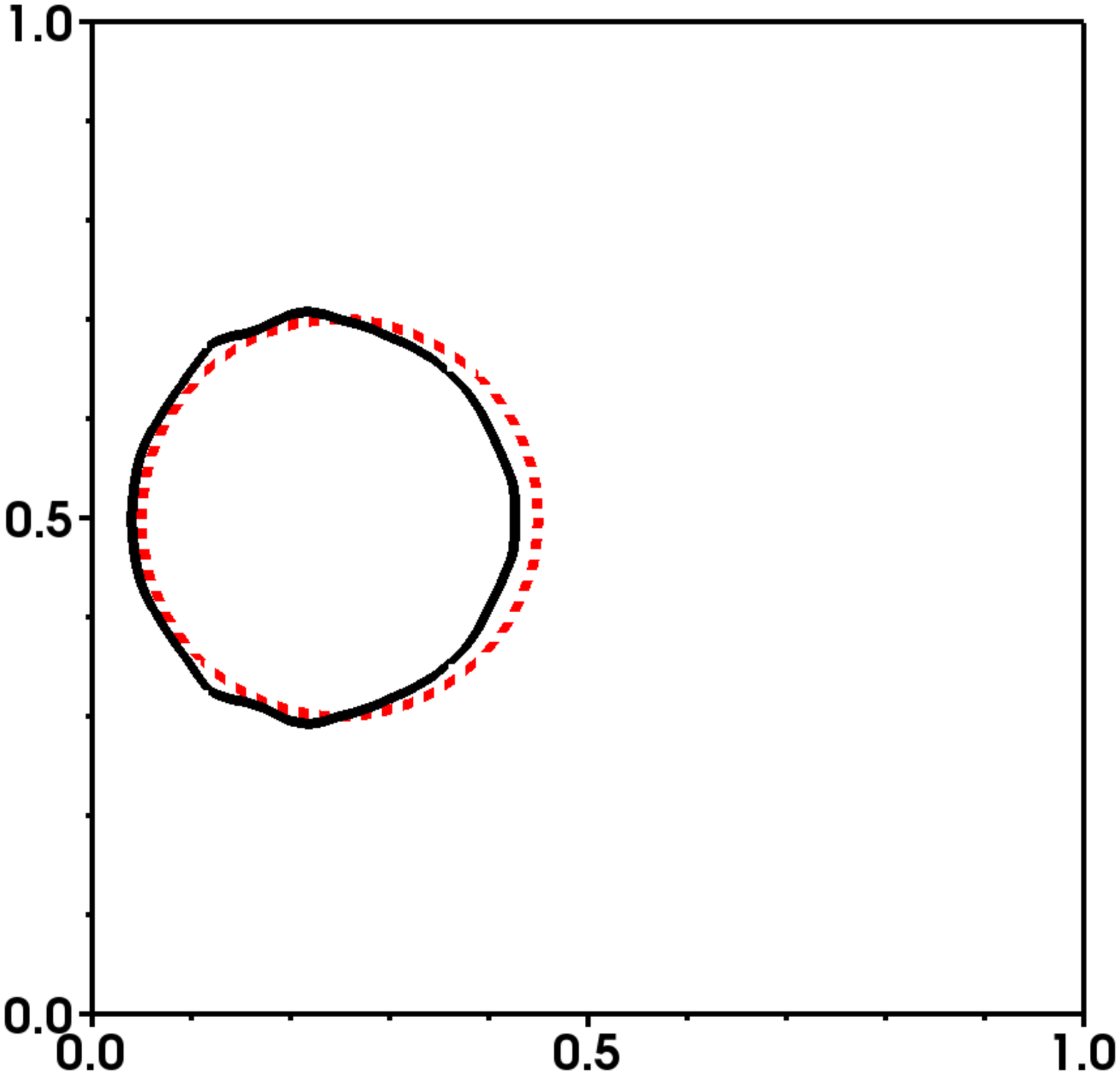}
    \label{c_PPM_bubble_N64}
  }
   \subfigure[$N = 128, t = 1$]{
    \includegraphics[scale = 0.22]{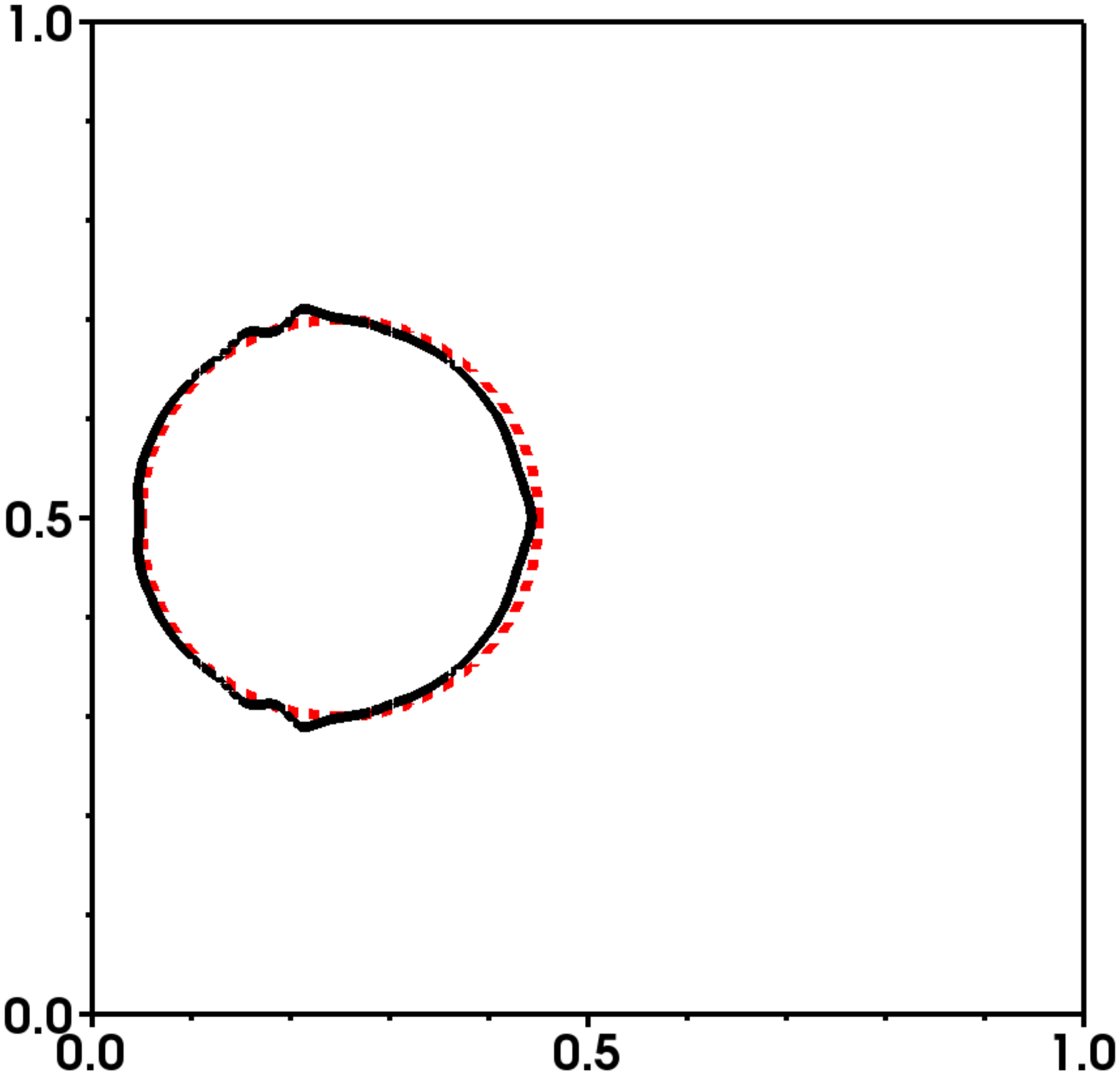}
    \label{c_PPM_bubble_N128}
  }
  \subfigure[$N = 256, t = 1$]{
    \includegraphics[scale = 0.22]{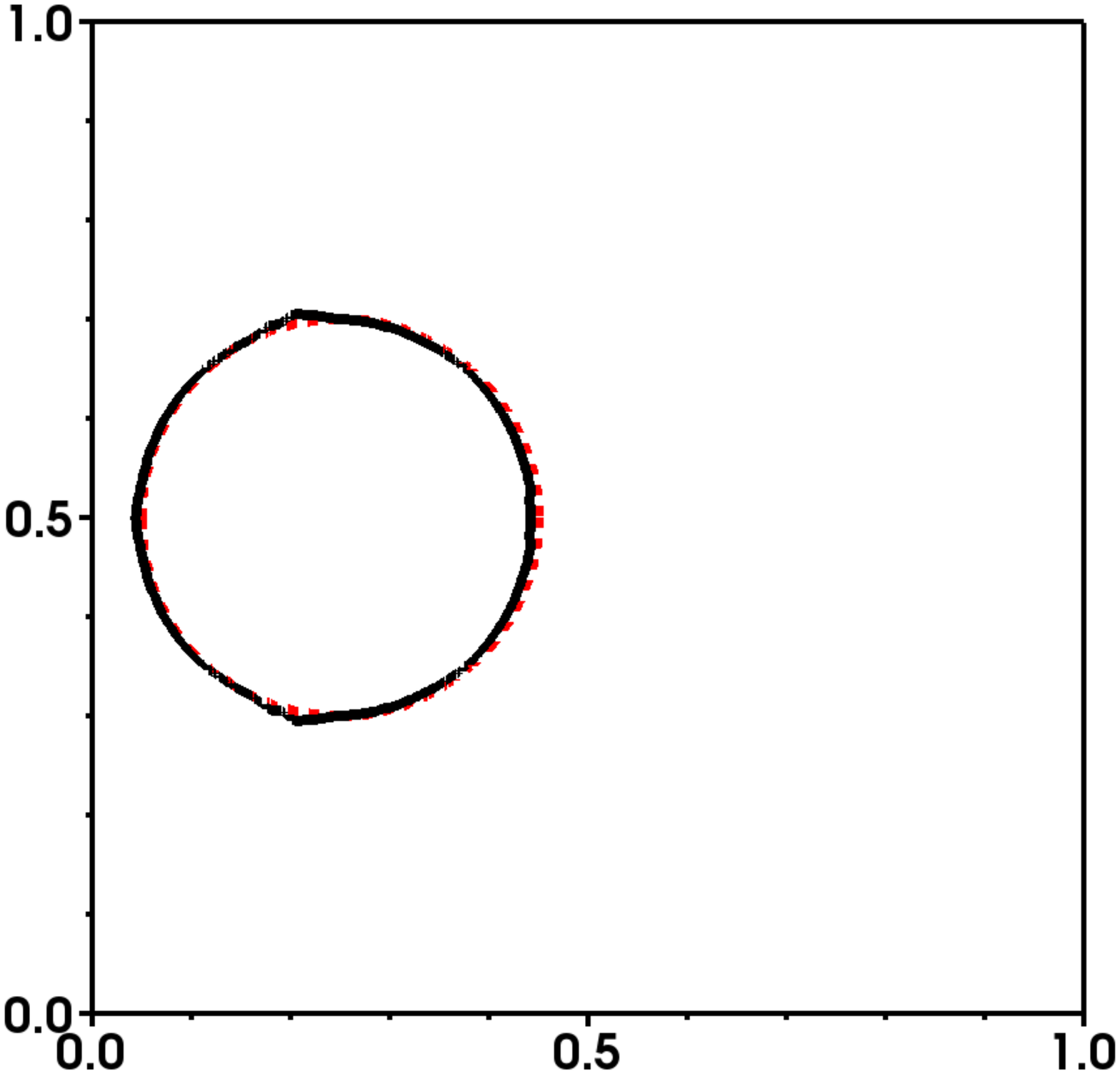}
    \label{c_PPM_bubble_N256}
  }
  \caption{Initial condition/exact solution (\texttt{---}, red) and numerical solution (---, black) 
  for convection of a high density droplet with density ratio $\rho_{\text{i}}/\rho_{\text{o}} = 10^6$ and initial
  velocity $\up(\x,0)$. Consistent transport of mass and momentum is used for these cases.
  PPM is used for advection of the level set and CUI limiters are used for mass and convective fluxes.
  \subref{c_PPM_bubble_N32}
  Grid size $N = 32$ at time $t = 1$.
  \subref{c_PPM_bubble_N64}
  Grid size $N = 64$ at time $t = 1$.
  \subref{c_PPM_bubble_N128}
  Grid size $N = 128$ at time $t = 1$.
  \subref{c_PPM_bubble_N256}
  Grid size $N = 256$ at time $t = 1$.}
  \label{fig_c_PPM_bubble}
\end{figure}

As a first test, we demonstrate the importance of consistent mass and momentum transport
for the stability of high density ratio flows. Fig.~\ref{fig_nc_PPM_bubble} shows the shape of the bubble
for four different grid spacings, $N = 32, 64, 128, 256$, when simulated with inconsistent mass and
momentum transport, i.e., by using a non-conservative momentum integrator. There is clear distortion 
in the bubble at time $t = 1$ at the
lower resolutions (Figs.~\ref{nc_PPM_bubble_N32} and~\ref{nc_PPM_bubble_N64}), whereas
the simulations quickly become unstable at higher grid resolutions 
(Figs.~\ref{nc_PPM_bubble_N128} and~\ref{nc_PPM_bubble_N256}). Lowering the time step
even further did not resolve these stability problems. However the simulations are stable when using
consistent mass and momentum transport (Fig.~\ref{fig_c_PPM_bubble}), and the bubble's shape converges
towards the analytical solution as the resolution increases. The bubble's distortion at lower resolutions can be
attributed to discretization errors from advection and interface tracking. For these cases, xsPPM7 is used to 
advect the level set, and CUI limiters are used for mass and convective fluxes. Therefore, these cases
also demonstrate that the level set and mass updates can have dissimilar limiters and still maintain numerical
stability for the consistent and conservative momentum integrator. This is in contrast to the numerical scheme of Patel and Natarajan~\cite{Patel2017}, which remains consistent only when same limiter is employed for the advection of algebraic VOF-scalar and momentum flux.

\begin{figure}[]
  \centering
  \subfigure[CUI limiter]{
    \includegraphics[scale = 0.16]{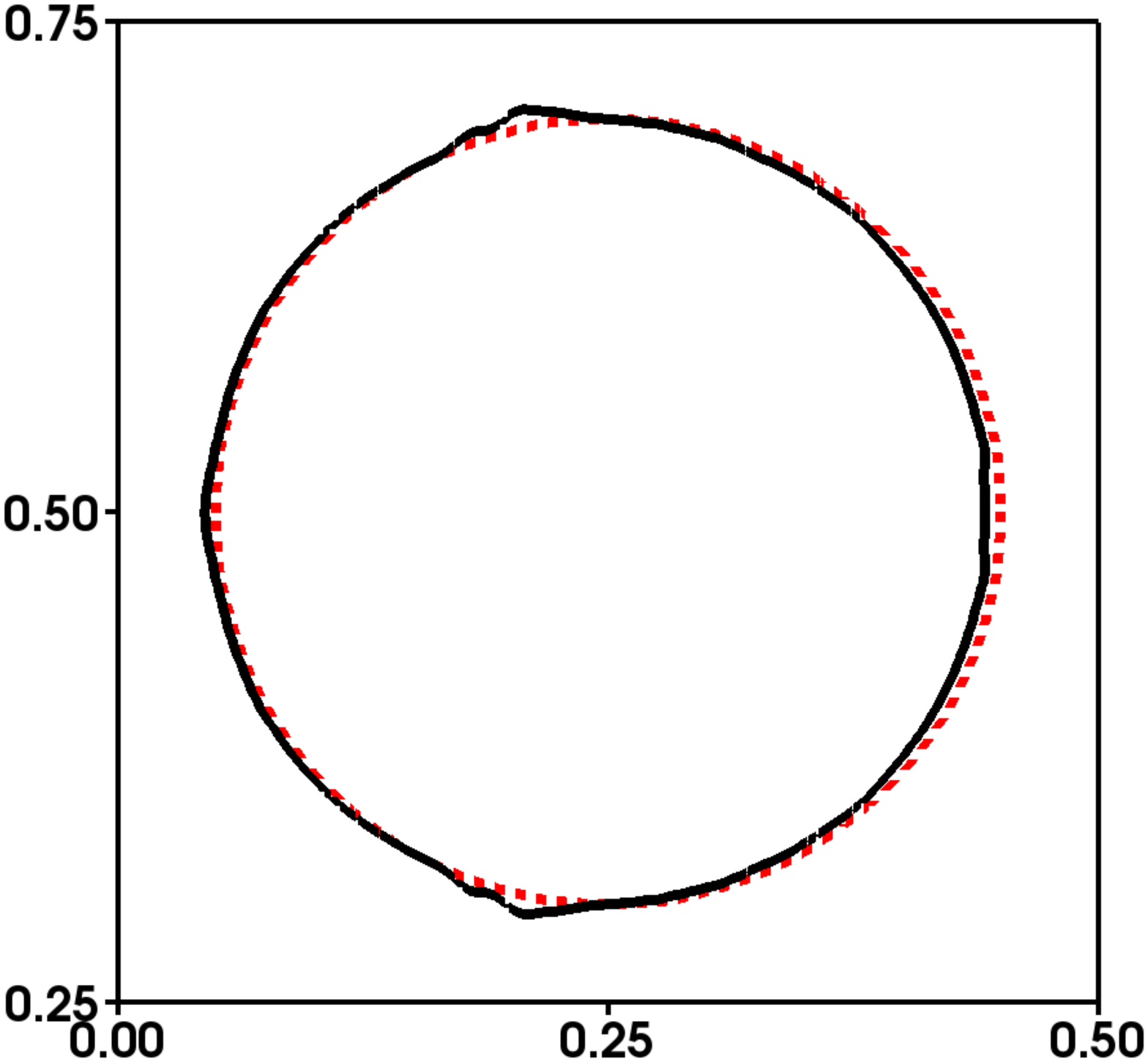}
    \label{c_PPM_CUI_zoom_bubble_N256}
  }
   \subfigure[M-Gamma limiter]{
    \includegraphics[scale = 0.16]{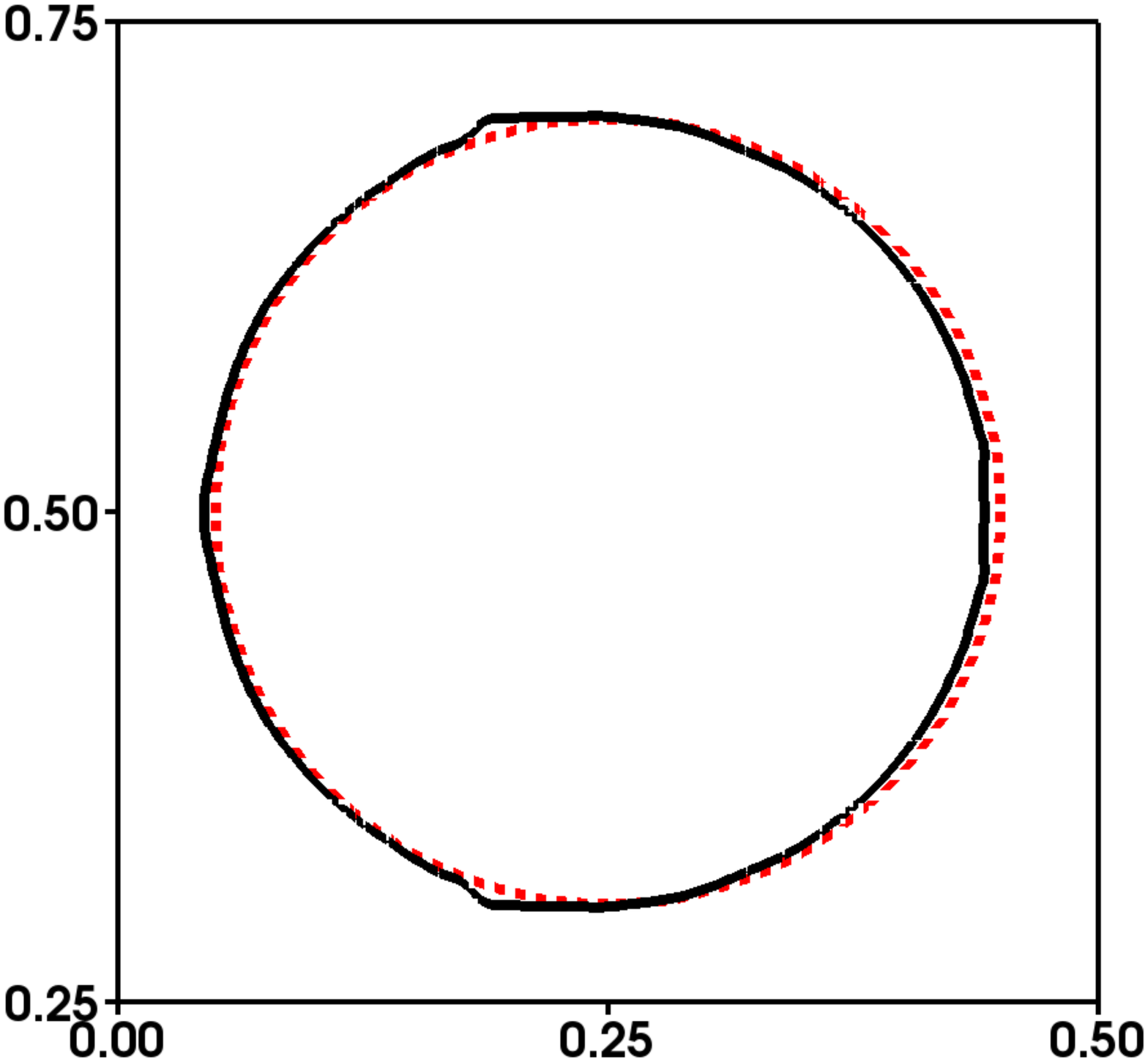}
    \label{c_PPM_MGAMMA_zoom_bubble_N256}
  }
   \subfigure[FBICS limiter]{
    \includegraphics[scale = 0.16]{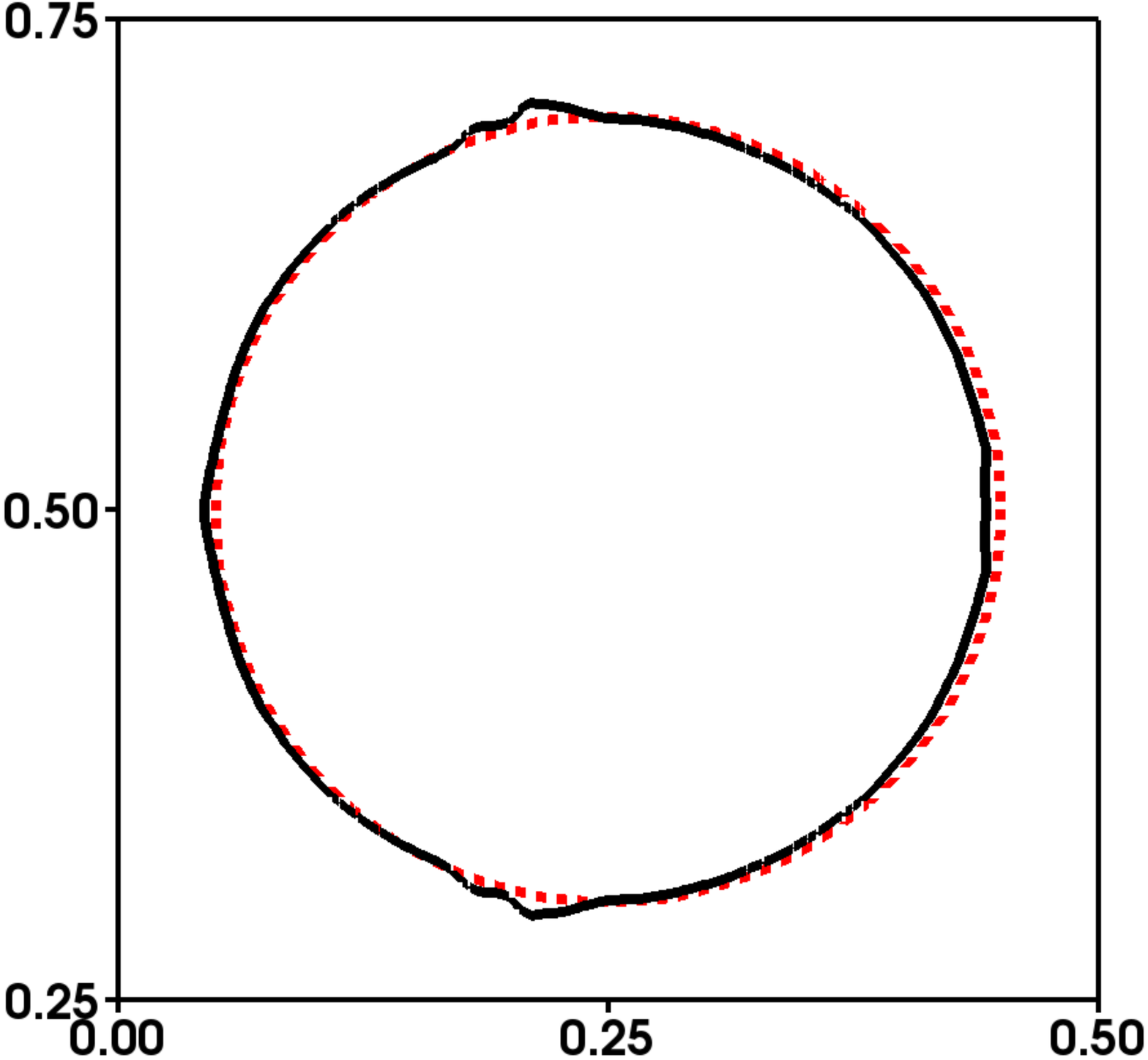}
    \label{c_PPM_FBICS_zoom_bubble_N256}
  }
  \subfigure[Minimum density vs. time step]{
    \includegraphics[scale = 0.18]{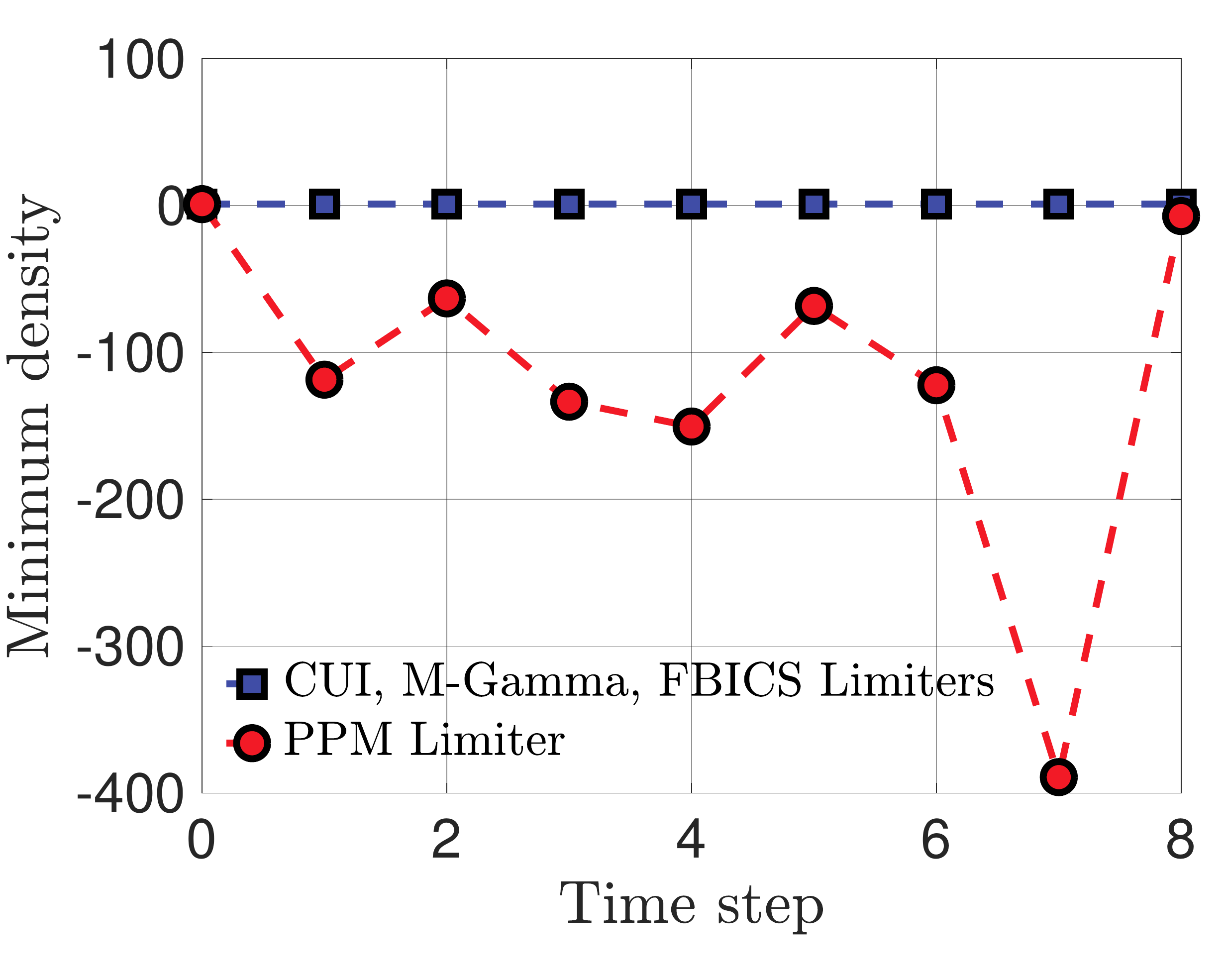}
    \label{min_density_vs_timestep}
  }
  \caption{Initial condition/exact solution (\texttt{---}, red) and numerical solution (---, black) 
  for convection of a high density droplet with density ratio $\rho_{\text{i}}/\rho_{\text{o}} = 10^6$ and initial
  velocity $\up(\x,0)$. Consistent transport of mass and momentum is used for these cases.
  PPM is used for advection of the level set and various limiters are used for mass and convective fluxes.
  The grid size is $N = 256$ for all cases.
  \subref{c_PPM_CUI_zoom_bubble_N256}
  CUI limiter at time $t = 1$.
  \subref{c_PPM_MGAMMA_zoom_bubble_N256}
  M-Gamma limiter at time $t = 1$.
  \subref{c_PPM_FBICS_zoom_bubble_N256}
    FBICS limiter at time $t = 1$.
  \subref{min_density_vs_timestep}
  Minimum density in the domain during the first eight simulation time steps.}
  \label{fig_c_PPM_bubble_zoom}
\end{figure}

We next demonstrate the importance of using a CBC satisfying limiter for the mass density update.
We consider the same simulation parameters above (for the consistent transport case) 
at a grid resolution of $N = 256$ and compare the shape of the bubble at $t = 1$ for four
different limiters: xsPPM7~\cite{Rider2007} along with high resolution versions of CUI~\cite{Patel2015}, M-Gamma~\cite{Patel2015}, and FBICS~\cite{Tsui2009}. Figs.~\ref{c_PPM_CUI_zoom_bubble_N256}--\ref{c_PPM_FBICS_zoom_bubble_N256} show the shape of the bubble for the CUI, M-Gamma, and
FBICS limiters. All of these limiters maintain numerical stability and produce reasonable results when
compared to the analytical solution. Results for the PPM limiter are not shown here because the simulation
quickly becomes unstable within ten time steps. Note that of these
four limiters, PPM is the only one that can produce undershoots and overshoots for advected quantities~\cite{Nonaka2011}.
It is required that the density remain positive throughout the computation to obtain a
physically accurate solution. Fig.~\ref{min_density_vs_timestep} shows the minimum updated density $\rho^{n+1}$
within each time step in the domain over the first eight time steps. It is clear that the CBC satisfying limiters maintain physically realistic
minimum (maximum) densities whereas PPM generates undershoots (overshoots) that eventually corrupt the simulation. Throughout the remainder of the simulation, the CBC satisfying limiters do not undershoot the minimum or overshoot the maximum initial density (data not shown for overshoots).

\begin{figure}[]
  \centering
  \subfigure[Inconsistent $t = 0.0$]{
    \includegraphics[scale = 0.22]{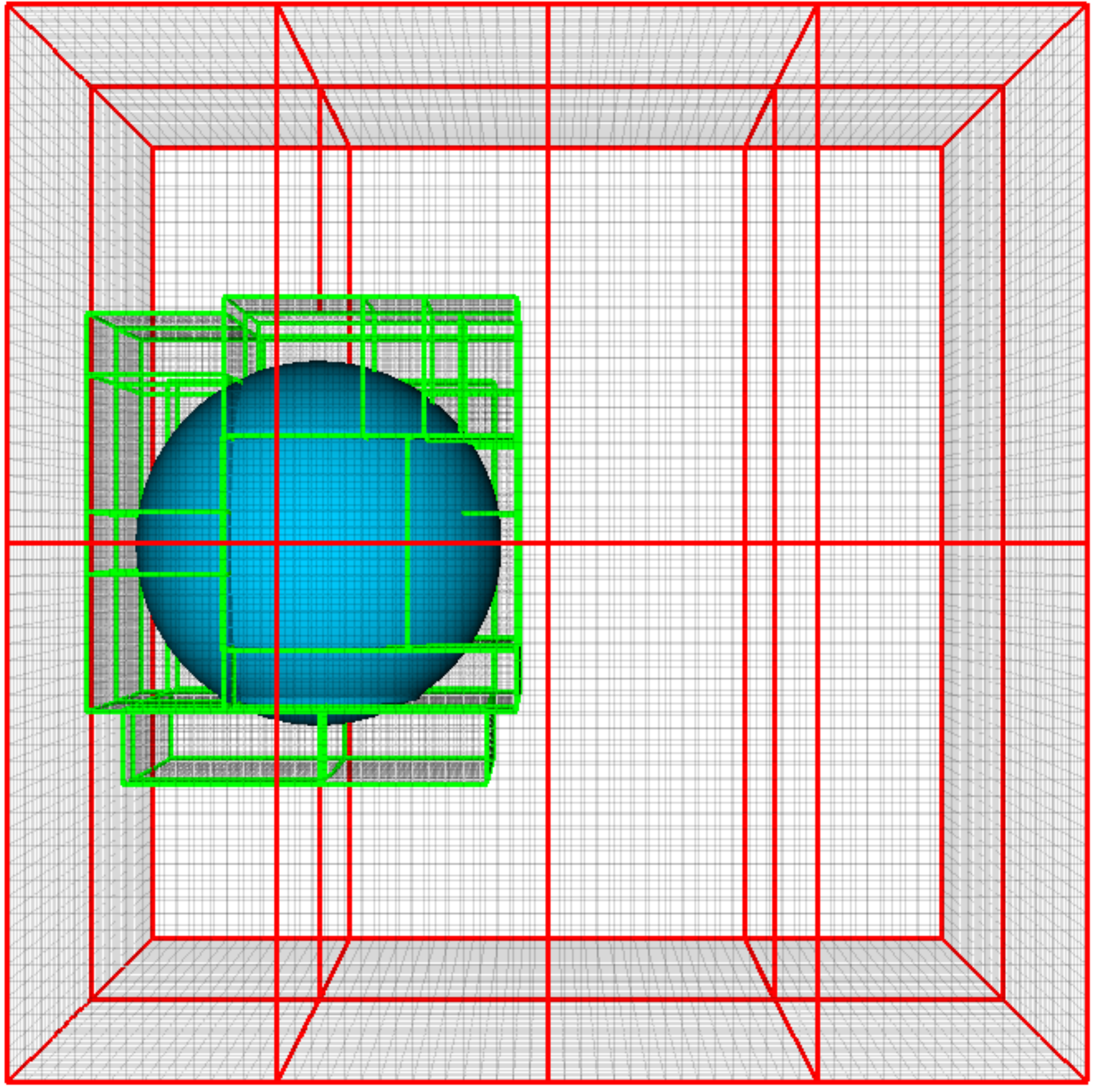}
    \label{nc_Bubble_3D_t0}
  }
   \subfigure[Inconsistent $t = 0.5$]{
    \includegraphics[scale = 0.22]{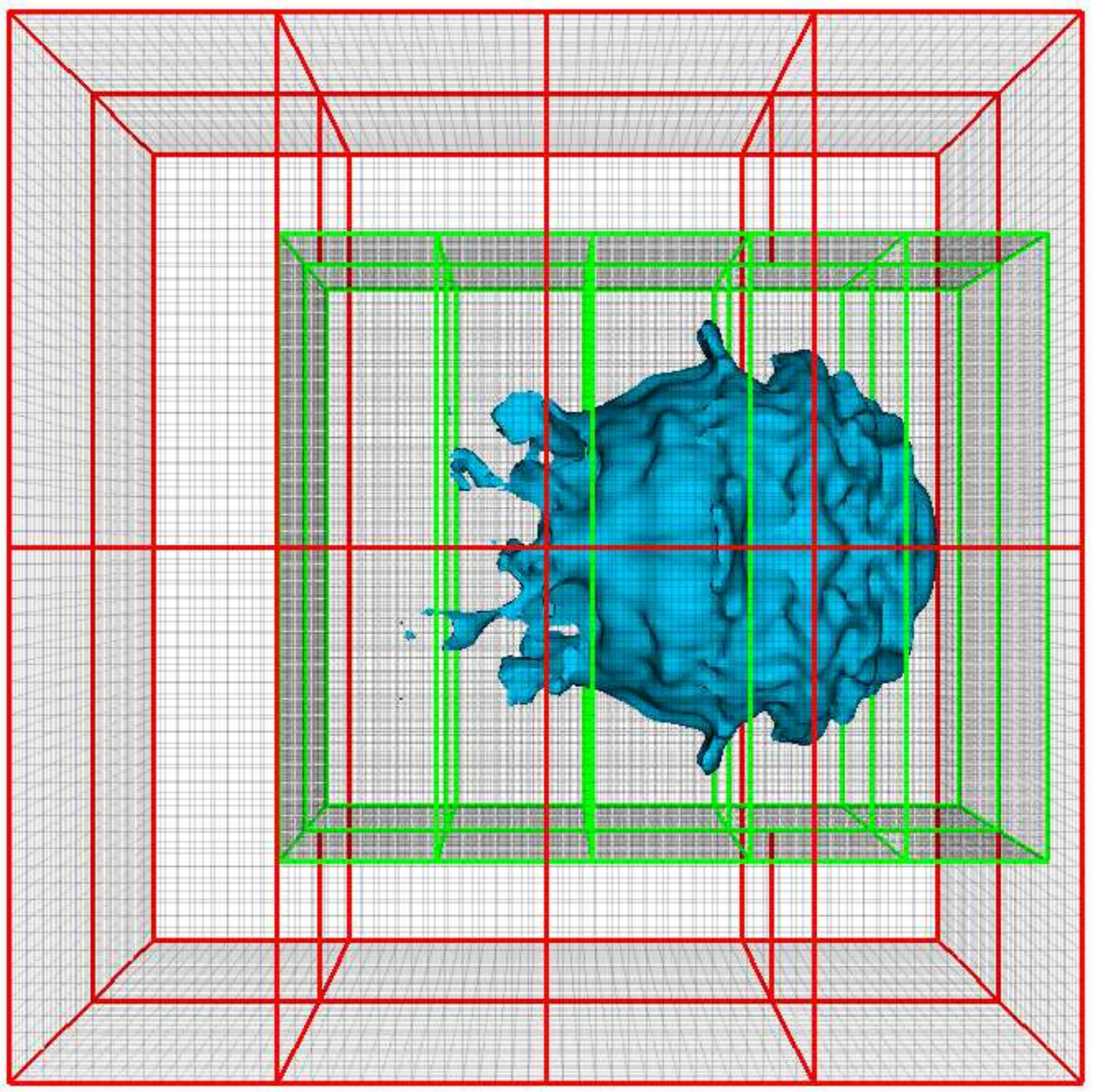}
    \label{nc_Bubble_3D_t05}
  }
   \subfigure[Inconsistent $t = 0.6$]{
    \includegraphics[scale = 0.22]{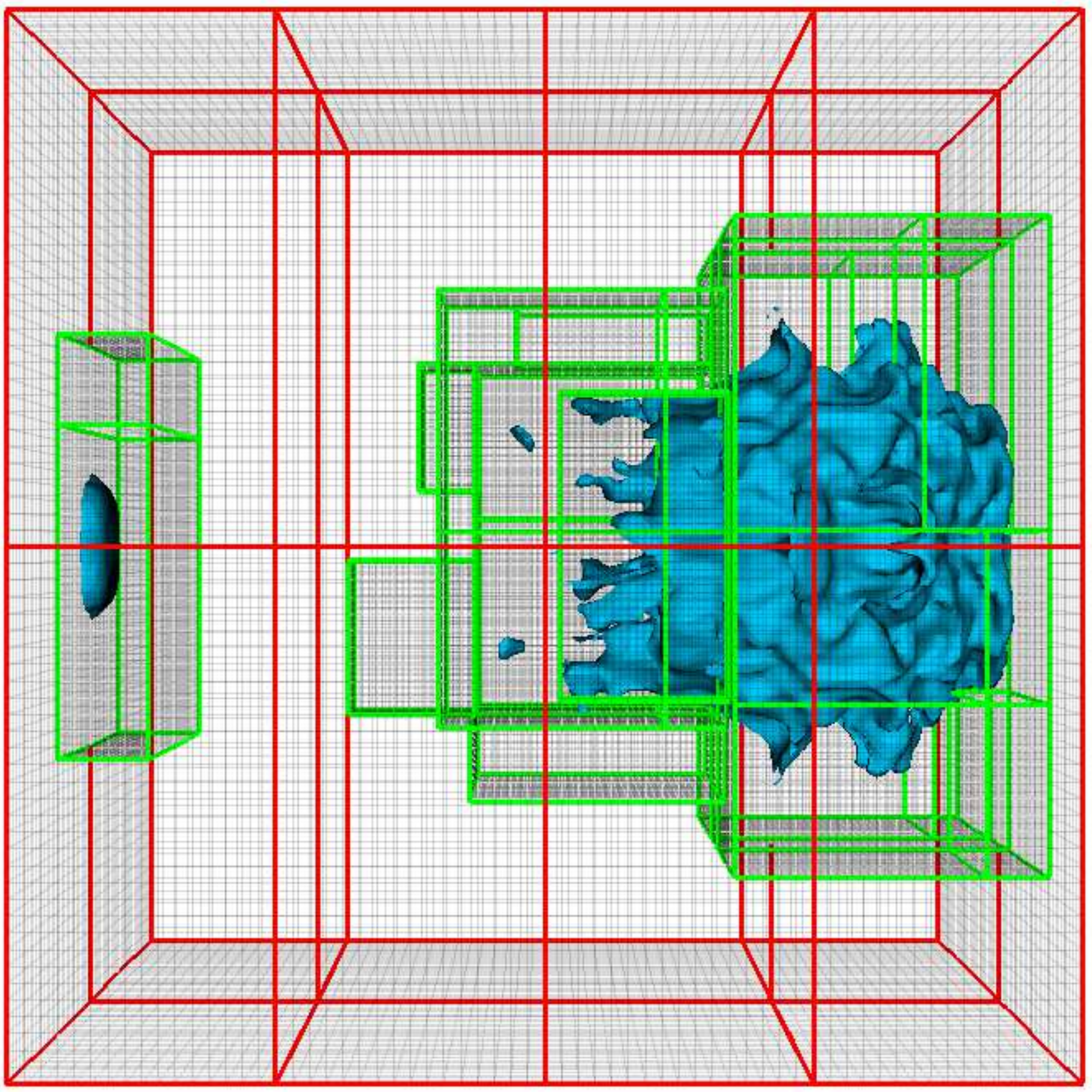}
    \label{nc_Bubble_3D_t06}
  }
    \subfigure[Consistent $t = 0.0$]{
    \includegraphics[scale = 0.22]{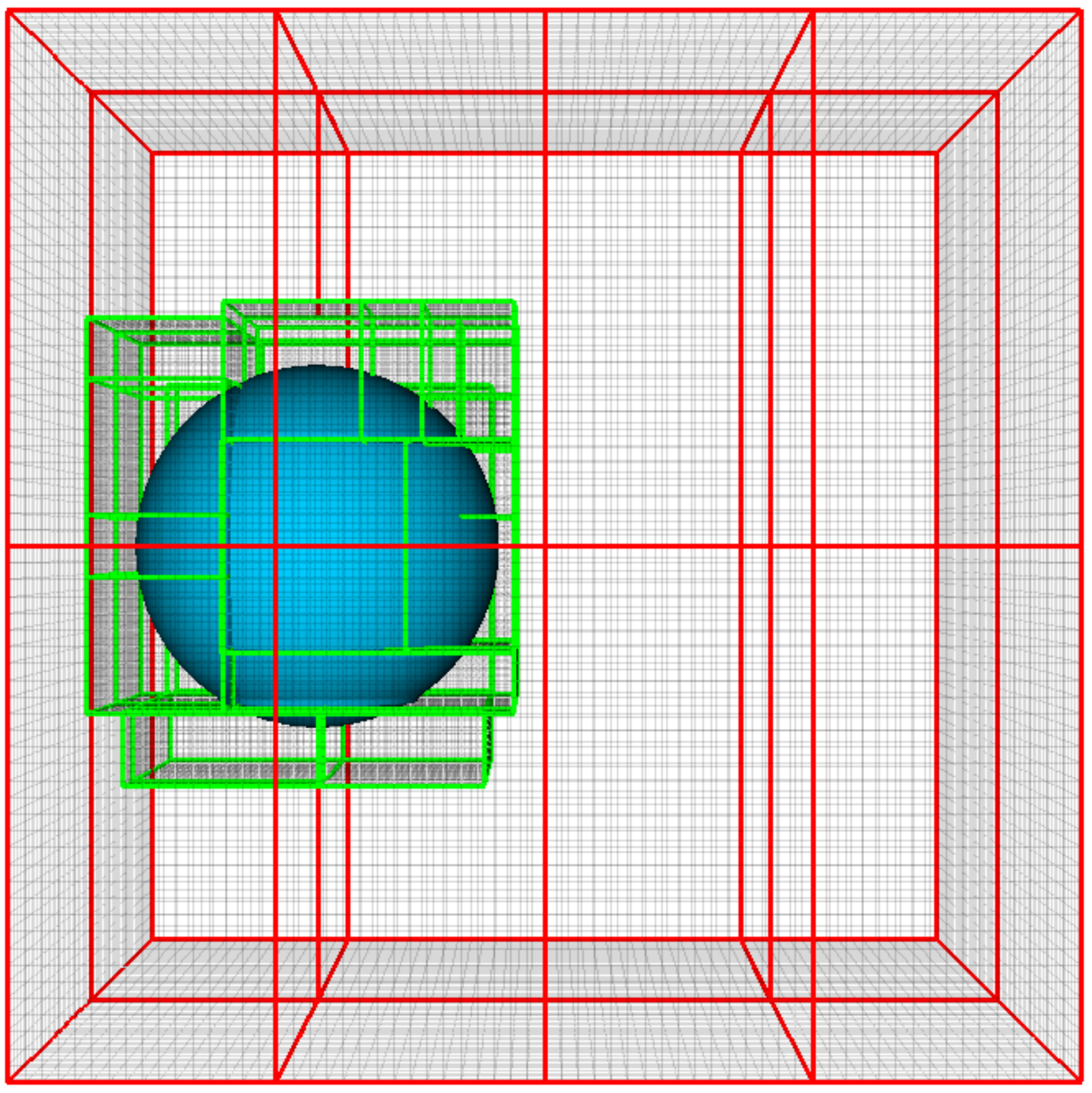}
    \label{c_Bubble_3D_t0}
  }
   \subfigure[Consistent $t = 0.5$]{
    \includegraphics[scale = 0.22]{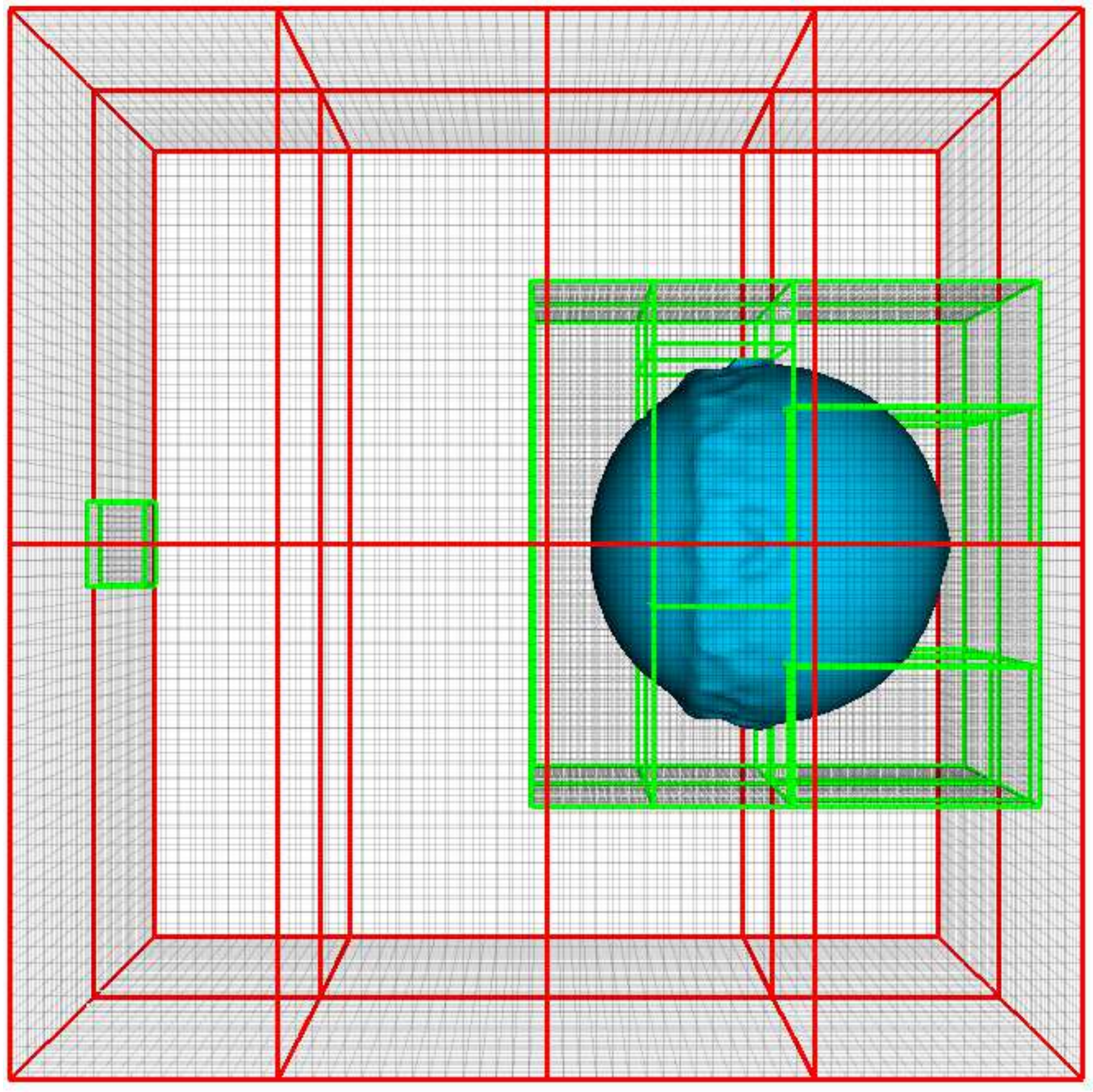}
    \label{c_Bubble_3D_t05}
  }
   \subfigure[Consistent $t = 1.0$]{
    \includegraphics[scale = 0.22]{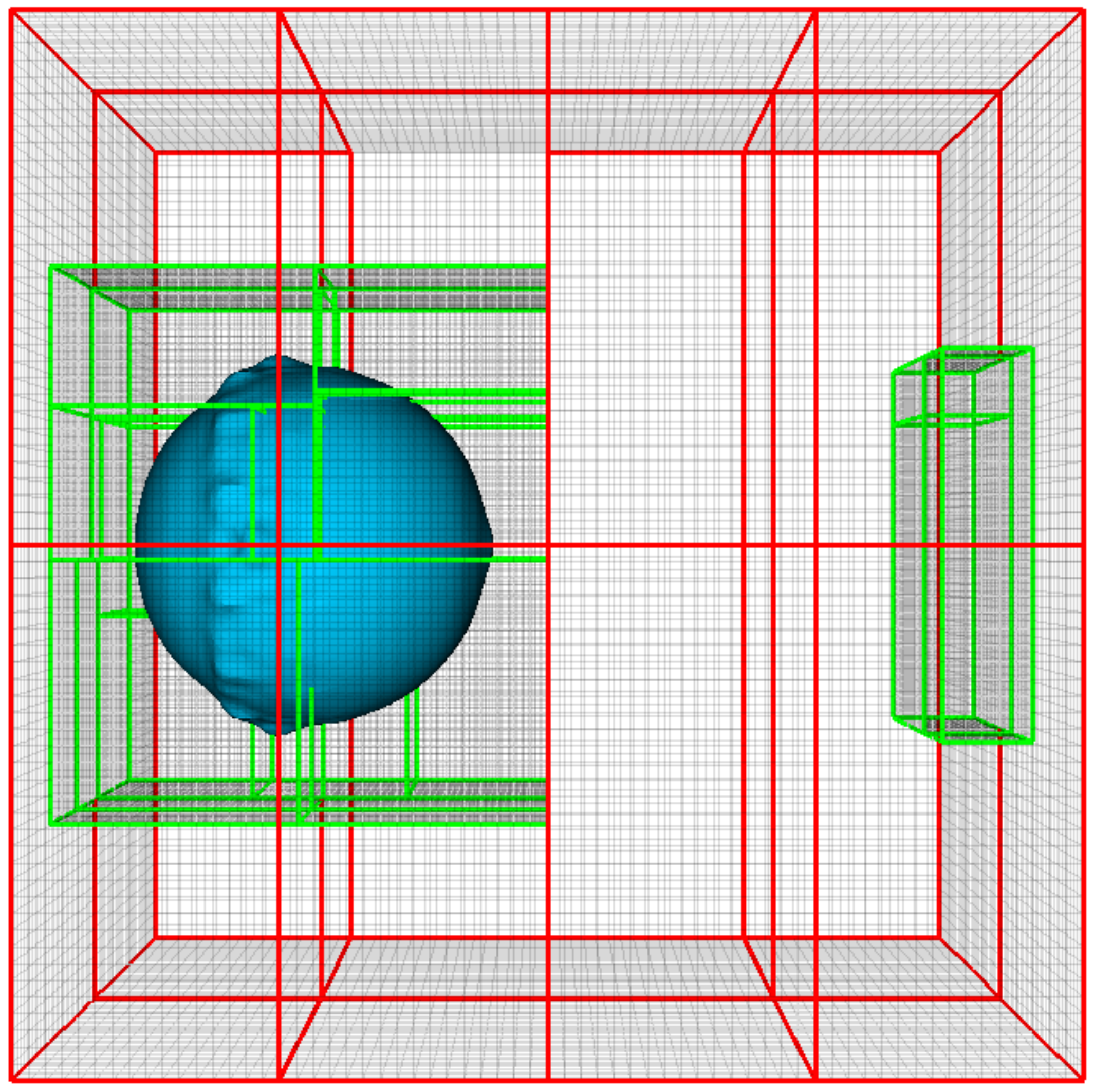}
    \label{c_Bubble_3D_t1}
  }
  \caption{Three-dimensional convection of a high density ratio droplet (blue)with density ratio
  $\rho_{\text{i}}/\rho_{\text{o}} = 10^6$ and initial velocity $\up(\x,0)$. The coarse grid has grid spacing $\dx_0 = \dy_0 = \dy_0 = 1/64$,
  with each simulation using $\ell = 2$ grid levels and refinement ratio $\nref = 2$. The coarse grid level is bounded by the red boxes, while the fine grid level is bounded by green boxes.
  \subref{nc_Bubble_3D_t0}-\subref{nc_Bubble_3D_t06} Three snapshots in time for inconsistent transport of mass and momentum, which becomes unstable shortly after $t = 0.6$.
  \subref{c_Bubble_3D_t0}-\subref{c_Bubble_3D_t1} Three snapshots in time for inconsistent transport of mass and momentum, which remains stable throughout the simulation time.}
  \label{fig_3D_bubble}
\end{figure}

Finally, we demonstrate that the consistent transport scheme produces stable results in three spatial dimensions
and with adaptive mesh refinement. Similar to the previous case, a dense bubble is placed in a fully periodic 
computational domain $\Omega = [0, L]^3$ with $L = 1$, which is discretized by a two-level ($\ell = 2$) locally refined grid with coarsest
grid spacing $\dx_0 = \dy_0 = \dy_0 = 1/64$ and refinement ratio $\nref = 2$.
At the finest level, the grid spacing is $\dx_\textrm{min} = \dy_\textrm{min} = \dz_\textrm{min} = 1/128$. %,
%which is equivalent to the grid size $N = 128$ for the 2D case. 
The radius of the bubble is $R = 0.2L$ with initial center 
position $(X_0, Y_0, Z_0) = (L/4, L/2, L/2)$. The density ratio between the bubble and the outer fluid is again 
$\rho_{\text{i}}/\rho_{\text{o}} = 10^6$, and the viscosity is set to $\mu = 0$ in the entire domain.
The initial velocity $\up(\x,0)$ is again set via density weighted projection, which
is zero for the non-horizontal velocity components. Each case is run until $t = 1$ with a constant time step size
$\dt = 2\dx_0/125$. Coarse grid cells are tagged for refinement where the local vorticity magnitude exceeds a relative threshold of 0.25, or where $\phi \le 2 \dx_0$, which ensures that the surface and interior of the bubble are always
placed on the finest grid level.
Fig.~\ref{fig_3D_bubble} shows evolution of the bubble at three snapshots in time. The top row shows the case
for inconsistent transport using the non-conservative integrator, which quickly becomes unstable, as in the two-dimensional case. The bottom row shows the case for consistent
transport, which remains stable and exhibits similar distortion because of numerical errors as the two dimensional case
with a comparable spatial resolution.

\subsection{Collapsing water column}
This section demonstrates the importance of consistent mass and momentum transport to achieve
stability for air-water density ratios. A water column of initial height and width of $a = 5.715 \times 10^{-2} $ m
is placed in a two-dimensional computational domain $\Omega = [7a, 1.75a]$ that is discretized using a $4N \times N$
grid. No-slip boundary conditions are imposed along
$\partial \Omega$. At the initial time, the bottom left corner of the water column coincides with the
bottom left corner of the computational domain. The density of water is $\rho_{\text{l}} = 1000$ kg/m$^3$ ,
the density of air is $\rho_{\text{g}} = 1.226$ kg/m$^3$, the viscosity of water is $\mu_{\text{l}} = 1.137 \times 10^{-3}$
kg $\cdot$ m/s, the viscosity of air is $\mu_{\text{g}} = 1.78 \times 10^{-5}$ kg $\cdot$ m/s, the surface tension coefficient
is $\sigma = 0.0728$ N/m, and the gravitational acceleration is $g = 9.81$ m/s$^2$ (directed in the 
negative $y$-direction). Both fluids are initially at rest. This problem has been studied numerically by 
Rezende et al.~\cite{Rezende2015} and by Patel and Natarajan~\cite{Patel2017}. We also consider the
analogous case in three spatial dimensions, in which a cubic water block with side length
$a = 5.715 \times 10^{-2} $ m is placed in a computational domain of size $\Omega = [7a, a, 1.75a]$
discretized by a $4N \times 9N/16 \times N$ grid. In this case, gravity is directed in the negative $z$-direction. This problem has been studied numerically
by Gu et al.~\cite{Gu2018} and experimentally by Martin and Moyce~\cite{Martin1952}.
For all of the following cases, the density is set via the level set function at the
beginning of each time step with one grid cell of smearing ($\ncells = 1$) on either side of the interface.

\begin{figure}[]
  \centering
    \subfigure[Inconsistent $t = 0.0$ s]{
    \includegraphics[scale = 0.28]{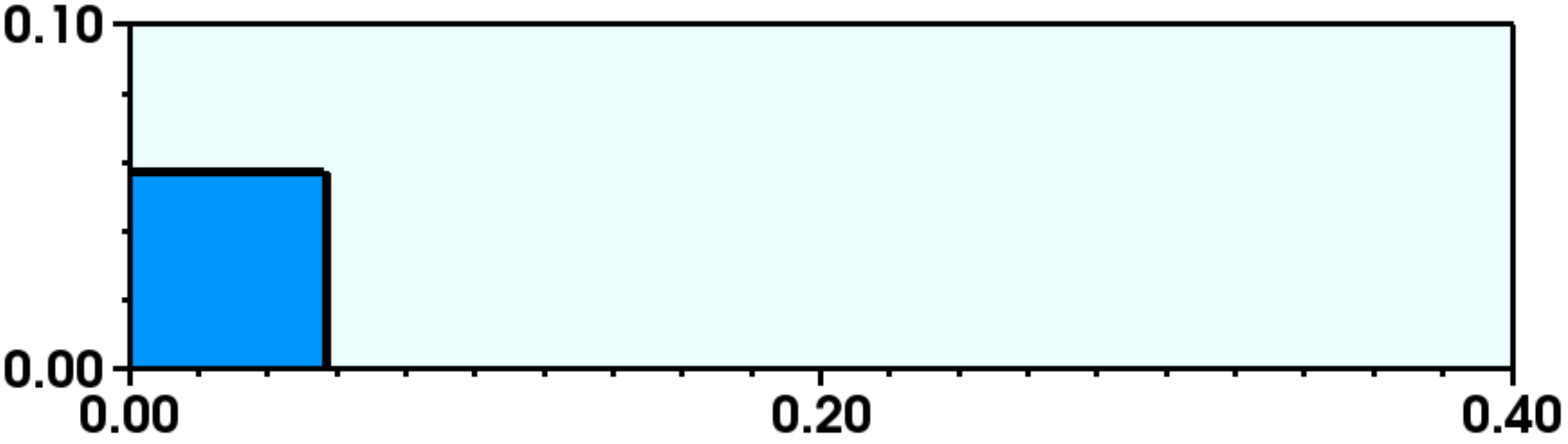}
    \label{nc_water_column_N128_t0}
  }
  \subfigure[Consistent $t = 0.0$ s]{
    \includegraphics[scale = 0.28]{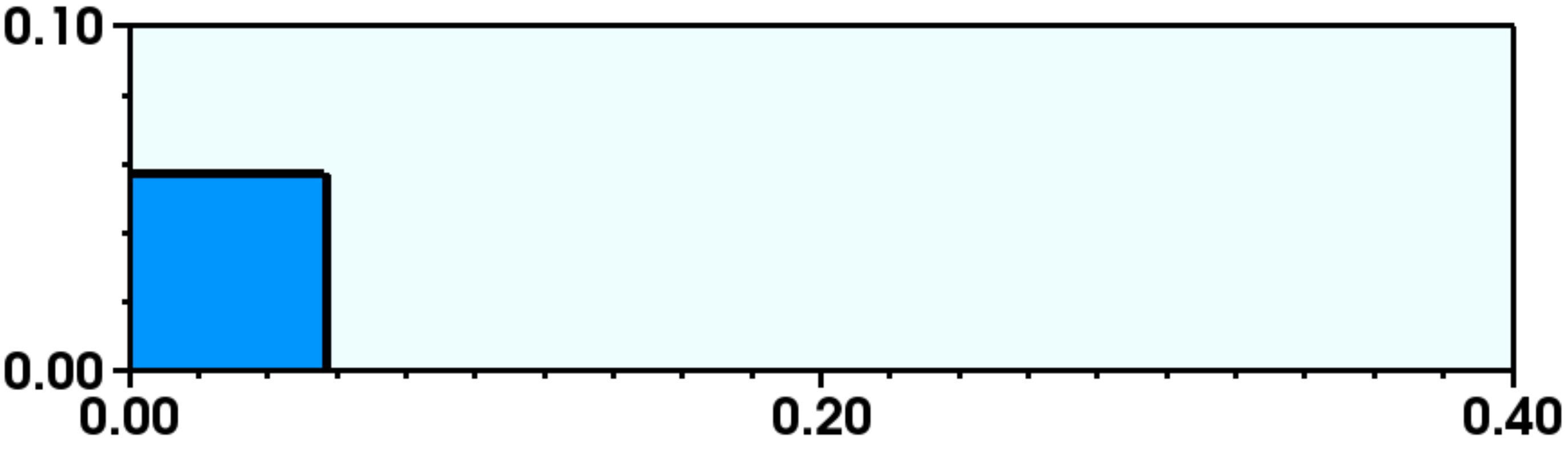}
    \label{c_water_column_N128_t0}
  }
    \subfigure[Inconsistent $t = 0.1$ s]{
    \includegraphics[scale = 0.28]{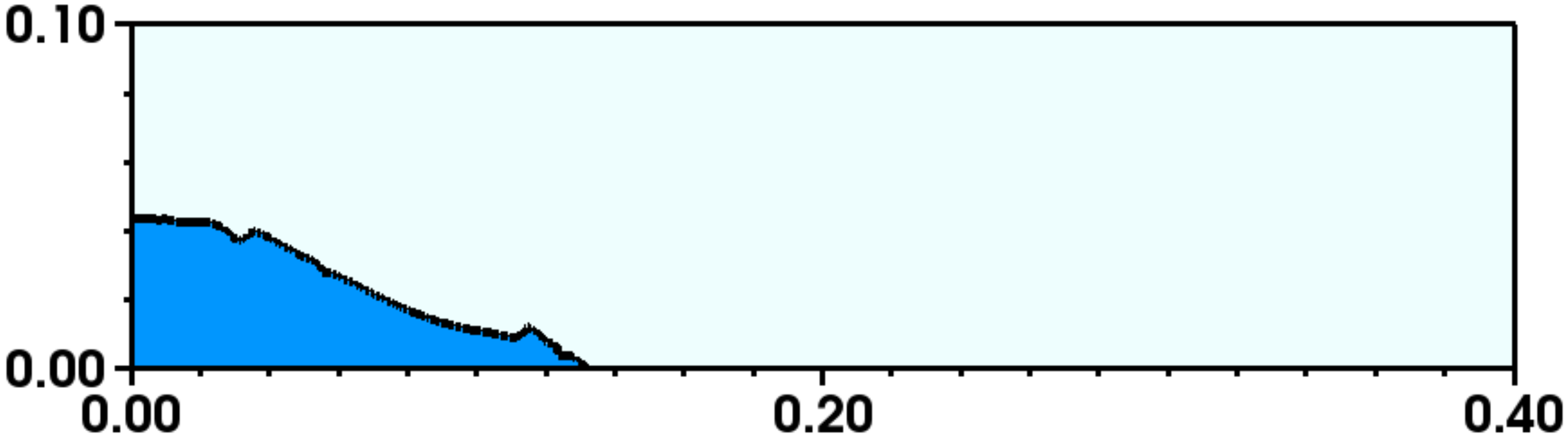}
    \label{nc_water_column_N128_t01}
  }
   \subfigure[Consistent $t = 0.1$ s]{
    \includegraphics[scale = 0.28]{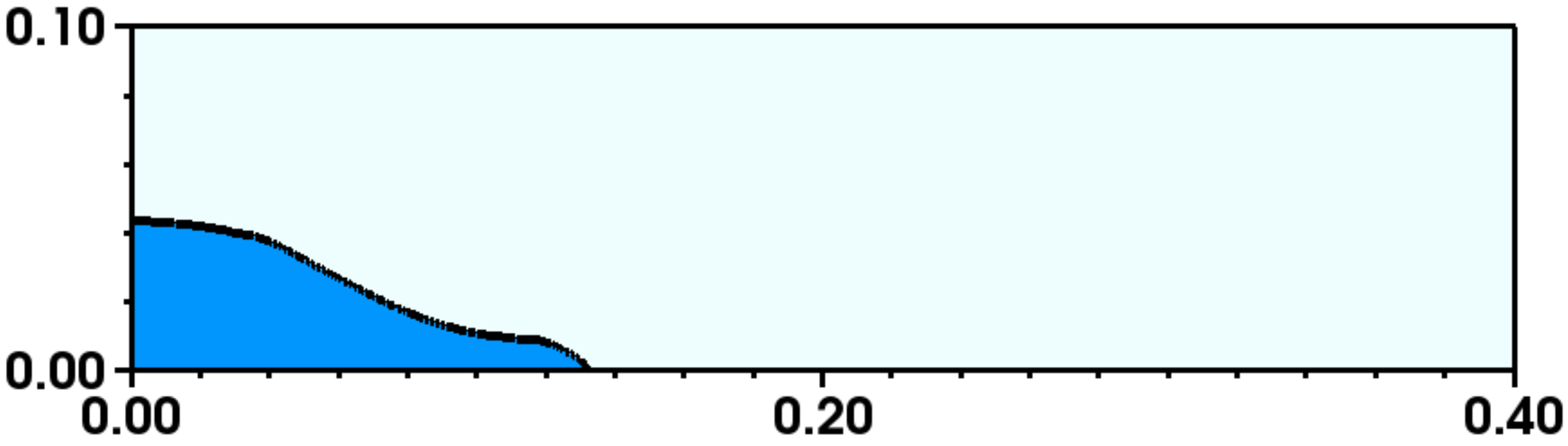}
    \label{c_water_column_N128_t01}
  }
  \subfigure[Inconsistent $t = 0.2$ s]{
    \includegraphics[scale = 0.28]{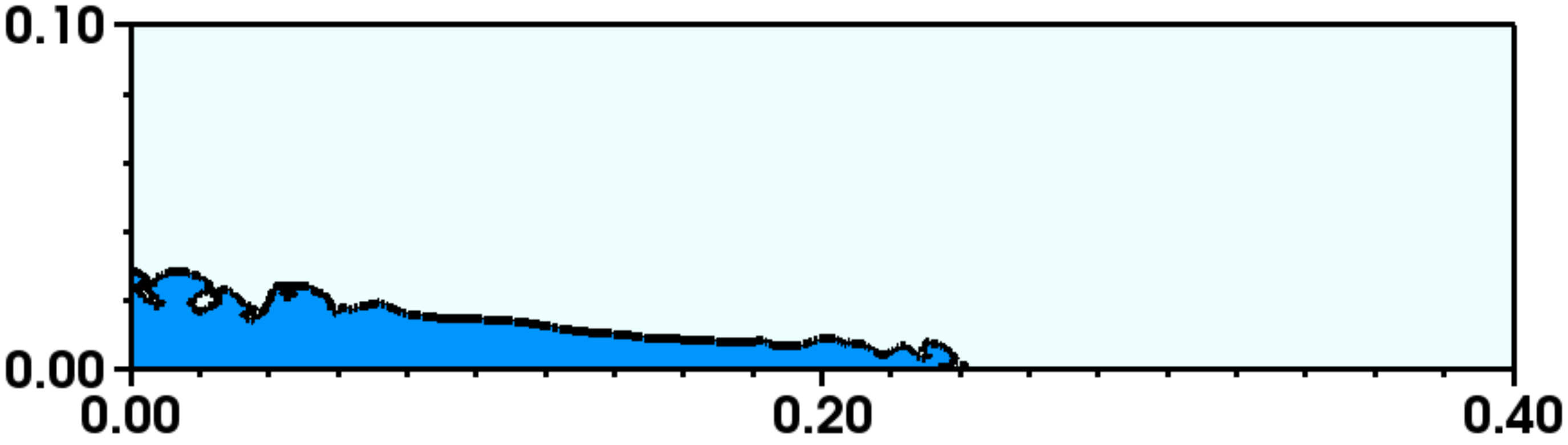}
    \label{nc_water_column_N128_t02}
  }
   \subfigure[Consistent $t = 0.2$ s]{
    \includegraphics[scale = 0.28]{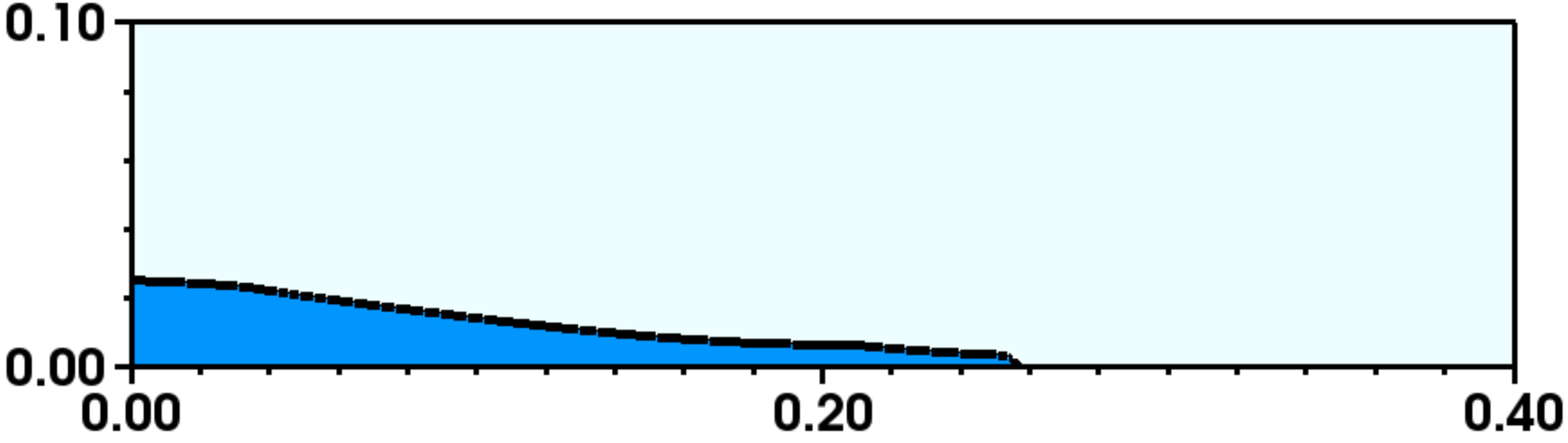}
    \label{c_water_column_N128_t02}
  }
  \subfigure[Inconsistent $t = 0.3$ s]{
    \includegraphics[scale = 0.28]{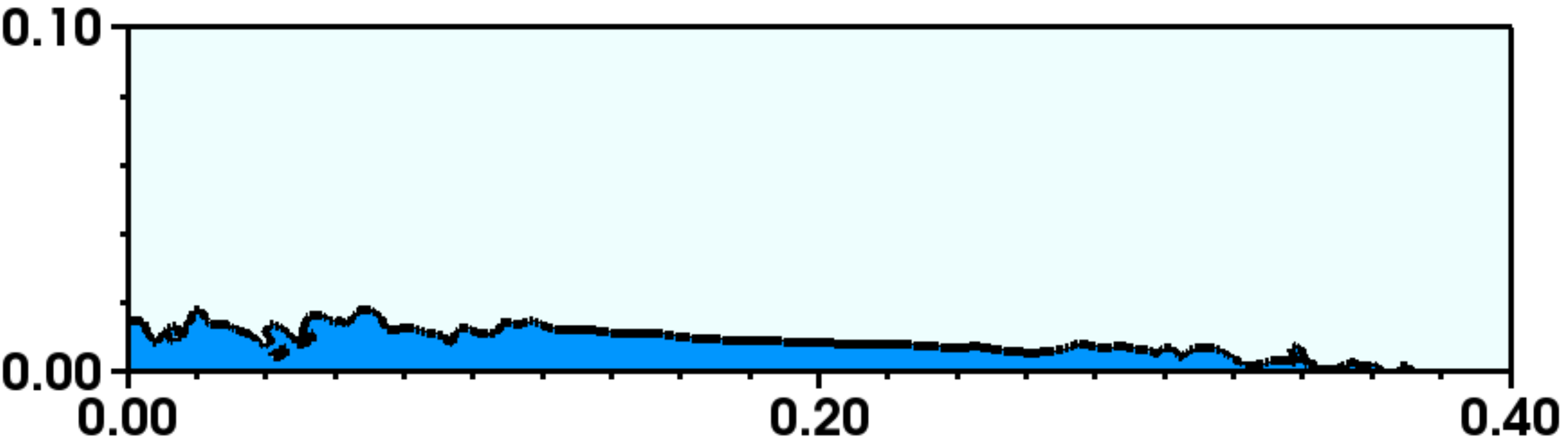}
    \label{nc_water_column_N128_t03}
  }
  \subfigure[Consistent $t = 0.3$ s]{
    \includegraphics[scale = 0.28]{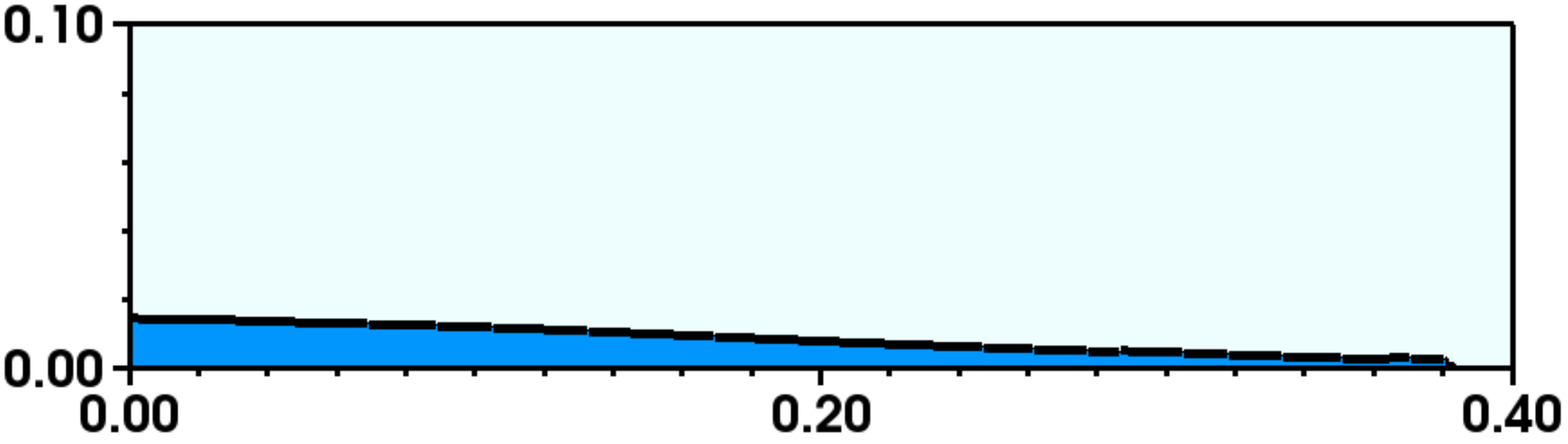}
    \label{c_water_column_N128_t03}
  }
  \caption{Two dimensional evolution of the spreading water column with density 
  ratio $\rho_{\text{l}}/\rho_{\text{g}} = 815.66$ and $\mu_{\text{l}}/\mu_{\text{g}} = 63.88$ at four different time instances: (left) Inconsistent and (right) consistent 
  transport of mass and momentum is 
  used with grid size $512 \times 128$.}
  \label{fig_2D_water_column_N128}
\end{figure}

\begin{figure}[]
  \centering
    \subfigure[$t = 0.0$ s]{
    \includegraphics[scale = 0.28]{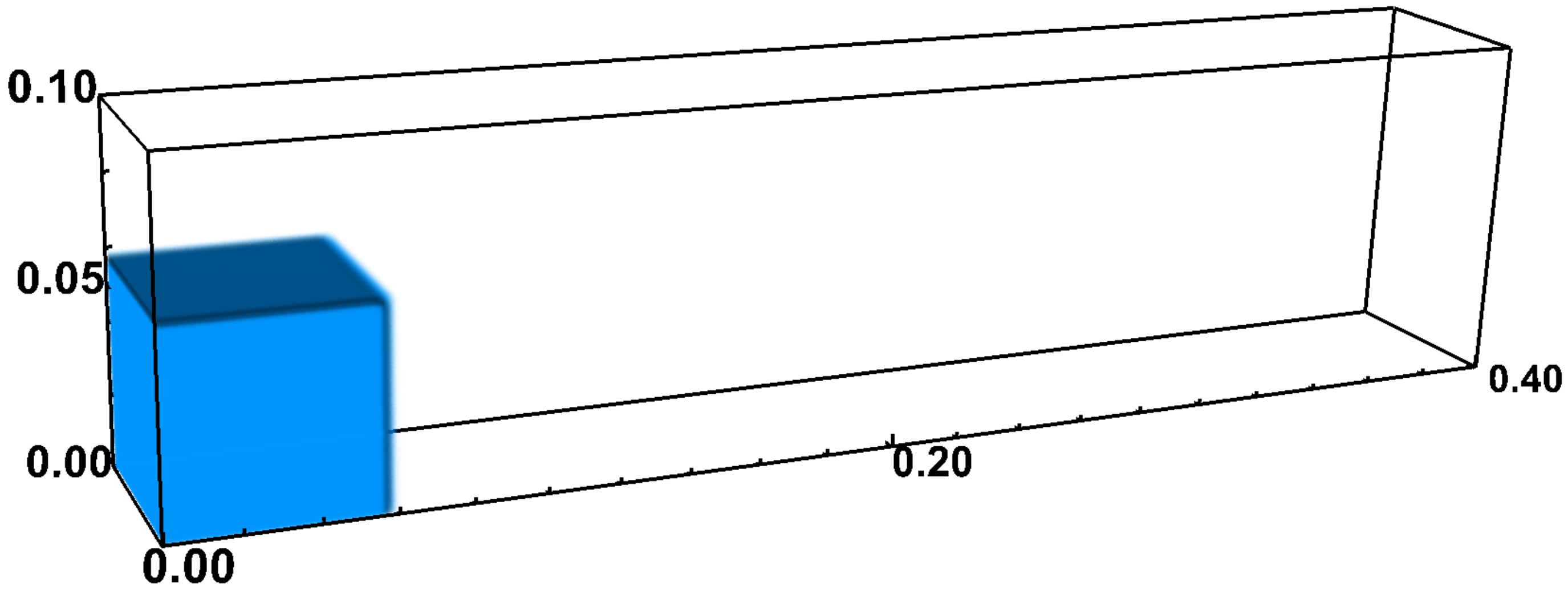}
    \label{c_3D_water_column_t0}
  }
  \subfigure[$t = 0.1$ s]{
    \includegraphics[scale = 0.28]{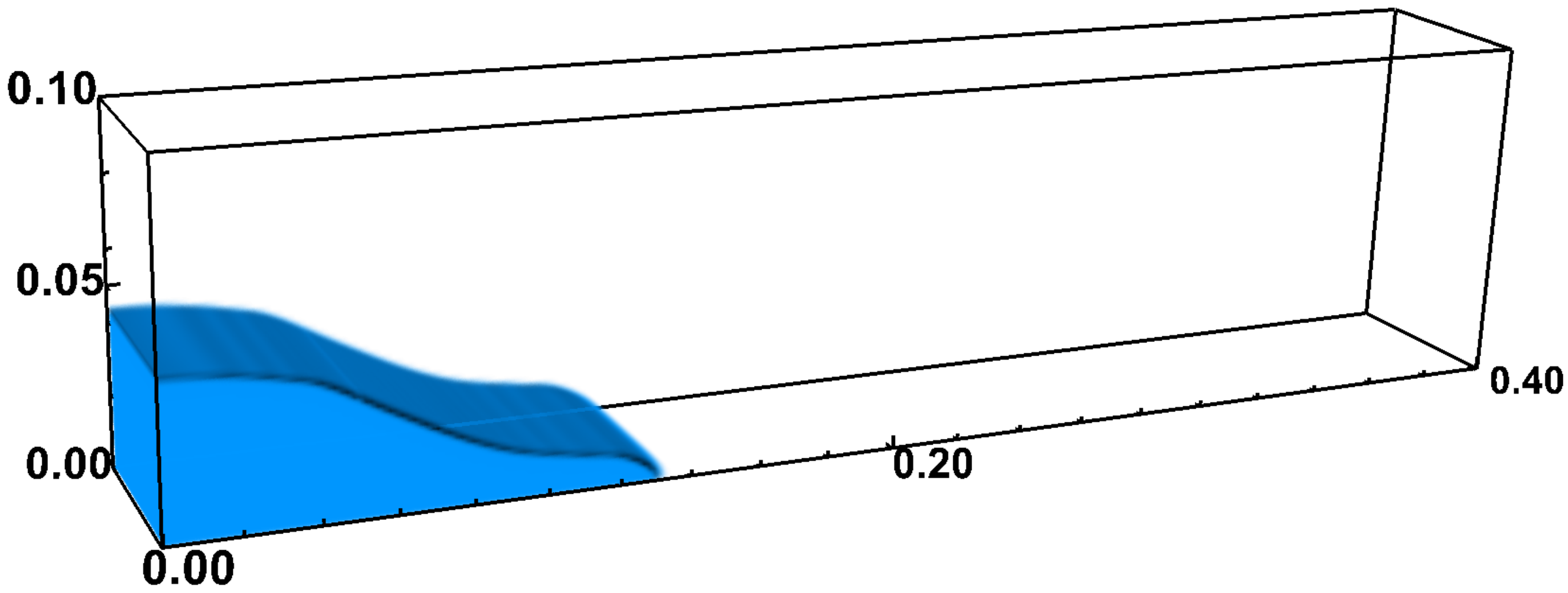}
    \label{c_3D_water_column_t01}
  }
    \subfigure[$t = 0.2$ s]{
    \includegraphics[scale = 0.28]{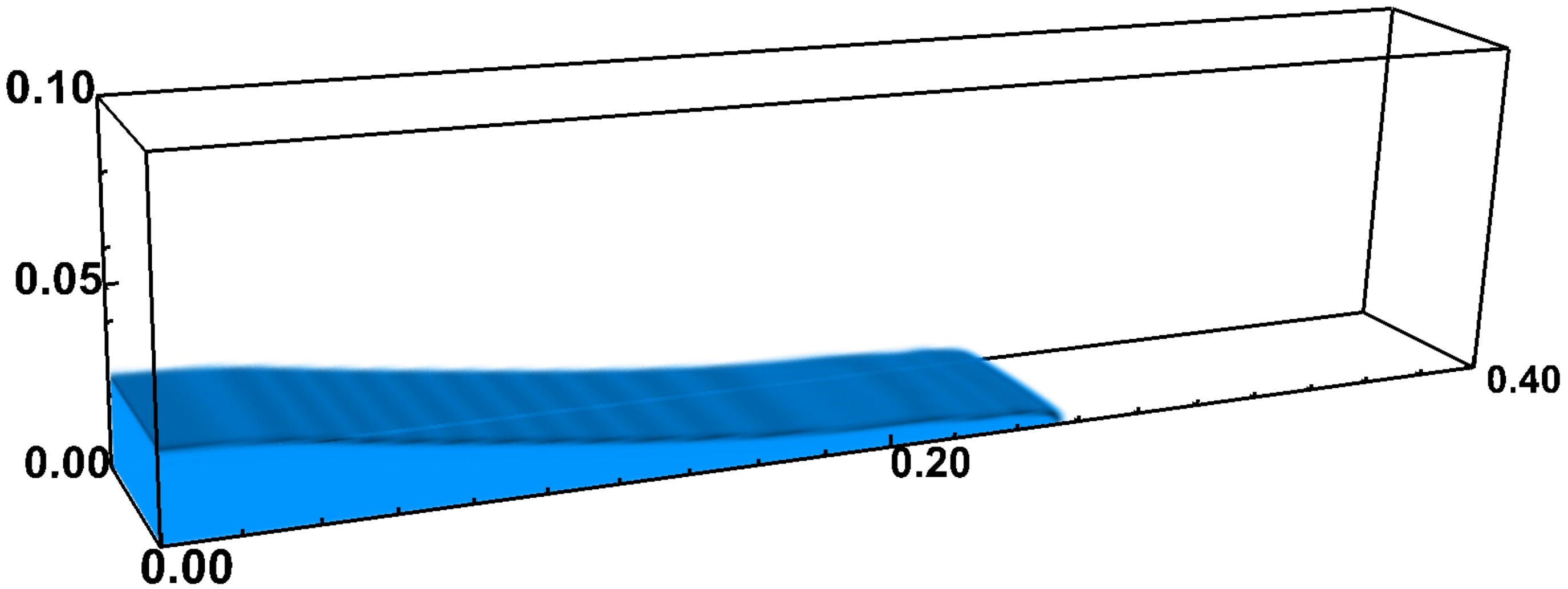}
    \label{c_3D_water_column_t02}
  }
   \subfigure[$t = 0.3$ s]{
    \includegraphics[scale = 0.28]{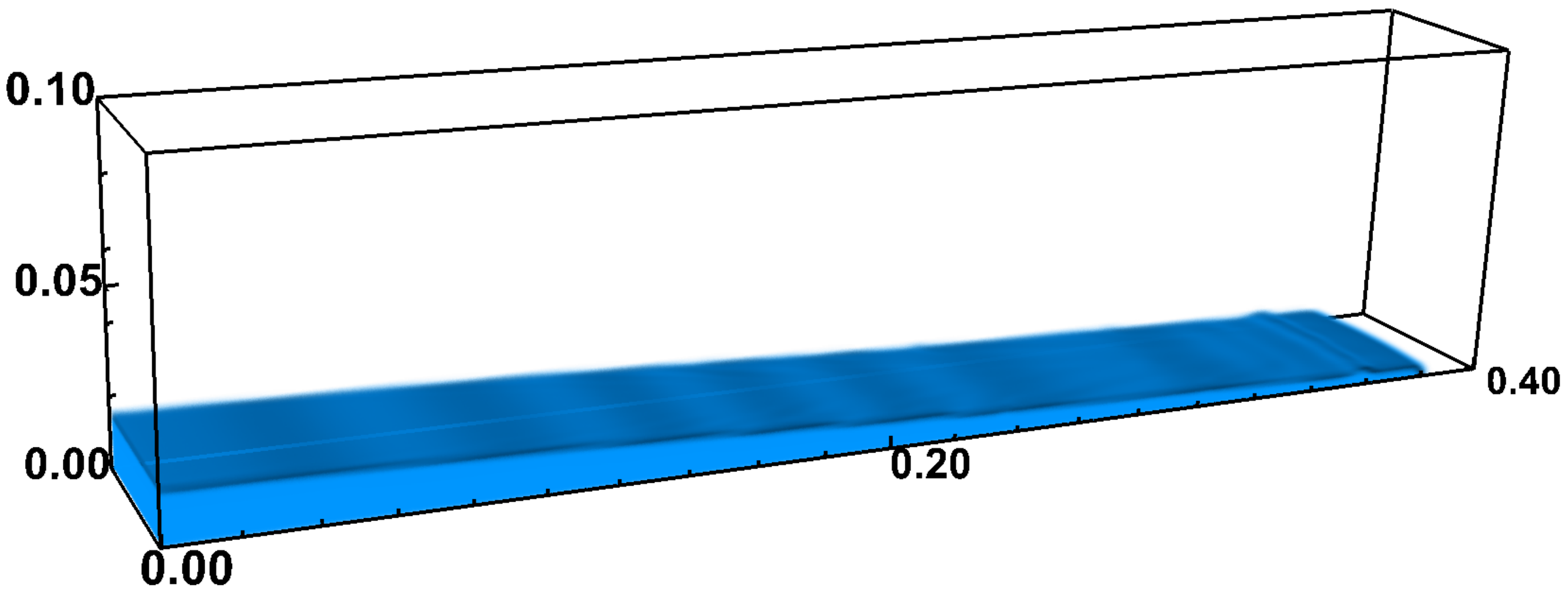}
    \label{c_3D_water_column_t03}
  }
 \caption{Three dimensional evolution of the spreading water column with 
 density ratio $\rho_{\text{l}}/\rho_{\text{g}} = 815.66$ and $\mu_{\text{l}}/\mu_{\text{g}} = 63.88$ 
 at four different time instances. Consistent transport of mass and 
 momentum is used for this case with
  grid size $256 \times 36 \times 64$.}
  \label{fig_3D_water_column_N64}
\end{figure}

Using $N = 128$ and uniform time step $\dt = 1/(62.5 N)$, we carry out the two-dimensional simulation using both 
inconsistent and consistent mass and momentum transport. Fig.~\ref{fig_2D_water_column_N128}
shows the evolution of the water column over time. Unphysical deformations and numerical instabilities 
plague the inconsistent approach whereas the consistent transport remains stable and produces a
physically accurate solution that compares favorably to the experimental study~\cite{Martin1952}. To demonstrate the qualitative accuracy of the present method in three
dimensions, we carry out the three-dimensional simulation using consistent transport with $N = 64$ 
(Fig.~\ref{fig_3D_water_column_N64}). The 3D case differs from the 2D case because that the water column makes contact with an additional pair of walls. It is seen that the numerical solution is again physically reasonable and stable, and that the three-dimensional results qualitatively agree with the two-dimensional simulation.

\begin{figure}[]
  \centering
    \subfigure[Front position]{
    \includegraphics[scale = 0.25]{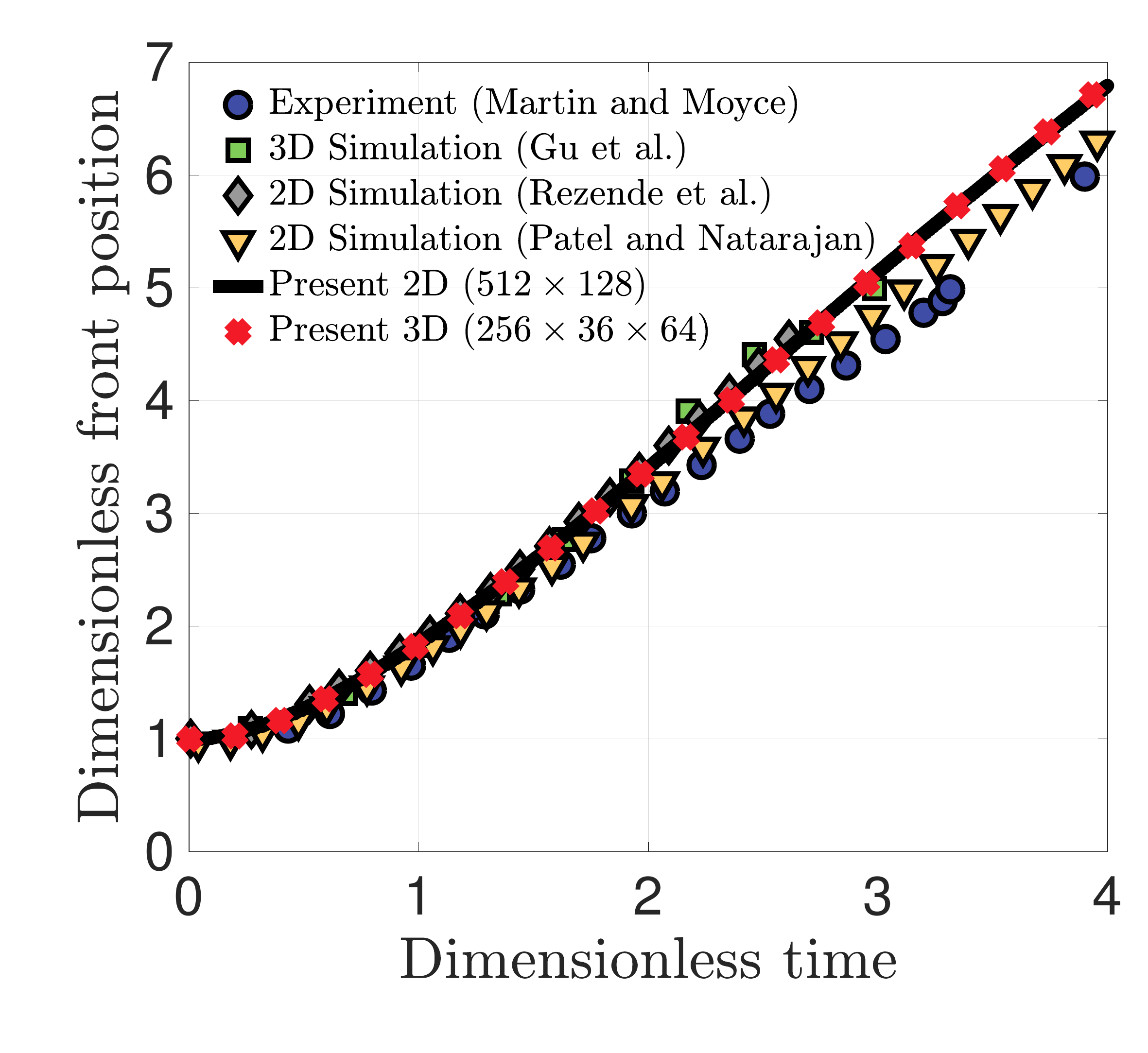}
    \label{dam_break_front}
  }
  \subfigure[Height]{
    \includegraphics[scale = 0.25]{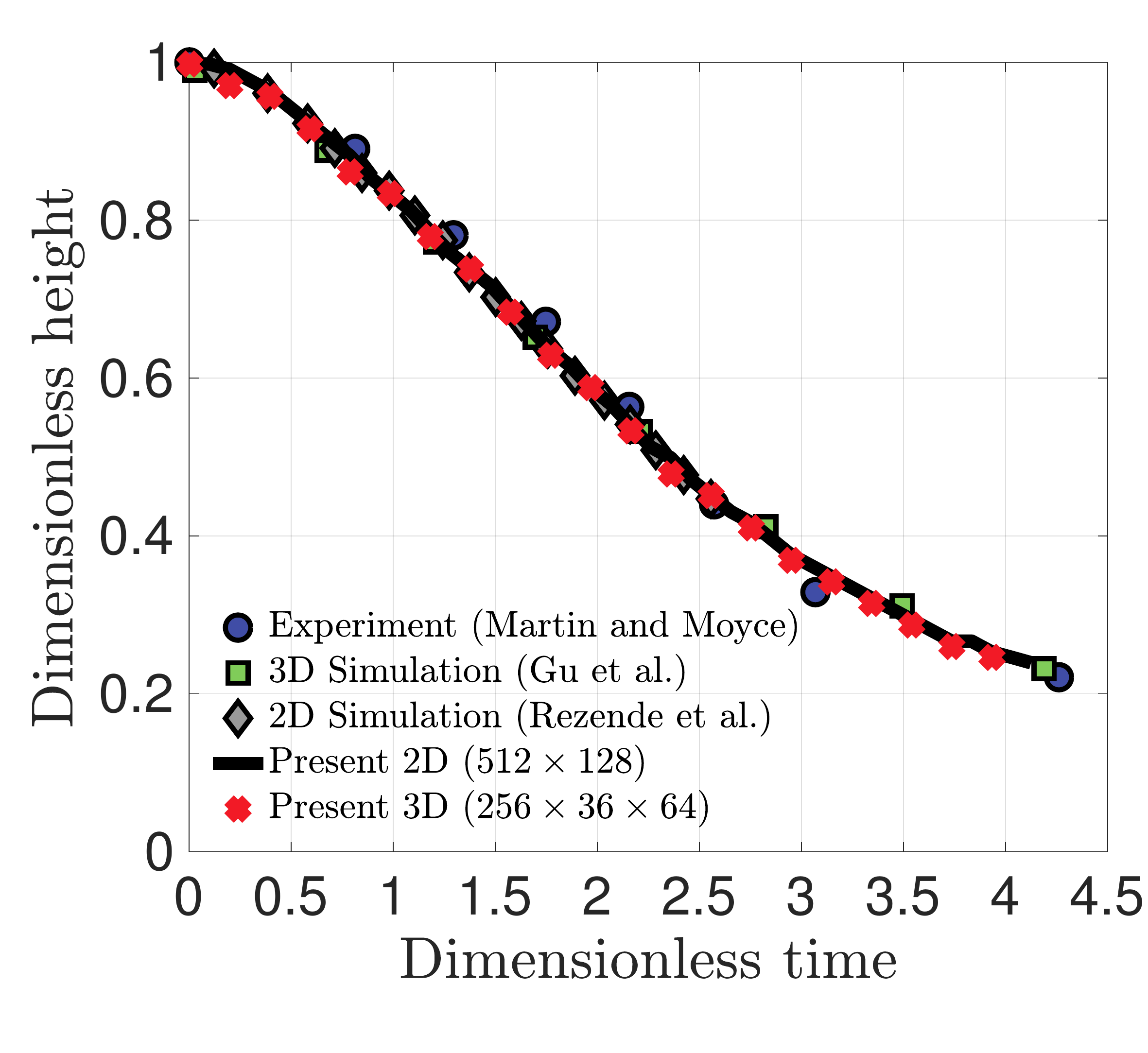}
    \label{dam_break_height}
  }
 \caption{
 Temporal evolution of \subref{dam_break_front} dimensionless front position and
 \subref{dam_break_height} dimensionless height for the present 2D (---, black) and
 3D ($\times$, red) consistent transport simulations, along with
 experimental data ($\bullet$, blue) from Martin and Moyce~\cite{Martin1952},
 3D simulation data ($\blacksquare$, green) from Gu et al.~\cite{Gu2018},
 2D simulation data ($\blacklozenge$, grey) from Rezende et al.~\cite{Rezende2015},
 and 2D simulation data ($\blacktriangledown$, yellow) from Patel and Natarajan~\cite{Patel2017}.
 }
  \label{fig_water_column_quantitative}
\end{figure}

Finally, we quantitatively compare our results to those of other numerical and experimental
studies. The front position and height (nondimensionalized by $a$) of the water column
is plotted against time (nondimensionalized by $\sqrt{a/g}$) in Fig.~\ref{fig_water_column_quantitative}.
Excellent agreement is demonstrated between the present study and previous works.
We again emphasize that this test case demonstrates how vitally important consistent mass and 
momentum transport is for stable simulation of practical multiphase flows, for which air-water density
ratios are ubiquitous.

\subsection{Droplet splashing on thin liquid film}

\subsubsection{Two spatial dimensions}

\begin{figure}[]
  \centering
    \subfigure[$\Re = 6.6$, $T = 0.2$]{
    \includegraphics[scale = 0.28]{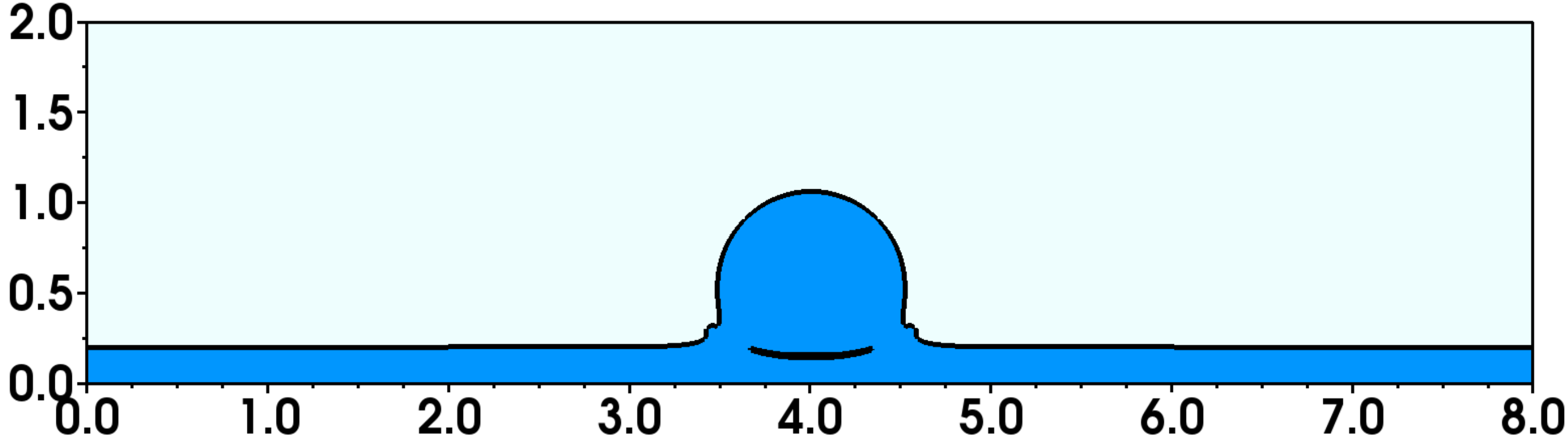}
    \label{Splash2D_LoRe_t02}
  }
  \subfigure[$\Re = 66$, $T = 0.2$]{
    \includegraphics[scale = 0.28]{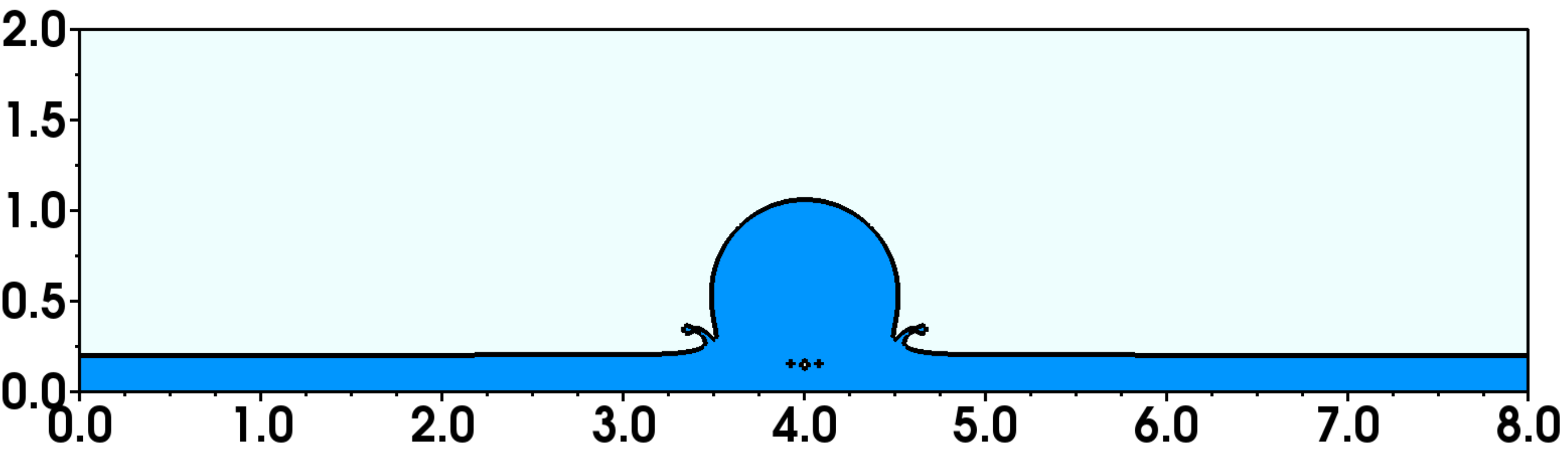}
    \label{Splash2D_HiRe_t02}
  }
    \subfigure[$\Re = 6.6$, $T = 0.5$]{
    \includegraphics[scale = 0.28]{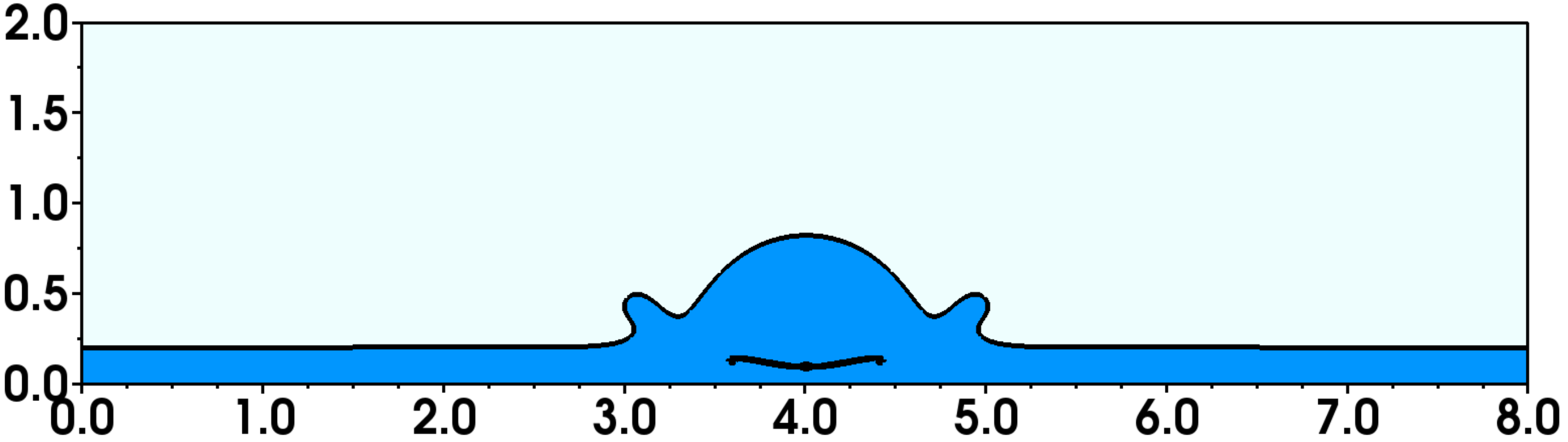}
    \label{Splash2D_LoRe_t05}
  }
   \subfigure[$\Re = 66$, $T = 0.5$]{
    \includegraphics[scale = 0.28]{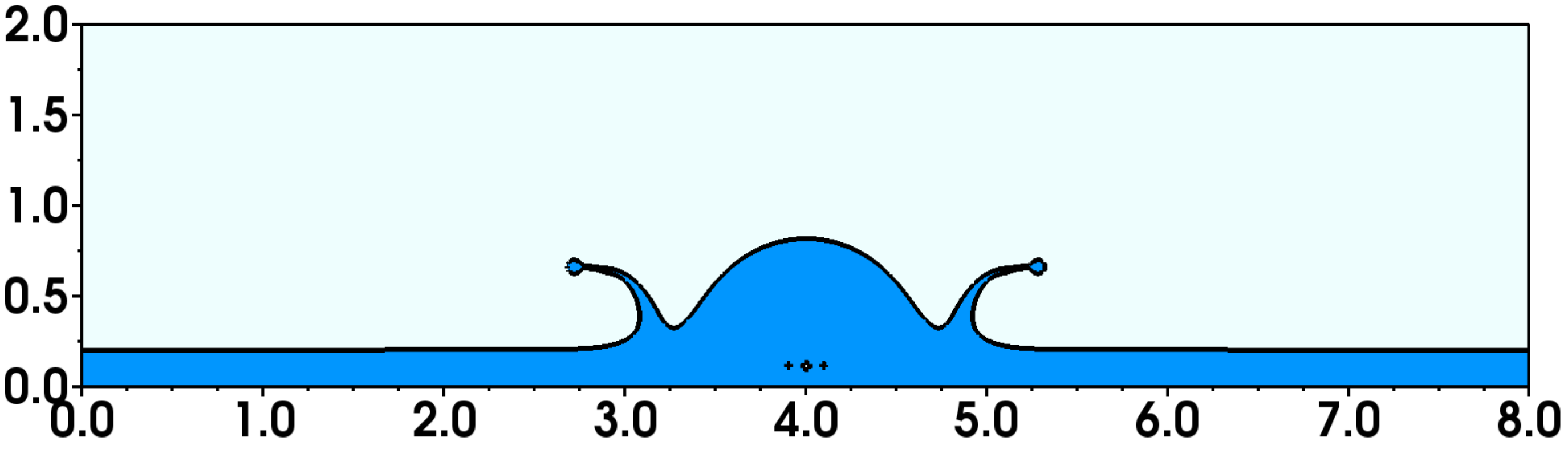}
    \label{Splash2D_HiRe_t05}
  }
  \subfigure[$\Re = 6.6$, $T = 0.7$]{
    \includegraphics[scale = 0.28]{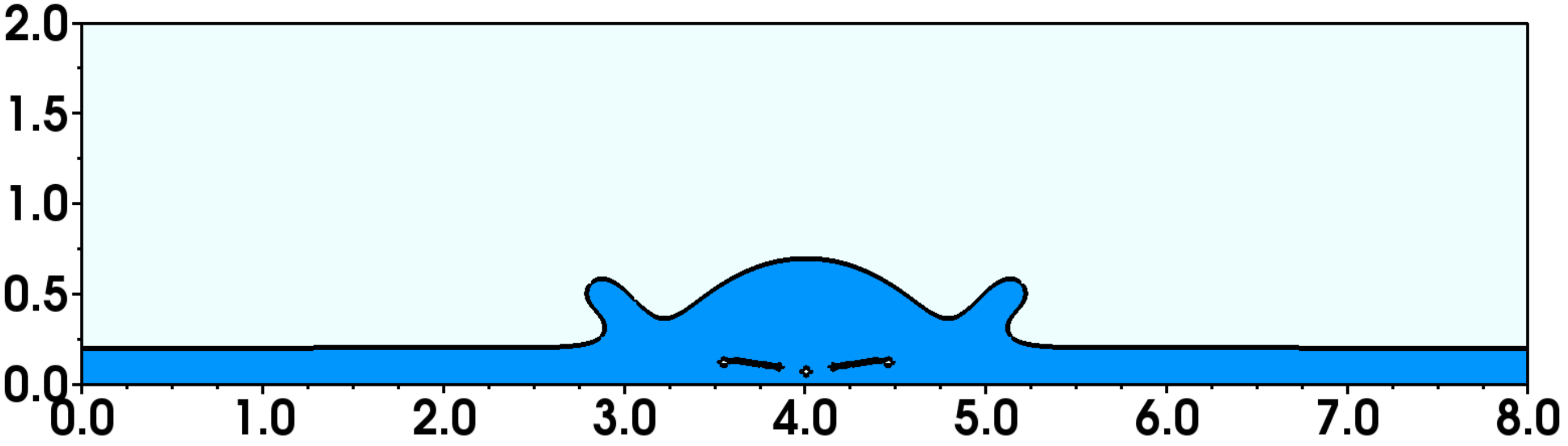}
    \label{Splash2D_LoRe_t07}
  }
   \subfigure[$\Re = 66$, $T = 0.7$]{
    \includegraphics[scale = 0.28]{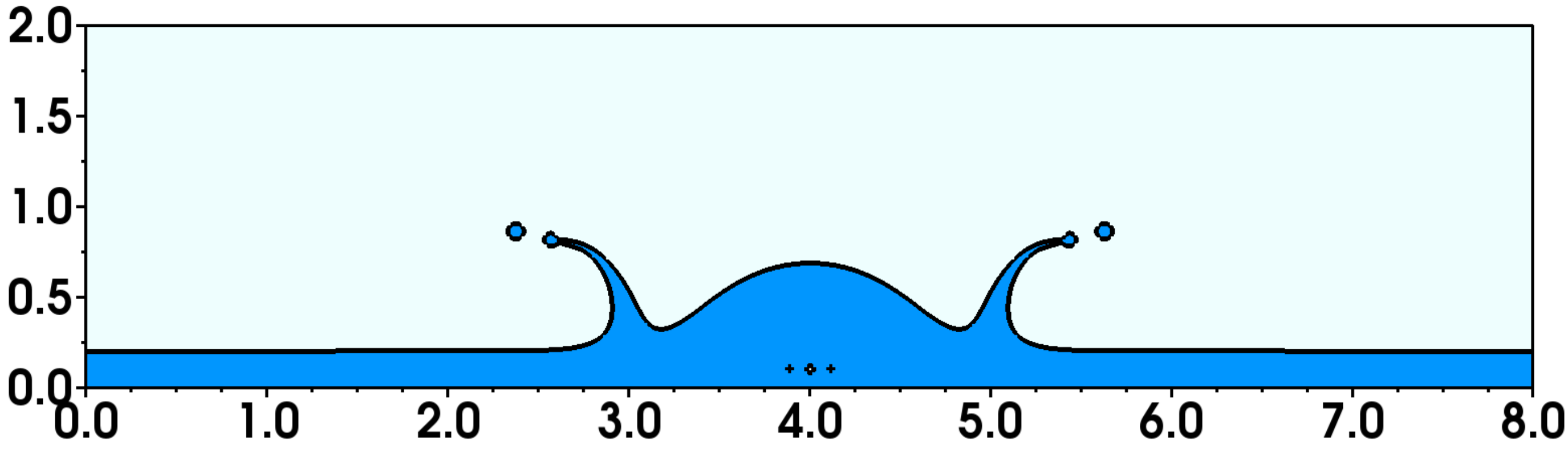}
    \label{Splash2D_HiRe_t07}
  }
  \subfigure[$\Re = 6.6$, $T = 1.0$]{
    \includegraphics[scale = 0.28]{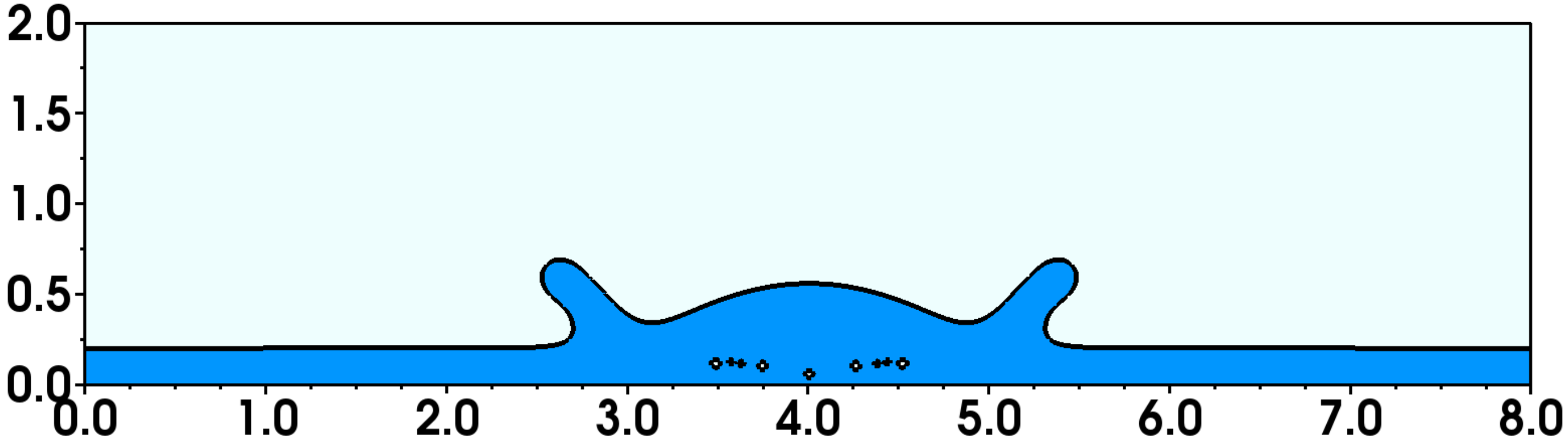}
    \label{Splash2D_LoRe_t1}
  }
  \subfigure[$\Re = 66$, $T = 1.0$]{
    \includegraphics[scale = 0.28]{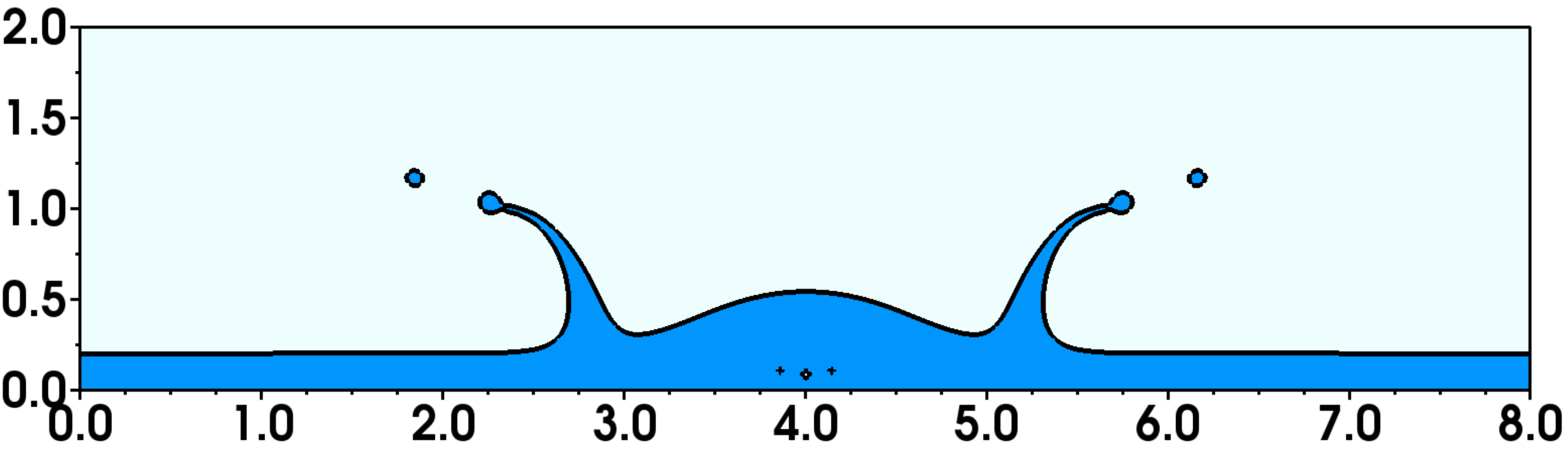}
    \label{Splash2D_HiRe_t1}
  }
  \subfigure[$\Re = 6.6$, $T = 1.5$]{
    \includegraphics[scale = 0.28]{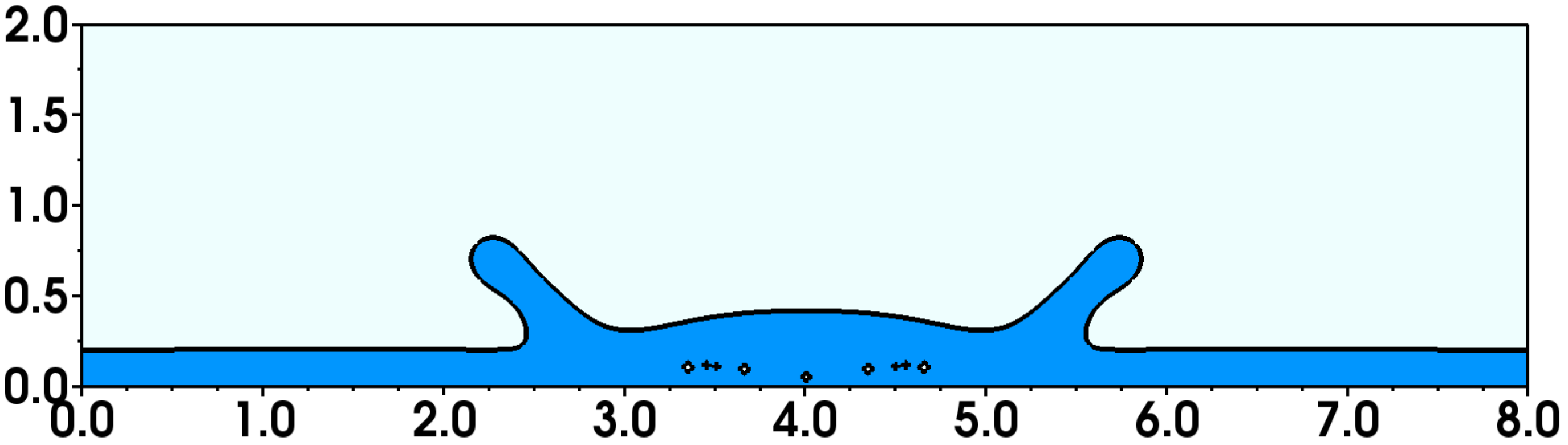}
    \label{Splash2D_LoRe_t15}
  }
  \subfigure[$\Re = 66$, $T = 1.5$]{
    \includegraphics[scale = 0.28]{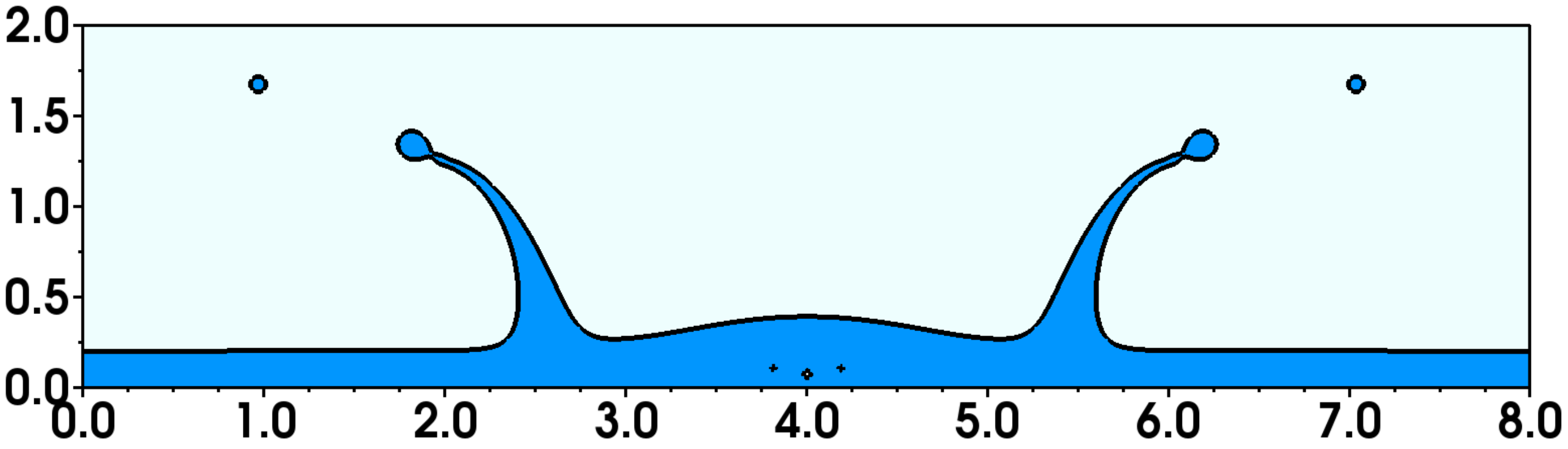}
    \label{Splash2D_HiRe_t15}
  }
  \caption{Temporal evolution of the droplet splashing on a thin liquid film with density 
  ratio $\rho_{\text{l}}/\rho_{\text{g}} = 815$ and $\mu_{\text{l}}/\mu_{\text{g}} = 55$ at five different time instances with grid
  size $1600 \times 400$: (left) $\Re = 6.6$ and (right) $\Re = 66$.}
  \label{fig_Splash2D}
\end{figure}

This section investigates the problem of a droplet splashing on a thin liquid film, which is 
representative of real-world applications such as spray cooling~\cite{Jia2003} and inkjet
printing~\cite{Van2004}. A circular droplet of initial diameter $D = 1$ is placed just above a liquid film in a two-dimensional
computational domain of size $\Omega = [0, 8D] \times[0, 2D]$, which is discretized
by an $4N \times N$ uniform grid with $N = 400$. The droplet has initial downward velocity
$\u(\x,0) = (0, -U)$, with $U = 1$, which is projected %using Eqs.~\eqref{eq_bubble_proj_poisson}
%and~\eqref{eq_bubble_proj_velocity} 
to produce a discretely divergence-free initial condition. % $\up(\x,0)$. 
The droplet has initial center $(X_0, Y_0) = (4D, 0.75D)$ and the liquid film has initial
height $0.2D$. The boundary normal and tangential velocities are set to zero,
imposing no-slip boundary conditions on $\partial \Omega$.
The density ratio between the liquid droplet (or the film phase) and the surrounding gas
phase is $\rho_{\text{l}}/\rho_{\text{g}} = 815$, with $\rho_{\text{g}} = 1$. The Reynolds number based on the gas phase,
$\Re = \rho_{\text{g}} U D/\mu_{\text{g}}$, is used to determine the gas phase viscosity, and the viscosity ratio between liquid and gas
is $\mu_{\text{l}}/\mu_{\text{g}} = 55$. The Weber number based on the gas phase, $\We = \rho_{\text{g}} U^2 D/\sigma = 0.126$,
is used to determine the surface tension coefficient $\sigma$. Gravity is neglected for this particular problem, which, because of the high impact velocity, 
is a convection and surface tension driven flow.
Time is nondimensionalized as $T = tU/D$, and each case is run until $T = 1.5$
with a constant time step size of $\dt = 1/(10 N)$.
The conservative discretization is used for all cases considered here,
but we found that the non-conservative discretization yields similar results
(data not shown). 
For all of the cases, the density is synchronized via the level set function at the beginning
of each time step with one grid cell of smearing ($\ncells = 1$) on either side of the interface. 
Similar problems have been
studied numerically using the VOF methods by Coppola et al.~\cite{Coppola2011}
and by Patel and Natarajan~\cite{Patel2017}, and using a lattice Boltzmann method by Li et al.~\cite{Li2013}.

\begin{figure}[H]
  \centering
    \subfigure[$\Re = 66$, $T = 0.2$]{
    \includegraphics[scale = 0.25]{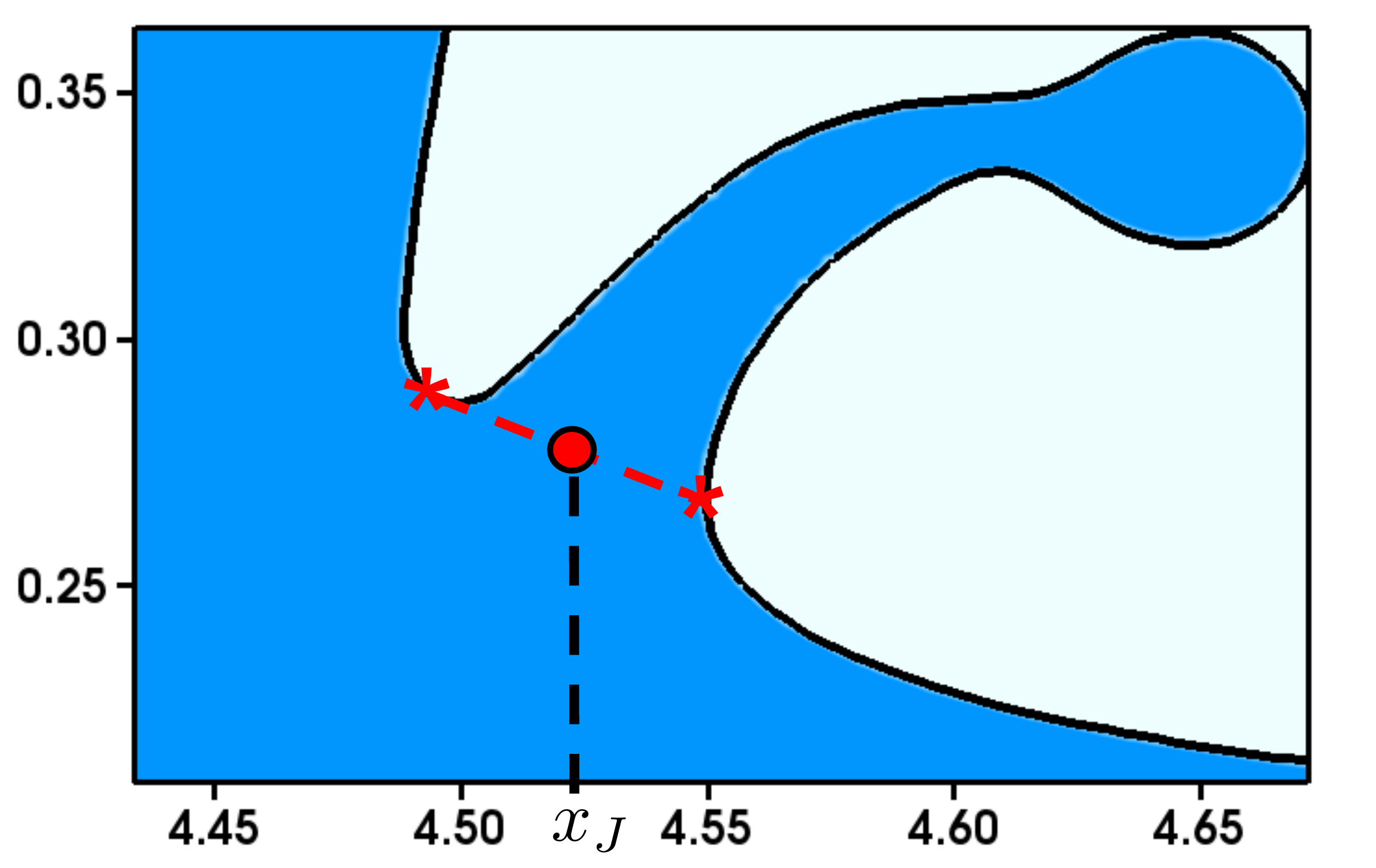}
    \label{jet_figure}
  }
  \subfigure[Jet base location vs. time]{
    \includegraphics[scale = 0.22]{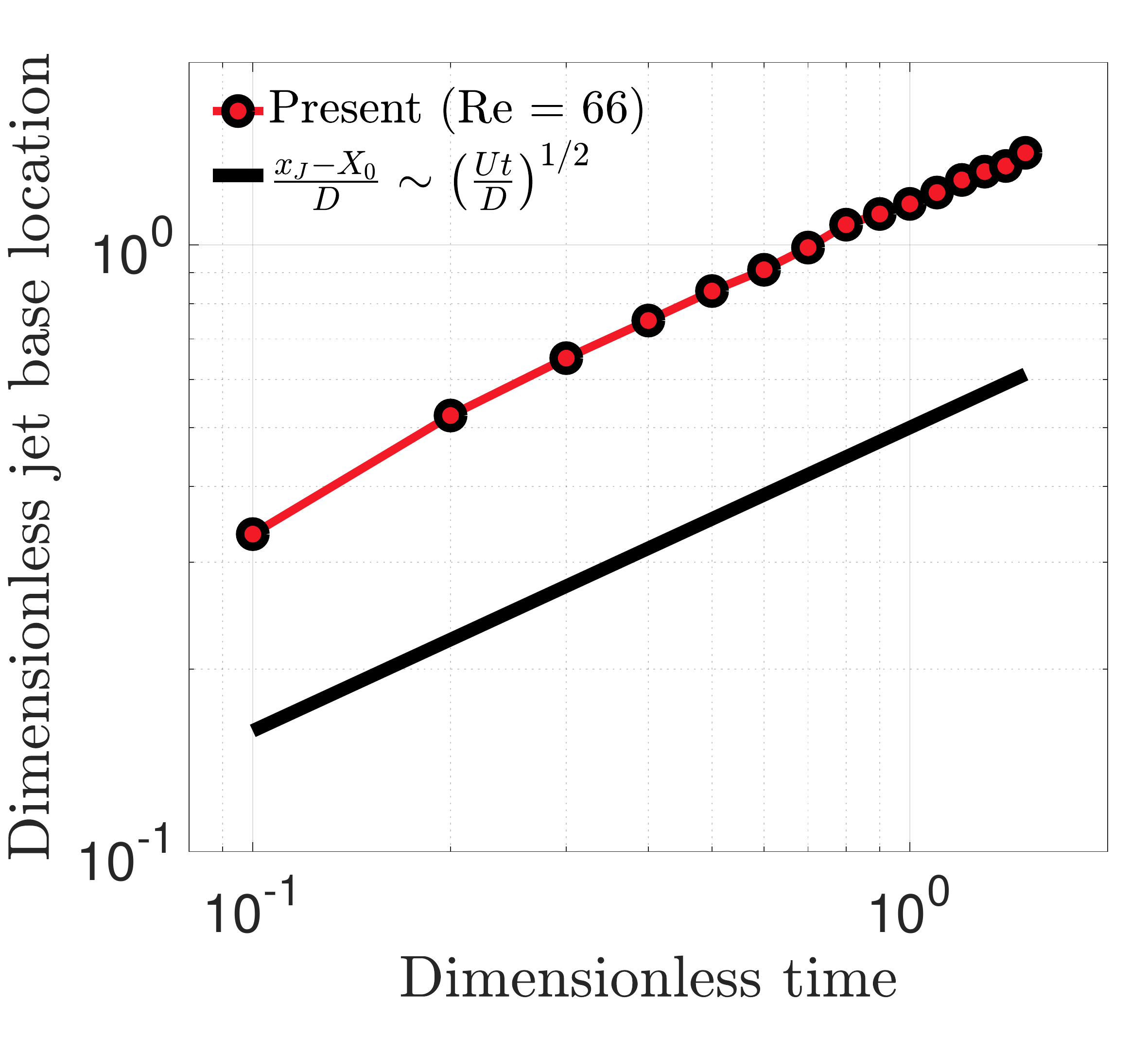}
    \label{splash_2d_spread}
  }
  \caption{\subref{jet_figure} Location of the jet base $x_\text{J}$ ($\bullet$, red), which is the midpoint of the two neck points ($*$, red) for the liquid sheet ($T = 0.2$, $\Re = 66$);
  \subref{splash_2d_spread} Temporal evolution of the dimensionless jet based location.}
  \label{fig_Re66_spread}
\end{figure}

As a first test we consider $\Re = 6.6$ (left panels in Fig.~\ref{fig_Splash2D}),
which can be directly compared to the results of Patel and Natarajan~\cite{Patel2017}.
It is seen that the droplet merges into the liquid film and produces a thick, symmetric
liquid sheet upon impact. Because of the relatively low Reynolds number, no satellite droplets are formed.
The results are in excellent qualitative agreement with prior work~\cite{Patel2017},
including resolution of the gas entrapment by the droplet into the liquid film. A second test case
is carried out at $\Re = 66$ (right panels in Fig.~\ref{fig_Splash2D}), which not only
produces a splash with higher vertical reach than the lower Reynolds number case but also
exhibits satellite droplets that break away from film. The results compare favorably to those 
of Patel and Natarajan~\cite{Patel2017}. As described by Coppola et al.~\cite{Coppola2011}, the jet base location $x_\text{J}$ can be computed by
averaging the $x$-position of the two neck points of the liquid sheet (see Fig.~\ref{jet_figure}).
According to the theoretical studies of Josserand and Zaleski~\cite{Josserand2003} and
Howison et al.~\cite{Howison2005}, the normalized jet base location $\left(x_\text{J}-X_0\right)/D$
should scale with the square root of dimensionless time $\left(U t/D\right)^{1/2}$. As shown in 
Fig.\ref{splash_2d_spread}, our simulation shows excellent quantitative agreement with this power law.
These cases demonstrate that surface tension flows with complex merging dynamics are adequately
simulated by the present numerical scheme.

\subsubsection{Three spatial dimensions}
For our final case, we consider the problem of a three-dimensional droplet splashing on a thin
liquid film. Experimental and theoretical consideration~\cite{Cossali1997, Yarin1995} of this 
particular problem has inspired numerous numerical studies of this phenomena, including
those of Nikolopoulos et al.~\cite{Nikolopoulos2005, Nikolopoulos2007} and
Manik et al.~\cite{Manik2018}. In particular, we follow a similar setup as described
in~\cite{Nikolopoulos2007}. A spherical droplet of initial diameter $D_0 = 1$ is placed in a computational 
domain of size $\Omega = [0,L]^3 = [0,8D_0]^3$, which is discretized by a base uniform grid of size $N^3$ (with $N = 20$) and four grid levels $\ell = 4$ with refinement ratio $\nref = 2$. Hence at the 
finest level, the grid spacing is $\dx_\textrm{min} = \dy_\textrm{min} = \dz_\textrm{min} = 8/160$.
The domain is filled with a liquid film of height $a$, with the dimensionless height set as
$\overline{a} = a/D_0 = 0.116$ and the droplet has initial center $(X_0, Y_0, Z_0) = (0.5L, 0.5L, 2.8D_0)$.
The density ratio between the liquid droplet (or the film phase) and the surrounding gas phase
is $\rho_{\text{l}}/\rho_{\text{g}} = 1000$, with $\rho_{\text{l}} = 1000$. The viscosity ratio between liquid and gas is 
$\mu_{\text{l}}/\mu_{\text{g}} = 40$, with $\mu_{\text{l}} = 8.9 \times 10^{-4}$. 
The droplet has initial downward velocity $\u(\x,0) = (0, -U_0)$, which is projected %using 
%Eqs.~\eqref{eq_bubble_proj_poisson} and~\eqref{eq_bubble_proj_velocity} 
to produce a discretely 
divergence-free initial condition $\up(\x,0)$. The Reynolds number based on the liquid phase,
$\Re = \rho_{\text{l}} U_0 D_0/\mu_{\text{l}} = 11294$, is used to specify the initial downward velocity.
The Weber number based on the liquid phase, $\We = \rho_{\text{l}} U_0^2 D_0/\sigma = 250$,
is used to specify the surface tension coefficient $\sigma$. The Froude number,
$\Fr = U_0^2/(g a) = 363$, is used to specify the gravitational constant $g$.
No-slip boundary conditions are applied on all computational
boundaries. A constant time step size of $\dt = 12.5 \dx_\textrm{min}$ is used, yielding an approximate
initial CFL number of $0.125$. The CFL number throughout the simulation remains less than $0.5$
for all time steps. The conservative discretization is used for this case.
The density is synchronized using the level set function at the beginning
of each time step with one grid cell of smearing ($\ncells = 1$) on either side of the interface.

\begin{figure}[H]
  \centering
    \subfigure[$T = -2.33$]{
    \includegraphics[scale = 0.23]{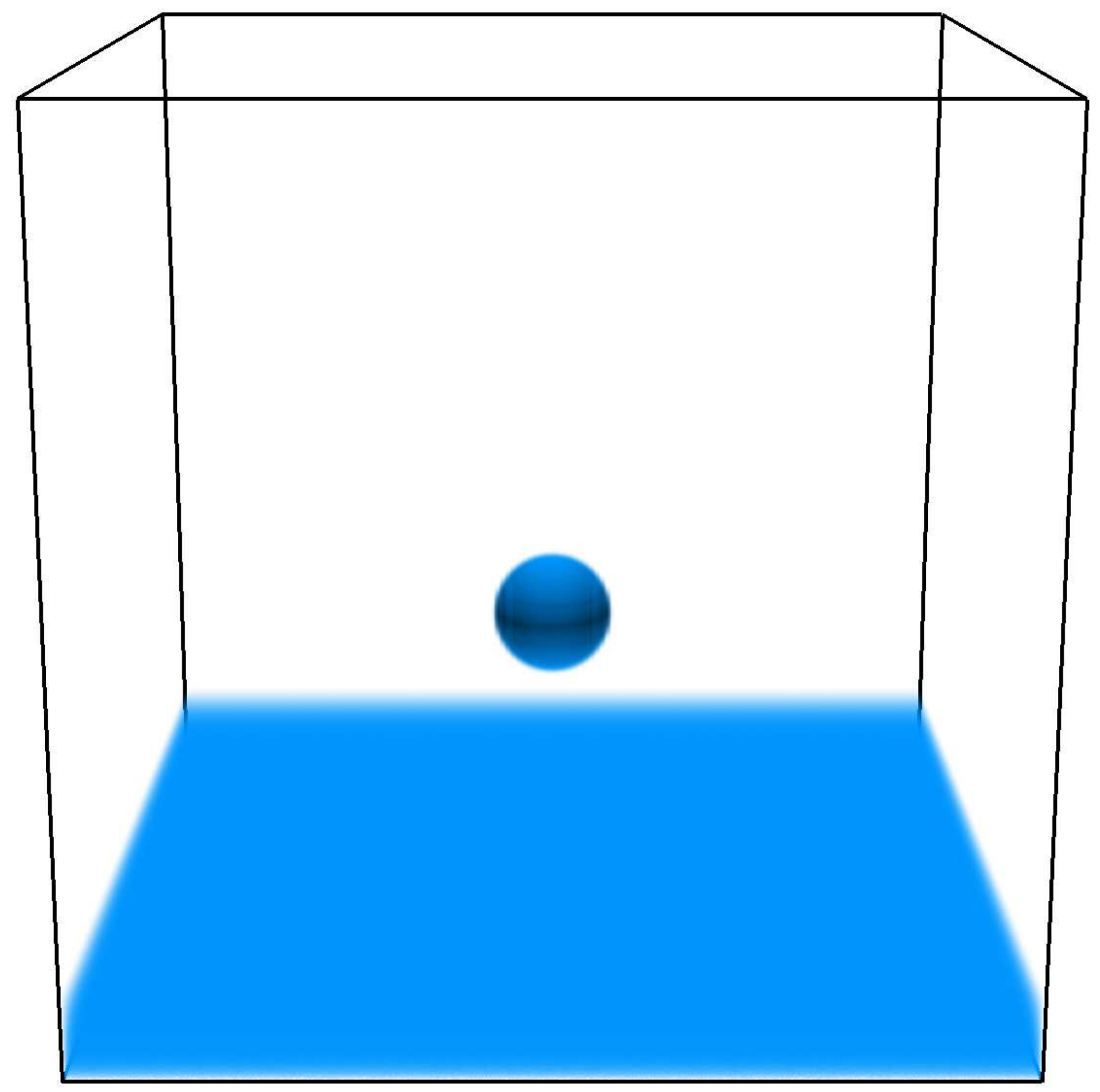}
    \label{3D_NC_Splash_amr_mT2p33}
  }
  \subfigure[$T = 0.00$]{
    \includegraphics[scale = 0.23]{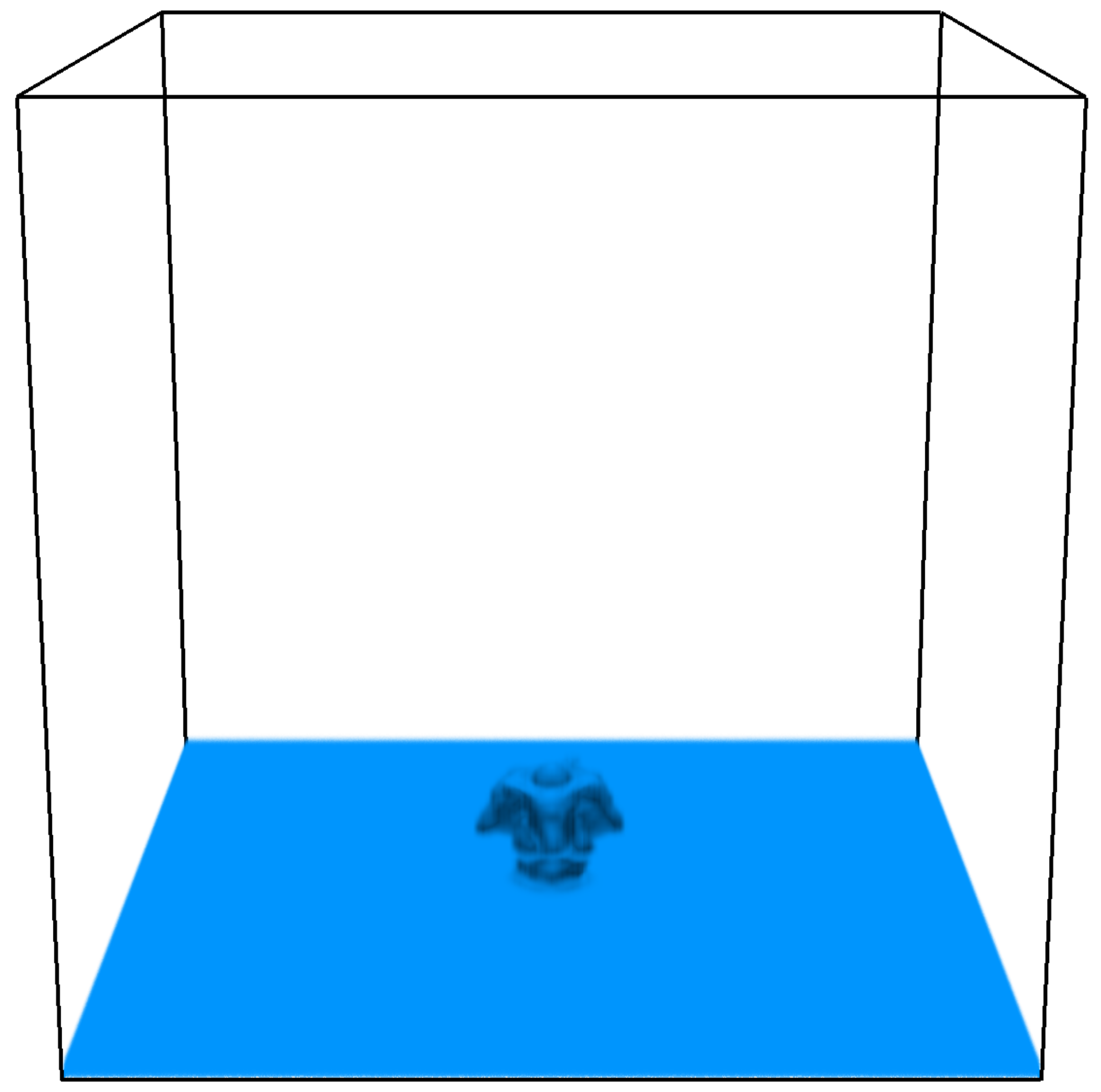}
    \label{3D_NC_Splash_amr_T0}
  }
  \subfigure[$T = 2.02$]{
    \includegraphics[scale = 0.23]{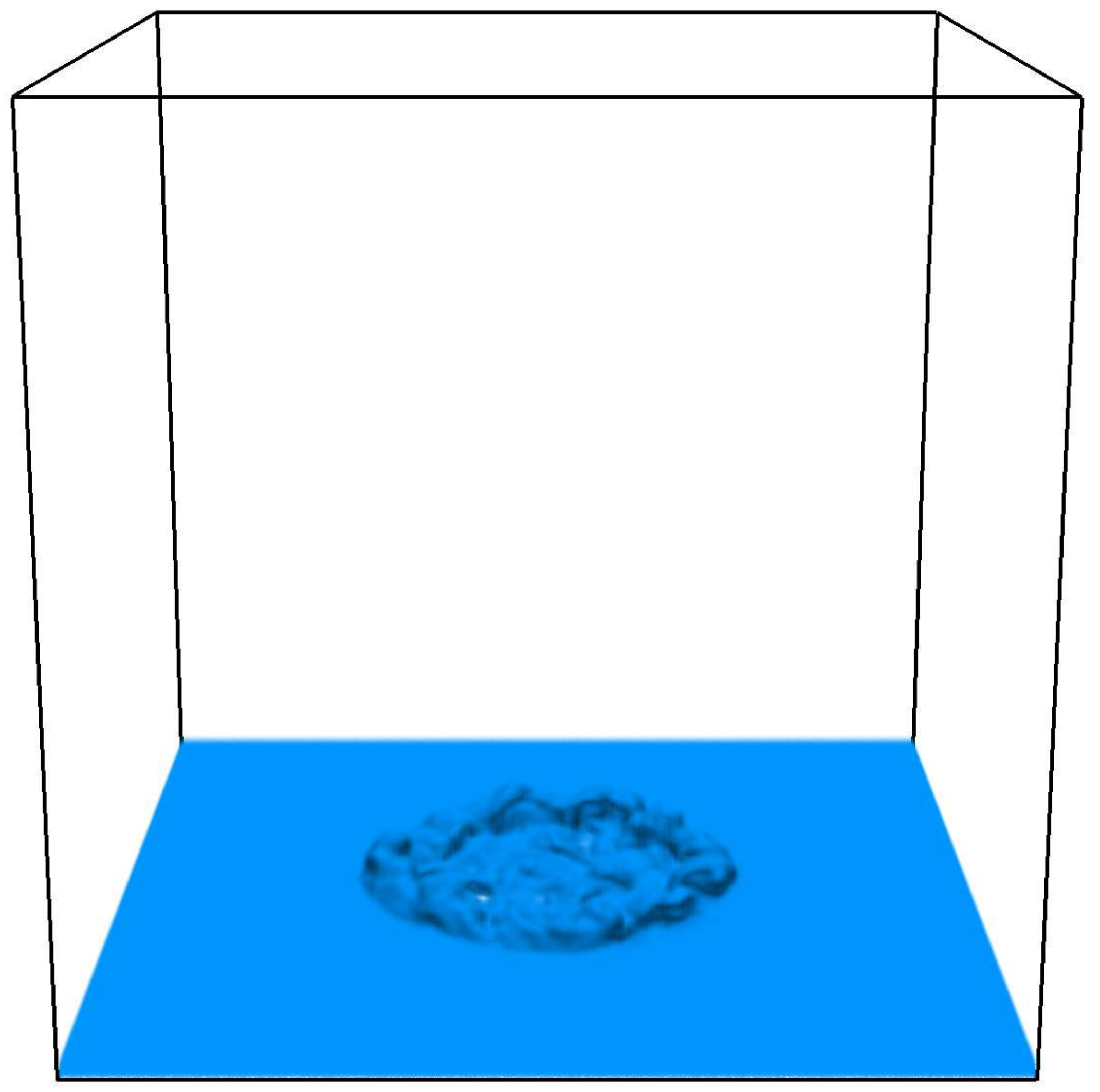}
    \label{3D_NC_Splash_amr_T2p02}
  }
  \caption{\subref{3D_NC_Splash_amr_mT2p33}-\subref{3D_NC_Splash_amr_T2p02}
  Temporal evolution of the 3D droplet splashing on a thin liquid film with density 
  ratio $\rho_{\text{l}}/\rho_{\text{g}} = 1000$ and $\mu_{\text{l}}/\mu_{\text{g}} = 40$ at three different time instances.
  Inconsistent transport of mass and momentum occurs while employing 
  a non-conservative momentum integrator.
  Contrast these results to those shown in Fig.~\ref{fig_Splash3D}, in which consistent transport is used.
  }
  \label{fig_Splash3D_NC}
\end{figure}

\begin{figure}[H]
  \centering
    \subfigure[$T = 0.00$]{
    \includegraphics[scale = 0.28]{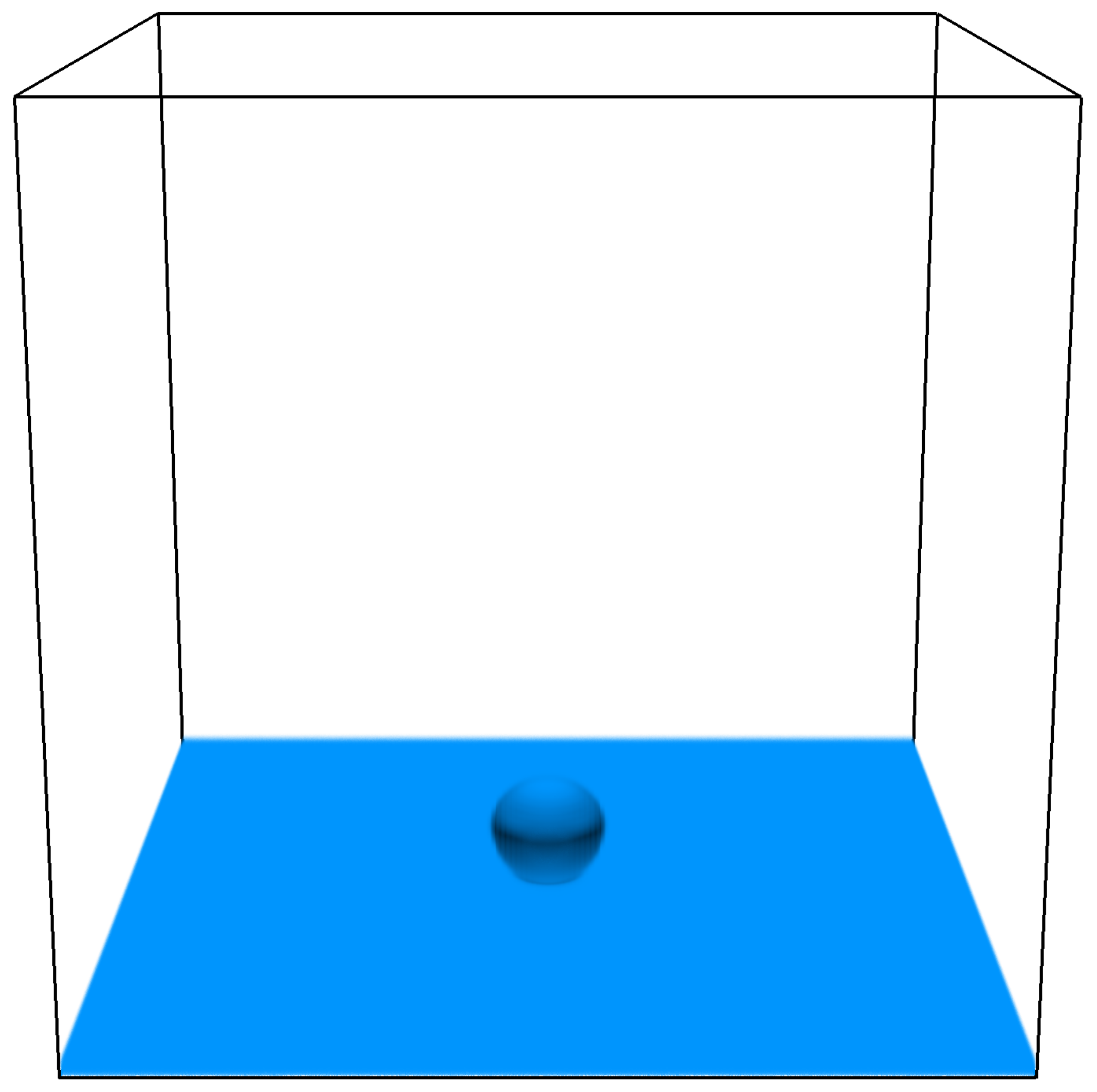}
    \label{3D_Splash_T0}
  }
  \subfigure[$T = 1.70$]{
    \includegraphics[scale = 0.28]{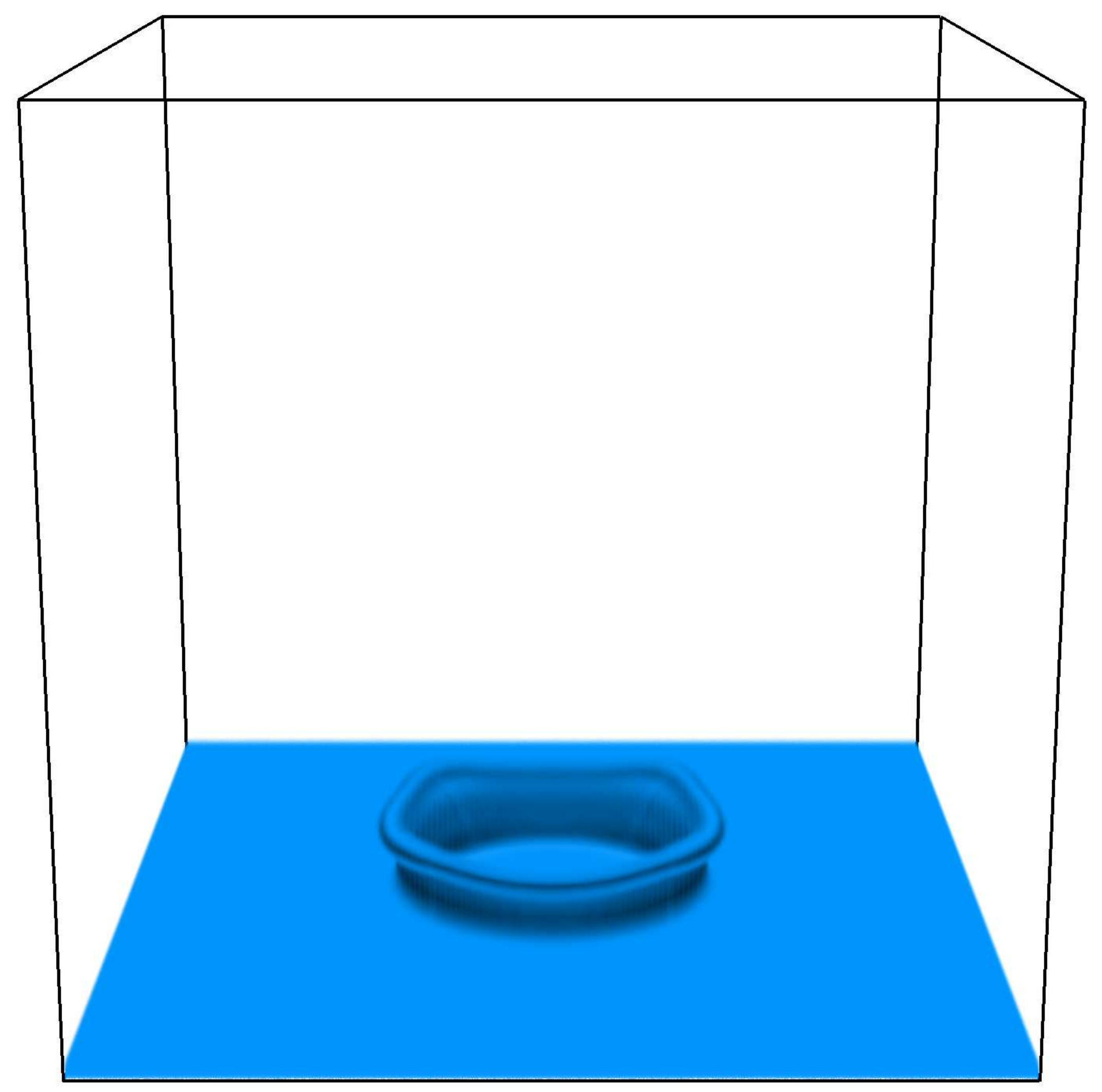}
    \label{3D_Splash_T17}
  }
  \subfigure[$T = 3.02$]{
    \includegraphics[scale = 0.28]{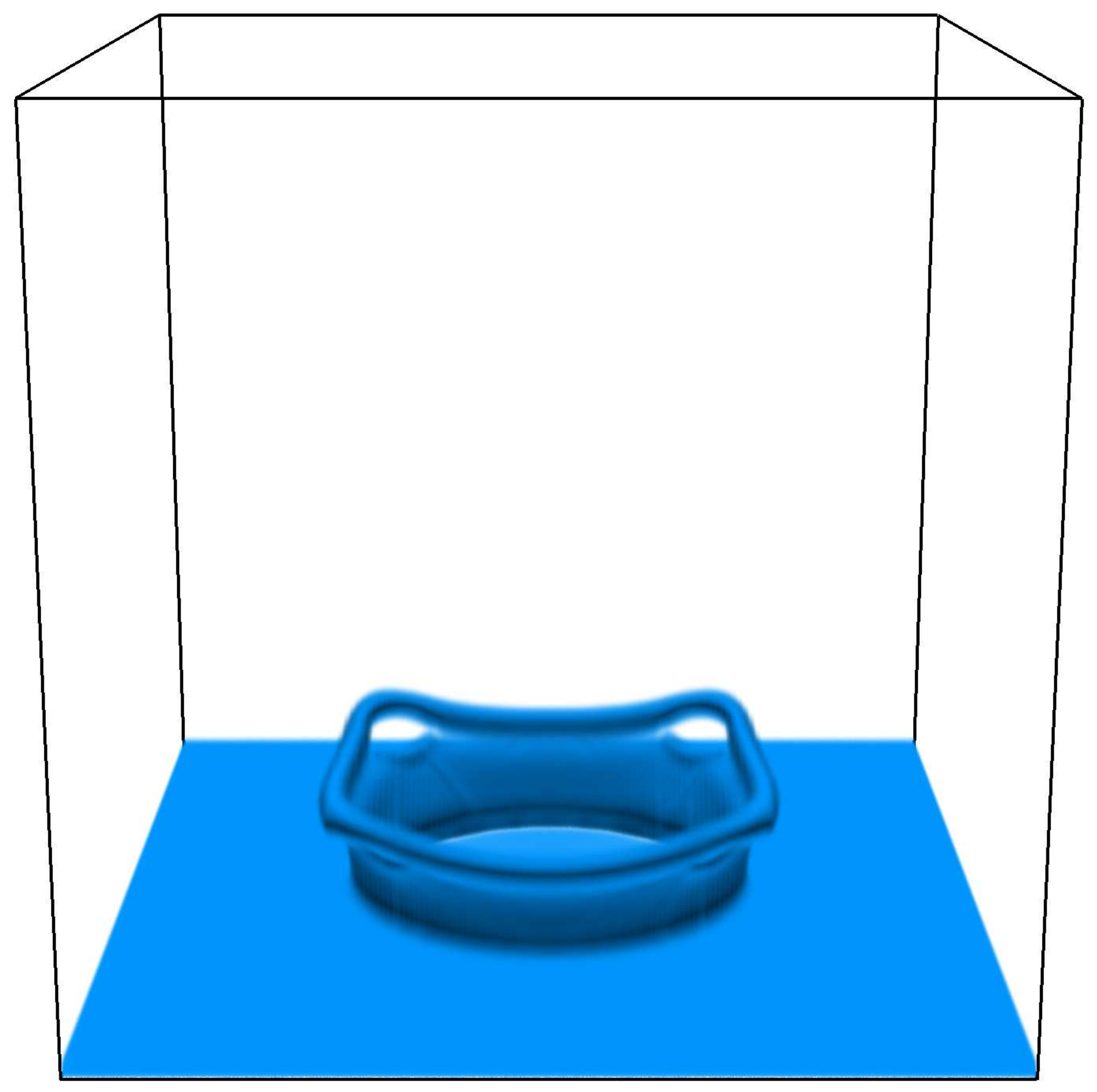}
    \label{3D_Splash_T302}
  }
  \subfigure[$T = 5.34$]{
    \includegraphics[scale = 0.28]{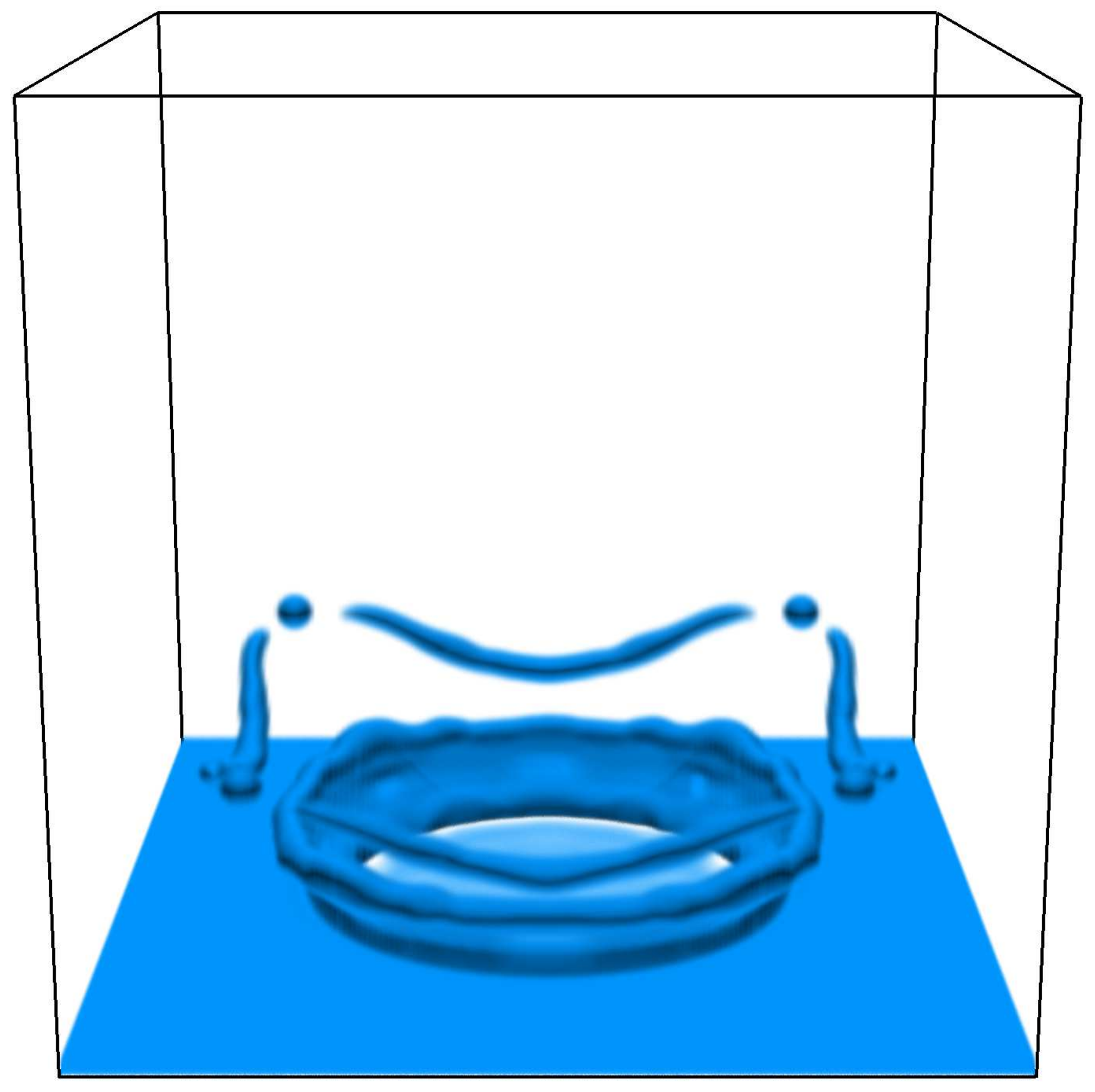}
    \label{3D_Splash_T534}
  }
  \subfigure[$T = 7.35$]{
    \includegraphics[scale = 0.28]{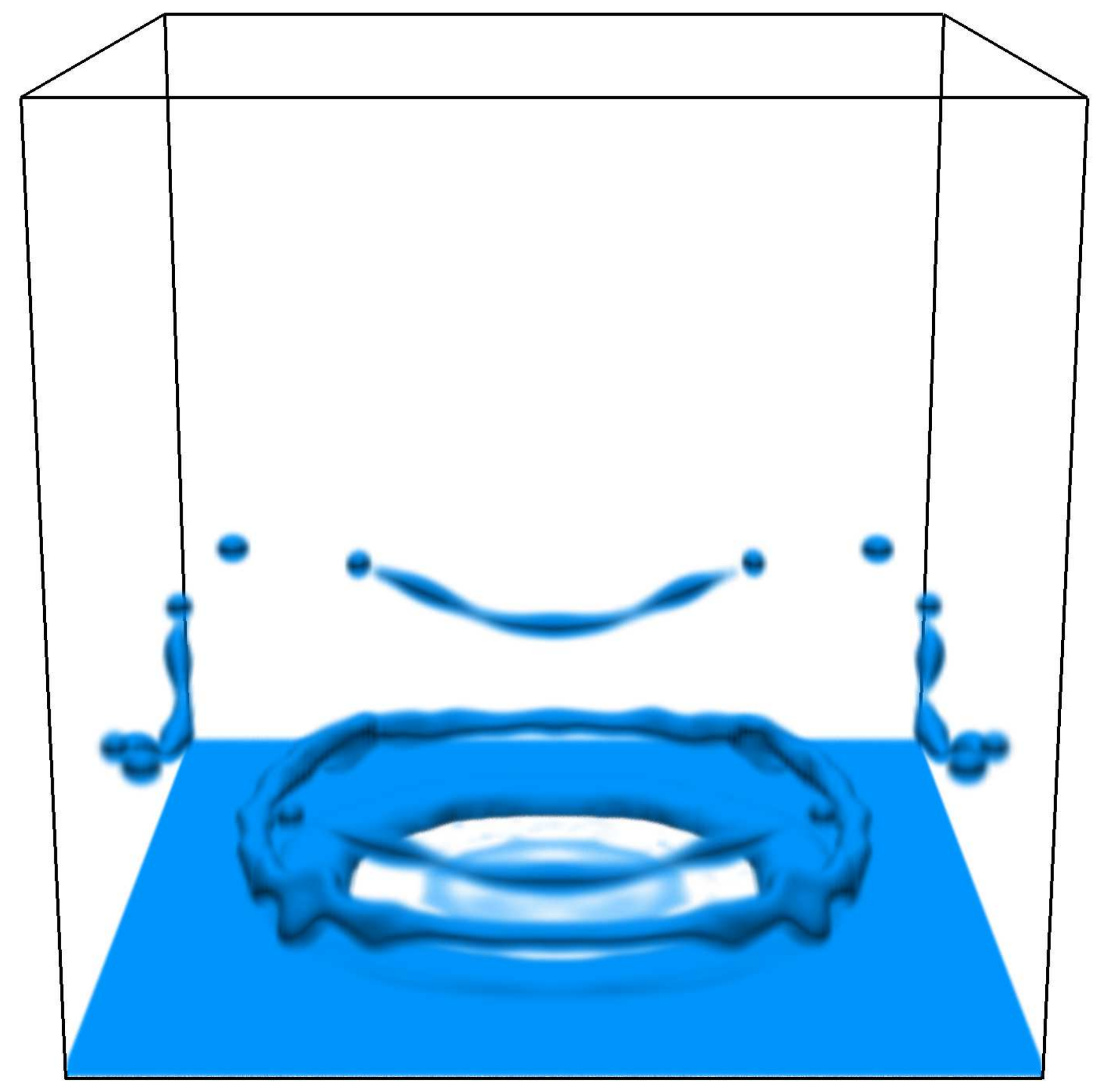}
    \label{3D_Splash_T735}
  }
  \subfigure[Mesh refinement at $T = 7.35$]{
    \includegraphics[scale = 0.28]{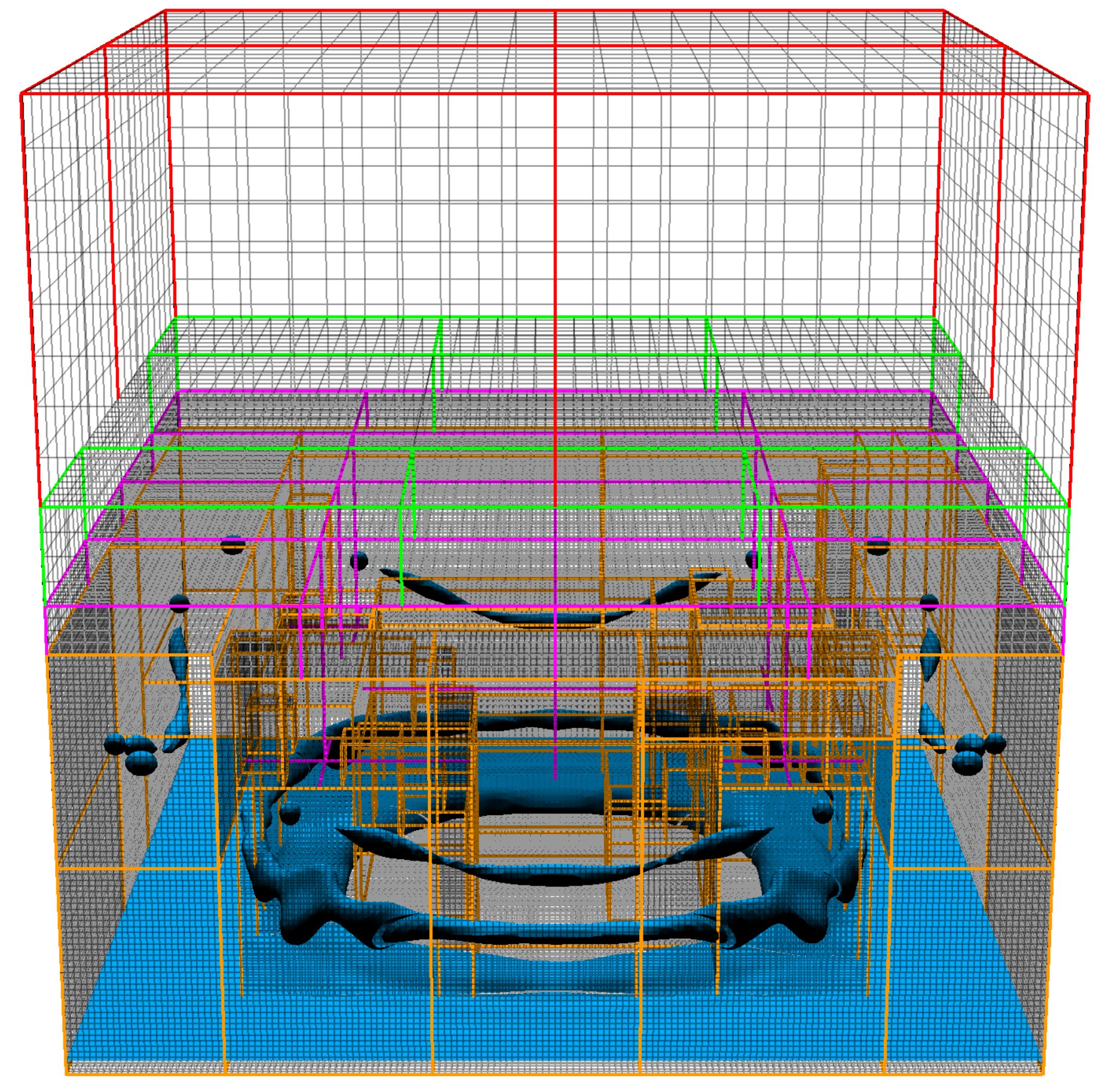}
    \label{3D_Splash_T735_amr}
  }
  \caption{\subref{3D_Splash_T0}-\subref{3D_Splash_T735}
  Temporal evolution of the 3D droplet splashing on a thin liquid film with density 
  ratio $\rho_{\text{l}}/\rho_{\text{g}} = 1000$ and $\mu_{\text{l}}/\mu_{\text{g}} = 40$ at five different time instances.
  \subref{3D_Splash_T735_amr} Locations of the different refined mesh levels from coarsest to finest:
  red, green, pink, orange.
  Consistent transport of mass and momentum is used for these cases.
  Unlike the results shown in Fig.~\ref{fig_Splash3D_NC}, here we observe a physical solution.
  }
  \label{fig_Splash3D}
\end{figure}

Time is made dimensionless via $T = (t - t_i)U_0/D_0$.
We again begin by demonstrating the importance of using consistent
transport of mass and momentum by attempting to run this case with the non-conservative
discretization. As seen in Fig.~\ref{fig_Splash3D_NC}, the droplet undergoes significant (unphysical)
deformation as it travels towards the liquid film. As a result, a coherent lamella does not form,
and we see unphysical impact dynamics.
In contrast, Fig.~\ref{fig_Splash3D} shows the post-impact splashing
dynamics of the liquid interface when this case is simulated with the conservative and consistent integrator.
As the droplet travels downwards, it deforms because of the resistance from the gas phase.
Upon impacting the liquid sheet at around $t_i = 218.75$ ($T = 0$), the droplet is no
longer a sphere. A lamella begins to emanate from the film and
forms a secondary ring thereafter. The ring eventually splits from the top of the lamella at around $T = 3.5$,
which then further disintegrates into smaller satellite droplets. The results are in very good qualitative agreement
with~\cite{Nikolopoulos2007}, with small dissimilarities in the formation of satellite droplets at later times.
This can be attributed to the differences in interface tracking methodology and computational set up of the problem.

\begin{figure}[H]
  \centering
    \subfigure[]{
    \includegraphics[scale = 0.23]{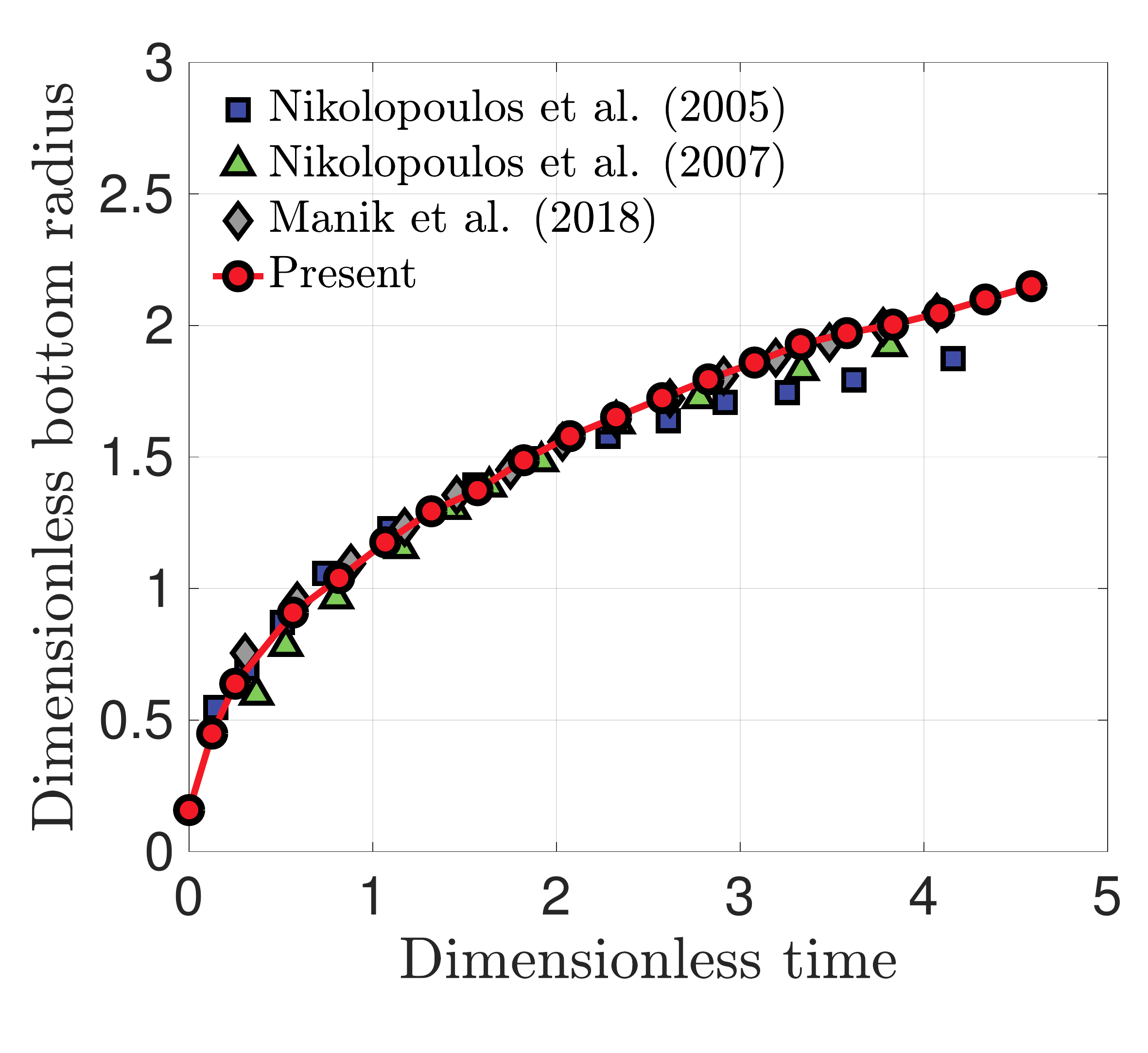}
    \label{splash_3d_radius}
  }
  \subfigure[]{
    \includegraphics[scale = 0.23]{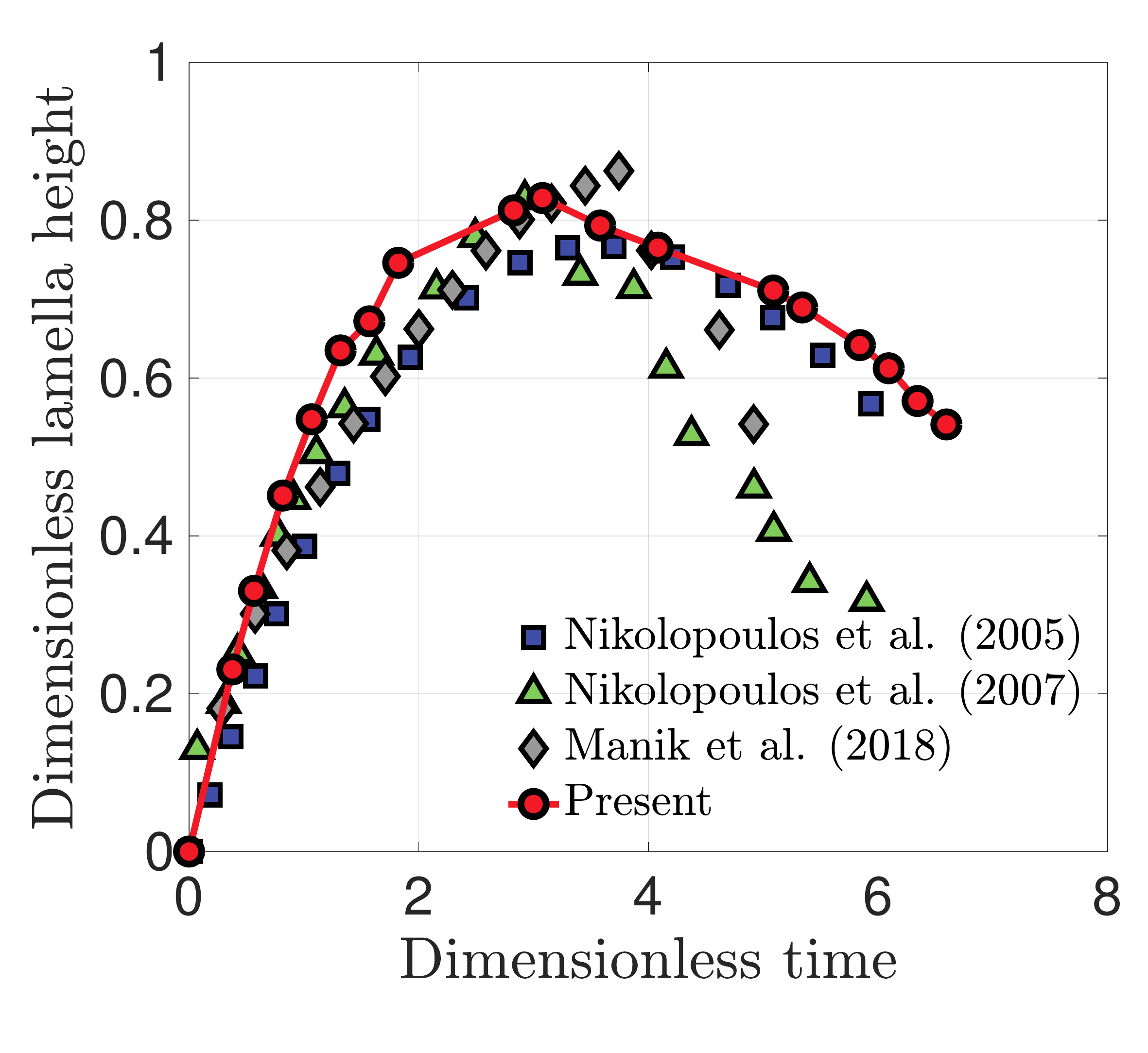}
    \label{splash_3d_height}
  }
   \caption{
 Temporal evolution of \subref{splash_3d_radius} dimensionless bottom radius and
 \subref{splash_3d_height} dimensionless lamella height for the present 3D ($\bullet$, red) simulations, along with
 3D simulation data ($\blacksquare$, blue) from Nikolopoulos et al. 2005~\cite{Nikolopoulos2005},
 3D simulation data ($\blacktriangle$, green) from Nikolopoulos et al. 2007~\cite{Nikolopoulos2005},
 and 3D simulation data ($\blacklozenge$, grey) from Manik et al.~\cite{Manik2018}.
 }
  \label{fig_3d_splash_quant}
\end{figure}

We plot the dimensionless radial distance at the bottom of the rim ($R_{b}/D_0$) 
and dimensionless height of the lamella ($a_l/D_0$) as a function of time in Fig.~\ref{fig_3d_splash_quant}.
The results for the present work compare favorably to previous computational 
studies~\cite{Nikolopoulos2005, Nikolopoulos2007, Manik2018} that use the VOF
method for interface tracking. This case demonstrates that the complex surface tension and gravity
driven splashing dynamics are accurately simulated by the present (consistent) numerical scheme,
even when significant adaptive mesh refinement is used.

\section{Conclusions}
This study extended the monolithic variable-coefficient incompressible 
Navier-Stokes solver of Cai et al.~\cite{Cai2014} and used it for practical high density ratio and high shear multiphase flow applications.
We demonstrated that our solver yields second-order convergence in a variety of conditions where analytic solutions are available, including with nontrivial physical boundary conditions, both on uniform and on locally refined computational grids.
Further, we demonstrated the importance of 
consistent mass and momentum transport to maintain numerical stability for convection-dominated high density
ratio flows. Achieving consistency between mass and momentum transport also necessitates the use of the conservative 
form of the momentum equation. 

The projection preconditioner described here, and by Griffith~\cite{Griffith2009} and Cai et al.~\cite{Cai2014}, has several
distinct advantages over the projection method solver. For variable-coefficient operators, there is an unavoidable commutator
error that is associated with using the projection method as a solver. By considering the coupled velocity-pressure Stokes
system, we avoid this source of error. Moreover, operator splitting approaches require the specification of artificial boundary
conditions for the velocity and pressure fields. These affect the overall solution accuracy of the projection method solver, but
only affect the convergence rate of the projection method preconditioner. Finally, we again emphasize that the using the
projection method as a preconditioner to the coupled Stokes system is no less efficient than using it as a solver.

We also demonstrated that the solver has scalable convergence properties at both low and high Reynolds numbers and for high density and viscosity ratios.
In our numerical experiments, we have seen that a number of factors can affect the convergence of
the FGMRES solver. Generally speaking, smoother problems (i.e., that use more grid cells of interface smearing) allow for more rapid linear solver convergence than problems with sharp interfaces.
Additionally, harmonic averaging of material properties leads
to fewer iterations than simple arithmetic averaging. We also see that larger Reynolds numbers
requires \emph{fewer} iterations whereas smaller Reynolds numbers requires \emph{more} iterations. Finally, the subdomain
solvers can be further optimized for higher or lower (than air-water combination) density and viscosity ratios by tweaking $\epssub$, or by using stronger Krylov subdomain solvers or multigrid preconditioners.

A key contribution of this paper is that it extends the first-order conservative discretization algorithm of Desjardins and 
Moureau~\cite{Desjardins2010} and Ghods and Hermann~\cite{Ghods2013} to a second-order accurate scheme. This is achieved here by 
using a SSP-RK3 time integrator for integrating the mass balance equation and employing higher-order CBC and TVD
satisfying limiters. Our implementation also allows for the use of a forward Euler or SSP-RK2 integrator for the mass balance equation, although we have found that these can become numerically unstable for high density ratio and convection dominated flows. In fact, Appendix~\ref{app_first_order_density_update} demonstrates that we revert to a first-order accurate solution by switching to the forward Euler mass integrator and upwind limiter as done originally in~\cite{Desjardins2010} and~\cite{Ghods2013}. Compatibility between mass and momentum transport is achieved at a discrete level by using the same mass 
flux in both the density evolution equation and the momentum equation.  The consistent discretization is also shown to be 
well-balanced with respect to the pressure gradient force and the gravitational and surface tension forces. By coupling this 
robust fluid solver with a level-set approach to interface tracking, we enable the simulation of complex multiphase flows such as spreading, 
merging, and splashing dynamics. We also demonstrated that level set function can be advected independently of the fluid solver, and can use a different (and possibly a non-CBC satisfying) convective limiter without affecting the overall stability of the scheme.  %The implementation is fully parallelized and open-source~\cite{IBAMR-web-page}.
Furthermore, we note that the present consistent mass and momentum update scheme remains stable even when conventional
projection method solvers~\cite{Chorin1968, Chorin1969} are used~\cite{HanLiu2019}.

The present numerical method can be easily extended to allow for fluid-structure interaction in presence of multiple phases.
In particular, future work involving the coupling between this solver and a constraint-based
immersed boundary method (CIB)~\cite{Bhalla13,Nangia17} is already underway. In addition to this, the
treatment of the viscous term described here allows for an implicit treatment of eddy viscosity,
which is a key ingredient to turbulence modeling using both RANS and LES formulations \cite{Spalart1992,Smagorinsky1963,Deardorff1970}.
Integrating this solver with CIB and turbulence modeling will enable simulation of many important industrial and engineering applications, including  high inertia vehicles,
wave-energy converter devices, and windmills. Finally, efforts are also underway to compute initial distance functions from CAD and STL files directly. This will allow complex geometries to be represented on Cartesian grids and further enhance our level set based solvers for realistic applications.

%%%%%%%%%%%%%%%%%%%%%%%%%%%%%%%%%%%%%%
\section*{Acknowledgements}
  
A.P.S.B.~acknowledges helpful discussions with Ganesh Natarajan for some of the example cases presented in this work. 
N.N., N.A.P., and A.P.S.B.~acknowledge computational resources
provided by Northwestern University's Quest high performance computing
service.  A.P.S.B.~acknowledges College of Engineering's Fermi high performance computing
service at the San Diego State University. N.N.~acknowledges research support from the National Science Foundation 
Graduate Research Fellowship Program (NSF award DGE-1324585).
B.E.G.~and N.A.P.~acknowledge support from the National Science Foundation's SI2 program (NSF awards OAC 1450327 and OAC 1450374).
B.E.G.~also acknowledges support from NSF awards OAC 1652541 and DMS 1664645.
A.P.S.B.~acknowledges research support provided by the San Diego State University.
This work also used the Extreme Science and Engineering Discovery Environment (XSEDE) Bridges at the
Pittsburgh Supercomputing Center through allocation TG-ASC170023, which is supported by National 
Science Foundation grant number ACI-1548562.

%%%%%%%%%%%%%%%%%%%%%%%%%%%%%%%%%%%%%%%%%%%
\appendix 

\renewcommand\thesection{\Alph{section}}

\section{Discretization of the viscous term in three spatial dimensions}
\label{app_3d_viscous}
\begin{figure}[]
  \centering
    \includegraphics[scale = 0.4]{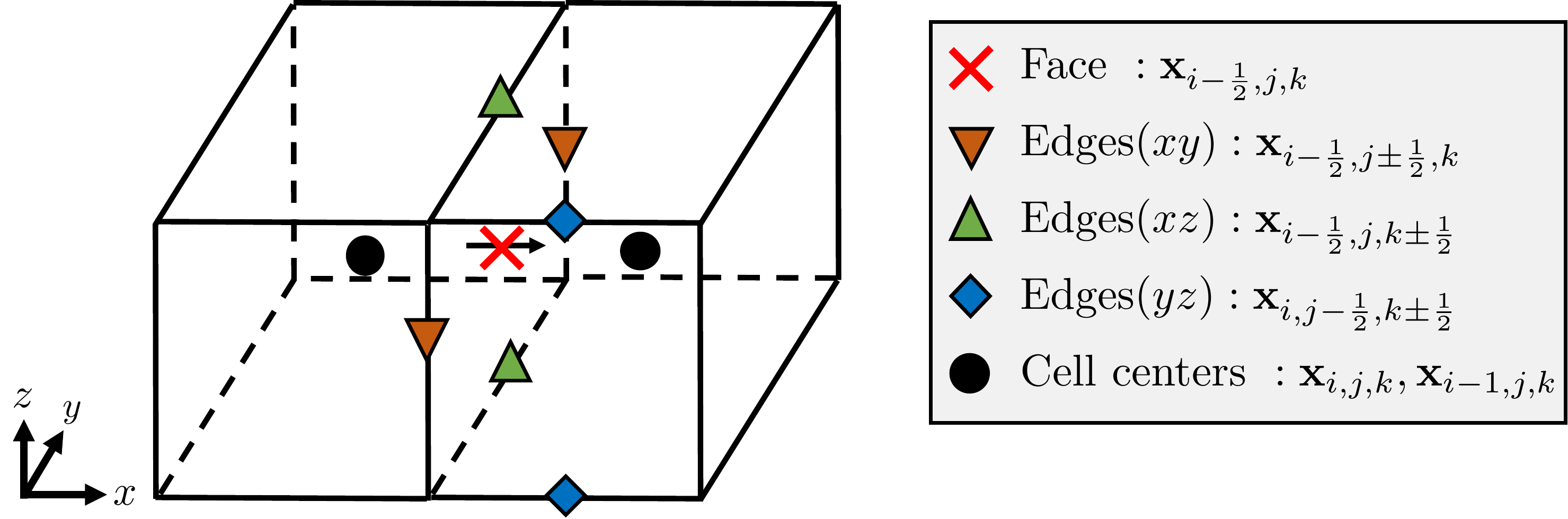}
  \caption{ Two adjacent 3D staggered grid cells on which the viscous component $(\Lmu \u)^x_{i-\half,j, k}$ is computed on
  a particular face ($\times$, red) using velocity derivatives and viscosity on
  cell centers ($\bullet$, black),
  $xy$-edges ($\blacktriangledown$, orange), and
  $xz$-edges ($\blacktriangle$, green). The $yz$-edges ($\blacklozenge$, blue) are \emph{not} used in the computation of $(\Lmu \u)^x_{i-\half,j, k}$, but would
  be required to compute $(\Lmu \u)^y_{i,j-\half, k}$ and $(\Lmu \u)^z_{i,j,k-\half}$}
  \label{fig_3d_staggered_grid}
\end{figure}

The numerical treatment of the strain rate tensor $\div \left[\mu \left(\grad \u + \grad \u^T\right) \right]$
in three spatial dimensions warrants additional discussion. 
For the velocity field $\u(\x,t) = (u(\x,t),v(\x,t),w(\x,t))$, the continuous strain rate tensor form of the viscous term is
\begin{equation}
\label{eq_visc_cont3d}
\div \left[\mu \left(\grad \u + \grad \u^T\right) \right] = 
\left[
\begin{array}{c}
 2 \D{}{x}\left(\mu \D{u}{x}\right) + \D{}{y}\left(\mu\D{u}{y}+\mu\D{v}{x}\right) + \D{}{z}\left(\mu\D{u}{z}+\mu\D{w}{x}\right) \\
 2 \D{}{y}\left(\mu \D{v}{y}\right) + \D{}{x}\left(\mu\D{v}{x}+\mu\D{u}{y}\right) + \D{}{z}\left(\mu\D{v}{z}+\mu\D{w}{y}\right) \\
 2 \D{}{z}\left(\mu \D{w}{z}\right) + \D{}{x}\left(\mu\D{w}{x}+\mu\D{u}{z}\right) + \D{}{y}\left(\mu\D{w}{y}+\mu\D{v}{z}\right) \\
\end{array}
\right],
\end{equation}
which leads to a discretization of the form
\begin{equation}
\label{eq_visc_discrete3d}
\Lmu \u= 
\left[
\begin{array}{c}
 (\Lmu \u)^x_{i-\half,j,k} \\
 (\Lmu \u)^y_{i,j-\half,k}  \\
 (\Lmu \u)^z_{i,j,k-\half} \\
\end{array}
\right].
\end{equation}
The two-dimensional discretization scheme required viscosity on both Cartesian cell centers, where $\D{u}{x}$ and $\D{v}{y}$ are naturally approximated,
and nodes, where $\D{u}{y}$ and $\D{v}{x}$ are naturally approximated. In three spatial dimensions, the off-diagonal
components of $\left[\mu \left(\grad \u + \grad \u^T\right) \right]$ are naturally defined on the \emph{edges} of
the grid cells,
\begin{align}
&\left(\mu\D{u}{y}+\mu\D{v}{x}\right)_{i-\half, j-\half,k} = \mu_{i-\half, j-\half,k} \left(\frac{u_{i-\half,j,k} - u_{i-\half,j-1,k}}{\dy} + \frac{v_{i,j-\half,k}-v_{i-1,j-\half,k}}{\dx}\right), \\
&\left(\mu\D{u}{z}+\mu\D{w}{x}\right)_{i-\half, j,k-\half} = \mu_{i-\half, j,k-\half} \left(\frac{u_{i-\half,j,k} - u_{i-\half,j,k-1}}{\dz} + \frac{w_{i,j,k-\half}-w_{i-1,j,k-\half}}{\dx}\right), \\
&\left(\mu\D{v}{z}+\mu\D{w}{y}\right)_{i, j-\half,k-\half} = \mu_{i, j-\half,k-\half} \left(\frac{v_{i,j-\half,k} - v_{i,j-\half,k-1}}{\dz} + \frac{w_{i,j,k-\half}-w_{i,j-1,k-\half}}{\dy}\right).
\end{align}
Fig.~\ref{fig_3d_staggered_grid} shows the location of these $xy$-, $xz$-, and $yz$-edges. The full discretization of the $x$-component of the strain rate tensor reads,
\begin{align}
 (\Lmu \u)^x_{i-\half,j, k} &= \frac{2}{\dx}\left[\mu_{i,j,k}\frac{u_{i+\half,j,k} - u_{i-\half,j,k}}{\dx} -
					        \mu_{i-1,j,k}\frac{u_{i-\half,j,k} - u_{i-\3half,j,k}}{\dx}\right] \nonumber \\ 
                    &+ \frac{1}{\dy}\left[\mu_{i-\half, j+\half,k}\frac{u_{i-\half,j+1,k} - u_{i-\half,j,k}}{\dy} - 
					         \mu_{i-\half, j-\half,k}\frac{u_{i-\half,j,k} - u_{i-\half,j-1,k}}{\dy}\right] \nonumber \\
	            &+ \frac{1}{\dy}\left[\mu_{i-\half, j+\half,k}\frac{v_{i,j+\half,k} - v_{i-1,j+\half,k}}{\dx} - 
					         \mu_{i-\half, j-\half,k}\frac{v_{i,j-\half,k} - v_{i-1,j-\half,k}}{\dx}\right] \nonumber \\
		   &+ \frac{1}{\dz}\left[\mu_{i-\half, j,k+\half}\frac{u_{i-\half,j,k+1} - u_{i-\half,j,k}}{\dz} - 
					         \mu_{i-\half, j,k-\half}\frac{u_{i-\half,j,k} - u_{i-\half,j,k-1}}{\dz}\right] \nonumber \\
	            &+ \frac{1}{\dz}\left[\mu_{i-\half, j,k+\half}\frac{w_{i,j,k+\half} - w_{i-1,j,k+\half}}{\dx} - 
					         \mu_{i-\half, j,k-\half}\frac{w_{i,j,k-\half} - w_{i-1,j,k-\half}}{\dx}\right] \label{eq_viscx_fd3d},		         
\end{align}
which again uses cell centered viscosities and requires approximations to the viscosity on edges. The edge viscosities can be computed
using arithmetic or harmonic averages of cell centered values, i.e., $\mu_{i-\half, j-\half,k}$ would be computed from 
$\mu_{i-1, j-1,k}$, $\mu_{i, j-1,k}$, $\mu_{i-1, j,k}$, and $\mu_{i, j,k}$. Approximations to the remaining components in Eq.~\eqref{eq_visc_discrete3d}
can be determined analogously. The other spatial discretizations described in Section~\ref{sec_discretized} can be straightforwardly
extended to three dimensions.

\begin{figure}[]
  \centering
  \subfigure[Velocity 3D]{
    \includegraphics[scale = 0.2]{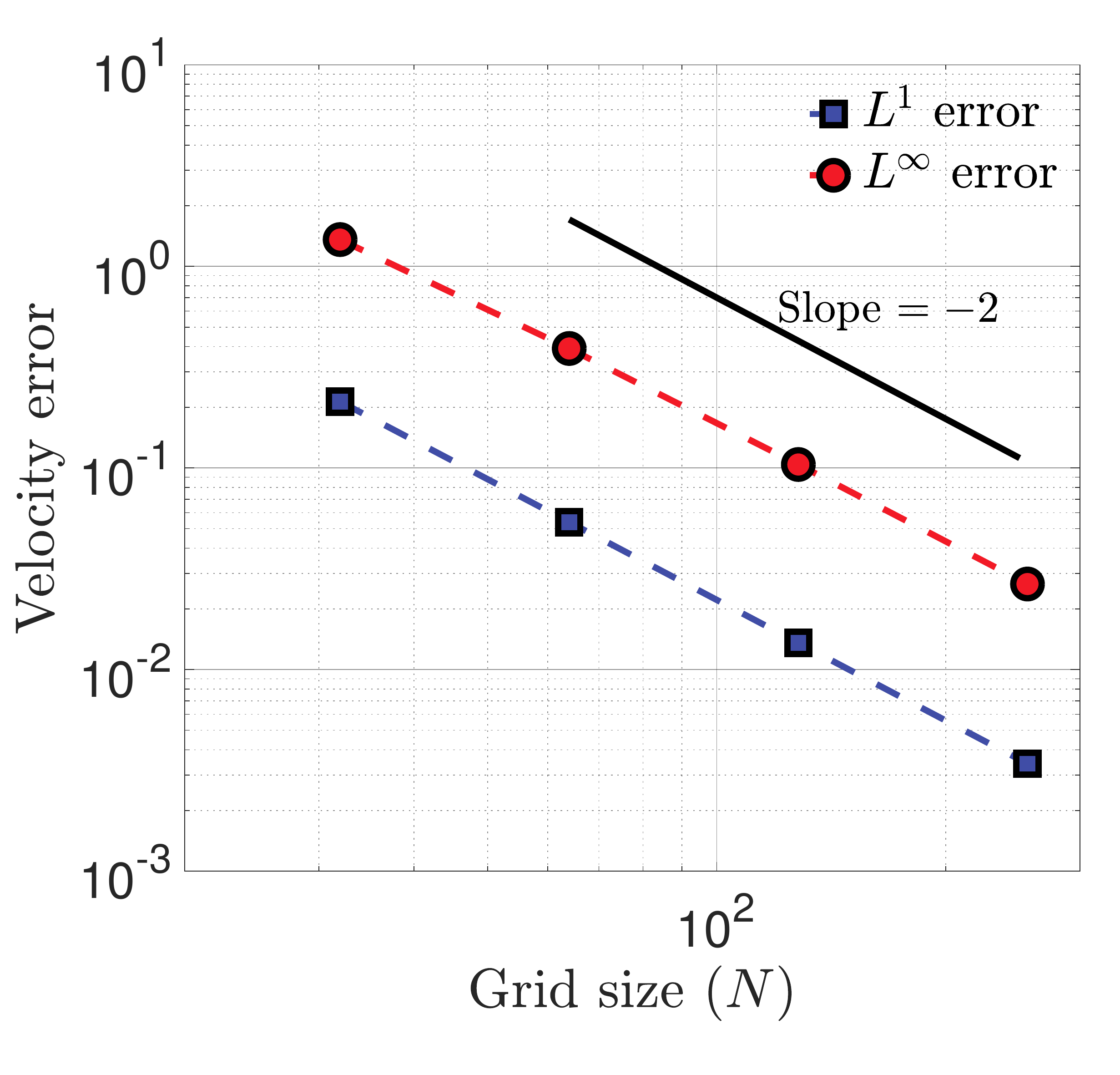}
    \label{fig_3D_nc_U_err_vel_tra}
  }
   \subfigure[Pressure 3D]{
    \includegraphics[scale = 0.2]{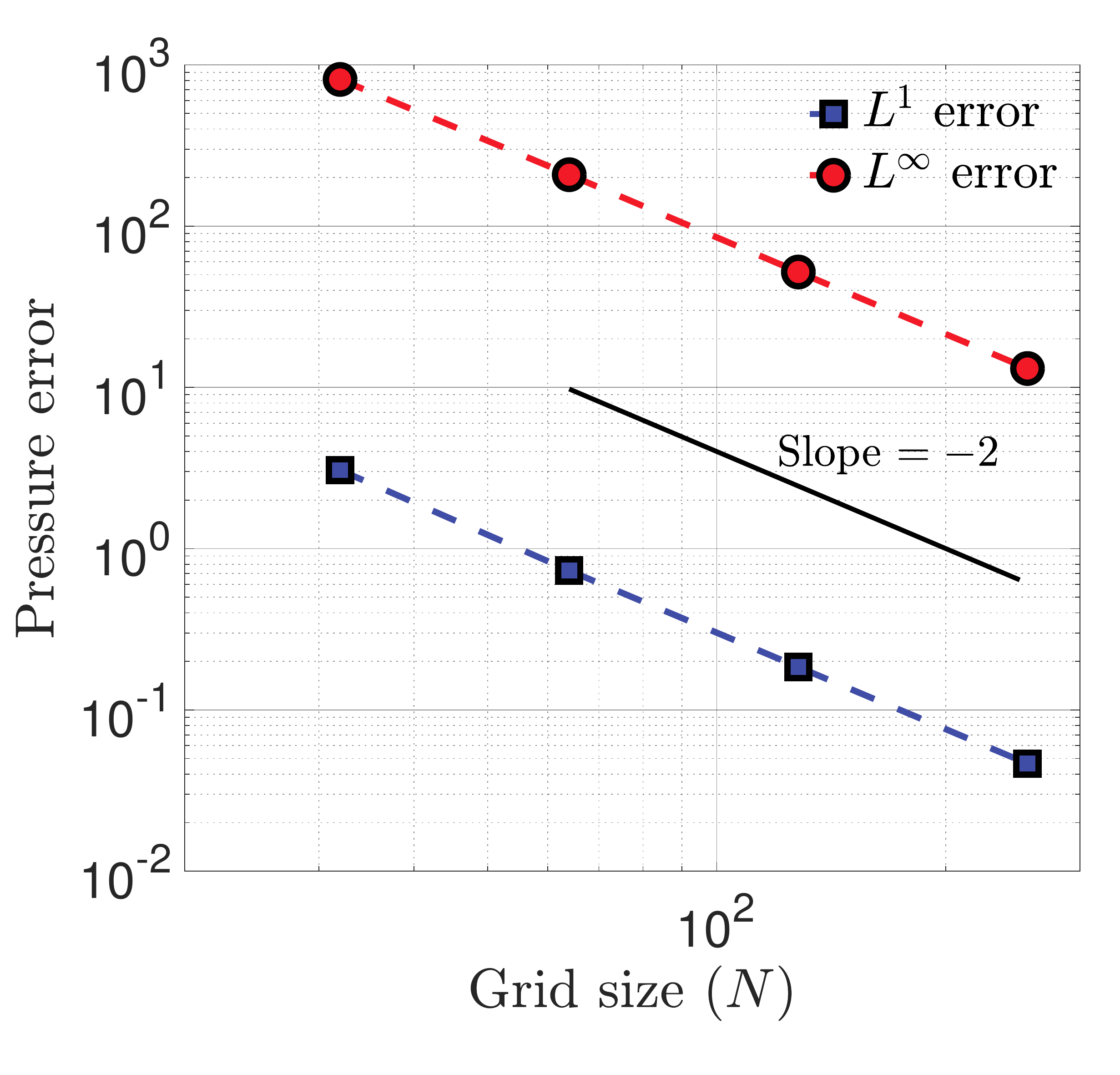}
    \label{fig_3D_nc_P_err_vel_tra}
  }
  \caption{ 
  $L^1$ ($\blacksquare$, blue) and $L^\infty$ ($\bullet$, red) errors as a function of grid size $N$ for the 3D non-conservative
  manufactured solution with specified normal velocity and tangential traction (vel-tra) boundary conditions:
  \subref{fig_nc_U_err_vel_tra}
  convergence rate for $\u$;
  \subref{fig_nc_P_err_vel_tra}
   convergence rate for $p$.
   }
  \label{fig_3D_nc_err_vel_tra}
\end{figure}

To complete our discussion, we consider the following manufactured solution in three spatial dimensions for the non-conservative set of equations,
\begin{align}
&u(\x,t) = 2 \pi  \cos (2 \pi  x) \sin (2 \pi  z) \cos (2 \pi  t-2 \pi  y), \label{eq_non_cons_ms_u3d}\\
&v(\x,t) = -2 \pi  \sin (2 \pi  x) \cos (2 \pi  z) \sin (2 \pi  t-2 \pi  y), \label{eq_non_cons_ms_v3d}\\
&w(\x,t) = \pi  \sin (2 \pi  x) \cos (2 \pi  t-2 \pi  y) (-2 \sin (2 \pi  z)-2 \cos (2 \pi  z)), \label{eq_non_cons_ms_w3d} \\ 
&p(\x,t) = 2 \pi  \sin (2 \pi  t-2 \pi  x) \cos (2 \pi  t-2 \pi  y) \sin (2 \pi  t-2 \pi  z), \label{eq_non_cons_ms_p3d}
\end{align}
together with time-independent density and viscosity fields of the form   
\begin{align}
\rho(\x) &= \rho_0 + \frac{\rho_1}{2}  \left(\tanh \left(\frac{0.1\,
   -\sqrt{(x-0.5)^2+(y-0.5)^2+(z-0.5)^2}}{\delta }\right)+1\right), \label{eq_non_cons_ms_rho3d}\\
\mu(\x) &= \mu_0+\mu_1+\mu_1 \sin (2 \pi  x) \cos (2 \pi  y) \sin (2 \pi  z). \label{eq_non_cons_ms_mu3d}
\end{align}
%Plugging Eqs.~\eqref{eq_non_cons_ms_u3d}-\eqref{eq_non_cons_ms_mu3d} 
%into the non-conservative momentum equation~\eqref{eqn_nc_momentum} yields a forcing term $\f(\x,t)$ that produces the desired 
%solution given by Eqs.~\eqref{eq_non_cons_ms_u3d}-\eqref{eq_non_cons_ms_p3d}.
The variations in density and viscosity are set to be similar to that of air and water:
$\rho_0 = 1$, $\rho_1 = -\rho_0 + 10^3$, $\mu_0 = 10^{-4}$, and $\mu_1 = -\mu_0 + 10^{-2}$.
The computational domain is the unit square $\Omega = [0,L]^2 = [0,1]^2$, which is discretized by $N$ grid cells in each direction.
The smoothing parameter is set to $\delta = 0.05L$.
The maximum velocities in the domain for this manufactured solution are $\BigO{1}$, hence a relevant  time scale is $L/U$ with $U = 1$.
Errors in the velocity and pressure are computed at time $T = tU/L = 0.1$ with a uniform time step $\dt = 1/(31.25 N)$, which
yields an approximate CFL number of $0.3$. A relative convergence tolerance of $\epsstokes = 10^{-12}$ is
specified for the FGMRES solver. We impose periodic boundary conditions in the $x$- and $z$-direction and consider
normal velocity and tangential traction boundary conditions in the $y$-direction (vel-tra).
Fig.~\ref{fig_3D_nc_err_vel_tra} shows the $L^1$ and $L^{\infty}$ errors for velocity and pressure
as a function  of grid size. Second-order convergence rates are achieved for both velocity and pressure in both
norms, which demonstrates that the solver is maintaining the expected order of accuracy in three spatial dimensions.

\section{Physical boundary conditions}
\label{app_boundary_conditions}
\begin{figure}[H]
  \centering
    \includegraphics[scale = 0.5]{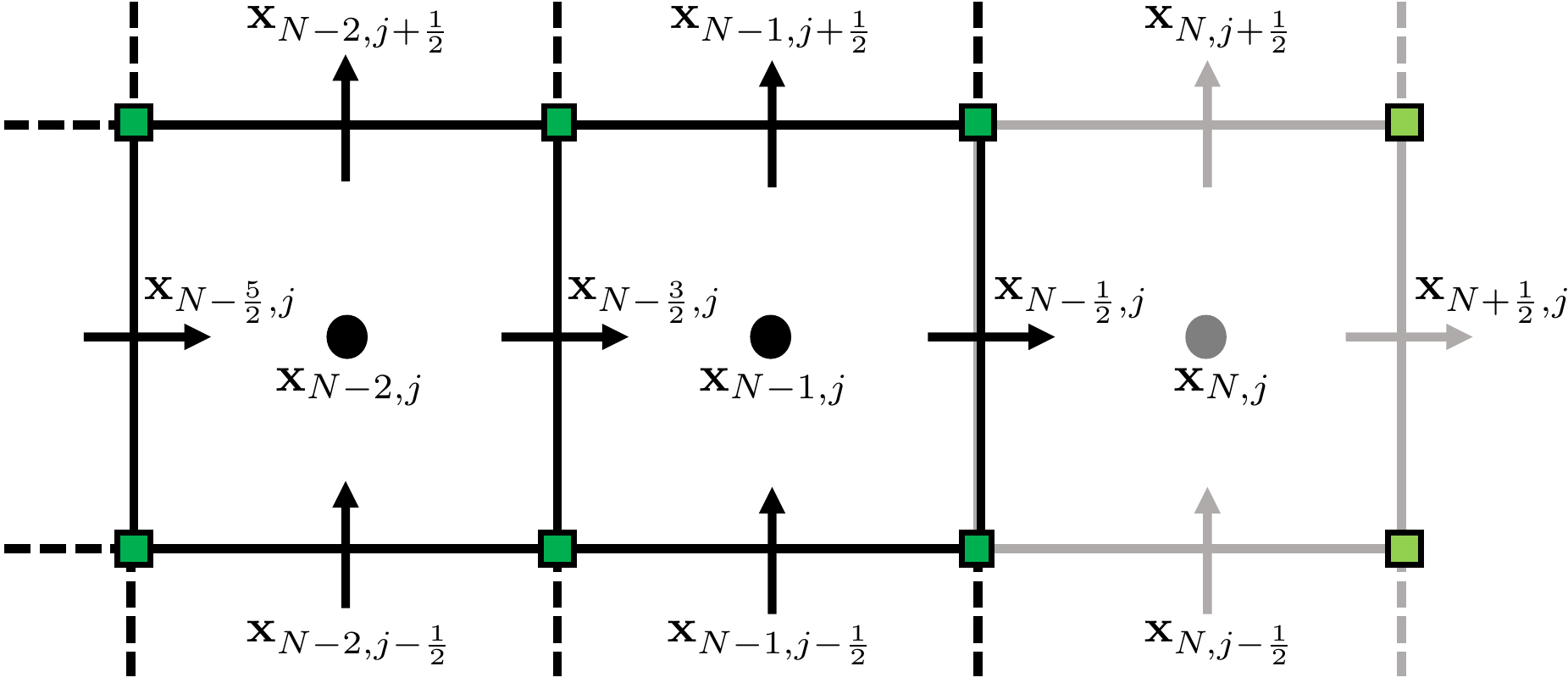}
    \label{bc_staggered_grid}
   \caption{The velocity, pressure, and viscosity degrees of freedom in the vicinity of cell $(N-1,j)$ required to 
   enforce physical boundary conditions. Interior and boundary values are colored in black, and ghost values
   are colored in grey.}
  \label{fig_bc_staggered_grid}
\end{figure}

This section details the physical boundary condition treatment used in this work. % along with and the determination of the ghost values for the fluid solver during a single
%time step $[t^n,t^{n+1}]$. 
The treatment of physical boundary conditions has been described in detail
by Griffith for the constant-coefficient case~\cite{Griffith2009}, although we note that the imposition of normal traction
boundary conditions used in this work differs slightly. We follow the same strategy as in Griffith~\cite{Griffith2009} and restrict
our attention to the vicinity of a single grid cell $(N-1,j), 0 \le j < N$ on the right side of the physical domain
(Fig.~\ref{fig_bc_staggered_grid}).
The treatment along other physical domain boundaries and in three dimensions 
is analogous.

\subsection{Scalar boundary conditions}
For the scalar fields, namely density $\rho$, viscosity $\mu$, and the signed distance function $\phi$, we restrict our attention to homogenous 
Neumann boundary conditions. % (although the implementation is flexible enough to use inhomogeneous Robin boundary conditions as well). 
For a given scalar field $\psi$, the boundary condition is
\begin{align}
\D{\psi}{x}\left(\x_{N-\half},t\right) = 0,
\end{align}
which is imposed using the standard ghost value treatment~\cite{Harlow1965},
\begin{equation}
%&\psi^{n}_{N,j} = \psi^{n}_{N-1,j}, \\
\psi^{n}_{N,j} = \psi^{n}_{N-1,j}.
\end{equation}
In cases where $n_\text{G} > 1$ ghost cell values are required for $\psi$, the condition is imposed by setting the $k^{\textrm{th}}$
ghost cell with the $k^{\textrm{th}}$ interior value, i.e., $\psi_{N+k-1} = \psi_{N-k}$ for $k = 1, \ldots, n_\text{G}$.

\subsection{Fluid boundary conditions}
Considering the right domain boundary once again, the outward unit normal is $\n = (1,0)$, and the unit tangent vector is $\vec{\tau} = (0, 1)$.
The viscous stress tensor for an incompressible fluid is
\begin{equation}
\label{eq_stress_tensor}
\vsigma = -p \I + \mu \left[\grad \u + \grad \u^T\right].
\end{equation}
Following Griffith~\cite{Griffith2009}, four types of physical boundary conditions for $\u$ and $\vsigma$ are considered in this work:
\begin{enumerate}
	\item The normal velocity is prescribed at face $\x_{N-\half, j}$: \\
	Suppose that the normal velocity at position $\x_{N-\half, j}$ is given by $u_{N-\half,j}^{\textrm{norm}}(t)$.
	The treatment of this boundary condition is straightforward because we can simply set
	\begin{equation}
	%&
	u^{n}_{N-\half, j} = u_{N-\half, j}^{\textrm{norm}}\left(t^n\right). %,  \nonumber \\
	%&u^{n+1}_{N-\half, j} = u_{N-\half, j}^{\textrm{norm}}\left(t^{n+1}\right).
	\end{equation}
	As in Griffith~\cite{Griffith2009}, no ghost values are required for $p_{N,j}$ or $u_{N+\half, j}$, nor may
	any pressure condition be prescribed at this boundary.
	
	\item The tangential velocity is prescribed at node $\x_{N-\half, j-\half}$: \\
	Suppose that the tangential velocity at position $\x_{N-\half, j-\half}$ is given by $v_{N-\half,j-\half}^{\textrm{tan}}(t)$. This boundary condition is imposed via a linear fit of the prescribed velocity
	and the closest internal value,
	\begin{equation}
	%&
	v^{n}_{N, j-\half} = 2 v_{N-\half, j-\half}^{\textrm{tan}}\left(t^n\right) - v_{N-1, j-\half}^n.%, \nonumber \\
	%&v^{n+1}_{N, j-\half} = 2 v_{N-\half, j-\half}^{\textrm{tan}}\left(t^{n+1}\right) - v_{N-1, j-\half}^{n+1} 
	\label{eq_tan_vel_ghost}
	\end{equation}
	
	\item The tangential traction is prescribed at node $\x_{N-\half, j-\half}$: \\
	Suppose that the tangential traction,
	%\begin{equation}
	%\label{eq_tan_trac}
	$(\vec{\tau} \cdot \vsigma \cdot \n) = \mu\left(\D{v}{x} + \D{u}{y}\right)$,
	%\end{equation}
	is given by $F_{N-\half,j-\half}^{\textrm{tan}}(t)$. In contrast with prior work~\cite{Griffith2009}, the
	spatially varying viscosity must be treated when imposing this boundary condition. Writing out a second-order finite
	difference approximation to this boundary condition %Eq.~\ref{eq_tan_trac} 
	at time $t^n$ % and setting it equal to the prescribed
	%tangential traction 
	yields
	\begin{equation}
	\label{eq_tan_trac_ghost}
	\mu^n_{N-\half, j-\half}\left(\frac{v^n_{N,j-\half}-v^n_{N-1,j-\half}}{\dx}+\frac{u^n_{N-\half,j}-u^n_{N-\half,j-1}}{\dy}\right)
	= F_{N-\half,j-\half}^{\textrm{tan}}(t^n),
	\end{equation}
	which is an expression for the ghost velocity $v_{N,j-\half}^{n}$ in terms of the interior values.
	%after minor algebraic rearrangement.
	Note that in the above expression the node-centered viscosity
	value is needed, which is already being computed and used for the discretization of the viscous operator
	(Eqs.~\eqref{eq_viscx_fd} and~\eqref{eq_viscy_fd}). %An analogous formula can be obtained for $v_{N,j-\half}^{n+1}$.
	
	\item The normal traction is prescribed at face $\x_{N-\half, j}$: \\
	Suppose that the normal traction,
	%\begin{equation}
	%\label{eq_norm_trac}
	$(\n \cdot \vsigma \cdot \n) = -p+2 \mu \D{u}{x}$,
	%\end{equation}
	is given by $F_{N-\half, j}^{\textrm{norm}}(t)$. Again, the spatially varying viscosity must be handled
	when imposing this boundary condition. This boundary condition requires ghost values $u_{N+\half,j}$ and $p_{N,j}$  
	to be set, which implies a second condition must be enforced in order to uniquely specify these values. As in prior work \cite{Griffith2009}, we
	enforce the divergence-free condition $\div \u(x_{N,j}, t) = 0$ in the ghost cell surrounding $x_{N,j}$.
	A second-order finite difference approximation to this condition at time $t^n$ is
	\begin{equation}
	\label{eq_div_free_bc}
	\frac{u^n_{N+\half,j} - u^n_{N-\half,j}}{\dx} + \frac{v^n_{N,j+\half}-v^n_{N,j-\half}}{\dy} = 0,
	\end{equation}
	which yields %the following expression for the horizontal ghost velocity value,
	\begin{equation}
	\label{eq_u_ghost_norm_trac}
	u^n_{N+\half,j} = u^n_{N-\half,j}-\frac{\dx}{\dy} \left(v^n_{N,j+\half}-v^n_{N,j-\half}\right).
	\end{equation}
	%Note that the above normal traction boundary condition will \emph{always} be specified in conjunction with one of the
	%tangential boundary conditions (Eqs.~\eqref{eq_tan_vel_ghost} or~\eqref{eq_tan_trac_ghost}) or periodic boundary
	%conditions. Therefore, the numerical implementation should ensure that $v^n_{N,j\pm\half}$ have been filled before
	%imposing the normal traction boundary conditions.
	Now that we have obtained the ghost value $u_{N+\half,j}$, we can write out the finite difference approximation to
	the traction boundary condition, % and set it equal to the prescribed normal traction
	\begin{equation}
	\label{eq_norm_trac_fd}
	-\frac{p_{N,j}^{n+\half} + p_{N-1,j}^{n+\half}}{2} + 
	\mu^{n+\half}_{N-\half,j}\left(\frac{3u^{n}_{N-\half,j} - 4 u^{n}_{N-\3half,j} +u^{n}_{N-\5half,j}}{2 \dx} 
	                                           +\frac{3u^{n+1}_{N-\half,j} - 4 u^{n+1}_{N-\3half,j} +u^{n+1}_{N-\5half,j}}{2 \dx}\right)
	= F_{N-\half, j}^{\textrm{norm}}(t^{n+\half}),
	\end{equation}
	in which we have used a one-sided second-order finite difference approximation to $\D{u}{x}$ at $\x_{N-\half, j}$.
	The viscosity on the face $\mu^{n+\half}_{N-\half,j}$ is computed using an arithmetic or harmonic average of
	the viscosity in the neighboring cells $\mu^{n+\half}_{N-1,j}$ and $\mu^{n+\half}_{N,j}$. %, which again requires the
	%ghost cell viscosity value to be set prior to imposing the normal traction.
	Rearranging Eq.~\eqref{eq_norm_trac_fd} yields an expression for the desired ghost value for pressure
	$p_{N,j}^{n+\half}$ based on interior values. %, thereby completing the treatment.
\end{enumerate}

We note that the numerical imposition of the normal traction boundary condition using (one-sided) second-order 
derivative here differs from the one-sided first-order approximation to $\D{u}{x}$ at $\x_{N-\half, j}$ previously used by Griffith~\cite{Griffith2009}. 

\section{Conservative form: Smooth density profile evolution}
\label{app_smooth_density}

\begin{figure}[H]
  \centering
  \subfigure[Velocity]{
    \includegraphics[scale = 0.2]{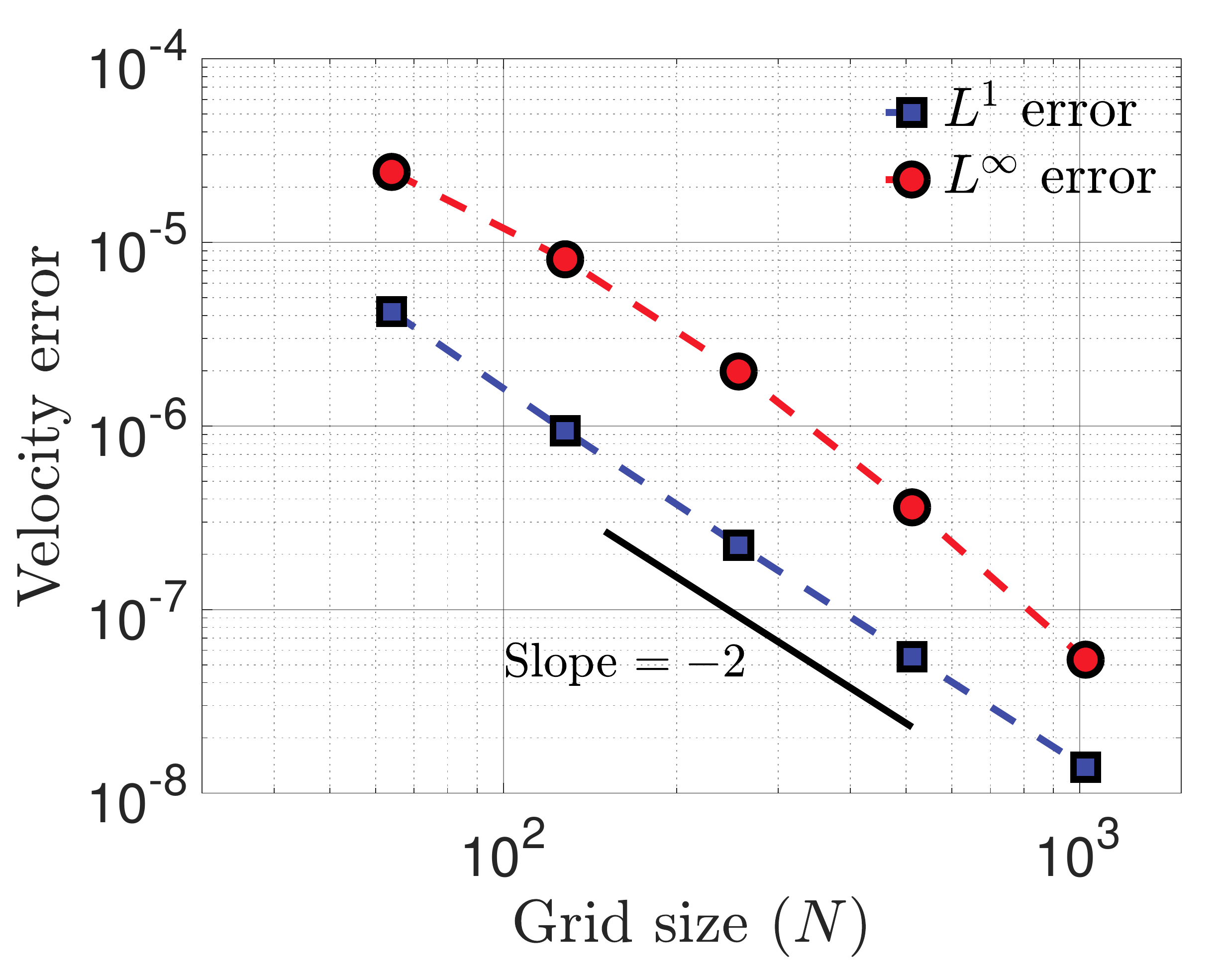}
    \label{fig_c_U_smooth_rho_err}
  }
   \subfigure[Pressure]{
    \includegraphics[scale = 0.2]{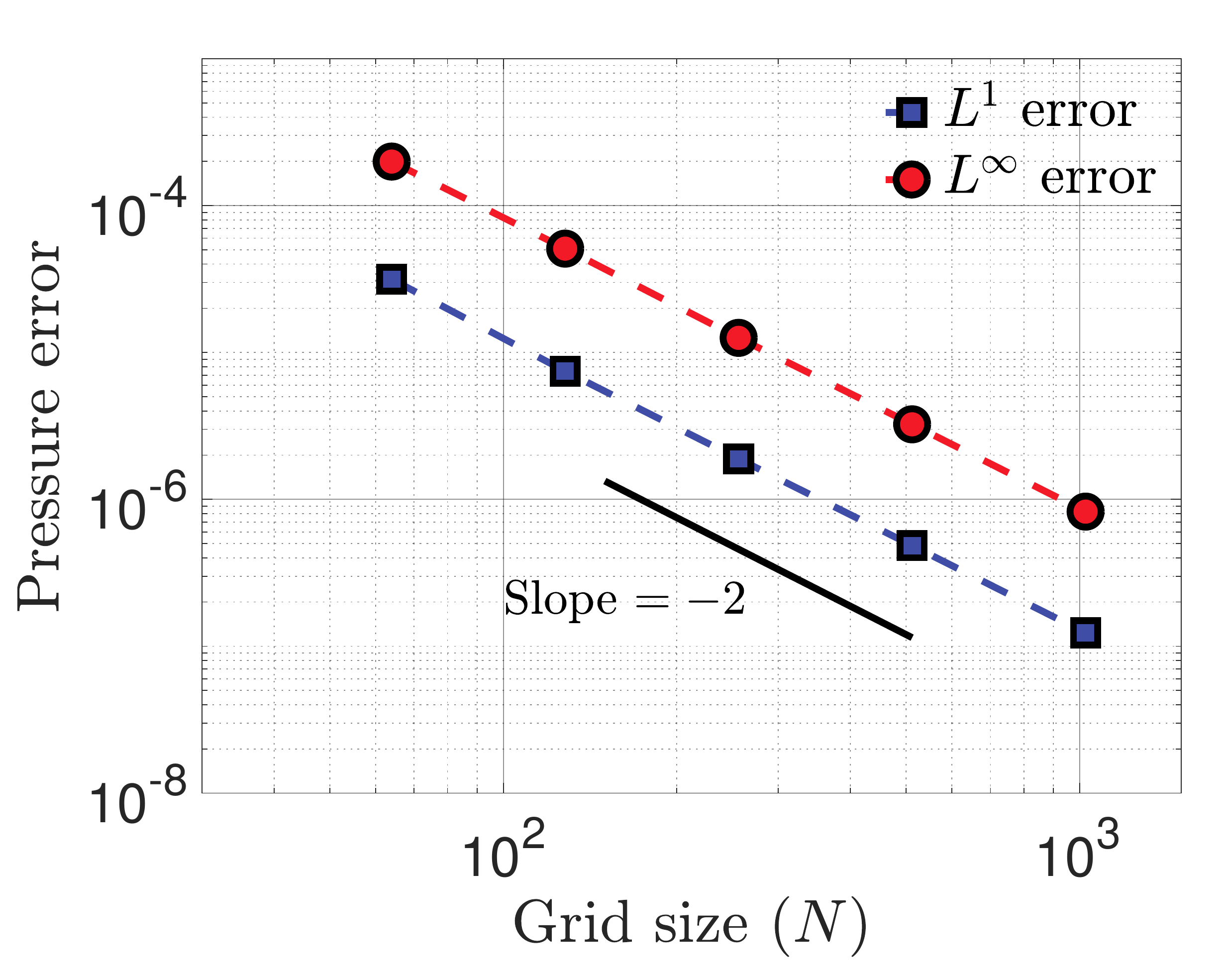}
    \label{fig_c_P_smooth_rho_err}
  }
   \subfigure[Density]{
    \includegraphics[scale = 0.2]{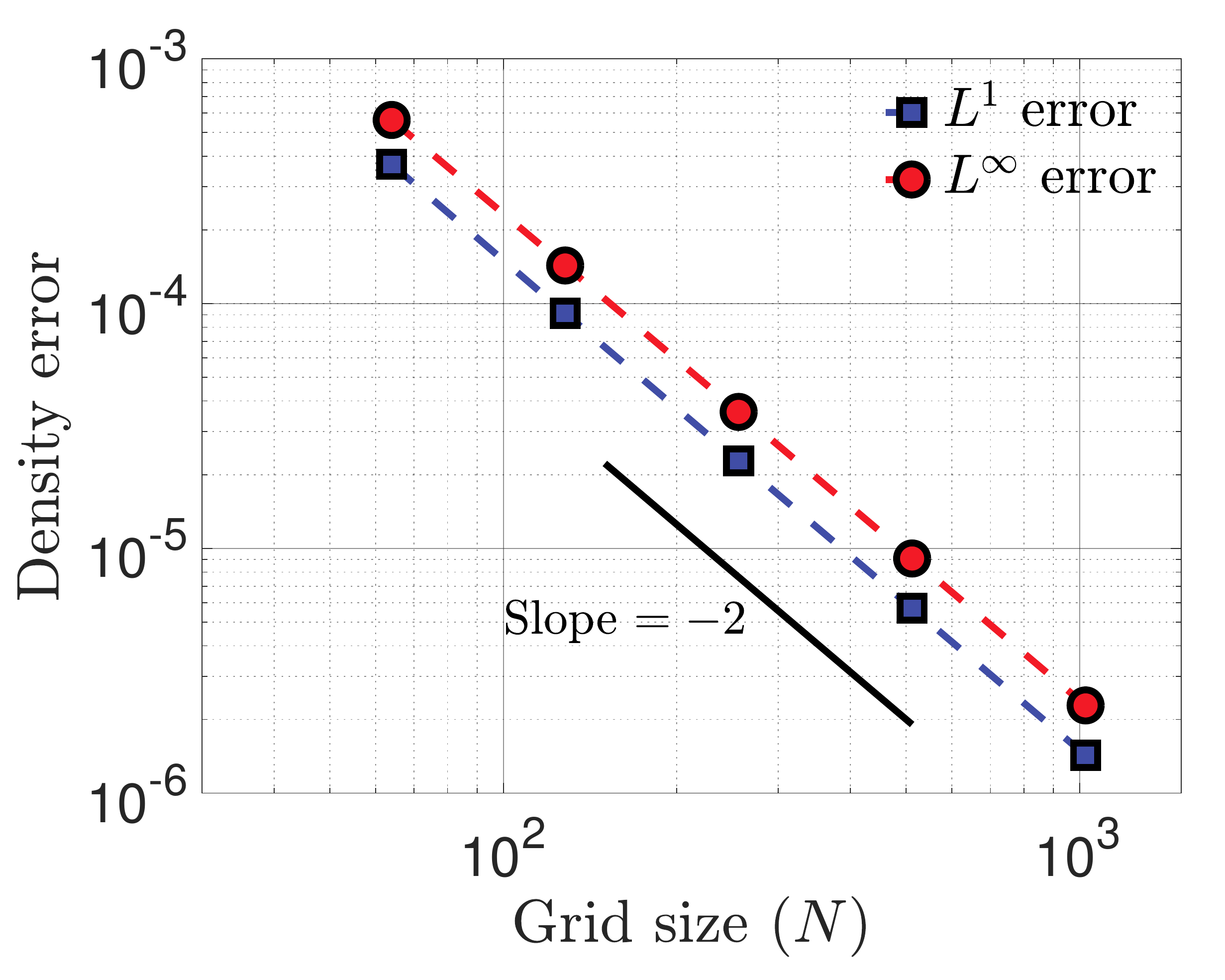}
    \label{fig_c_RHO_smooth_rho_err}
  }
  \caption{ 
  $L^1$ ($\blacksquare$, blue) and $L^\infty$ ($\bullet$, red) errors as a function of grid size $N$ for the conservative
  manufactured solution with specified normal velocity and normal velocity boundary conditions on all boundaries, and smooth
  density profile given by Eq.~\eqref{eq_cons_ms_smooth_rho}. In these cases,
  the density $\rho$ is \emph{not} reset between time steps.
  \subref{fig_c_U_smooth_rho_err}
  convergence rate for $\u$;
  \subref{fig_c_P_smooth_rho_err}
   convergence rate for $p$;
   \subref{fig_c_RHO_smooth_rho_err}
   convergence rate for $\rho$.
   }
  \label{fig_c_err_smooth_rho}
\end{figure}

In Sec.~\ref{sec_cons_ms}, we note that the $C^0$ spatial continuity of the manufactured solution
density field can lead to reductions in the pointwise convergence rate. This section
demonstrates that using the smooth density field,
\begin{equation}
\rho(\x,t) = 2 + x \cos (\sin(t))+y \sin (\sin (t)), \label{eq_cons_ms_smooth_rho}
\end{equation}
yields second-order pointwise convergence rates for $\u$, $p$, and $\rho$. The same velocity
(Eqs.~\eqref{eq_cons_ms_u}--\eqref{eq_cons_ms_v}), pressure (Eq.~\eqref{eq_cons_ms_p}),
and viscosity (Eq.~\eqref{eq_cons_ms_mu}) fields are used to produce the desired forcing term.
Note that this density field also satisfies the conservative mass balance equation~\eqref{eqn_cons_of_mass}.
For \emph{all} physical boundaries, specified normal velocity and tangential velocity boundary conditions are used.
The domain $\Omega = [-L,L] = [-1,1]^2$ is discretized by an $N \times N$ grid.
The maximum velocities in the domain for this manufactured solution are $\BigO{1}$, hence a relevant  time scale is $L/U$ with $U = 1$.
Errors in the velocity and pressure
are computed at time $T = tU/L = 0.6$ with a uniform time step size $\dt = 1/(1.042N)$, which yields an approximate CFL
number of $0.5$. A relative convergence tolerance of $\epsstokes = 10^{-12}$ is specified for the FGMRES solver.
For these cases, density is evolved and \emph{not} reset between time steps.
Fig.~\ref{fig_c_err_smooth_rho} shows the $L^1$ and $L^{\infty}$ errors for velocity, pressure, and density as a function of grid size.
Second-order convergence rates are achieved for velocity, pressure, and density in both norms.

\section{First-order density update scheme} \label{app_first_order_density_update}

\begin{figure}[H]
  \centering
  \subfigure[Velocity]{
    \includegraphics[scale = 0.2]{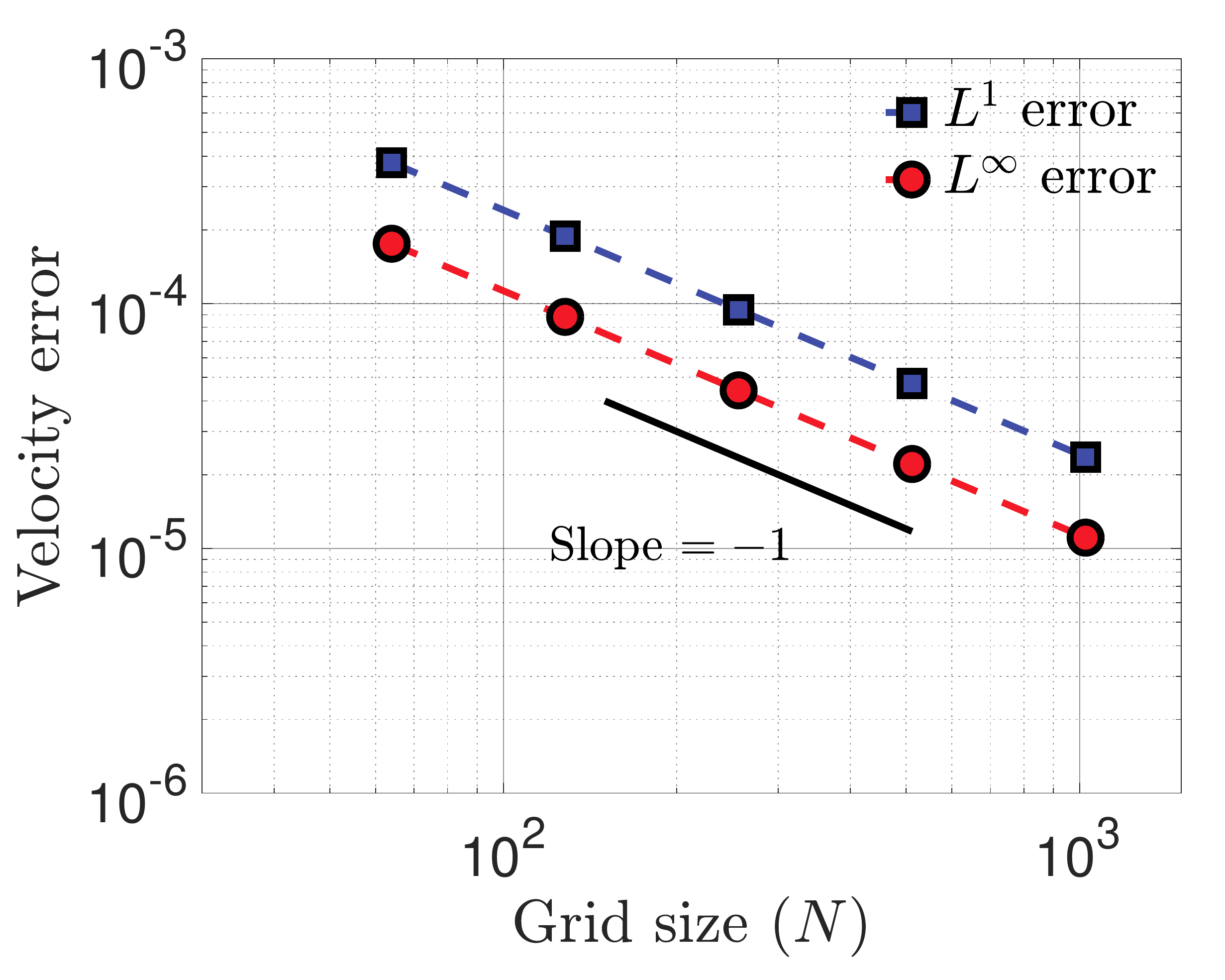}
    \label{fig_c_U_smooth_rho_err_FE_Upwind}
  }
   \subfigure[Pressure]{
    \includegraphics[scale = 0.2]{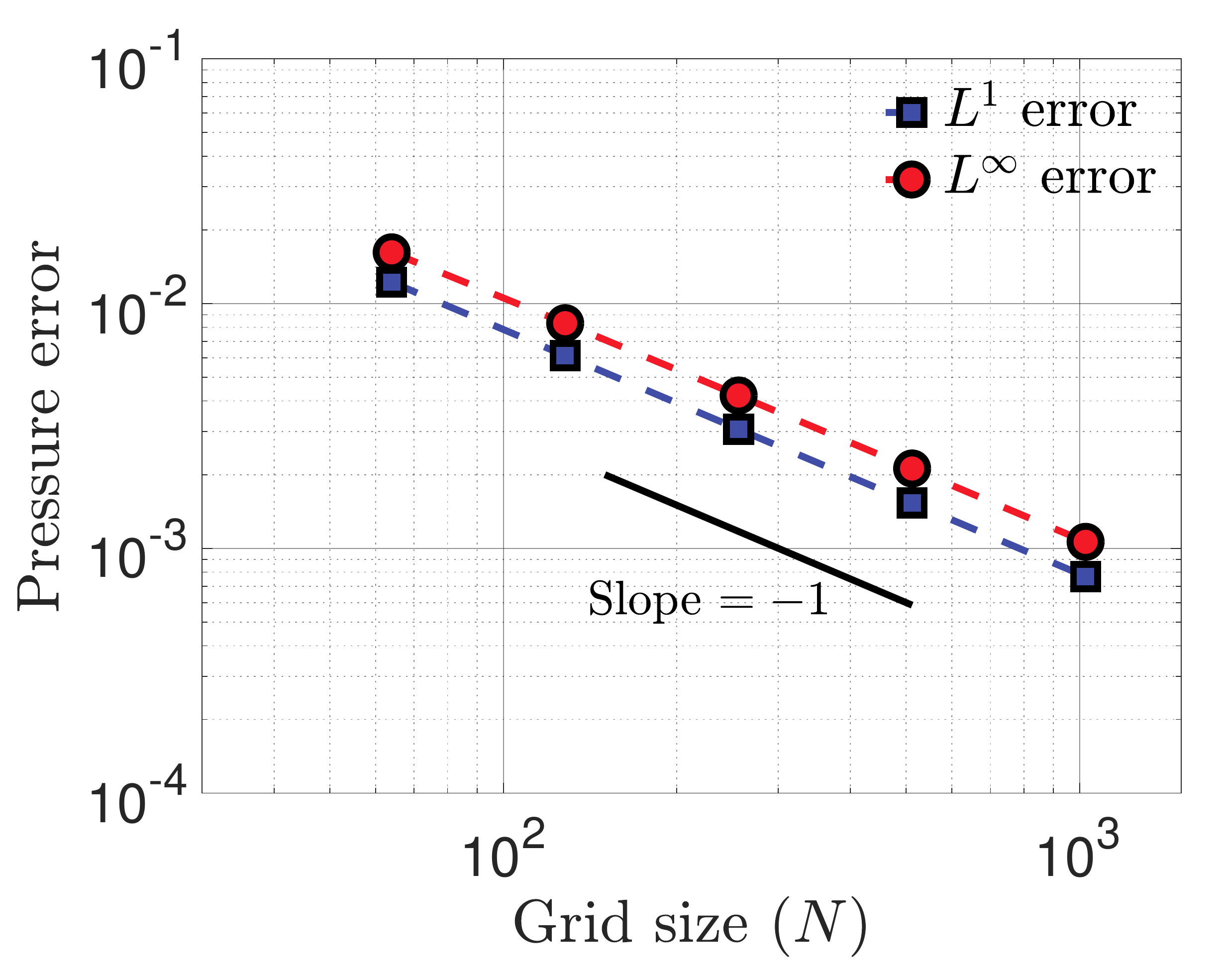}
    \label{fig_c_P_smooth_rho_err_FE_Upwind}
  }
   \subfigure[Density]{
    \includegraphics[scale = 0.2]{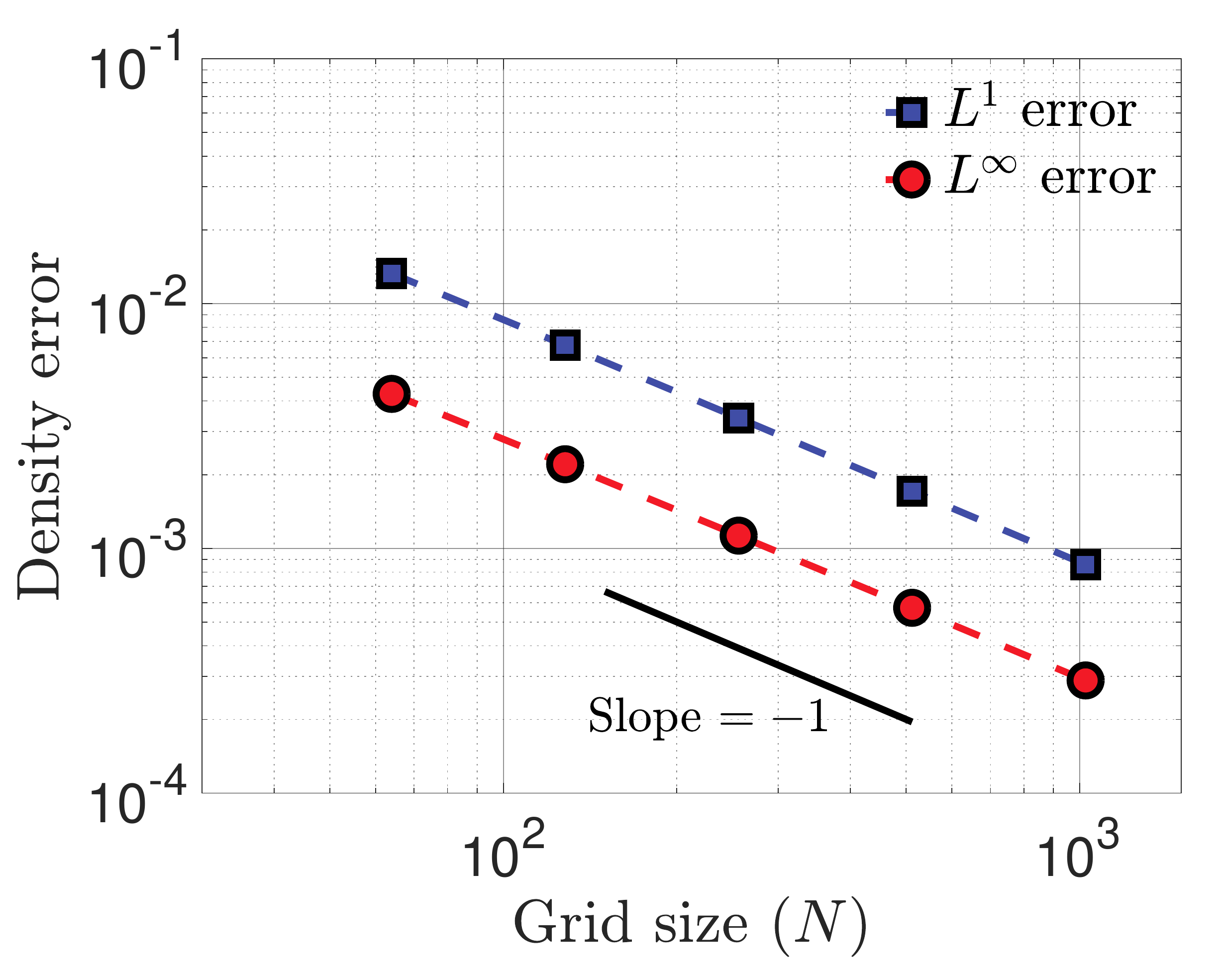}
    \label{fig_c_RHO_smooth_rho_err_FE_Upwind}
  }
  \caption{ 
  $L^1$ ($\blacksquare$, blue) and $L^\infty$ ($\bullet$, red) errors as a function of grid size $N$ for the conservative
  manufactured solution with specified normal velocity and normal velocity boundary conditions on all boundaries, and smooth
  density profile given by Eq.~\eqref{eq_cons_ms_smooth_rho}. In these cases,
  the density $\rho$ is \emph{not} reset between time steps. Forward Euler timestepping is used to update $\rho$,
  while an upwind scheme is used to approximate $\rho^{\text{lim}}$ on the faces of the shifted control volumes.
  \subref{fig_c_U_smooth_rho_err_FE_Upwind}
  convergence rate for $\u$;
  \subref{fig_c_P_smooth_rho_err_FE_Upwind}
   convergence rate for $p$;
   \subref{fig_c_RHO_smooth_rho_err_FE_Upwind}
   convergence rate for $\rho$.
   }
  \label{fig_c_err_smooth_rho_FE_Upwind}
\end{figure}

Here, we demonstrate the importance of the SSP-RK3 and CUI limited density update for maintaining
second-order convergence rates. Using the same manufactured solution of Appendix~\ref{app_smooth_density}, we instead use forward Euler
timestepping for the density update and an first-order upwind scheme to approximate the shifted control volume density $\rho_{\text{lim}}$.
We still use CUI to approximate the shifted control volume velocity $\u_{\text{lim}}$ for the convective derivative.
A similar scheme was considered in~\cite{Desjardins2010}.
Fig.~\ref{fig_c_err_smooth_rho_FE_Upwind} shows that first order convergence rates are obtained not only for the
density field, but also for pressure and velocity. Therefore, the additional complexity of the primary discretization scheme is justified.

%%%%%%%%%%%%%%%%%%%%%%%%%%%%%%%%%
%%%%%%%%%%%%%%%%%%%%%%%%%%%%%%%%%

\section*{Bibliography}
\begin{flushleft}
 \bibliography{VCINS_SOLVER}
\end{flushleft}

\end{document}